\documentclass[12pt]{article}
\usepackage{cite}
\usepackage{epsf}
\newcommand{\gsim}{\lower.7ex\hbox{$\;\stackrel{\textstyle>}{\sim}\;$}}
\newcommand{\lsim}{\lower.7ex\hbox{$\;\stackrel{\textstyle<}{\sim}\;$}}

\begin{document}

\begin{titlepage}
\begin{flushright}
TPI-MINN-99/08-T\\
UMN-TH-1743/99\\
hep-th/9902018
\end{flushright}

\vspace{1.3cm}
\bigskip
\begin{center} \Large
{\bf  Instantons Versus Supersymmetry: \\
Fifteen Years Later }

\end{center}

\vskip 12pt

\begin{center} {\Large
Mikhail Shifman  and  Arkady Vainshtein}
\vspace{.4cm}

{\normalsize {\it   Theoretical Physics Institute, University  of 
Minnesota,
Minneapolis, MN 55455}}

\vspace{2cm}

Abstract

\end{center}

An introduction to the instanton formalism in supersymmetric gauge theories is
given. We explain how the instanton calculations, in conjunction with analyticity in
chiral parameters and other general properties following from
supersymmetry, allow one to establish exact results in the weak and strong
coupling regimes. Some key applications are reviewed, the main emphasis is put
on the mechanisms of the dynamical breaking of supersymmetry.

\end{titlepage}

\newpage

\tableofcontents

\newpage

\section{Introduction and Outlook}
\label{sec1}

\renewcommand{\theequation}{1.\arabic{equation}}
\setcounter{equation}{0}

Most theorists firmly believe that  the underlying theory of 
fundamental interactions is supersymmetric. If so, at energies below
100 GeV supersymmetry (SUSY) must be broken, since no traces of the
Fermi-Bose degeneracy were detected so far. One can speak of two
alternative patterns of the SUSY breaking: explicit or spontaneous.
Although the former pattern is sometimes discussed in the
literature, this direction is obviously peripheral. For aesthetical and
theoretical  reasons the main trend is the spontaneous breaking of
supersymmetry. Various mechanisms of the spontaneous SUSY breaking
were worked out; a common feature of all of them is the  occurrence
of the massless Goldstino. When the supergravity is switched on the
phenomenon analogous to the Higgs mechanism occurs --
the Goldstino mixes with the gravitino and is ``eaten up"
by the latter -- the gravitino becomes massive. Therefore, phenomenology
of the spontaneous SUSY breaking cannot be considered in isolation
from  gravity.

To us, the most interesting question is {\em dynamics} lying behind
the spontaneous SUSY breaking. More exactly, our prime topic
is the nonperturbative gauge dynamics which, in certain
supersymmetric theories, creates non-supersymmetric vacua.
Why this happens in some models and does not in others? Which features
of the supersymmetric Yang-Mills theories are crucial
for this phenomenon and which are secondary?

These and other similar questions were first posed in the
beginning of the 1980's. Many breakthrough discoveries were made then,
approximately fifteen years ago.   In the subsequent decade
the issue was in a dormant state. It was revived recently in connection
with new breakthrough discoveries in nonperturbative gauge dynamics.
The present review is an attempt to give a unified ``big picture"
of the development of the subject spanning over 15 years.

Several reviews and lectures published recently are devoted
to the dynamical SUSY breaking. Usually the main emphasis
is made on general aspects and phenomenological consequences.
Not much attention is paid to theoretical tools 
developed over these years, which allow one
to obtain exact results in strongly coupled gauge theories, in certain
instances. The formalism of the supersymmetric instantons is one of
such tools. A significant part of this review is devoted to
in-depth studies of the supersymmetric instantons.
More exactly, we focus on the aspects of superinstanton calculus
that are important for the mechanisms of the dynamical SUSY breaking.
The second part of the review presents a survey of such mechanisms.

\subsection{Dynamical SUSY breaking: what does that mean?}

Logically, there are two possibilities for the 
spontaneous breaking of supersymmetry. It could be broken at the 
tree (classical) level by virtue of one of two known 
mechanisms: the Fayet-Iliopoulos mechanism or the O'Raifeartaigh 
mechanism (see Sec.\ \ref{sec51}). Both were most popular in the 1970's and 
early 1980's. They are out of fashion now. If one believes that
the only genuine mass scale in the fundamental theory is that of 
Grand Unification, $\sim 10^{16}$ GeV, or the Planck scale,
$\sim 10^{19}$ GeV, then it is natural to expect that
the tree-level breaking would produce the splittings between the 
masses of ordinary particles and their superpartners of the same 
order of magnitude. Then, next-to-nothing is left of supersymmetry
as far as physics of our low-energy world is concerned. In particular,
SUSY  will have nothing to do with the hierarchy problem --
its  main {\em raison d'etre} in the model-building. 

If supersymmetry is unbroken at the tree level, it remains unbroken 
at any finite order of perturbation theory.  If the SUSY breaking still 
occurs in this case, it must be exponential in the inverse coupling 
constant, $\sim e^{-C/g^2}$, where $C$ is a positive constant. A  natural very
small parameter  appears in the theory.  
The exponent $ e^{-C/g^2}$ suppresses all SUSY breaking effects.  Starting 
from the fundamental scale  $\sim 10^{19}$ GeV, one can  
expect to  get in this case the mass splitting
between the superpartners of order of the ``human" scale, $\sim 
10^{2}$ GeV. Then supersymmetry may be instrumental in the 
solution of the hierarchy problem and in  shaping the
major regularities of the electroweak theory.  It is just this scenario 
that is  called  the dynamical SUSY breaking, that will be in the focus 
of  the present review. The term  was put into circulation by Witten 
\cite{Witten1}.  The models where the phenomenon occurs
are rather sophisticated in their structure and are not so abundant.
The searches that eventually led to the discovery of such models 
present a dramatic chapter in the history of supersymmetry. We will 
dwell on various aspects of this story and on  nuances of the 
dynamical SUSY breaking in due course.  Whether the models
developed so far may be relevant to nature is a big question mark.
Since supersymmetry is not even yet observed, it seems premature 
to submerge too deeply into the phenomenological aspects.
Our prime emphasis will be on the underlying dynamics which is 
beautiful by itself. It is inconceivable that such an elegant 
mechanism will not be a part of the future theory in this or 
that form. 

From the  ample flow of the literature devoted to the dynamical 
supersymmetry breaking we have chosen only one little stream. 
The early works \cite{Witten2,NSVZ1,RoVe,ADS2}  established the 
basic principles of the construction and set the framework and formalism 
 that are 
universally used at present in this range of questions. The ideas 
elaborated in these works were further advanced and supplemented 
by new discoveries, of which most important are two mechanisms:
one of them is due to specific features of confinement in SUSY gauge 
theories \cite{ISS} and another is due to a quantum deformation of 
the moduli
space \cite{IT}. It may well happen that only one of several 
competing  dynamical scenarios will survive in the future. But which 
one?
Since the answer is unknown, we should get acquainted with all of 
them. 

\subsection{Hierarchy problem}

As we have mentioned, the dynamical SUSY breaking could explain
how supersymmetry could survive down to energies constituting a 
tiny 
fraction of  the Planck mass $M_{\rm Pl}$. This is only one aspect of 
the
hierarchy problem. One can ask oneself why $W$ and $Z$ bosons are 
so light compared to $M_{\rm Pl}$? Or, in other words, why
the expectation value of the Higgs field triggering 
 the SU(2) breaking in the Standard Model is so small?

One needs the Higgs doublet to be essentially  massless on the scale 
of $M_{\rm Pl}$. The elementary charged scalars are not protected 
against large quadratically divergent corrections pushing their 
masses to $M_{\rm Pl}$, except in supersymmetric theories. 
Supersymmetry pairs elementary scalars with elementary spinors 
that could be naturally massless because of their chirality. 
In fact, SUSY propagates the notion of chirality to the scalar fields.
It is quite possible then that the same exponentially small
nonperturbative effects, which are responsible for the dynamical 
SUSY breaking,  provide the Higgs field with an exponentially small
(on the scale $M_{\rm Pl}$) expectation value and mass. We do not 
know 
exactly how this could happen but {\em a priori} this is conceivable.

\subsection{Instantons}

Many mechanisms of the supersymmetry 
breaking to be discussed below refer to the weak coupling regime.
In a sense, the smallness of the coupling constant is a prerequisite in 
our explorations.  In the weak coupling regime
the mystery surrounding nonperturbative effects in the gauge 
theories fades away -- we know that  certain nonperturbative  phenomena are due
to instantons \cite{BPST}.  Thus, the instantons are the most important technical
element of the whole construction.

The instantons were discovered in \cite{BPST} in the context of 
quantum chromodynamics (QCD). In QCD they are instrumental in revealing a 
topologically  nontrivial structure in  the space of fields. They can not 
be  used,  however, for  quantitative analyses of the nonperturbative 
effects since the  instanton-based  approximations in QCD are not closed: 
typically the instanton  contributions are saturated at large distances, 
where the coupling constant  becomes of order  unity, and the 
quasiclassical (weak coupling) methods become  inapplicable. 
As was noted by 't Hooft \cite{Hooft1}, in the models where the 
gauge  symmetry is spontaneously broken, e.g. in the Standard Model 
(SM),  the  instanton  calculations become 
well-defined, the integrals over the instanton size $\rho$  are cut off 
at $\rho\sim v^{-1}$ where $v$ is the expectation value of the Higgs 
field. The  running coupling constant is frozen at distances of order 
$v^{-1}$,  and if it is  small at such distances, as is the case in the Standard 
Model, all  approximations  that go with the instanton analyses are 
justified. 

From the calculational side everything is okey with the instantons in 
the Higgs regime. However, the exponentially 
small instanton contribution is usually buried in the background
of much larger perturbative contributions, which mask it and make it 
totally negligible. The only exception known so far is the baryon number 
violating  effects in the Standard Model. These effects identically vanish 
to any finite order in perturbation theory, but are generated by instantons 
\cite{Hooft1}.

Instantons in supersymmetric theories are very peculiar. Many SUSY 
models possess the so called {\em vacuum valleys}, or {\em flat directions}, 
or {\em moduli spaces of vacua}. There are infinitely many physically 
inequivalent degenerate vacuum states. As a matter of fact, the 
degeneracy is continuous. The degeneracy is protected by supersymmetry 
from being lifted in perturbation theory to any finite order.
The instantons may or may not lift the vacuum degeneracy. If they do,
a drastic restructuring of the vacuum state occurs. The instantons are 
the leading driving force in the vacuum restructuring; the perturbative 
background is just absent. Thus, we encounter here with the  
situation similar to
 the baryon number violation in the Standard Model, except that 
the  baryon number violation is an exotic process, 
while the instanton-generated superpotentials in the vacuum valleys of 
the SUSY models are quite typical. 

The distinguished  role of the instantons in the phenomena to be 
discussed below  is 
not the only special circumstance. Another remarkable feature of the 
supersymmetric instantons, with no parallels in nonsupersymmetric 
models,  is the fact that their contribution can be found {\em exactly} 
\cite{NSVZ1}. 
In conventional theories a perturbative expansion is built around the 
instanton  solution, so that  the corresponding contribution takes the 
following generic form
\begin{equation}
e^{-C/\alpha }(C_0 + C_1 \alpha  + C_2 \alpha^2 + ...)\, , \,\,\, 
\alpha\equiv 
\frac{g^2}{4\pi}\, .
\end{equation}
The  series in $\alpha $ never terminates, although practically, of 
course, it is  hard  to go beyond the one-loop approximation. The 
supersymmetric instanton  is unique:  the general loop  expansion 
(i.e. the expansion in powers of $\alpha$)  trivializes.  No $\alpha$ series  
emerges in the instanton-generated quantities. All quantum 
corrections around the instanton solution cancel each other 
\cite{DADV,NSVZdop}, and the problem reduces to consideration of 
the  classical  solution itself, and the zero modes it generates 
\cite{NSVZ4,NSVZ3}.  The zero  modes are associated with the 
symmetries of the problem that are  nontrivially realized on the 
instanton solution. The cancellation of the  quantum  corrections is 
due to the fact that the instanton field is a very special  field 
configuration: each given instanton  preserves one half of the original 
supersymmetry. 

Even though the $\alpha$ series does not appear one may pose a question
of the iteration of the exponential terms, i.e. multi-instantons.
It turns out that in many cases these iterations do not take place -- the
one-instanton results are exact. In other instances the multi-instantons can be
summed up exactly.

Quite often the instanton calculations are performed by ``brute force", by using
methods that are essentially non-supersymmetric. Such an approach is
quite possible, especially in those instances when the authors
are not interested in  overall numerical constants, while the
general structure of the result is known {\em a priori}, say, from $R$ 
symmetries.  Our approach is different. About one half of the review
is devoted to building a superinstanton formalism which takes
maximal advantage of SUSY, at every stage. Although it takes
some effort to master it, once this is done various instanton 
calculations can be carried out with ease. In this way the reader gets 
a much better insight in  technical and conceptual aspects
of nonperturbative gauge dynamics. 

\subsection{The Higgs mechanism {\em en route}}

At first sight, the supersymmetric gauge theories look very similar to 
ordinary  QCD.  In the simplest case of SUSY  gluodynamics the only 
difference  refers to  the representation of the color group to which the 
fermions belong.  In  SUSY   gluodynamics the fermions (gluinos) 
transform according to the  adjoint  representation of the group, as 
the  gluons, and are the Majorana  fields. If one  adds the matter 
superfields  in the fundamental representation, one  gets the  quarks, 
and, in  addition, their scalar partners, squarks. Keeping in  mind this 
similarity, it was tempting to interpret the supersymmetric gauge  
dynamics in  parallel to that of QCD. As a matter of fact, in some of 
the  first works  (e.g. \cite{RV})  devoted to nonperturbative effects 
in  supersymmetric QCD  (SQCD) attempts were made to closely follow 
the  pattern of QCD,  which gave  rise to paradoxes. The paradoxes 
are  resolved as follows: in many instances 
the spontaneous  breaking of the gauge symmetry takes place. The theory
one deals with is  in  the Higgs phase:  the gauge 
bosons are heavy which  freezes  the gauge coupling constant. 
This is in clear
contradistinction to what  happens in QCD. The Higgs  regime is a common 
phenomenon in the SUSY  gauge  theories with  matter. 

The Higgs phenomenon is very well known from
the Standard Model. In SM the
potential  energy associated with the Higgs field is
\begin{equation}
U_H (\phi ) = C(\bar \phi \phi - v^2 )^2\, .
\label{SMpot}
\end{equation}
It is obvious that the minimum of the energy corresponds to a 
nonzero  expectation value 
\begin{equation}
\left\langle\bar \phi \phi\right\rangle_{\rm vac} = v^2\, .
\label{phi2}
\end{equation}
If $v$ is large, we  deal with the classical field and can speak 
of the  average value not only of the square of the field (\ref{phi2}), 
but of  the field  itself.

In contrast to the Standard Model, however,  where the potential 
(\ref{SMpot}) is usually postulated, in the SUSY gauge theories of 
interest,  similar  potentials are  generated dynamically, by 
instantons, and are  fully calculable in the models where the 
expectation  values of the  scalar fields are large. One may ask how 
large  parameters can appear. 
The set of dimensional parameters in the ``microscopic" SQCD is the 
same as in  QCD. The basic parameter is $\Lambda$, the scale 
parameter which  determines the value of the running gauge 
coupling constant. In  addition, the  set of the dimensional 
parameters includes the masses of the matter  fields, 
quarks and squarks,
\begin{equation}
\Lambda\, , \,\,\, m_1\, , \,\,\,  m_2\, , ... \, .
\end{equation}
In the simplest case of one flavor the vacuum expectation value of 
the scalar field can be estimated as
\begin{equation}
\left\langle \phi\right\rangle_{\rm vac} \sim \Lambda 
\left(\frac{\Lambda}{m_1}\right)^{1/4}\, ,
\end{equation}
and $\langle \phi \rangle_{\rm vac}$ is large
provided $m_1 \ll \Lambda$. There is  nothing  of the kind in QCD where the
masses of the $u$ and $d$ quarks  can be  assumed to vanish and no vacuum
condensates  become  infinite in this  limit.

The crucial distinction between QCD and SQCD arises due to the 
existence of  the flat directions in the latter. The vacuum state can be 
at any point  from the  bottom of the valley. Classically, all these 
points are degenerate. That  is why  small perturbations (say, the 
mass terms, Yukawa couplings or instanton-generated
superpotentials) can  drastically change the  values of  the vacuum 
condensates. 

In the supersymmetric theories with the fundamental matter
there is no phase transition between the Higgs and confinement
regimes~\cite{BRFS}.
By continuously changing some parameters (for instance, the mass
parameters $m_i$) one passes from the weak coupling Higgs regime to the
strong coupling confinement regime. SUSY has a remarkable property:
it enforces holomorphic dependences of chiral quantities on chiral parameters.
Moreover, general features of the theory can often prompt one
the structure of the analytic function. Then, performing
calculations in the weak coupling regime allows one to establish exact 
results referring to the strong coupling regime. 
This is the reason why the Higgs phenomenon in SUSY gauge 
theories is of paramount importance in the technical aspect.

\subsection{Chiral versus nonchiral models}

All gauge theories with matter can be divided in two distinct classes:
chiral and non-chiral matter. The second class includes 
supersymmetric generalization of QCD, and all other models 
where each matter multiplet is accompanied by the corresponding 
conjugate representation. In other words,  the gauge symmetry does 
not forbid a mass term  for  {\em all} matter fields. 

Models with the chiral matter are those where mass terms are 
impossible (more exactly, at least for some of the matter fields
the quadratic terms in the superpotential are forbidden by gauge 
invariance). The matter sector in such theories is severely 
constrained by the absence of the internal anomalies in the theory.

For a long time SUSY practitioners were convinced
that the dynamical SUSY breaking   can occur  only in rather exotic 
models where the  matter sector is chiral. The  conclusion that the 
chirality of the matter is a necessary condition seemingly followed  
from  {\em  Witten's index} \cite{Witten2}. Needless to say that this 
constraint drastically  narrows down the class of models to be 
examined. 

Recently it was realized, however, that nonchiral models
may also experience the dynamical SUSY breaking. The 
Intriligator-Thomas-Izawa-Yanagida (ITIY) mechanism \cite{IT}
and its derivatives \cite{MDDGR} are nonchiral.
The fact that nonchiral models of the dynamical SUSY breaking
exist was an unexpected discovery. It is highly probable that
the future searches for new mechanisms will focus on the nonchiral 
models. The vast majority of the mechanisms established so far are 
chiral, though. In particular,
 all models found in the 1980's are chiral. 

\subsection{What will not be discussed}

At the early stages it was  believed that a dynamical SUSY breaking 
could be built in directly into the supergeneralization of the Standard 
Model (Minimal Supersymmetric Standard Model, or MSSM), or into 
the SU(5)-based theory of grand unification  (GUT), with a minimal 
set of quintets and decuplets. It was realized rather quickly that
a reasonable  pattern of the mass splittings between the 
superpartners was not attainable in this way.  In particular, gauginos 
came out unacceptably light. At present no model is known
where the dynamical SUSY breaking would act
 directly in the more or less established SM-related sector and would 
lead
to no contradictions with phenomenology. The word ``established"
is used above as opposed to hypothetical. To solve the problem 
theorists invented a totally new sector -- a part of the world which is 
pure fantasy so far -- and dubbed it {\em supersymmetry breaking 
sector}.  It consists of some  hypothetical fields, none of which 
is discovered, whose sole role in today's theory is to break 
supersymmetry.  Then this breaking is communicated to our world
through yet another sector, a {\em messenger} sector. 
Two alternative scenarios of how this communication could be 
achieved are elaborated in the literature. In the first scenario, which 
was most popular in the 1980's, the contact between the two worlds 
-- ours and the supersymmetry breaking sector -- is realized only 
through gravity. The gravity is universal. Once it is switched on,
the supersymmetry breaking sector generates a mass $m_{3/2}\neq 
0$ to the originally massless gravitino,
\begin{equation}
m_{3/2} \sim \frac{m_{\rm br}^2}{M_{\rm Pl}}\, ,
\end{equation}
where $m_{\rm br}$ is the scale of supersymmetry breaking. The 
coupling of gravitino to the fields belonging to our world then
generates SUSY breaking terms in the Lagrangian, which from the 
point of view of the human observer look as an  explicit 
supersymmetry  breaking by soft terms. The mass splittings between 
the  superpartners is
\begin{equation}
\Delta m \sim m_{3/2} \sim \frac{m_{\rm br}^2}{M_{\rm Pl}}
\, .
\end{equation}
The SUSY breaking scale  is the geometrical average between $\Delta 
m $ and $M_{\rm Pl}$;  if we want $\Delta m$ to be of order 100 
GeV, $m_{\rm br} \sim 10^{11}$ GeV.  This scenario is called {\em 
gravity-mediated}.

In the second scenario gauge fields play the role of the messenger 
interaction. Messenger (heavy) quarks and leptons are introduced, 
that are coupled directly to the supersymmetry breaking sector on 
the one hand, and to our gauge bosons, on the other. Since the gauge 
coupling constant is much stronger than the gravitational constant,
the scale of supersymmetry breaking need not be so large as in the 
previous case. In the {\em gauge-mediated} scenario $m_{\rm br}$ is 
of order 10 TeV.

Both scenarios have their virtues and shortcomings. In the 
gravity-mediated models the suppression of the flavor changing 
neutral currents does not come out naturally while in the 
gauge-mediated models one typically ends up with a  horrible  
messenger sector. Moreover, the gauge-mediated approach
is not closed, one needs supergravity anyway. Indeed,
the spontaneous SUSY breaking inevitably produces a massless
fermion, the Goldstino. No such massless fermions seem to exist;
therefore, one should make the Goldstino massive 
which can only be achieved
through its coupling to gravitino.

It seems that neither of the approaches is fully
phenomenologically acceptable and aesthetically appealing.  
The issue remains  a challenge for  the future theory.  
The question of  implementation of the 
dynamical SUSY breaking in realistic models 
will not be discussed at all. It will just
assumed that this dynamical  phenomenon  is somehow 
realized in a consistent way, either through a supersymmetry 
breaking sector and messengers, or somehow else. We will limit 
ourselves to the internal structure of possible  mechanisms of  the 
dynamical SUSY breaking. Many excellent reviews are devoted to 
phenomenological aspects of
gravity-mediated and gauge-mediated scenarios. The reader is 
referred to e.g. Ref.\ \cite{GRM}. 

\subsection{Instantons vs. supersymmetry:  literature travel guide} 

The subject we are going to discuss is vast, its development spans 
almost two decades. The advancement is not always straightforward.
Therefore,  it is convenient to 
provide a brief literature travel guide. 

The  importance of the dynamical SUSY breaking was first 
realized by Witten in 1981 \cite{Witten1} who started considering 
general aspects of the phenomenon and provided many insights.
The first instanton calculations in supersymmetric gauge theories 
ascend to 1982: in Ref.\ \cite{AHW} certain three-dimensional
models were considered, while in Ref.\ \cite{1} 
the four-dimensional theories were addressed for the first time. 
In the latter work
the question was raised as to the compatibility of 
 the instanton-induced 't~Hooft vertex with supersymmetry.
 Shortly after,  the problem was solved in \cite{NSVZ1} where basic elements of 
super-instanton calculus were introduced. The formalism was 
applied in calculation of appropriately chosen $n$-point functions of
chiral superfields (all fields of one and the same chirality).
It was shown that, thanks  to supersymmetry, the result
 could only be a constant, and in many instances this constant 
turned out to be non-vanishing due to the instanton contribution. 
By exploiting the 
cluster 
decomposition at large distances the gluino condensate was found.
As a byproduct, the exact Novikov-Shifman-Vainshtein-Zakharov
(NSVZ) $\beta$ function emerged \cite{NSVZdop}. 
Then, this approach was generalized \cite{RoVe}
to include SUSY gauge theories 
with matter. This line of research culminated in the very 
beginning of 1984  when the SU(5) model with
$M$ (anti)decuplets and $M$ quintets was considered \cite{MV}.
The instantons were shown to generate the gluino condensate, 
which under certain conditions on the Yukawa couplings is incompatible with
 supersymmetry.
The conflict  was interpreted as a dynamical SUSY breaking.

In fact, a few weeks before this work, the SU(5) model with
one  decuplet and one anti-quintet was analyzed in 
Ref.\ \cite{ADS1}. The model is  strongly coupled; 
the analysis was based on indirect (symmetry) arguments and the 
't~Hooft matching. This was the first example ever where the
dynamical SUSY breaking was indicated.

Meanwhile, it was realized \cite{ADS2,ADS3} that in a wide class
of models with matter and classically flat directions
the gauge symmetry was implemented in
the Higgs regime.  By an appropriate adjustment of parameters 
(masses, Yukawa couplings, etc.)
 one could guarantee
the theory to be weakly coupled. Then the heavy gauge bosons could 
be  integrated out  and the vacuum structure could be 
inferred
from the effective low-energy theory of the moduli.
The equivalence between the two approaches --
calculation of the condensates and the low-energy Lagrangian for 
the moduli -- was established in \cite{NSVZ2}.
Being formally equivalent, the method of the effective Lagrangians
for the moduli turned out to be more ``user-friendly". 
The underlying physical picture is  transparent, an advantage that 
can hardly be overestimated. 
In the quest for the dynamical SUSY breaking 
the effective Lagrangian approach of Affleck, Dine and Seiberg became 
standard; some ideas and technical elements of the condensate
calculations were later incorporated, though. 

Affleck, Dine and Seiberg were the first to realize \cite{2} that in SU(5) 
model  with two or more generations (i.e.\ $M$ (anti)decuplets and $M$ 
quintets  where $M\geq 2$) the dynamical SUSY breaking 
occurs in  the weak  coupling regime, provided the Yukawa couplings 
are small,  and the  model, being fully calculable,  presents a great 
theoretical  laboratory. Their  paper  \cite{2} was submitted for 
publication on   February 10, 1984  --  the next day after \cite{MV},  
where the  very  same model was  analyzed, with the  same 
conclusion of the dynamical SUSY breaking. 

The basic elements of  the superinstanton formalism
(the one-instanton problem in the theories with matter)
were worked out by  1985~\cite{NSVZ3}.  Some additional 
elements were
elaborated in~\cite{novfu}. Significant improvements were later made
in Ref.\ \cite{Yung}, in connection with the instanton
calculations in extended supersymmetries.  A continuous progress
in the multi-instanton calculations$\,$\footnote{The problems where
the multi-instanton calculations are needed will not 
be discussed in this review.}
culminated in the explicit construction~\cite{DKM} of the $n$-instanton
measure for arbitrary $n$. Much work was done to make  this achievement
possible;
the interested reader can turn  to the review
paper~\cite{DKMS} for references and comments.

Drawing the picture in broad touches one may say that in 1984 and 
1985 the first stage was 
completed -- although work 
continued and some novel important results were obtained in the 
following two  or  three years (e.g.~\cite{seib88}),
the interest to the issue have been  rapidly drying out. 
The subject was in a dormant state for about a decade; it was revived 
in 1993, after  breakthrough observations due to Seiberg
\cite{Nati1,Nati2}. New models exhibiting the dynamical SUSY 
breaking were  constructed. Some of them continue the trend 
established in the 1980's,  others  are  based on new ideas. The first 
class is  represented  by the  so-called {\em 4-1} model and its 
relatives \cite{DNNS,PT}. In the second class the most prominent is 
the ITIY model \cite{IT}, which is nonchiral and, nevertheless, 
ensures the SUSY breaking.  The model is strongly coupled, by 
default, and is noncalculable.  Another novel mechanism is the 
Intriligator-Seiberg-Shenker (ISS) model \cite{ISS}, which also takes 
place in the strong coupling regime. Both have numerous 
descendants. 

\section{Supersymmetric Theories: Examples and Generalities}
\label{sec3}

\renewcommand{\theequation}{2.\arabic{equation}}
\setcounter{equation}{0}

 Our first task  is  to reveal general features of the supersymmetric  theories 
instrumental in the  dynamical SUSY breaking.  We  start   from  a brief
review of the  simplest  models,  intended mainly in   order  to introduce our 
notations and highlight some basic formulas.  
The feature we will focus on is the existence of continuum of degenerate classical 
vacua, the so called $D$-flat directions (in the absence of the superpotential). These
vacua define physically inequivalent theories.  The fermion-boson cancellation
inherent to supersymmetry guarantees  that no superpotential is generated to 
any finite order  of perturbation theory, i.e. the
$D$-flatness is  maintained. The vacuum degeneracy  can be broken only by
nonperturbative effects which will be discussed in Sec.\ \ref{secSA}. Traveling
over the $D$-flat directions one finds oneself, generally speaking, 
 in the Higgs phase.  We  will dwell on peculiarities of the Higgs mechanism in  
supersymmetric  gauge theories. Anomalous and non-anomalous $R$ symmetries
play a special role in the analysis, and we will dwell on them too.
 Finally,  we will consider  issues related
to   Witten's index.  

\subsection{Superspace and superfields}
\label{sec30}

The four-dimensional space $x^\mu$ can be promoted to superspace
by adding four Grassmann coordinates $\theta_\alpha$ and $\bar
\theta_{\dot\alpha}$, ($\alpha,\,\dot\alpha=1,2$). The coordinate transformations
\begin{equation}
\{x^\mu ,\theta_\alpha\,, \bar\theta_{\dot\alpha}\}:\qquad
\delta\theta_\alpha=\varepsilon_\alpha\,,\quad 
\delta\bar\theta_{\dot\alpha}=\bar\varepsilon_{\dot\alpha}\,, \quad 
\delta x_{\alpha\dot\alpha}= -2i\,\theta_{\alpha}\bar\varepsilon_{\dot\alpha}
-2i\,\bar\theta_{\dot\alpha}\varepsilon_{\alpha}
\label{susytr}
\end{equation}
add SUSY to the translational and Lorentz transformations$\,$\footnote
{Our notation is close but not identical to that 
of Bagger and Wess \cite{BW}. The main distinction is the 
conventional choice of the metric tensor $g_{\mu\nu} 
=\mbox{diag}(+---)$ as opposed to the $\mbox{diag}(-+++)$
version of Bagger and Wess. For further details see Appendix in Ref.\ \cite{CS}.
Both, the spinorial and vectorial indices will be denoted by the Greek letters.
To differentiate between them we will use the letters from the beginning of the
alphabet for the spinorial indices, e.g. $\alpha$, $\beta$ and so on, reserving
those from the end of the alphabet (e.g. $\mu$, $\nu$, {\em etc.}) for the vectorial
indices.}.

Here  the Lorentz vectorial indices are transformed into  spinorial according to the
standard rule
\begin{equation}
A_{\beta\dot\beta} = A_\mu (\sigma^\mu )_{\beta\dot\beta}\,,\qquad
  A^\mu = \frac{1}{2} 
A_{\alpha\dot\beta}(\bar\sigma^\mu )^{\dot\beta\alpha}\, ,
\end{equation}
where
\begin{equation}
 (\sigma^\mu )_{\alpha\dot\beta} = \{ 1, 
\vec\tau\}_{\alpha\dot\beta} 
\,,\qquad   (\bar\sigma^\mu
)_{\dot\beta\alpha}=(\sigma^\mu)_{\alpha\dot\beta}
  \, . 
\end{equation}
We use the notation $\vec \tau$ for the Pauli matrices throughout the paper.
The lowering and raising of the indices is performed by virtue of the
$\epsilon^{\alpha\beta}$ symbol
($\epsilon^{\alpha\beta}=i(\tau_2)_{\alpha\beta}$, $\epsilon^{12}=1$). For
instance, 
\begin{equation}
(\bar\sigma^\mu
)^{\dot\beta\alpha}=\epsilon^{\dot\beta\dot\rho}\,\epsilon^{\alpha\gamma}\,
(\bar\sigma^\mu )_{\dot\rho\gamma} =\{1, -\vec\tau\}_{\dot\beta\alpha}
 \, . 
\end{equation}

Two invariant subspaces $\{x^\mu_L\,,\theta_\alpha\}$ and
$\{x^\mu_R\,,\bar\theta_{\dot\alpha}\}$ are spanned on 1/2 of the
Grassmann coordinates,
\begin{eqnarray}
\{x^\mu_L\,,\theta_\alpha\}:&\qquad&\delta\theta_\alpha =\varepsilon_\alpha\, ,
\quad \delta({x}_L)_{\alpha\dot\alpha}=-
4i\, \theta_{\alpha}\bar\varepsilon_{\dot\alpha}\,;\nonumber\\[0.2cm]
\{x^\mu_R\,,\bar\theta_{\dot\alpha}\}:&\qquad&\delta\bar\theta_{\dot\alpha}
=\bar\varepsilon_{\dot\alpha}\, ,
\quad \delta({x}_R)_{\alpha\dot\alpha}=-
4i\, \bar\theta_{\dot\alpha}\varepsilon_{\alpha}\,,
\label{sutra}
\end{eqnarray}
where 
\begin{equation}
({x_{L,R}})_{\alpha\dot{\alpha}} = {x}_{\alpha\dot{\alpha}} \mp
2i\, \theta_{\alpha}\bar{\theta}_{\dot{\alpha}}\, .
\label{chcoor}
\end{equation}
The minimal supermultiplet of fields includes one complex scalar field $\phi (x)$ 
(two bosonic states) and one complex Weyl spinor $\psi^\alpha 
(x)\, , \,\,\alpha = 1,2$ (two fermionic states).
Both fields are united in one {\em chiral superfield},
\begin{equation}
{\Phi ({x}_L,\theta )} = \phi ({x}_L) + \sqrt{2}\theta^\alpha 
\psi_\alpha ({
x}_L) +  \theta^2 F({x}_L)\, ,
\label{chsup}
\end{equation}
where $F$
is an auxiliary component. The  field  $F$ 
appears in the Lagrangian without
the  kinetic term. 

In the gauge theories one also uses a {\em vector superfield},
\begin{eqnarray}
&&V(x,\theta,\bar\theta)= C+i\theta\chi -i \bar\theta \bar\chi
+\frac{i}{\sqrt{2}} \theta^2 M -\frac{i}{\sqrt{2}} \bar\theta^2 {\bar M}
\nonumber \\
&&-2\theta_\alpha \bar\theta_{\dot \alpha } v^{\dot\alpha\alpha} +\left\{
2i \theta^2 \bar\theta_{\dot \alpha} \left[\bar\lambda^{\dot \alpha} - \frac{i}{4}
\partial^{\alpha \dot \alpha} \chi\right] +\mbox{H.c.}\right\}   + \theta^2
\bar\theta^2 \left[D - \frac 1 4 \partial^2 C\right]\,.
\label{vecsf}
\end{eqnarray}
The superfield $V$ is real, $V=V^\dagger$, implying that the bosonic fields
$C$, $D$ and $v^\mu=\sigma^\mu_{\alpha\dot\alpha}v^{\dot\alpha\alpha}/2$
are real. Other fields are complex, and the bar denotes the complex conjugation.

The transformations~(\ref{sutra}) generate the SUSY transformations
of the fields which can be written as
\begin{equation}
\delta V= i \left(Q\varepsilon +  \bar Q\bar\varepsilon\right)\,V              
\end{equation}
where $V$ is a generic superfield (which could be chiral as well). The
differential operators $Q$ and $\bar Q$ can be written as
\begin{equation}
Q_\alpha=-i \frac{\partial}{\partial \theta^\alpha} +\partial_{\alpha\dot\alpha}
\bar\theta^{\dot \alpha}\,,\quad
\bar Q_{\dot\alpha}=i\frac{\partial}{\partial \bar\theta^{\dot\alpha}} -
\theta^{\alpha}\partial_{\alpha\dot\alpha}\,, \quad \left\{Q_\alpha\,,\bar
Q_{\dot\alpha}\right\}=2i\partial_{\alpha\dot\alpha}\,.
\label{diffq}
\end{equation}
These differential operators give the explicit realization of the SUSY algebra,
\begin{equation}
\left\{Q_\alpha\,,\bar Q_{\dot\alpha}\right\}=2P_{\alpha\dot\alpha}\,,\quad
\left\{Q_\alpha\,, Q_{\beta}\right\}=0\,,\quad \left\{\bar Q_{\dot\alpha}\,,
\bar Q_{\dot\beta}\right\}=0\,,\quad
\left[Q_\alpha\,,P_{\beta\dot\beta}\right]=0\,,
\label{susyalgebra}
\end{equation}
where  $Q_\alpha$ and $\bar Q_{\dot\alpha}$ are the supercharges while
$P_{\alpha\dot\alpha}=i\partial_{\alpha\dot\alpha}$ is the energy-momentum
operator.  The {\em superderivatives} are defined as the differential operators 
$\bar D_\alpha$, $D_{\dot\alpha}$ 
anticommuting with $Q_\alpha$ and $\bar
Q_{\dot\alpha}$,
\begin{equation}
D_\alpha=\frac{\partial}{\partial \theta^\alpha} -i \partial_{\alpha\dot\alpha}
\bar\theta^{\dot \alpha}\,,\quad
\bar D_{\dot\alpha}=-\frac{\partial}{\partial \bar\theta^{\dot\alpha}} +i
\theta^{\alpha}\partial_{\alpha\dot\alpha}\,, \quad \left\{D_\alpha\,,\bar
D_{\dot\alpha}\right\}=2i\partial_{\alpha\dot\alpha}\,.
\end{equation}

\subsection{The generalized Wess-Zumino models}
\label{sec31}

The generalized Wess-Zumino model describes interactions of an arbitrary number
of the  chiral superfields. Deferring the discussion of the general case, we start from
the simplest  original Wess-Zumino model \cite{WZ1} (sometimes referred to as
the minimal model).

\subsubsection{The minimal model}
\label{sec310}

The model contains one chiral superfield $\Phi (x_L, \theta)$ and its complex
conjugate $\bar \Phi (x_R, \bar\theta)$, which is  antichiral.
The action of the model is
\begin{equation}
S = \frac{1}{4}\int\! {\rm d}^4 x \,{\rm d}^4\theta  \,\Phi\bar{\Phi} +  
 \frac{1}{2}\int \! {\rm d}^4 x \,{\rm d}^2 \theta\, {\cal W}(\Phi) +
\frac{1}{2}\int \! {\rm d}^4 x \,{\rm d}^2 \bar\theta\, \bar{\cal W}(\bar\Phi )
\, .
\label{lagrwz}
\end{equation}
Note that the first term is the integral over the full superspace, while the second
and the third run over the chiral subspaces. The holomorphic function 
$ {\cal W}(\Phi )$ is called the {\it superpotential}. 
In components the Lagrangian  has the form
\begin{equation}
{\cal L} = (\partial^{\mu} \bar\phi)(\partial_{\mu} \phi) + 
\psi^{\alpha} i \partial_{\alpha\dot\alpha}\bar{\psi}^{\dot\alpha} + 
\bar F F
 + \left\{ F\,{\cal W}'(\phi ) -  \frac{1}{2} {\cal W}''(\phi )\psi^2 
  + {\rm H.c.}\right\} \, .
\label{wzcomp}
\end{equation}
From Eq.\ (\ref{wzcomp}) it is obvious that $F$ can be eliminated by
virtue of the classical equation of motion,
\begin{equation}
\bar F = -\,\frac{\partial \,{\cal W}(\phi)}{\partial \phi}\, ,
\label{feom}
\end{equation}
so that the {\em scalar potential} describing self-interaction of the
field $\phi$ is
\begin{equation}
V(\phi , \bar\phi) = \left| \frac{\partial \,{\cal W}(\phi)}{\partial 
\phi}\right|^2\, .
\label{scalpo}
\end{equation}
In what follows we will often denote the chiral superfield and its 
lowest (bosonic) component by one and the same letter, making no 
distinction  between capital and small $\phi$. Usually it is clear from 
the context what is meant in each particular case.

If one  limits oneself to renormalizable theories, the superpotential 
${\cal W}$ must be a polynomial function of $\Phi$ of power not 
higher than 
three. In the model at hand, with one chiral superfield, the generic
superpotential  can be always reduced to the following ``standard" form
\begin{equation}
{\cal W}(\Phi) = \frac{m^2}{\lambda} \,\Phi - \frac{\lambda}{3} 
\Phi^3\, .
\label{spot}
\end{equation}
The quadratic term can be always  eliminated by a redefinition of the 
field $\Phi$. Moreover, by using the $R$ symmetries (Sec.\ \ref{sec34}) one can
always choose the phases of  the constants $m$ and $\lambda$ at will. 

Let us study the set of classical vacua of the theory, {\it the vacuum manifold}.
In the simplest case of the vanishing superpotential, ${\cal W}=0$, any
coordinate-independent field
$\Phi_{\rm vac}=\phi_0$ can serve as a vacuum\footnote{By coordinates we mean
here coordinates in superspace, $x_\mu$, $\theta_\alpha$,
$\theta_{\dot\alpha}$.}.  The vacuum manifold is then the
one-dimensional (complex) manifold
$C^1=\{
\phi_0\}$. The continuous degeneracy is due to the absence of the potential
energy, while the kinetic energy vanishes for any constant $\phi_0$.

This continuous degeneracy  is lifted by the superpotential.
In particular, the superpotential~(\ref{spot}) implies two
 classical
vacua,
\begin{equation}
\phi_{\rm vac} = \pm \frac{m}{\lambda}\, .
\label{vacua}
\end{equation}
Thus, the continuous manifold of vacua $C^1$ reduces to  two points.
 Both vacua are physically equivalent. This equivalence
 could be explained by the spontaneous 
breaking of 
$Z_2$ symmetry, $\Phi \to -\Phi$, present in the 
action~(\ref{lagrwz}) with the superpotential~(\ref{spot}).

\subsubsection{The general case}
\label{sec3100}

In many instances generalized Wess-Zumino models emerge at low 
energies  as effective theories describing the low-energy behavior of 
``fundamental" gauge theories, much in the same way as the pion 
chiral  Lagrangian presents a low-energy limit of QCD. 
In this case they need not be renormalizable, the superpotential need 
not be polynomial, and the kinetic term need not be canonic.
The most general action compatible with SUSY and containing not more than two
derivatives $\partial_\mu$ is
\begin{equation}
S = \frac{1}{4}\int\! {\rm d}^4 x \,{\rm d}^4\theta \, {\cal K}(\Phi_i ,
\bar{\Phi}_j )+  
\left\{ \frac{1}{2}\int\! {\rm d}^4 x \,{\rm d}^2\theta \,{\cal W}(\Phi_i) +{\rm
H.c.}\right\}\, ,
\label{klagrwz}
\end{equation}
where $\Phi_i$ is a set of the chiral superfields, the superpotential ${\cal 
W}$ is an analytic function of all chiral variables $\Phi_i$, 
while the kinetic term is determined by the function   ${\cal K}$ depending on
both chiral $\Phi_i$ and antichiral $\bar \Phi_j$ fields. Usually ${\cal K}$ is
referred to as the {\em K\"{a}hler potential} (or the  K\"{a}hler 
function).  The K\"{a}hler potential is real.

In components the Lagrangian takes the form
\begin{equation}
{\cal L} =\sum_{i,j=1}^n \left\{G^{i\bar j}\,
\partial_\mu\phi_i\,\partial^\mu\bar\phi_{\bar j} -\left[ G^{-1}\right]_{i\bar j}
\frac{\partial {\cal W}}{\partial \phi_i }\,\frac{\partial \bar{\cal W}}{\partial
\bar\phi_{\bar j} } \right\} + \mbox{fermions}
\end{equation}
where 
 \begin{equation}
G^{i\bar j}=\frac{\partial^2\,{\cal K}}{\partial \phi_i \partial \bar \phi_{\bar j}}
\label{metric}
\end{equation}
plays the role of the metric on the space of fields (the {\it target space}),
and $G^{-1}$ is the inverse matrix.

What is the vacuum manifold in this case? In the absence of the superpotential,
${\cal W}=0$, any set $\phi^0_i$ of constant fields is a possible vacuum.
Thus, the vacuum manifold  is the K\"{a}hler manifold of the complex dimension
$n$  and the metric $G^{i\bar j}$ defined in Eq.\ (\ref{metric}). If ${\cal W}\neq 0$
the conditions of the $F${\em -flatness}
\begin{equation}
\frac{\partial {\cal W}}{\partial \phi_i }=0
\label{Fflat}
\end{equation}
single out some submanifold of the original K\"{a}hler manifold. This submanifold
may be continuous or discrete. If no solutions of Eq.\ (\ref{Fflat}) exist,
supersymmetry is spontaneously broken, see examples in Sec.\ \ref{sec5}.

To illustrate this general construction let us 
consider the   model with two  superfields $\Phi$ and $X$, and 
\begin{equation}
{\cal K}=\bar \Phi \Phi + \bar X X\,,\quad
{\cal W}(\Phi , X) = \frac{m^2}{\lambda}\Phi  - \frac{\lambda}{3}\Phi^3 
-\alpha\Phi 
X^2\, .
\end{equation}
In this simple case the  K\"{a}hler manifold is two-dimensional complex space
$C^2$.  If $m^2/\lambda=0$ and $\lambda=0$ but $\alpha\neq 0$
the vacuum manifold is one-dimensional space, $X=0$ and $\Phi$ arbitrary.
Switching on all three coefficients in ${\cal W}$ reduces the vacuum manifold
to four points. The first pair is at 
$\Phi =\pm m/\lambda\, , \,\,\, X= 0$; another pair is at $\Phi =0 \, , \,\,\, X= \pm
m/\sqrt{\alpha\lambda}$. Inside each pair the vacua are equivalent due to
$Z_2\times Z_2$ symmetry of the model. 

\subsection{Simplest gauge theories}
\label{secGT}

Now let us proceed to the gauge models, which constitute the main 
contents  of this review.

\subsubsection{Supersymmetric quantum electrodynamics } 
\label{sec311}

Supersymmetric quantum electrodynamics (SQED) is the simplest and, 
historically, the first  \cite{GL}  supersymmetric gauge theory. 
This model supersymmetrizes QED.
In QED the electron  is described by the
Dirac  field. One Dirac field contains two chiral (Weyl) fields:
 left-handed and  right-handed, both with the electric charge~1.
Its complex conjugate contains  left-handed and  right-handed Weyl spinors
with the electric charge~$-1$. So, in terms of Weyl spinors the electron is described
by two left-handed fields, one with charge $+1$, another with charge
$-1$, complex conjugated fields are right-handed.
 In  SQED each Weyl  field is accompanied by a complex scalar field,
selectron. 
Thus, we get two  chiral  superfields, $S$  and $T$ of the opposite 
electric charges. 

Apart from the matter sector there  exists the gauge sector 
which includes the photon and photino. In the superspace one uses 
the vector superfield $V$, see Eq.~(\ref{vecsf}).
The SQED  Lagrangian is
\begin{equation}
{\cal L } = \left\{ \frac{1}{8\, e^2}\int\!{\rm d}^2\theta \, W^2 + {\rm
H.c.}\right\} +
\frac{1}{4}\int \!{\rm d}^4\theta \left(\bar{S}e^V S + \bar{T}e^{-V} T \right)
 + \left\{ \frac{m}{2}\int\!{\rm d}^2\theta \, ST + {\rm H.c.} \right\}\, ,
\label{sqed}
\end{equation}
where $e$ is the electric charge, $m$ is the electron/selectron mass, and
$W_\alpha$ is the supergeneralization of the photon field strength 
tensor,
\begin{equation}
{W}_{\alpha} = \frac{1}{8}\;\bar{D}^2\, D_{\alpha } V =
  i\left( \lambda_{\alpha} + i\theta_{\alpha}D - \theta^{\beta}\, 
F_{\alpha\beta} - 
i\theta^2{\partial}_{\alpha\dot\alpha}\bar{\lambda}^{\dot\alpha} 
\right)\, . 
\label{sgpfst}
\end{equation}
This superfield is chiral, $W_{\alpha}=W_{\alpha}(x_L, \theta)$.
Its form  as well as the form 
of interaction is fixed by the SUSY generalization of the
gauge  invariance \cite{JWBZ},
\begin{eqnarray}
&&S(\!x_L,\theta)\to { e}^{i\Lambda(x_L,\theta)}\, S(\!x_L,\theta)\,,
\quad 
~~T(\! x_L,\theta)\to { e}^{-i\Lambda(x_L,\theta)}\, T(\! x_L,\theta)\,;
\nonumber\\[0.2cm]
&&\bar S(\! x_R,\bar\theta)\to { e}^{-i\bar\Lambda(x_R,\bar\theta)}\,
\bar S(\! x_R,\bar\theta)\,,
\quad 
\bar T(\! x_R,\bar\theta)\to { e}^{i\bar\Lambda(x_R,\bar\theta)}\,
\bar T(\! x_R,\bar\theta)\,;
\nonumber\\[0.2cm]
&&V(\! x,\theta, \bar\theta) \to V(\! x,\theta, \bar\theta)
-i\left[\Lambda(\! x_L,\theta) -\bar\Lambda(\! x_R,\bar\theta)\right]\,.
\end{eqnarray}
The gauge parameter which was a function of $x$ in QED is now promoted to the
chiral superfield $\Lambda(x_L,\theta)$. Using this gauge freedom one eliminates,
for example, all terms in the first line in Eq.\ (\ref{vecsf}),
\begin{equation}
V_{\rm WZ} = -2\theta^{\alpha}\bar\theta^{\dot\alpha}v_{\alpha\dot\alpha} - 
2i\bar\theta^2(\theta\lambda ) + 2i 
\theta^2(\bar\theta\bar{\lambda}) + 
\theta^2\bar\theta^2 D\, .
\label{vecsf1}
\end{equation}
This is called the {\em Wess-Zumino gauge}.

If we take into account the  rules of integration over the Grassmann 
numbers
we immediately see that
the integration ${\rm d}^2\theta$ singles out the $\theta^2$ component
of the chiral superfields $W^2$ and $ST$, i.e. the   $F$ terms.  
Similarly, the integration ${\rm d}^2\theta {\rm d}^2\bar \theta$ singles out the
$\theta^2\bar \theta^2$ component of the real superfields $\bar Se^VS$ and $\bar 
Te^{-V} T$ , i.e. the $D$ terms. The fact that the electric charges of $S$ and $T$ are 
opposite is explicit in Eq.\ (\ref{sqed}). The theory describes 
conventional electrodynamics of one Dirac and two complex
scalar fields. In addition, it includes  photino-electron-selectron 
couplings and self-interaction of the selectron fields of a very special 
form, to be discussed below.  

In the Abelian gauge theory one is allowed to add to the Lagrangian
the so-called $\xi$ term,
\begin{equation}
\Delta{\cal L}_\xi = \frac{\xi }{4} \int\! {\rm d}^2\theta {\rm d}^2\bar \theta 
\,V(\! x,\theta , \bar\theta ) \equiv \frac{\xi }{2} \int\! {\rm d} \theta^\alpha
W_\alpha \equiv  \xi D\, .
\label{xiterm}
\end{equation}
It plays an important role in the Fayet-Iliopoulos mechanism of the tree-level
spontaneous SUSY  breaking (Sec.\ \ref{sec51}). Although this term is specific
for the  Abelian theories, one can find 
some analogs in the non-Abelian gauge theories too. 

The $D$ component of $V$ is an auxiliary field (similarly to $F$); it enters the 
Lagrangian without derivatives,
\begin{equation}
{\cal L} =
\frac{1}{2e^2}D^2 +D\,(\bar S S-\bar T T) +\xi D +\dots
\label{dLd}
\end{equation}
where the ellipses denote $D$-independent terms. Thus, $D$
 can be eliminated by substituting the classical equation of 
motion. In this way we get  the so-called 
$D$-potential, describing the self-interaction of selectrons,
\begin{equation}
U_D = \frac{1}{2e^2}\,D^2\, , \,\,\,  D=-e^2 (\bar S S-\bar T T +\xi) \, .
\end{equation}
This is only a part of the potential energy.
The full  potential is obtained by adding the part generated by 
the $F$ terms of the matter fields, see Eq.\ (\ref{scalpo}) with ${\cal 
W}\to mST$,
\begin{equation}
U(S,T) = \frac{e^2}{2}(\bar S S-\bar T T+ \xi)^2 +|mS|^2 +|mT|^2\, .
\label{polpot} 
\end{equation}
This expression is sufficient for examining the structure of the vacuum manifold
(we do not give here the full component expression for the SQED 
Lagrangian, deferring this task till  Sec.\ \ref{sec312}, where
the transition to components is elaborated in more complicated 
non-Abelian gauge theories). 

The  energy of any field configuration in
supersymmetric theory is positive-definite. Thus, any configuration with the
zero energy is automatically a  vacuum, i.e. the vacuum manifold is determined by
the condition $U(S,T)=0 $. Assume at first that the mass term and the $\xi$ term
are absent, $m=\xi =0$, i.e. we deal with massless SQED. Then,
\begin{equation}
U(S,T) = \frac{e^2}{2}(\bar S S-\bar T T)^2\equiv 0\, ,
\end{equation}
Modulo gauge transformations the general solution is
\begin{equation}
S=\Phi\,, \quad T=\Phi\,,
\end{equation}
where $\Phi$ is a complex parameter.
 One can think of the potential $U$
as of a mountain ridge; the $D$-flat directions then present the flat 
bottom of the valleys. This explains the origin of the term vacuum valleys.
The (classical) vacuum 
manifold is one-dimensional complex line, parametrized by  $\Phi$. 
Each point at the manifold can be viewed as a vacuum of a particular
theory.
 
Considering the parameter $\Phi$ as a chiral superfield $\Phi(x_L,\theta)$,
we arrive at  the Wess-Zumino model with
the K\"ahler potential $\bar \Phi \Phi$. The model describes the supermultiplet
containing one massless scalar and one Weyl fermion. 

It is not difficult  to verify that there is no other light excitations
at the generic point
on  the vacuum manifold. Indeed, at $\Phi\neq 0$ the theory is  in the Higgs phase:
the photon supermultiplet becomes massive.  The photon  field ``eats up" one of the
real scalar fields residing in 
$S,T$, and becomes massive,  along with another real scalar field 
which acquires the very same mass. The photino teams up  with a linear
combination of two  Weyl spinors in $S,T$, and becomes a massive Dirac field, with
the same  mass as the photon. One Weyl spinor and one complex scalar (two 
real fields) remain massless. 

The consideration above was carried out in the Wess-Zumino gauge.
The  gauge invariant
parametrization  of the vacuum manifold is given by the product $ST$. This
product is a chiral  superfield, of zero charge, so it is gauge invariant. Every point
from  the bottom of the valley is in one-to-one correspondence with the 
value of $ST=\Phi^2$. 

The occurrence of the flat directions is the  single most crucial 
feature of  the SUSY gauge  theories  instrumental in  the dynamics 
of  SUSY breaking. The issue will be discussed in more detail in
Secs.\ \ref{vmg} and  \ref{vme}.   We started from  the simplest 
example, SQED, to
get  acquainted  with the phenomenon. 

Introducing the mass term $m\neq 0$ one  lifts the vacuum 
degeneracy, making the bottom of the valley non-flat.
The mass term pushes the theory towards the origin of the valley.
Indeed, with the mass term switched on the only solution 
corresponding to the vanishing energy is $S=T=0$. The vacuum 
becomes unique. If, in addition, $\xi\neq 0$, supersymmetry is 
spontaneously broken, see Sec.\ \ref{sec51}. 

\subsubsection{Supersymmetric QCD with one flavor}
\label{sec312}

As the next step,  we  consider SUSY generalization of  QCD (to be referred to as
SQCD). Here we limit ourselves to the gauge group  SU(2)
with the matter sector consisting of one flavor. The gauge sector consists of 
 three gluons and their superpartners -- gluinos. The corresponding superfield 
now is a matrix in the color space,
\begin{equation}
V=V^a T^a\,, 
\end{equation}
where $T^a$ are the matrices of the color generators. In the SU(2) 
theory  $T^a=\tau^a/2$ where $\tau^a$ are the Pauli matrices, $a=1,2,3$.

Similarly to SQED, the matter sector is  built from two superfields. Instead of the
electric charges  now we must pick up  certain representations of SU(2). In SQED 
the fields $S$ and $T$ have the opposite electric charges. Analogously, in SQCD
one superfield must be in the fundamental representation and another in
anti-fundamental. The specific feature of SU(2) is the equivalence of doublets and
anti-doublets. 
Thus, the matter is described by the set of superfields $Q_f^\alpha$
where $\alpha =1,2$ is the color index  and $f= 1,2 $
is a ``subflavor" index. Two subflavors comprise one flavor.

The Lagrangian of the model is
\begin{equation}
{\cal L} = \left\{ \frac{1}{4g^2}\, 
\int \! {\rm d}^2 \theta\, \mbox{Tr}\,W^2 +{\rm H.c.}\right\} +
\frac{1}{4} \int \! {\rm d}^2\theta {\rm d}^2\bar\theta\,
\bar Q^f e^V Q_f +
\left\{ \frac{m}{4}\int\! {\rm d}^2\theta\, Q_\alpha^f Q_{f}^\alpha
+{\rm H. c.}\right\}\, .
\label{su2lagr}
\end{equation}
 The chiral superfield $W_\alpha$ which includes the gluon 
field strength tensor, is the non-Abelian generalization of Eq.\ 
(\ref{sgpfst}),
\begin{equation}
{W}_{\alpha} = \frac{1}{8}\;\bar{D}^2\, \left( {e}^{-V}\! D_{\alpha } {e}^V\right) =
  i\left( \lambda_{\alpha} + i\theta_{\alpha}D - \theta^{\beta}\, 
G_{\alpha\beta} - 
i\theta^2{\cal D}_{\alpha\dot\alpha}\bar{\lambda}^{\dot\alpha} 
\right)\, . 
\label{sgfst}
\end{equation}
Unlike the situation in the Abelian case, now  ${W}_{\alpha}$ is not
 (super)gauge
invariant, Eq.~(\ref{sgfst}) refers to the Wess-Zumino gauge.

Note that the SU(2) model under consideration, with  one 
flavor, possesses a global SU(2) (subflavor) invariance
allowing one to freely rotate the superfields $Q_f$.  All indices corresponding 
to the
SU(2) groups (gauge, Lorentz and subflavor) can be lowered and 
raised by  means of the $\epsilon^{\alpha\beta}$ symbol, according to the 
general rules.

The Lagrangian presented in Eq.\ (\ref{su2lagr}) is unique if the requirement of
renormalizability is imposed. Without this requirement the last term in Eq.\
(\ref{su2lagr}), the superpotential, could be supplemented, e.g., by the quartic 
color
invariant $( Q^{\alpha f}Q_{\alpha f})^2$.  The cubic term is not
allowed in SU(2).  In general, the 
renormalizable models with a richer matter
sector may allow  for the cubic in $Q$ terms in the superpotential.

It is instructive to pass from the superfield notations to components.
We will do this exercise once.
Start from $W^2$. The $F$ component of $W^2$ includes the kinetic 
term of the gluons and  gluinos, as well as the square of the $D$ term,
\begin{equation}
 \frac{1}{4g^2}\, 
\int \! {\rm d}^2 \theta\, \mbox{Tr}\,W^2 = 
-\frac{1}{8 g^2} \left( G_{\mu\nu}^a G^{a\mu\nu} -i G_{\mu\nu}^a\tilde 
G^{a\mu\nu}\right)  +\frac{1}{4g^2}D^a D^a 
+\frac{i}{2g^2}
\lambda^a\sigma^\mu {\cal D}_\mu\bar \lambda^a \, .
\label{ld}
\end{equation}
Note that the inverse coupling constant $1/g^2$ can be treated as a complex
parameter,
\begin{equation}
\frac{1}{g^2}\to \frac{1}{g^2}- i\,\frac{\vartheta}{8\pi^2}
\label{itheta}
\end{equation}
where $\vartheta$ is the vacuum angle. For the time being the occurrence of
 the $\vartheta$ angle is not important.

The next term to be considered is $\int\!{\rm  d}^2\theta {\rm 
d}^2\bar\theta\,\bar  Qe^V\! Q$. Calculation of the $D$ component of $\bar Qe^V Q$
is a more  time-consuming  exercise since we must take into account the fact 
that $Q$ depends  on $x_L$ while $\bar Q$ depends on $x_R$; the 
both arguments differ from  $x$. Therefore, one has to expand in this 
difference. The factor $e^V$  sandwiched  between $\bar Q$ and $Q$ 
covariantizes all derivatives. Taking the field $V$ in the Wess-Zumino gauge
one gets
\begin{eqnarray}
\frac{1}{4} \int \!{\rm  d}^2\theta {\rm 
d}^2\bar\theta\,\bar  Q^f e^V\! Q_f &\!\!=&\!\! {\cal D}^\mu\bar\phi^f\,{\cal
D}_\mu
\phi_f + \bar F^f F_f+ D^a \,\bar\phi^f T^a \phi_f \nonumber\\[0.2cm] 
&& + i\psi_f
\sigma^\mu{\cal D}_\mu
\bar\psi^f + \left[ i\sqrt{2}
(\psi_f\lambda )\bar\phi^f +\mbox{H.c.}\right]\,.
\label{barss}
\end{eqnarray}
Finally, we present the superpotential term,
\begin{equation}
\frac{m}{4}\int\! {\rm d}^2\theta\, Q_\alpha^f Q^\alpha_f=
m \,\phi_\alpha^f F^\alpha_f -\frac{m}{2} \psi_\alpha^f\psi^\alpha_f\,.
\end{equation}

The fields $D$ and $F$ are auxiliary and can be eliminated by  virtue of the
equations of motion. In this way we get the potential energy,
\begin{equation}
U= U_D +U_F\,,\quad U_D=\frac{1}{2g^2} D^a D^a\,,\quad U_F= \bar F_\alpha^f 
F^\alpha_f \,,
\label{dplusf}
\end{equation}
where 
\begin{equation}
D^a = - g^2\,
\bar\phi_f T^a \phi^f \,,\quad F^\alpha_f =-\bar m \,\bar\phi^\alpha_f\,.
\label{vd}
\end{equation}

The $D$ potential $U_D$  represents
a quartic self-interaction of the scalar fields, of a very peculiar form.
Typically in the $\phi^4$ theory the potential has one -- at most 
several -- minima. The only  example  with a continuous manifold of 
points of minimal energy which was well studied in the context of
non-supersymmetric theories
is the spontaneous breaking of a global  continuous symmetry, say, 
U(1).  In this case all points belonging to the vacuum manifold  are 
physically equivalent. The $D$ potential (\ref{dplusf}) has a specific 
structure -- there is a continuous vacuum degeneracy, the minimal (zero) energy is
achieved on an infinite set of the field configurations which are {\em  not}
physically equivalent. 

To examine the vacuum manifold let us start again from the case of the vanishing
superpotential, i.e. $m=0$.  From Eq.\ (\ref{vd}) it is clear that the classical
space of vacua is defined by the $D$-flatness  condition
\begin{equation}
D^a =  - g^2\, \bar\phi_f T^a \phi^f =0\,,\quad a=1,2,3\,. 
\label{dflat}
\end{equation}
The notion of  $D$-flatness  is specific for  the Wess-Zumino gauge description of the
vacuum manifold. Later on we will present a more general (and more geometrical)
construction of the vacuum manifold (Sec.\ \ref{vmg}).
 
In the case at hand  it is not difficult to find
the $D$ flat direction explicitly. Indeed, 
consider the scalar fields of the form
\begin{equation}
\phi^\alpha_f = v \left( \begin{array}{cc}
1 & 0\\  0 & 1
\end{array} \right)\, ,
\label{dflat2}
\end{equation}
 where $v$ is an arbitrary complex constant. It is obvious that
for any value of $v$ all $D$'s vanish. $D^1$ and $D^2$ vanish 
because
$\tau^{1,2}$ are off-diagonal matrices; $D^3$ vanishes after 
summation over
two subflavors. 

It is quite obvious that if  $v\neq 0$ the original  gauge symmetry 
SU(2) is totally Higgsed.
Indeed, in the vacuum field  (\ref{dflat2}) all three gauge bosons 
acquire mass  $M_V= g|v|$. Needless to say that supersymmetry is not broken. 
It is instructive to trace the reshuffling of the degrees of freedom before and after 
the Higgs phenomenon. In the unbroken phase, corresponding to $v=0$,
we have three massless gauge bosons (6 degrees of freedom),
three massless gauginos (6 degrees of freedom), four matter Weyl
fermions (8 degrees of freedom), and
four complex matter scalars (8 degrees of freedom).
In the broken phase three matter fermions combine with
the gauginos to form three massive Dirac fermions (12 degrees of 
freedom). Moreover, three matter scalars combine with the gauge fields
to form three {\em massive} vector fields (9 degrees of freedom) 
plus three massive (real) scalars.
What remains massless? One complex scalar field,
corresponding to the motion along the  bottom of the valley, $v$,
and its fermion superpartner, one Weyl fermion. The
balance between the fermion and boson degrees of freedom is 
explicit.

Thus, we see that in the effective low-energy theory only one chiral superfield
$\Phi$ survives. This chiral superfield can be introduced as a supergeneralization
of Eq.\ (\ref{dflat2}),
\begin{equation}
Q^\alpha_f = \Phi \left( \begin{array}{cc}
1 & 0\\  0 & 1
\end{array} \right)\, .
\label{dflat3}
\end{equation}
Substituting this expression in the original Lagrangian~(\ref{su2lagr}) we get 
\begin{equation}
{\cal L}_{\rm eff}=\frac{1}{2} \int \!{\rm  d}^2\theta {\rm 
d}^2\bar\theta\,\bar \Phi \Phi + \left\{
\frac{m}{2}\int\! {\rm d}^2\theta\, \Phi^2
+ {\rm H.c.}\right\}\,.
\label{lel}
\end{equation}
Here we also included the superpotential term assuming that
$|m|\ll g|v|$. Thus, the low-energy theory is that of the free chiral superfield with
the mass $m$. We hasten to add that Eq.\ (\ref{lel}) was obtained at the classical
level. The quantum corrections do modify it as we will see later. In particular, the
kinetic term receives perturbative corrections. The expansion parameter is
$1/\log |\Phi |/\Lambda$. The superpotential term is also renormalized but only
at the nonperturbative level, see Sec.\ \ref{sec471}. In this way we arrive at an
effective low-energy Lagrangian of the general form~(\ref{klagrwz}) with one
chiral superfield.

\subsection{The vacuum manifold: generalities}
\label{vmg}

In this section we present a general approach to the construction of the vacuum
manifold in the gauge theories. Particular applications will be given in the
subsequent sections. We start from a historical remark. A gauge invariant
description of the system of the vacuum valleys  was suggested in Ref.\
\cite{BDSF} and extensively used in Refs.\ \cite{TT,ADS2};  recently the issue was
revisited in  Ref.\ \cite{Luty1}.
In these  works it is explained that the set of  proper coordinates 
parametrizing the
space  of the classical  vacua is  nothing else but the set of  independent  gauge
invariant polynomials constructed from  the  chiral matter fields.  Another name
for these coordinates, often used in the literature, is  the {\it moduli}. The vacuum
manifold is referred to as the moduli space. 

In the previous sections we considered the U(1) and SU(2) theories in the
Wess-Zumino gauge. This gauge is extremely convenient in the unbroken phase.
At the same time, 
for the general analysis of  the Higgs phase, a superanalog of the unitary gauge is
more suitable. To illustrate the statement we turn again to the same SU(2) theory
with one flavor.  
As was just explained,
 in this theory only one physical Higgs superfield $\Phi$
survives. Correspondingly, Eq.\  (\ref{dflat3}) can be viewed as the unitary gauge
condition, rather than the parametrization of the vacuum manifold; any 
field configuration $Q^\alpha_f$ can be cast in the form~(\ref{dflat3}) by an
appropriate gauge transformation. In this gauge the Lagrangian becomes
\begin{equation}
{\cal L}= \left\{ \frac{1}{4g^2}\, 
\int \! {\rm d}^2 \theta\, \mbox{Tr}\,W^2 +{\rm H.c.}\right\} +
\frac{1}{4} \int \! {\rm d}^2\theta {\rm d}^2\bar\theta\, \bar \Phi \Phi\, {\rm
Tr} \,e^V \,.
\end{equation}
(We omit the mass term for the time being.)

To verify that this is indeed the analog of the unitary gauge one can rewrite the
Lagrangian in components keeping the terms up to quadratic in $V$,
\begin{eqnarray}
{\cal L}_{\rm quad}\!&=&\! -\frac{1}{4g^2}
G_{\mu\nu}^aG^{a\mu\nu}+\frac{1}{2g^2}D^aD^a +
\frac{|\phi |^2}{2}\left[ v_\mu^a v^{a\mu} + C^a D^a 
+\frac{1}{4}\partial_\mu C^a
\partial^\mu C^a
\right.\nonumber\\[0.2cm]
&&\left.+\frac 1 2 \bar M^a M^a\right] +2\partial_\mu \bar\phi \,\partial^\mu \phi
+\mbox{fermionic~part}\,.
\label{arkone}
\end{eqnarray}
Note that the term linear in $V$ drops out in Tr$\,e^V$.
Eliminating the auxiliary fields $D$ and $M$ we arrive at the theory containing  the
massive vector triplet plus the scalar triplet (their common mass is $g|\phi |$),
 plus
the massless modulus field $\phi$, plus their fermionic partners.  

The following general construction extends this example.
Consider a generic gauge theory, based on the gauge group $G$, with  matter
$Q=\{Q_i\}$ in the representation $R$, which can be reducible. The index $i$
runs from 1 to $n$ where $n$ is the dimension of the representation $R$. 
The Lagrangian has the form
\begin{equation}
{\cal L} = \left\{ \frac{1}{8T(R)g^2}\, 
\int \! {\rm d}^2 \theta\, \mbox{Tr}\,W^2 +{\rm H.c.}\right\} +
\frac{1}{4} \int \! {\rm d}^2\theta {\rm d}^2\bar\theta\,
\bar Q e^V Q 
\, ,
\label{Glagr}
\end{equation}
where the group theory coefficient $T(R)$
is defined after Eq.~(2.90). In particular, $T(R)=1/2$
if the generators are in the fundamental representation
of SU($N$), which is usually assumed.
For the time being the superpotential is set to zero.  The theory is invariant under 
the (super)gauge transformations
\begin{eqnarray}
{e}^{V(\! x,\theta, \bar\theta)}\!&\to& \!\!{e}^{i\bar\Lambda(\!
x_R,\bar\theta)}\,{ e}^{V(\! x,\theta, \bar\theta)}\,
{ e}^{-i\Lambda(\! x_L,\theta)} \,,\quad
W_\alpha(\!x_L,\theta) \to 
{e}^{i\Lambda(\!
x_L,\theta)}\,W_\alpha(\!x_L,\theta)\,
{ e}^{-i\Lambda(\! x_L,\theta)} 
\,, \nonumber\\[0.2cm]
Q(\!x_L,\theta)\!\! &\to &\!\! { e}^{i\Lambda(x_L,\theta)}\, Q(\!x_L,\theta)
\,,\quad \quad \quad ~~~~
\bar Q(\! x_R,\bar\theta)\to 
\bar Q(\! x_R,\bar\theta)\,{ e}^{-i\bar\Lambda(x_R,\bar\theta)}\,.
\label{gaugetr}
\end{eqnarray}
It is seen that for the spatially constant fields these gauge transformations elevate
the original group $G$  to  its complex extension $G_c$. The group $G_c$ acts in the
$n$-dimensional complex space $C^n$. All points of $C^n$ belonging to one and the
same gauge orbit of $G_c$ are physically identical. After this identification is done,
we get the space ${\cal M}$ of physically distinct classical vacua as a quotient 
\begin{equation}
 {\cal M}=C^n/G_c\,.
\label{quot}
\end{equation}

In fact, the space ${\cal M}$ is not a manifold but, rather,  a sum of manifolds,
\begin{equation}
 {\cal M}=\sum_i {\cal M}_i\,,
\label{quotsum}
\end{equation}
where each ${\cal M}_i$ is characterized by a subgroup $H_i$ of $G$ which remains
unbroken. For instance, in the SU(2) model with one flavor discussed above the
original complex space is $C^4$,  its complex dimension is four. The group $G_c$
which is a complexified SU(2) is three-dimensional. Moreover, the space ${\cal M}$ 
is the sum of two manifolds: the zero-dimensional ${\cal M}_1$ consisting of one
point, $\Phi=0$, and the one-dimensional manifold ${\cal M}_2$ which is
$\Phi\neq 0$.  The stability group $H_1$ coincides with SU(2) (the entire SU(2)
remains unbroken). For ${\cal M}_2$ the stability group is trivial (all vector bosons
are Higgsed).

Returning to the generic gauge theory let us consider the case when all vector bosons
are Higgsed. It implies that $n\ge d_G$,  where $d_G$ is the
dimension of the group $G$ (the number of the generators). We pick up such ${\cal
M}_i$ in Eq.\ (\ref{quotsum}) whose stability group is trivial. The
$d_G$ degrees of freedom are eaten up in the process of Higgsing the gauge bosons.
Then the complex dimension of ${\cal M}_i$ is $n-d_G$.

To determine the K\"ahler metric on ${\cal M}$ we do the following. First, introduce
$n-d_G$ complex coordinates on ${\cal M}$ in some way,
\begin{equation}
Q= Q(\tau_1,\dots , \tau_{n- d_G})\,.
\label{parametr}
\end{equation}
One of the possible ways of  parametrizing  ${\cal M}$ is exploiting  the set of
gauge invariant chiral polynomials constructed from the fields $Q$. Generally
speaking, their number is larger than $n- d_G$, but the number of independent
invariants is equal to $n- d_G$.

Second, the condition of the vanishing energy, (i.e. $D^a=0$, cf. Eq.\ (\ref{arkone})) 
is
\begin{equation}
\frac{\partial}{\partial C^a} \bar Q (\bar \tau) e^C Q(\tau)=0\,.
\label{eqonc}
\end{equation}
Here we get $d_G$ equations for $d_G$ quantities $C^a$, so the solution
$C^a(\tau , \bar\tau)$ is unique$\,$\footnote{If the solution is non-unique
it means that $Q(\tau)$ belongs to ${\cal M}_i$ with a nontrivial stability group
$H_i$.}.
Once the solution is found the 
 K\"ahler metric is obtained,
\begin{equation}
{\cal K}=\bar Q (\bar \tau) e^{C(\tau , \bar\tau)} \, Q(\tau)\,.
\label{Kmetric}
\end{equation}
In the mathematical literature the procedure of 
constructing the kinetic term for the moduli, after integrating out
all heavy gauge bosons, goes under the name the {\em K\"{a}hler
quotient}, e.g. \cite{HKLR}.

Although the construction described above solves the issue of the K\"{a}hler
 metric on the moduli space
in principle, in practice solving Eqs.\ (\ref{eqonc}) is a
difficult technical task. Therefore, it is instructive to see how the general
procedure is related to the
$D$-flatness conditions in the Wess-Zumino gauge. To pass to this gauge we
perform the gauge transformation~(\ref{gaugetr}) with $\Lambda = -iC/2$,
\begin{equation}
\hat Q={ e}^{C/2}\, Q \,,\quad
\bar {\hat Q}=  \bar Q \,{ e}^{C/2}\,,\quad
{ e}^{\hat V}={e}^{-C/2}\,{ e}^{ V}{e}^{-C/2}\,.
\label{gaugeC}
\end{equation}
Note that, after this transformation, the fields $\hat Q$ and $\bar {\hat Q}$
depend on the parameters $\tau_i$ in a non-holomorphic way, unlike $ Q$ and
$\bar {Q}$, whose dependence was holomorphic. 
The gauge transformed $\hat C$ vanishes, and Eqs.\ (\ref{eqonc}) take the form
\begin{equation}
\left. \frac{\partial}{\partial \hat C^a} \bar{\hat Q} e^{\hat C} \hat
Q\right|_{\hat C^a=0}=\bar{\hat Q} \,T^a \hat
Q=0\,.
\label{eqDflat}
\end{equation}
This is precisely the $D$-flatness
conditions in the Wess-Zumino gauge.

What happens if  $n< d_G$? In this case it clear that the group $G$ cannot be fully
Higgsed. A part of the group $G$ can be realized in the Higgs mode, however, while 
a subgroup $H$ of $G$ remains unbroken. Then the consideration above can be
repeated with the substitution of $d_G$ by  $d_G-d_H$. The gauge orbit which
identifies the points on $C^n$ is that of the quotient $G/H$. Even at $n> d_G$ for
each manifold ${\cal M}_i$ with nontrivial $H_i$
the situation is similar.

The technical difficulty of solving Eqs.\ (\ref{eqonc}) explains
why the  K\"{a}hler metric on the moduli space is explicitly found
only in several relatively simple models. Usually one analyzes the $D$-flatness
conditions in the Wess-Zumino gauge, rather than the general relations
(\ref{eqonc}). This strategy proves to be simpler. One tries to find a particular
solution of the $D$-flatness
conditions containing a sufficiently large number of parameters. Once found, the
particular solution is then promoted to the general solution by virtue of the
flavor symmetries of the model under consideration.  An instructive example
is discussed in Sec.\ \ref{su52}.
It is customary to use the gauge invariant polynomials as the  moduli parameters.
This corresponds to a particular choice of the moduli parameters $\tau$
introduced above. The use of the gauge invariant polynomials makes absolutely
transparent the realization of the quotient space~(\ref{quot}). On the other hand 
using more general parametrizations~(\ref{parametr}) may result in algebraically
simpler expressions for the K\"ahler potentials.

\subsection{The vacuum manifold: examples}
\label{vme}

\subsubsection{ SU(5) model with one quintet and one 
(anti)decuplet}
\label{secSU5}

The approach based on the chiral polynomials is very convenient
for establishing the fact of the existence (non-existence)
of the moduli space of the classical vacua, and in counting the
dimensionality of this space.  As an example, consider the 
SU(5) model with one quintet and one 
(anti)decuplet. This is the simplest chiral model with no internal anomalies.
It describes  Grand Unification,
with one generation of quarks and leptons. 
This example of the non-chiral matter is singled out historically -- 
the dynamical supersymmetry breaking was 
first found in this model \cite{ADS1,MV}.

The matter sector consists of
one quintet field 
$V^\alpha$, and one (anti)decuplet  antisymmetric field 
$X_{\alpha\beta}$.
One can see that in this case there are no chiral invariants at all. For instance, 
$V^\alpha V^\beta X_{\alpha\beta}$, vanishes due to antisymmetricity of 
$X_{\alpha\beta}$. Another candidate,
$\epsilon^{\alpha\beta\gamma\delta\rho}X_{\alpha\beta}X_{\gamma\delta}
X_{\rho\sigma}V^\sigma$, vanishes too.  This means that no
$D$ flat directions exist. The same conclusion can  be reached by explicitly
parametrizing $V$ and $X$; inspecting then  the $D$-flatness conditions one can
conclude that they have no  solutions, see e.g. Appendix A in Ref.\ \cite{Amati}.

Thus, the classical vacuum manifold reduces to the point $V=X=0$.

\subsubsection { The {\em 3-2} model of Affleck, Dine, and Seiberg }
\label{sec3-2}

The  Affleck-Dine-Seiberg (ADS) model~\cite{ADS3}, also known as the {\em 3-2}
model,  is based  on the direct 
product of two gauge groups, SU(3)$\times$SU(2) (that is where the 
name comes from). It can be obtained from the Standard Model by 
eliminating from the latter certain elements inessential for the 
dynamical SUSY breaking. Following \cite{ADS3} we retain only
one generation of matter, discard the hypercharge U(1) and
the field $\bar e_L$, singlet with respect to SU(3)$\times$SU(2) 
(``color" and ``weak isospin"). Thus, altogether we will be dealing 
with  14 Weyl fermions.

In terms of the  SU(3) color group we deal with SQCD with three colors and 
two flavors, $u$ and $d$. The quark sector includes
the following chiral (left-handed) superfields:
\begin{equation}
Q^{\alpha f}\equiv \{ u^\alpha , d^\alpha\}  \,,\quad q_{\alpha \bar f}\equiv \{\bar
u_\alpha\, , \, \bar d_\alpha\}\,, \quad (\alpha = 1,2,3;\,\,\,
f,\bar f=1,2)\, .
\label{qusect}
\end{equation}
The flavor SU(2) (the superscript $f$) of the left-handed particles, $Q^{\alpha
f}$,  is gauged -- this is our weak interaction. As for the antiparticles $q_{\alpha
\bar f}$, the subscript
$\bar f$ is the index of a global SU(2) which remains ungauged.

The  gauge bosons $W^\pm$  and $W^0$, and their superpartners,
transform according to the  adjoint representation of the group SU(2) of the weak
isospin. It is here that the asymmetry appears between the right- and 
left-handed matter. 
In addition, to avoid the internal (global) anomaly \cite{Witten3} we 
must add to the matter sector one more doublet of chiral superfields, 
the lepton doublet
\begin{equation}
L^f = \{ \nu , \, e\}\, .
\label{leptdou}
\end{equation}

No mass term that would be invariant with respect to both SU(3) 
and SU(2) can be built. Thus, the model is indeed chiral.
In  search of the valleys we will first count the dimension of the 
moduli space. To this end one must construct all independent chiral 
invariants. There are no bilinear invariants, as was just mentioned.
Two cubic invariants exist, however,
\begin{equation}
I_{\bar f} = Q^{\alpha f }q_{\alpha \bar f} L^g\,\varepsilon_{fg}\, ,
\quad (\bar f=1,2)\, .
\label{cuinv}
\end{equation}
They form a doublet of the global SU(2). 
One more chiral invariant  is quartic,
\begin{equation}
J = (Q^{\alpha f }q_{\alpha \bar f} )\, (Q^{\beta g }q_{\beta \bar g} 
)\,\epsilon_{fg} \,\epsilon^{\bar f\bar g}\equiv \, \mbox{det}\, \{ Q q\}\, .
\label{quainv}
\end{equation}
We conclude that the moduli space  has complex dimension 
three. Out of 14 fields 11 (the dimension of the SU(3)$\times$SU(2)) are eaten up
by the Higgs mechanism.

In the case at hand it is not difficult to find an  explicit
parametrization of the $D$-flat directions.  
The two-parametric family of field configurations
for which $\bar Q T^a Q=0$ for all $a$ has the form
\begin{equation}
Q^{\alpha f} = \left(\begin{array}{cc}
\tau_1 & 0\\
0 & \tau_2\\
0 & 0\\
\end{array}\right)\, ,\quad
q_{\alpha \bar f} = \left(\begin{array}{cc}
\tau_1 & 0\\
0 & \tau_2\\
0 & 0\\
\end{array}\right)\, ,\quad L=\left( 0\,, \,\sqrt{|\tau_1|^2-|\tau_2|^2}\right)\,.
\label{arktwo}
\end{equation}
In terms of the general construction of Sec.\ \ref{vmg} the fields above
correspond to $\hat Q$. They satisfy the Wess-Zumino $D$-flatness conditions
$\bar Q T^a Q=0$.

 For this configuration  
the gauge invariants $I_{\bar f}$ and $J$ are
\begin{equation}
I_{\bar 1} =-\tau_1^2\sqrt{|\tau_1|^2-|\tau_2|^2}\,,\quad 
I_{\bar 2} =0\,,\quad 
J=\tau_1^2\tau_2^2\,.
\label{ginv}
\end{equation}
The expressions above demonstrate that the vacuum family (\ref{arktwo})  is
general modulo the global SU(2) rotations,
\begin{equation}
q^\prime_{\alpha \bar g}=  U_{\bar g}^{\bar
f}\, q_{\alpha \bar f}\,,
\end{equation}
where $U$ is a matrix from SU(2). The U(1) part of this SU(2) is irrelevant (it
changes the phases of $\tau_{1,2}$ which are arbitrary anyway); therefore, 
we deal with the SU(2)/U(1) quotient in the flavor space. This quotient is
equivalent to the sphere which is  parametrized by one complex parameter.
Altogether, we have three complex parameters -- exactly the number we
need. We do not write out the flavor parametrization explicitly since we will use a
round-about way to account for the flavor symmetry.

Now we are ready to get the K\"ahler potential,
\begin{equation}
{\cal K}=\bar{\hat Q} \hat Q =3\bar\tau_1\tau_1+\bar\tau_2\tau_2\,,
\label{Ktau}
\end{equation}
where $\tau_i$ are the moduli fields depending on $x_L$ and $\theta$
(correspondingly, $\bar\tau_i$ depend on $x_R$ and $\bar\theta$), 
and we set the matrix of the global SU(2) rotation to unity, $U=1$.

As was already mentioned, in the literature it is customary to use the gauge
invariants as the moduli fields. In the given problem the gauge invariants
$I_{\bar f}$ and
$J$ are related to 
$\tau_i$ by virtue of Eq.\  (\ref{ginv}).  We can rewrite the K\"ahler
potential~(\ref{Ktau}) in terms of $I_{\bar f}$ and $J$.
Because of the flavor SU(2), the invariants $I_1$ and 
$I_2$  can only enter in the combination
\begin{equation}
A =\frac{1}{2}\left(\bar I_1 I_1 + \bar I_2 I_2\right) 
= \frac{1}{2}\, 
(\bar\tau_1 \tau_1)^2 \left( \bar\tau_1 \tau_1 - \bar\tau_2 \tau_2\right) 
\, .
\label{comba}
\end{equation} 
As for $J$,  it is convenient to introduce
\begin{equation}
B = \frac{1}{3}\sqrt{\bar J J}= \frac{1}{3}\bar\tau_1 \tau_1\bar\tau_2 \tau_2
\,  . 
\label{combb}
\end{equation}
Then
\begin{equation}
\bar\tau_1 \tau_1= \left(A- \sqrt{A^2 - B^3}\right)^{1/3}
+
\left(A +\sqrt{A^2 - B^3}\right)^{1/3}\,,\quad
\bar\tau_2 \tau_2 = \frac{3B}{\bar\tau_1 \tau_1}
\, .
\label{combbb}
\end{equation}
Thus, the  K\"ahler potential takes the form
\begin{eqnarray}
{\cal K}&=&
3\left[\left(A- \sqrt{A^2 - B^3}\right)^{1/3} 
+
\left(A +\sqrt{A^2 - B^3}\right)^{1/3}\right]\nonumber\\[0.1cm]
&&+\frac{3B}{\left(A- \sqrt{A^2 - B^3}\right)^{1/3}
+\left(A +\sqrt{A^2 - B^3}\right)^{1/3} 
}
\, .
\label{kkt}
\end{eqnarray}
For an alternative derivation of the K\"{a}hler potential in the {\em 
3-2} 
model see \cite{BPR}. 

\subsubsection{SU(5) model with two quintets and two 
(anti)decuplets}
\label{su52}

This  model was the 
first  example of the instanton-induced
supersymmetry breaking  in the weak coupling regime 
\cite{MV,2}.
It presents another example of the anomaly-free chiral 
matter sector. Unlike the one-family model (one quin\-tet and one 
(anti)decu\-plet)  the $D$-flat directions do exist. Generically, the gauge SU(5)
symmetry is completely broken, so that 24 out of 30 chiral matter 
superfields are eaten up in the super-Higgs mechanism. Therefore, 
the vacuum valley should be parametrized by six complex moduli.

Denote two quintets present in  the  model as
$V_f^\alpha$ $(f=1,2)$, and two (anti)decu\-plets as 
$(X_{\bar g})_{\alpha\beta}$ where ${\bar g}=1,2$  and 
the matrices $X_{\bar g}$ are antisymmetric in color indices 
$\alpha ,\beta$. Indices $f$ and ${\bar g}$ reflect the 
SU(2)$_V \times$ SU(2)$_X$ flavor symmetry of the model.
Six independent chiral  invariants are
\begin{equation}
M_{\bar g}= V_k X_{\bar g} V_l \varepsilon^{kl}\,  ,
\qquad
B_{{\bar g} f}= X_{\bar g} X_{\bar k} X_{\bar l} V_f 
\varepsilon^{{\bar k}{\bar l}} \, ,
\label{MBinv}
\end{equation}
where the gauge indices in $M$ are convoluted in a 
straightforward manner $V^\alpha X_{\alpha \beta} V^\beta$, while 
in $B$ one uses the $\epsilon$ symbol, 
$$
X_{\bar g} X_{\bar k} X_{\bar l} V_f =
\epsilon^{\alpha\beta\gamma\delta\rho}
(X_{\bar g})_{\alpha\beta}(X_{\bar k})_{\gamma\delta}(X_{\bar
l})_{\rho\kappa}
(V_f)^\kappa\, .
$$
The choice of invariants above  implies
 that there are no moduli transforming as 
$\{4,\,2\}$ under the flavor group (such moduli vanish).

In this model the explicit parametrization of the valley is far from 
being obvious, to put it mildly. The most convenient strategy of the 
search is analyzing the five-by-five matrix
\begin{equation}
D^{\alpha}_{\beta} = V^{\alpha}_f {\bar V}^f_{\beta} +
({\bar X}^{\bar g})^{\alpha\gamma}(X_{\bar g})_{\gamma\beta}\,.
\label{55M}
\end{equation}
If this matrix 
is proportional to the unit one, the vanishing of the $D$ terms is 
guaranteed.  (Similar strategy based on analyzing analogs of 
Eq.\ (\ref{55M}) is applicable in other cases as well). 

A solution of $D$-flatness conditions~(\ref{55M}) containing three real
parameters  was suggested long ago in Ref.\ \cite{IVS}. Recently, 
four-parametric solution was found~\cite{Veldhuis1}. It has the form
\begin{equation}
V_1 = \left(
\begin{array}{c}
c \\ 0 \\ v^1_3 \\ 0\\ d
\end{array}\right) , \qquad
V_2 = \left(
\begin{array}{c}
0 \\ v^2_2 \\ 0 \\ v^2_4 \\ 0
\end{array}\right) ,
\label{MM6}
\end{equation}
$$
 X_1 =
\left(
\begin{array}{ccccc}
0 & 0 & x^1_{13} & 0 & 0 \\
0 & 0 & 0 & 0 & 0 \\
-x^1_{13} & 0 & 0 & 0 & 0\\
0 & 0 & 0 & 0 & x^1_{45} \\
0 & 0 & 0  & -x^1_{45} & 0 
\end{array}\right)\! , \;\;
X_2 =
\left(
\begin{array}{ccccc}
0 & a & 0 & 0 & 0 \\
-a & 0 & x^2_{23} & 0 & x^2_{25} \\
0 & -x^2_{23} & 0 & b & 0 \\
0 & 0 & -b & 0 & x^2_{45} \\
0 & -x^2_{25} & 0  & -x^2_{45} & 0 
\end{array}\right)\! ,
$$
where
\begin{eqnarray}
&& v_3^1  = -\frac{a}{\sqrt{a^2-c^2}} \,\sqrt{b^2-(a^2-c^2)}\,,\nonumber\\[0.2cm]
&& v_2^2  = \frac{c}{b} \sqrt{b^2-(a^2-c^2)} \,\sqrt{
\frac{b^2}{a^2-c^2}+\frac{d^2}{a^2}}\,,\quad v_4^2  = - \sqrt{a^2-c^2} \,
            \sqrt{ \frac{b^2}{a^2-c^2}+\frac{d^2}{a^2}}\,,\nonumber\\[0.2cm]
&& x_{13}^1  = \frac{c}{b}  
        \sqrt{a^2-c^2} \,\sqrt{\frac{b^2}{a^2-c^2}+\frac{d^2}{a^2}}\,,\quad
x_{45}^1  = \frac{a}{b} 
        \sqrt{a^2-c^2} \,\sqrt{
\frac{b^2}{a^2-c^2}+\frac{d^2}{a^2}}\,,\nonumber\\[0.2cm]
&& x_{23}^2 = \frac{c}{\sqrt{a^2-c^2}}\, \sqrt{b^2-(a^2-c^2)}\,,\quad
x_{25}^2 = -\frac{c d}{a}\,,\nonumber\\[0.2cm]
&& x_{45}^2 = \frac{d}{b a}\, \sqrt{a^2-c^2} \sqrt{b^2-(a^2-c^2)}\,.
\end{eqnarray}

The most general valley parametrization depends on 12 real 
parameters. Nevertheless, the four-parametric solution above is sufficient 
for the full reconstruction of the K\"ahler potential provided that  the flavor
symmetry is taken into account. Indeed, the flavor symmetry requires 
the K\"ahler
potential to depend on the following four  combinations:
\begin{equation}
I_1=\bar M^{\bar g} M_{\bar g}\,,\;\, I_2=\frac 1 2 \bar B^{\bar g f} B_{\bar g
f}\,,\;\, I_3=\bar M_{\bar g}\bar B^{\bar g f}B_{\bar k f}M^{\bar k}\,,\;\,
I_4=\frac 1 2 \bar B^{\bar k h} \bar B^{\bar g f}  B_{\bar g h} B_{\bar k f}\,.
\label{miful}
\end{equation}

Substituting the solution~(\ref{MM6}) in the original kinetic term
\begin{equation}
\bar V^f V_f +\frac 12 \bar X^{\bar g} X_{\bar g}\,,
\end{equation}
one finds the K\"ahler potential in terms of four parameters $a,b,c,d$, which can
then, in turn, be  expressed  in terms of four invariants $I_i$. In this way
one arrives at
\begin{equation}
{\cal K}(I_i)=\frac{3}{5}\sqrt{B} \left[\cos \left( \frac{1}{3} \arccos
\frac{A}{B^{3/2}}\right) +\frac{1}{4}\,\frac{1 }{\cos \left(
\frac{1}{3} \arccos
\frac{A}{B^{3/2}}\right)}\right]\,,
\label{kaposu5}
\end{equation}
where 
\begin{equation}
A  =  125 I_1\,,\qquad
B  = \frac{25}{9} 
\left[ \sqrt{ I_2 +  \sqrt{ I_4 - I_2^2}} +
       \sqrt{I_2 -\sqrt{ I_4 - I_2^2}} \right]\,.
\end{equation}
This K\"ahler potential was obtained in Ref.\ \cite{Veldhuis1}. The remarkable
feature of the result is the absence of the invariant $I_3$. The fact that ${\cal K}$
 does not depend on $I_3$ implies
some extra SU(2) flavor  symmetry of the moduli space, in addition to the obvious
SU(2)$_V
\times$ SU(2)$_X$. The extra flavor symmetry of ${\cal K}$
was not expected {\em a priori}. 

\subsection{The impact of the superpotential}
\label{sec33}

So far we discussed the structure of the classical vacua in the theories without
superpotential. We saw that in a large class of such theories there is a continuous 
manifold of physically inequivalent vacua -- moduli space. What is the physical
way to pick up a particular point in this space? 

To this end one can introduce a small perturbation in the form of a superpotential
${\cal W}$ which lifts the vacuum degeneracy. A few distinct scenarios of what
happens then exist. For a sufficiently general superpotential no continuous
degeneracy survives. The vacuum manifold shrinks to several isolated points
determined by the extremal points of ${\cal W}$. In some particular cases
it may happen that
no supersymmetric vacua  exist. Then, logically there are two possibilities:
there may exist a non-SUSY vacuum with a positive energy
density (supersymmetry  is spontaneously broken, see Sec.\ \ref{sec51}), 
or there
may be no vacuum at all at finite values of the  fields. The latter case 
is called the
{\em run-away vacuum}. Let us discuss  in more detail this  phenomenon, which is
an interesting animal in the zoo of the supersymmetric models.

Consider, e.g.,  the minimal Wess-Zumino model with the superpotential
${\cal W} = \mu^3\,\ln (\Phi /\mu)$. Certainly, this is a nonrenormalizable 
model. Never mind; assume that this  model is 
nothing but a low-energy effective description of some fundamental
theory which is renormalizable and well-defined.

The scalar potential emerging from the logarithmic superpotential
is
\begin{equation}
U = \frac{\mu^6}{|\phi|^2}\, .
\end{equation}
For any finite $\phi$ the vacuum energy density ${\cal E}$ is 
positive; the supersymmetric state ${\cal E}=0$ is achieved only 
asymptotically,  at $|\phi | \to\infty$.  The theory has no vacuum state in the
regular  sense  of the  word. To trace the fate of the sliding vacuum we 
can stabilize the theory by introducing extra terms 
in  the superpotential, for instance, $\Delta {\cal W}=m\Phi^2/2$ with an arbitrarily
small $m$. The scalar potential becomes
\begin{equation}
U = \left|\, \frac{\mu^3}{\phi}+m\phi\,\right|^2\, ,
\end{equation}
and the supersymmetric vacuum is located at $\phi^2=-\mu^3/m$. 
The limit $m\to 0$ makes the notion of the run-away vacuum explicit.
The  run-away vacuum could be of interest  in the cosmological context, but
we do not touch this subject here.

The most famous example of the gauge theory with the run-away 
vacuum is SQCD with $N_f = N_c-1$ in the limit of the strictly 
massless  matter (Sec.\ \ref{sec471}).  The stabilization can be readily achieved 
in  this  case  by adding a mass term to the matter fields. Other 
examples with a  similar behavior will be considered too.

\subsection{The impact of quantum effects}
\label{seciqe}
In the discussion above we did not touch  yet the quantum effects. They can be of
two types, perturbative and nonperturbative. Let us discuss them in turn.

The perturbative corrections do not renormalize the superpotential -- this is the
essence of the non-renormalization theorem for the $F$-terms~\cite{GRS}.
Therefore, the $F$-flatness conditions, $F=0$, remain intact. 
What the perturbative corrections affect  is the K\"ahler potential. In the Higgs
phase the parameter governing the amplitude of corrections is the gauge coupling
constant $\alpha(\phi)\propto (\log \phi/\Lambda)^{-1}$ where $\phi$ is the
scale of the moduli fields and $\Lambda$ is the scale parameter of the gauge
theory. For  large moduli, $\phi\gg\Lambda$, the coupling is weak and the
corrections to the  K\"ahler potential are calculable. If $\phi\sim\Lambda$,
however, the corrections explode, and the K\"ahler potential is not calculable.

The crucial role of the nonperturbative effects is that they can show up in the
superpotential. Thus, they can affect  the $F$-flatness conditions. In particular,
even if the tree-level superpotential vanishes, a superpotential can be generated
nonperturbatively, lifting the vacuum degeneracy. That is how the run-away
vacuum occurs in  SQCD with $N_f = N_c-1$. 

Similarly to the perturbative corrections, in the Higgs phase in the weak coupling
range the nonperturbative corrections to the superpotential are calculable. 
The tool allowing one to do the calculation is instantons.  Many interesting
phenomena occur  at this level, for instance, the dynamical SUSY breaking,
see Sec.\ \ref{sec54}. This happens because the nonperturbatively generated 
superpotentials have a different structure compared to the tree-level
superpotentials allowed by renormalizability.

Moreover, since we deal with the $F$-terms which are severely constrained by the
general SUSY properties, some results can be propagated from the weak to the
strong coupling regime. This is apparently the most interesting aspect of the
supersymmetric gauge dynamics.

\subsection{Anomalous and non-anomalous  U(1) symmetries}
\label{sec34}

An important role in the analysis of SUSY gauge theories belongs to 
global symmetries. We encountered  some examples
of the flavor symmetries in  the discussion of the
$D$-flat directions in the {\em 3-2} and SU(5) models. 
In this section we focus on U(1) symmetries.

One global  U(1) symmetry, usually called the $R$ symmetry,
is inherent to any supersymmetric theory because of its geometrical
nature. In the superspace this
$R$ transformation is expressed by the phase rotation of $\theta$,
\begin{equation}
R:\qquad \theta\to e^{i\alpha}\,\theta\,, \quad  \bar\theta\to
e^{-i\alpha}\,\bar\theta\,,
\quad x_\mu \to x_\mu\,.
\label{chiraltr}
\end{equation}
The commutator of this transformation with the SUSY
transformation~(\ref{susytr}) is
\begin{equation}
[Q_\alpha , \Pi]=Q_\alpha\,,\qquad [\bar Q_{\dot \alpha} , \Pi]=-\bar Q_{\dot
\alpha}\,,
\label{compiq}
\end{equation}
where $\Pi$ is the generator of the transformation~(\ref{chiraltr}).
These transformations generate the following transformations of the superfields:
\begin{equation}
\Phi(x,\theta,\bar\theta) \to e^{-ir\,\alpha}\,\Phi(x,
e^{i\alpha}\theta,e^{-i\alpha}\bar\theta) \,,
\label{chiralsup}
\end{equation}
where $r$ is the corresponding $R$ charge of the field $\Phi$.

To get acquainted more closely with this U(1) symmetry let us consider first
supersymmetric gluodynamics, the simplest non-Abelian 
gauge theory. The Lagrangian is obtained from Eq.\ (\ref{su2lagr})
by omitting the part with the matter fields,
\begin{equation}
{\cal L} =  \frac{1}{4g^2}  \int\!{\rm d}^2\theta \,\mbox{Tr}\, W^2 + \, 
\mbox{H.c.}
\, ,
\label{SFYML}
\end{equation}
where
$ W = W^aT^a$, and $T^a$ are the generators of $G$ in the 
fundamental representation.  The 
gauge group $G$  can be arbitrary. In components
\begin{equation}
{\cal L} =  \frac{1}{g^2} \left\{ -\frac 14 
G_{\mu\nu}^aG^{a\mu\nu} + 
i\lambda^{a \alpha} 
{\cal D}_{\alpha\dot\beta}\bar\lambda^{a\dot\beta}
\right\} \,  .
\label{SUSYML}
\end{equation}

With the massless gluino field,   the Lagrangian
(\ref{SUSYML}) is invariant under the chiral rotations
$\lambda\to \lambda e^{-i\alpha}$. This corresponds to the chiral transformation 
of superfields~(\ref{chiralsup}) with the $R$ charge of $V$  equal to zero and that
of $W$ equal to one.  The classically conserved $R$ current is~\cite{SSF}
\begin{equation}
R_\mu= \frac 12 \,(\sigma_\mu)_{\alpha\dot\alpha}  R^{\alpha\dot\alpha}=-
\frac{1}{g^2}\, \lambda^a \sigma_\mu \,\bar\lambda^a\,,\quad  \Pi=\int\!{\rm
d}^3x
\,R_0\, . 
\label{Rcur}
\end{equation}
From the commutation relations~(\ref{compiq}) it follows that this current
enters  the same supermultiplet as the supercurrent and the
energy-momentum tensor~\cite{FEZU}. In other words, the axial current
$R_\mu$ is the lowest component of the  superfield ${\cal
J}_{\alpha\dot\alpha}$
\begin{eqnarray}
{\cal J}_{\alpha\dot\alpha} &\!\!=&\!\! -\frac{4}{g^2}\,\mbox{Tr} \left[e^V
W_\alpha e^{-V}\bar W_{\dot \alpha}\right]\\
 &\!\!=&\!\!  R_{\alpha\dot\alpha} -
\left\{  i\theta^{\beta} J_{\beta\alpha\dot\alpha} 
 + 
\mbox{H.c.}    \right\} -
 2\, \theta^{\beta}\bar{\theta}^{\dot\beta}  \,
     \vartheta_{\alpha\dot\alpha\beta\dot\beta} +\frac 12\left\{
\theta_\alpha \bar\theta_{\dot \beta}\,i\,\partial^{\gamma\dot \beta}
R_{\gamma\dot \alpha} +\mbox{H.c.}\right\} +\dots\;,
\nonumber
\label{decom2}
\end{eqnarray}
where 
$J_{\beta\alpha\dot\alpha} $ is the 
supercurrent,  and ${\vartheta_{\alpha\dot\alpha\beta\dot\beta}}$ is 
the  energy-momentum tensor,
\begin{eqnarray}
J_{\beta\alpha\dot\alpha}&\!\!=&\!\!(\sigma_\mu)_{\alpha\dot\alpha}\,
J^\mu_\beta=\frac {4i}{g^2}\,\mbox{Tr}\, G_{\alpha\beta}\,\bar
\lambda_{\dot \alpha}\,,\\
\vartheta_{\alpha\dot\alpha\beta\dot\beta}&\!\!=&\!\!
(\sigma^\mu)_{\alpha\dot\alpha}(\sigma^\nu)_{\beta\dot\beta}\,
\vartheta_{\mu\nu}=\frac{2}{g^2}\,\mbox{Tr}\left[i\,\lambda_{\{\alpha}{\cal
D}_{\beta\}\dot\beta} \bar \lambda_{\dot \alpha} 
-i\,{\cal D}_{\beta\{\dot\beta}\lambda_\alpha \,\bar \lambda_{\dot \alpha\}}
+ G_{\alpha\beta}
\bar G_{\dot\alpha\dot\beta}\right]\,.
\nonumber
\end{eqnarray}
The symmetrization over $\alpha,\beta$ or $\dot \alpha,\dot\beta$
is marked by the braces. Note that all component expressions we presented 
above
refer to the Wess-Zumino gauge. 

The classical equation for ${\cal J}_{\alpha\dot\alpha}$
\begin{equation}
\bar{D}^{\dot\alpha}{\cal J}_{\alpha\dot\alpha} =0\,,
\end{equation}
besides the conservation of all three currents, contains also the relations
\begin{equation}
\vartheta_{\mu}^\mu=0\,,\qquad (\sigma_\mu)_{\alpha \dot\alpha}
J^{\mu\alpha}=J^\alpha_{\alpha\dot\alpha}=0\,,
\end{equation}
which express the classical conformal and superconformal symmetries.
At the quantum level these symmetries are broken.  The
conservation  of the $R_\mu$ current is also lost at the quantum level
-- this is the
celebrated chiral anomaly.  In particular,  the one-loop result for $\partial^\mu
R_\mu $ is
\begin{equation}
\partial^\mu R_\mu = \frac{T_G }{16\pi^2}\,  
G_{\mu\nu}^a\tilde{G}^{a \mu\nu}\, .
\label{3anom}
\end{equation}
%
The group factors $T_G$ (and $T(R)$  to be used below) are defined as 
follows. Let $T^a$ be the generator of the group $G$ in the 
representation $R$. Then Tr$\,(T^aT^b) = T(R)\delta^{ab}$. Moreover, 
$T(R)$ in the adjoint representation is denoted by $T_G$. According 
to the terminology used in the mathematical literature $T(R)$ is one 
half of the Dynkin index for the representation $R$; another name is the dual
Coxeter number. For SU($N$) one has $T_G= N$ and $T$(fund) =1/2.

The superfield generalization of Eq.\ (\ref{3anom}) is 
\begin{equation}
\bar{D}^{\dot\alpha}{\cal J}_{\alpha\dot\alpha} =
-\frac{T_G}{8\pi^2}\,D_\alpha {\rm Tr}\,W^2\, .
\label{sganom}
\end{equation}
Equation (\ref{3anom}) is  nothing but 
one  of the components of this superrelation. What are other components?
They present the anomalies in $(\sigma_\mu)^{\alpha \dot\alpha}
J^\mu_\alpha$ and $\vartheta_{\mu}^\mu$,
\begin{equation}
(\sigma_\mu)_{\alpha \dot\alpha}
J^{\mu\alpha}=J^\alpha_{\alpha \dot\alpha}=  i\,
\frac{T_G }{4\pi^2}\,
 \mbox{Tr}\left[ \bar G_{\dot\alpha\dot \beta}\bar \lambda^{\dot
\beta}\right]\,,\qquad
\vartheta^\mu_{\mu}=
\frac{T_G }{16\pi^2}\,\mbox{Tr} \,\left[G_{\rho\sigma}G^{\rho\sigma}\right]
\,.
\label{confan}
\end{equation}
Besides, $\vartheta_{\mu\nu}$ and $ J^\mu_\alpha$
cease to be conserved. Indeed,  acting by $D^\alpha$ on Eq.\ (\ref{sganom}) and
combining it with complex conjugate one arrives at
\begin{equation}
\partial^{\mu}{\cal J}_{\mu} =i\,
\frac{T_G}{32\pi^2}\,{\rm Tr}\left[ D^2 \,W^2- \bar D^2 \,\bar W^2\right]\, .
\label{diverg}
\end{equation}
In perfect parallel with the axial current the right-hand side is in fact  a full
derivative,
\begin{equation}
\partial^\mu R_\mu = \frac{T_G }{16\pi^2}\, \partial^\mu K_\mu\,,\qquad 
\partial^{\mu}{\cal J}_{\mu}  =
\frac{T_G}{16\pi^2}\,\partial^{\mu}{\cal K}_{\mu}\,,
\label{Kanom}
\end{equation}
where the superfield ${\cal K}_{\mu}$ generalizes the Chern-Simons current
$K_\mu$,
\begin{equation}
{\cal K}_{\mu}=4\,\epsilon^{\mu\nu\rho\sigma}\,{\rm Tr}\left\{
{\cal A}_\nu \partial_\rho {\cal A}_\sigma -\frac{i}{3} {\cal A}_\nu [{\cal A}_\rho
{\cal A}_\sigma ]\right\}\,.
\end{equation}
Here we introduced the superfield ${\cal A}_\mu$ which generalizes the standard
vector potential $A_\mu$,
\begin{equation}
{\cal A}_\mu= - \frac {1}{8}\,(\sigma_\mu)^{\alpha\dot\alpha}\, [D_\alpha , \bar
D_{\dot \alpha}] V\,.
\end{equation}
We can return to the conserved currents by defining
\begin{equation}
\tilde R_\mu=R_\mu -\frac{T_G}{16\pi^2}\,{ K}_{\mu}\,,\qquad
\tilde{\cal J}_\mu={\cal J}_\mu -\frac{T_G}{16\pi^2}\,{\cal K}_{\mu}\,.
\label{tildecur}
\end{equation}
The price we pay is that the corrected current $\tilde{\cal J}_\mu$ is not a
supergauge invariant object. Nevertheless, in the Wess-Zumino gauge it is only the
lowest component of
${\cal K}_{\mu}$ (equal to $K_\mu$) which is explicitly gauge non-invariant. 

Following a standard route one defines the conserved supercurrent $\tilde
J_{\beta\alpha\dot \alpha}$ and the  energy-momentum tensor
$\tilde\vartheta_{\mu\nu}$ as the $\theta$ and $\theta\bar \theta$ components
of $\tilde{\cal J}_\mu$,
\begin{eqnarray}
\tilde J^\beta_{\alpha \dot\alpha}&\!\!=&\!\! J^{\beta}_{\alpha \dot\alpha}-i\,
\frac{T_G }{2\pi^2}\,\delta^\beta_\alpha\,
 \mbox{Tr}\left[ \bar G_{\dot\alpha\dot \beta}\bar \lambda^{\dot
\beta}\right]\,,\nonumber\\[0.2cm]
\tilde\vartheta_{\mu\nu}&\!\!=&\!\! \vartheta_{\mu\nu}-
\frac{T_G }{16\pi^2}\,g_{\mu\nu}\,
\mbox{Tr} \,\left[G_{\rho\sigma}G^{\rho\sigma}\right]
\,.
\end{eqnarray}
These currents are conserved,  but the conformal and superconformal
anomalies are still there. 
\begin{equation}
\tilde J^\alpha_{\alpha \dot\alpha}= -3 i\,
\frac{T_G }{4\pi^2}\,
 \mbox{Tr}\left[ \bar G_{\dot\alpha\dot \beta}\bar \lambda^{\dot
\beta}\right]\,,\qquad
\tilde\vartheta^\mu_{\mu}=-3\,
\frac{T_G }{16\pi^2}\,\mbox{Tr} \,\left[G_{\rho\sigma}G^{\rho\sigma}\right]
\,.
\label{confan1}
\end{equation}
Compared with Eq.\ (\ref{confan}) we got an extra coefficient $-3$. We stress
once more 
that Eq.\ (\ref{confan1}) gives the anomalies in the {\em conserved} supercharge
and energy-momentum tensor.

The supermultiplet structure of the  anomalies in $\partial^\mu
R_\mu$, the trace of the  energy-momentum tensor $\tilde\theta^\mu_\mu$
and in 
$\tilde J^\alpha_{\alpha\dot\alpha} $ (the three 
``geometric" anomalies) was revealed in \cite{Grisa}.

Inclusion of the matter fields in the model-building typically results 
in  additional global symmetries, and, in particular, in additional U(1) symmetries.
Some of them act  exclusively in the matter sector. These are
usually quite evident and are immediately detectable. Somewhat less obvious are
 U(1) symmetries which act on both, the matter and gluino fields.  They
 play a distinguished role in the  analysis of possible instanton-induced
effects.  Here we intend to present a classification of the anomalous and
non-anomalous U(1) symmetries.

The general Lagrangian of the gauge theory with matter is given in Eq.\
(\ref{Glagr}) in the absence of the superpotential. The matter field $Q$ consists of
some number of irreducible representations of the gauge group. Every irreducible
representation will be referred to as ``flavor'', $\{Q\}=\{Q_1, \dots Q_{N_f}\}$.
It is clear that additionally to U(1) discussed above one can make the phase
rotations of each of $N_f$ matter fields independently.  Thus, altogether we have 
$N_f+1$ chiral rotations. Adding a general superpotential ${\cal W}$ eliminates, in
principle, all these U(1) symmetries. However, if the classical conformal symmetry
is unbroken, at least one U(1) survives \cite{FEZU}. The conformal invariance
implies the absence of dimensional parameters; in other words, it limits the
superpotential to be cubic in
$Q$.  Then the action is invariant under the following  transformation:
\begin{equation}
V(x,\theta,\bar\theta)\to V(x,e^{i\alpha}\theta,e^{-i\alpha}\bar\theta)\,,\quad
Q(x_L,\theta)\to e^{-2i\alpha/3}\, Q(x_L,e^{i\alpha}\theta)\,.
\end{equation}
The cubic form of the superpotential fixes the $R$ charge of the matter
to be $2/3$. In components the same transformations look as 
\begin{equation}
A_\mu\to A_\mu\,,\,\,\,\lambda_\alpha \to e^{-i\alpha}\lambda_\alpha\, , \,\,\,
\psi_\alpha^f \to e^{i\alpha/3}\psi_\alpha^f \, , \,\,\,
\phi^f \to e^{-2i\alpha/3}\phi^f \, .
\label{rnot}
\end{equation}
The corresponding chiral current, which can be viewed as a generalization of
the current~(\ref{Rcur}), is denoted $R^0_\mu$ and has the form
\begin{equation}
R_\mu^0 = -\frac{1}{g^2}\, 
\lambda^a\sigma_\mu\bar\lambda^a
+\frac{1}{3}\sum_f \left( \psi_f\sigma_\mu\bar\psi_f
-2i\, \phi_f \stackrel{\leftrightarrow}D_\mu \bar\phi_f\right)\, .
\label{rnotcur}
\end{equation}
This current is the lowest component of the superfield ${\cal
J}_{\alpha\dot\alpha}^0$,
\begin{eqnarray}
{\cal J}_{\alpha\dot\alpha}^0&\!\!=&\!\!
\frac{4}{g^2}\,\mbox{Tr} \left[\bar W_{\dot \alpha}\, e^V\, 
W_\alpha\,  e^{-V}\, \right] \nonumber\\[0.2cm]
&\!\!-&\!\! \frac{1}{3}\sum_f \bar Q_f \left(
\stackrel{\leftarrow}{\bar \nabla}_{\dot \alpha}\,e^{V} \nabla_{ \alpha} 
-
e^{V} {\bar D}_{\dot \alpha}\,\nabla_{ \alpha} +
 \!\stackrel{\leftarrow}{\bar \nabla}_{\dot
\alpha}\stackrel{\leftarrow}D_{ \alpha} e^{V}
\right)Q_f \,,
\end{eqnarray}
where the background-covariant derivatives are introduced,
\begin{equation}
\nabla_{ \alpha} Q=e^{-V}D_{ \alpha}\left( e^{V} Q\right)\,,\quad
\bar\nabla_{\dot \alpha} \bar Q=e^{V}\bar D_{\dot \alpha}\left( e^{-V} \bar
Q\right)
\,.
\end{equation}
The superfield current ${\cal
J}_{\alpha\dot\alpha}^0$ plays the same geometrical role as ${\cal
J}_{\alpha\dot\alpha}$ in  SUSY gluodynamics. In particular, the
$\theta\bar\theta$ component contains the total energy-momentum tensor of the
theory.

The remaining $N_f$ currents are due to the phase rotations of each flavor
superfield independently,
\begin{equation}
Q_f(x_L,\theta)\to e^{-i\alpha_f}\, Q_f(x_L,\theta)\,.
\end{equation}
Note that $\theta$ is not touched by these transformations -- we deal with
the 
genuinely flavor symmetry. The corresponding chiral currents are
\begin{equation}
R_\mu^f = 
- \psi_f\sigma_\mu\bar\psi_f
- \phi_f \stackrel{\leftrightarrow}D_\mu \bar\phi_f\, .
\label{rfcur}
\end{equation}
In the superfield language,  $R_\mu^f$ is the $\theta\bar\theta$ component of
the so called Konishi currents~\cite{Konishi}
\begin{equation}
{\cal J}^f=\bar{Q}_f e^{V} Q_f\,,
\end{equation}
To make the situation similar to ${\cal J}_{\alpha\dot\alpha}$ we can form
another superfield ${\cal J}_{\alpha\dot\alpha}^f$,
\begin{equation}
{\cal J}_{\alpha\dot\alpha}^f=-\frac 12\,[D_\alpha , \bar D_{\dot\alpha}] \,{\cal
J}^f=-\frac 12\, [D_\alpha , \bar D_{\dot\alpha}]\,\bar{Q}_f\, e^{V} Q_f\,,
\label{jalal}
\end{equation}
of which  $R_\mu^f$  is the lowest component. There is a deep difference between
the flavor current ${\cal J}_{\alpha\dot\alpha}^f$ and the geometric current
${\cal J}_{\alpha\dot\alpha}^0$: 
the latter contains the supercurrent and the energy-momentum tensor in the higher
components while higher components of the former are conserved trivially.

With all these definitions in hands we are ready to discuss the non-conservation of
the currents both due to the classical superpotential and the quantum anomalies.
For the geometric current ${\cal J}_{\alpha\dot\alpha}^0$  one has
\begin{eqnarray}
\bar{D}^{\dot\alpha}J_{\alpha\dot\alpha}^0 &\!\!= &\!\!\frac{2}{3} 
D_{\alpha}\left\{
\left[ 3{\cal W } - \sum_f Q_f\, \frac{\partial{\cal W }}{\partial Q_f}
\right] \right.
 \nonumber\\[0.2cm]
&\!\!- &\!\!
\left.
\left[ \frac{3T_G- \sum_f T(R_f)}{16\pi^2}\,{\rm Tr}\,W^2 + 
\frac{1}{8}\sum_f\gamma_f 
\bar{D}^2 (\bar{Q}_{\!f}\, e^{V} Q_f) \right]
\right\}\, ,
\label{geom}
\end{eqnarray}
where $\gamma_f$ are the anomalous dimensions$\,$\footnote{For 
the definition
of $\gamma_f$'s see Eq.\ (\ref{defgamma}).}  of the matter 
fields $Q_f$. 
The first line is purely classical. It is seen that the classical part vanishes for the
cubic in $Q$ superpotential, as it was discussed above. 
The second line is the quantum anomaly.  Being understood in the operator form, 
this anomaly is exact~\cite{anomaly}. Higher loops enter through the anomalous
dimensions $\gamma_f$. Let us memorize this relation --
it will play an important role in what follows.

The  anomaly  in the Konishi currents ${\cal J}^f$ is expressed by the  formula
\cite{Konishi}
\begin{equation}
\bar{D}^2 {\cal J}^f=  \bar{D}^2\, (\bar{Q}_f e^{V} Q_f) = 
4 \,Q_f \frac{\partial{\cal W }}{\partial Q_f} +
\frac{T(R_f)}{2\pi^2}\,{\rm Tr}\, W^2\, .
\label{ka1}
\end{equation} 
The first term on the right-hand side is classical, the second term is the anomaly.
Note that in this operator relation there are no higher-order
corrections, in contrast with the situation with
the geometric anomalies, Eq.\ (\ref{geom}).

By combining the Konishi currents with the $R^0_\mu$ current, with 
the appropriate coefficients, one establishes all conserved anomaly-free
$R$ currents of the
theory under consideration (provided that they exist, of course). To this end it is
convenient to write the anomaly relations in the form of divergences of  
${\cal J}^0_{\alpha\dot\alpha}$ and  ${\cal J}^f_{\alpha\dot\alpha}$,
\begin{eqnarray}
\partial^{\alpha\dot\alpha}{\cal J}_{\alpha\dot\alpha}^0 &\!\!= &\!\!-\frac{i}{3} 
D^2\left\{
\left[ 3{\cal W } - \sum_f \left(1+\frac{\gamma_f}{2}\right)Q_f\,
\frac{\partial{\cal W }}{\partial Q_f}
\right] \right.
 \nonumber\\[0.2cm]
&\!\!- &\!\!
\left.
\frac{1}{16\pi^2}\left[ 3T_G- \sum_f \left(1-\gamma_f\right)T(R_f)\right]{\rm
Tr}\,W^2  
\right\}+{\rm H.c.}\, ,\label{geom1}
\label{konish11}
\end{eqnarray}
and 
\begin{equation}
\partial^{\alpha\dot\alpha}{\cal J}_{\alpha\dot\alpha}^f= 
i\,D^2\left\{\frac 12
\,Q_f \frac{\partial{\cal W }}{\partial Q_f} +
\frac{T(R_f)}{16\pi^2}\,{\rm Tr}\, W^2\right\} +{\rm H.c.}
\, ,\label{konish1}
\end{equation}
where we used Eqs.\ (\ref{jalal}), (\ref{geom}) and (\ref{ka1}) plus the algebraic
relation
\begin{equation}
\partial^{\alpha\dot\alpha}[D_\alpha ,\bar D_{\dot \alpha}]=-\frac{i}{4}\left(D^2
\bar D^2 -\bar D^2 D^2 \right)\, .
\end{equation}

From the equations above it is clear that there exist $N_f$ linear combinations
of $N_f +1$ chiral currents which are free from the gauge anomaly. The 
choice is not unique, of course. In particular, we can choose one of such currents in
the form 
\begin{equation}
{\cal J}_{\alpha\dot\alpha}^{0f}={\cal J}_{\alpha\dot\alpha}^0 -\frac{3T_G -\sum_f
\left(1-\gamma_f\right)T(R_f)}{3\sum_f T(R_f)}\,\sum_f {\cal
J}_{\alpha\dot\alpha}^f
\,,
\label{0fcur}
\end{equation}
The coefficient in front of the second term is proportional to the $\beta$ function,
see Sec.\ \ref{sec46}. In the extreme ultraviolet, where $\alpha\to 0$, the
anomalous dimensions $\gamma_f(\alpha)$ vanish, and we are left with the
one-loop expression for the current ${\cal J}_{\alpha\dot\alpha}^{0f}$.  The
formula~(\ref{0fcur}) is valid, however, for any $\alpha$. Thus, it smoothly
interpolates to the strong coupling range~\cite{KSV}. In particular, if the theory is
conformal in the infrared the second term vanishes and the current ${\cal
J}_{\alpha\dot\alpha}^{0f}$ coincides with the geometrical current  ${\cal
J}_{\alpha\dot\alpha}^{0}$.

The remaining
$N_f -1$ currents can be chosen as
\begin{equation}
{\cal J}_{\alpha\dot\alpha}^{fg}= T(R_g)\,{\cal J}_{\alpha\dot\alpha}^f
-T(R_f)\,{\cal J}_{\alpha\dot\alpha}^g
\,, 
\label{fgcur}
\end{equation}
where one can fix $g$ and consider all $f\neq g$.
For nonvanishing superpotential the classical part in the divergence of currents
(\ref{0fcur}) and  (\ref{fgcur}) can be simply read off from Eqs.\ (\ref{geom1}) and
(\ref{konish1}). Generically, no conserved current survive. For specific
superpotentials 
it happens that 
one can  find  conserved combinations of currents. It is certainly the case
if the theory with the given superpotential has flat directions. The surviving
moduli can be classified according to the conserved $R$ charges. 
The examples will
be given below.

\subsection{Effective Lagrangian and the anomalous U(1)}
\label{secEL}

Let us return for a while  to  SUSY gluodynamics, in this theory a single  chiral
current  (\ref{Rcur}) exists. As well-known, the anomaly in this current, see
Eq.~(\ref{3anom}), does {\em not} lead to  the breaking of U(1) in perturbation
theory. Indeed, one can build a conserved  (but gauge non-invariant) current
$\tilde R_\mu$ given in  Eq.\ (\ref{tildecur}). The corresponding charge is gauge
invariant in perturbation theory. 

 At the nonperturbative level,  the U(1) symmetry
is lost~\cite{Hooft1}. The only remnant of the continuous chiral symmetry that
survives 
\cite{Witten2} is a discrete subgroup $Z_{2T_G}$,
\begin{equation}
\lambda\to e^{-i\pi k/T_G}\,\lambda\, , \quad  k = 1,2, ..., 2T_G\, .
\label{large}
\end{equation}
The fact that SUSY gluodynamics is invariant under the discrete
$Z_{2T_G}$ can be verified, for instance, by analyzing the instanton-induced
't~Hooft interaction -- the number of the gluino zero modes on the instanton
is $2T_G$, as will be discussed in more detail in Sec.\ \ref{sec4}.

One can visualize all  anomalies, as well as
the discrete invariance $Z_{T_G}$ via the
Veneziano-Yankielowicz effective Lagrangian~\cite{veneziano}, 
\begin{equation}
{\cal L}_{VY}=\frac{T_G}{3} \int \! {\rm d}^2\theta\,  \Phi  \ln \left(\frac{\Phi
}{\sigma}\right)^{T_G}  + \mbox{H.c.} +\mbox{invariant terms}
\,  ,
\label{VYL}
\end{equation}
where $\Phi$ is a composite color-singlet chiral superfield, 
$$
\Phi = \frac{3}{32\pi^2 T_G }\,\mbox{Tr}\,W^2\, ,
$$
and $\sigma$ is expressed via the  scale parameter of the theory
$\Lambda$ as
\begin{equation}
\sigma = {\rm e}\Lambda^3 \, ,\quad {\rm e}=2.718\dots\,.
\end{equation}
The superpotential term in ${\cal L}_{VY}$ is the only one non-invariant under
the 
geometrical transformations discussed in the previous section. The omitted terms,
including the kinetic one, must be invariant. To see how
the superpotential term generates the 
anomalies
let us consider its variation under the chiral transformation~(\ref{chiralsup})
\begin{equation}
\Phi(x_L,\theta) \to e^{-2i\alpha}\Phi (x_L, e^{i\alpha}\theta)\, . 
\label{phitran}
\end{equation}
Then
\begin{equation}
\delta {\cal L}_{VY} = -i\delta\alpha\, \frac{2T_G^2}{3}
 \int {\rm d}^2\theta\, \Phi +{\rm H.c.}\,,
\end{equation}
in full accord with Eq.\ (\ref{3anom}). One can easily check that all other
geometrical anomalies are reproduced as well.

Note that the logarithmic term in the  Lagrangian~(\ref{VYL}) is not fully 
defined since the logarithm is a multivalued function. The differences between the
branches is not important for the generation of
the  anomalies -- this difference resides
in the invariant terms of ${\cal L}_{VY}$. This difference is important, however, 
for the ``large'' chiral transformations~(\ref{large}) forming the $Z_{2T_G}$
subgroup.

The proper definition was suggested in Ref.\ \cite{kovner2}. For the $n$-th 
branch one must define a corresponding Lagrangian,
\begin{equation}
{\cal L}_n=\frac{T_G}{3} \int \! {\rm d}^2\theta\,  \Phi  \left[\ln \left(\frac{\Phi
}{\sigma}\right)^{T_G} + 2i\pi n\right] + \mbox{H.c.} 
\label{Ln}
\,  ,
\end{equation}
where a specific branch is ascribed to  the logarithm. In terms of the original
theory the parameter $n$ shifts the vacuum angle $\theta \to \theta+2\pi n$.
The discrete transformations~(\ref{large}) act on $\Phi$ as  in Eq.\
(\ref{phitran}), with the substitution $\alpha\to \pi k/T_G$. Thus, these
transformations convert ${\cal L}_n\to {\cal L}_{n-k}$. This implies the
$Z_{T_G}$ invariance of the theory provided that the partition function sums over
all $n$,
\begin{equation}
{\cal Z}=\sum_{n=-\infty}^\infty \int\!{\cal D}\,\Phi \,e^{\,i\!\int\! {\rm d}^4 x
\,{\cal L}_n}\,.
\end{equation}
The invariance group is $Z_{T_G}=Z_{T_G}/Z_2$ because $\Phi$ is quadratic in
$\lambda$, thus  identifying $\lambda$ and $-\lambda$. 

The construction of Ref.\ \cite{kovner2} determines the number of the vacuum
states to be
$T_G$. The zero energy states are obtained from the stationary of points of the
superpotential. They lie at
\begin{equation}
\langle \Phi\rangle_k = \Lambda^3 e^{2\pi i k/ T_G}\,,\qquad k=0,\dots, T_G -1\,.
\label{condF}
\end{equation}
Summarizing, we see that the theory consists of $T_G$ sectors exhibiting the
spontaneous breaking of $Z_{2T_G}\to Z_2$. 

One should keep in mind that the Lagrangian (\ref{VYL})
is not Wilsonean, it cannot be used for obtaining  the scattering amplitudes and
other information of a similar   nature. Its only {\em raison d'etre} is the explicit
realization of the anomalous and non-anomalous symmetries of SUSY 
gluodynamics and the vacuum structure compatible with 
these symmetries.  

The point $\Phi=0$ requires a special considerations. If one takes the
Lagrangians (\ref{Ln}) literally, there exists~\cite{kovner2} a chirally symmetric
vacuum at $\Phi=0$. It was  overlooked  in the original analysis~\cite{veneziano}.
If this is indeed the case, the chirally symmetric vacuum drastically affects various
mechanisms of the dynamical SUSY breaking. We will dwell on  this issue in 
Sec.\ \ref{sec7}.

\subsection{Witten's index: where to look for the dynamical  supersymmetry 
breaking?}
\label{sec36}

The answer to this question  experienced a dramatic evolution over the last 
decade. In the 1980's people believed that only the chiral gauge theories
are suitable candidates for the dynamical SUSY breaking. 
Correspondingly, the searches were limited to this rather narrow 
class of models.  At present it became clear that, under certain conditions,
some nonchiral gauge theories can do the job too. 

First, we will introduce the notion of {\em Witten's index}, one of the most 
important theoretical tools in this range of questions, and 
explain why initially theorists'  attention was attracted exclusively to the 
chiral  gauge theories.  We also discuss  a deeper theoretical 
understanding achieved in  the 1990's. Novel elements, introduced into 
circulation recently, allow one to construct mechanisms
of the dynamical SUSY breaking,  based on the nonchiral gauge 
theories, although the chiral ones still remain the most  important 
supplier of such mechanisms. 

Witten's index  is defined as  
\begin{equation}
I_W = n^b_{E=0} - n^f_{E=0} 
\end{equation}
where $n^b_{E=0}$  and $n^f_{E=0} $ are the numbers of the bosonic and
fermionic zero-energy states, respectively.

As was emphasized in~\cite{Witten2}, $I_W$ is an invariant that does not 
change under any continuous deformations of the theory --
such as the particle masses, the volume in which the theory is 
defined, the values of the coupling constants and so on. Under 
such deformations the levels of the system breathe,
they can come to and leave zero, but as long as the Hamiltonian is 
supersymmetric, once, say,  a bosonic state comes to  zero, it must be 
accompanied by its fermionic counterpartner, so that $I_W$ does not 
change. If $I_W\neq 0$ the theory does have at least $I_W$ 
zero-energy states.  The existence of the zero-energy 
vacuum state is the necessary and sufficient condition for 
supersymmetry to be realized linearly, i.e. stay unbroken. Thus, only 
the $I_W = 0$ theories could produce  the dynamical SUSY 
breaking. 

Witten's index can be calculated for the gauge theories based on arbitrary Lie
group  (for  the time being let us forget about matter). Its value is given by
$T_G$. The values of $T_G$ for the semi-simple Lie groups are collected in
Table \ref{tabdynkin}. In  the  theories where the gauge group is a product of
semi-simple groups, $G=G_1\times G_2\times\dots$, Witten's index $I_W =
T_{G_1}\times T_{G_2}\times\dots$\,.

\begin{table}
\begin{center}
\begin{tabular}{|c|c|c|c|c|c|c|c|c|}
\hline
~ & ~  & ~  &~ & ~ & ~ & ~ & ~ &~\\[-0.1cm]
 $ \mbox{Group}$  &  SU$(N)$  &  SO$(\!N) $  &Sp$\,(\!2N)$  &~$G_2$~&~$F_4$~
&~$E_6$~&~$E_7$~&~$E_8$~\\[0.2cm]\hline
\vspace*{-0.2cm}
~ & ~  & ~  &~ & ~ & ~ & ~ & ~&~ \\
$T_G$ & $ N$  & $ N-2$  &$N+1$ & 4 & 9&12&18 &30\\[0.2cm] 
\hline
\end{tabular}
\caption{The dual Coxeter number (one half of the Dynkin index) for various 
groups.}\label{tabdynkin}
\end{center}
\end{table}

Two alternative calculations of $I_W$ are known in the literature. 
The first was Witten's original calculation who deformed the theory by putting it in
a finite three-dimensional volume $V=L^3$. The size $L$ is such that the coupling 
$\alpha(L)$ is weak, $\alpha(L)\ll 1$.  The field-theoretical problem of counting
the number of the  zero-energy states becomes, in the limit $L\to 0$, a 
quantum-mechanical problem of counting  the gluon and gluino  zero modes. 
In practice, the problem is still quite  tricky because of subtleties associated with
quantum mechanics on  the group spaces.  

The story has a dramatic
development. The result obtained in the first paper  in~\cite{Witten2} was $I_W=
r+1$ where $r$ stands for the rank of  the group. For the unitary and simplectic 
groups $r+1$  coincides with $T_G$. However, for the orthogonal (starting from 
SO(7)) and exceptional groups $r+1$  is smaller than  $T_G$. 
The overlooked zero-energy states  in the SO($N$) quantum mechanics of the
zero modes were found by the same author  only 15 years later! (See the second
paper in \cite{Witten2}).
Further  useful comments can be found in \cite{SMILGA} where 
additional states in  the exceptional groups were exhibited.

An alternative calculation of $I_W$ \cite{SVMO} resorts to another 
deformation which, in a sense, is an opposite extreme.
A set of auxiliary matter fields with small mass terms is introduced  in such a
way that the theory becomes completely Higgsed and weakly coupled. Moreover,
for a certain ratio of the mass parameters the pattern of the gauge symmetry
breaking  is  step-wise, e.g. 
$$
\mbox{SU($N)\to $ SU($N-1) \to \dots\to $SU(2)$\to$ nothing}, 
$$
In the weakly coupled  theory
everything is calculable. In particular, one can  find the vacuum states and count
them. This was done in~\cite{SVMO}. It turns out that the gluino condensate is a
convenient marker of the  vacua --  it takes distinct  values in the various
vacua$\,$\footnote{Actually, using the gluino condensate as an order parameter
was suggested by Witten~\cite{Witten2}, he realized the mismatch for the
orthogonal groups.}. 
The gluino condensate 
$\langle\lambda\lambda\rangle$ was
{\em exactly} calculated in \cite{SVMO}; 
the result 
is  multiple-valued, 
\begin{equation}
\langle\lambda\lambda\rangle \propto e^{ 2\pi i k/T_G}\,, \qquad k=0,1, ... , T_G-1
\,,
\end{equation}
cf. Eq.\ (\ref{condF}). All vacuum states are, of course, bosonic in this 
theory, implying that $I_W =T_G$.   In the limit when all mass parameters
tend to infinity the auxiliary matter fields decouple and we return to the
strongly coupled SUSY gluodynamics. Since $I_W$ is invariant under 
this deformation, the result $I_W =T_G$ stays valid irrespective of how large or
small the mass parameters are. 

The crucial element of the index analysis is the assumption that no
vacuum state runs away to infinity in the space of fields  in the process of
deformation. For instance, in Witten's analysis~\cite{Witten2} it was tacitly
assumed that  at $L\to
\infty$ no  fields develop infinitely large expectation values. The analysis based
on Higgsing  of the theory by
virtue of the auxiliary matter~~\cite{SVMO} confirms this assumption. However, as
it was found recently, 
 when physical (rather than the auxiliary)  matter fields are introduced, it is
not always true that  in the process of
deformation no 
fields develop infinitely large expectation values.

If all matter fields are massive, nothing changes in the calculation of Witten's
index, it remains the same.  Thus, what remains to study is the massless matter
sector.  (Remember: technically, such models are divided in two classes: chiral and
nonchiral. The matter  which allows one to introduce a mass term for every
matter field is called  nonchiral, otherwise the model is called 
chiral.) 

A natural deformation in the nonchiral models is the introduction of the mass
term~\cite{Witten2}. Then the index is the same as in SUSY gluodynamics,
$I_W=T_G\neq 0$. For this reason,
 for a long time it was believed that the
nonchiral models do not provide for the opportunity of the dynamical SUSY
breaking. Recently it was demonstrated~\cite{IT}, however, that in some particular
models the limit $m\to 0$ leads to the situation where some vacua run away to
infinity in the space of fields (Sec.\ \ref{sec33}).  This means that the index
determined by the
$m\neq 0$ deformation is unphysical. The physically relevant index must count
only those zero energy states which lie in a finite domain in the space of fields.
The index defined in such a way could be zero even in the nonchiral
models~\cite{IT}. Thus, with this remark in mind, there is no conceptual
distinction between the nonchiral and chiral models. In addition to the formal
Witten's index one should check whether there are run-away vacua in the limit of
vanishing deformation. For instance, it is conceivable that in some chiral theory
with the formal index zero the {\em physical index}  (i.e. $n^b_{E=0} - n^f_{E=0}$
calculated for the field configurations  belonging to  a
finite domain in the space of fields) could be nonzero, provided that some
fermionic states run away.

Historically, the first examples of the dynamical SUSY breaking were found 
in the chiral theories. The class of such models is strongly constrained. 
Building the chiral matter sector, one has to proceed with caution,
in order to avoid {\em internal} chiral triangle anomalies  (in the theories with
the vector-like couplings, i.e. the nonchiral theories, the  anomalous
triangles do not appear, automatically). Since matter fermions have 
both the vector
and axial-vector couplings, the set of the fermion fields  must be arranged in such
a way that all chiral triangles coupled to the  gauge  bosons of the theory  cancel in 
the total  sum.  Otherwise,  the gauge invariance is lost, and the theory becomes 
ill-defined, intractable. The best-known  example is the Standard Model. If one
considers, say the $Z Z\gamma$ effective vertex, arising due to fermionic loops, the
$u$-quark contribution is anomalous. The anomaly is  canceled after summation
over all fermions belonging to the first  generation.

The easiest  way to build a chiral theory that will be automatically 
anomaly-free is to start from a larger anomaly-free  theory, and to 
pretend  that  the gauge  symmetry of the original model is somehow 
spontaneously broken  down to a  smaller group. The gauge bosons 
corresponding to the broken   generators are  frozen out. The matter 
fields that are singlets with  respect to the  unbroken 
subgroup can be discarded. The remaining matter sector may well  
be chiral, but there will be no internal anomalies. For instance, to get 
the  SU(5)  theory, we may start from SO(10), where all representations 
are (quasi)real, so  this  theory is  automatically anomaly-free. 
Assume,  we introduce  the  matter  in  the  representation {\bf 16} of SO(10). 
Now, we break  SO(10) down to SU(5). The representation  {\bf 16}
can be decomposed with respect to  SU(5) as a singlet, a quintet plus 
an  (anti)decuplet. Drop the singlet. We are left with the SU(5) model
with one quintet and one  (anti)decuplet. We can further break SU(5)  
down to SU(3)$\times$SU(2), a cascade which eventually leads us to 
the {\em 3-2} model of Affleck, Dine and Seiberg.

In some of the chiral models obtained in this way one finds a 
self-consistent weak coupling regime --  under a certain choice of 
parameters  the original gauge group is completely Higgsed (Sec.\ \ref{sec54}).
Others are intrinsically strongly coupled (Sec.\ \ref{sec61}).  The latter case could
be connected to the former one by introducing auxiliary nonchiral matter. This
program (similar to that of Ref.\ \cite{SVMO}) was carried out in several  models,
 see Secs.\ \ref{sec56} and \ref{sec61}. 

The discovery of the nonchiral models of the dynamical SUSY breaking  is
extremely important since  it expands the class of the SUSY-breaking theories. 
The initial findings~\cite{IT} were further  generalized in Ref.\ \cite{MDDGR}.
The development was welcome by  builders of the SUSY-breaking mechanisms. 

Let us comment on another recent development. In Ref.\ \cite{kovner2}
arguments were given in favor of the existence  of the chirally symmetric phase
in SUSY gluodynamics. If such phase does exist, it is  not  ruled out that  in some
chiral gauge models  that are known to have  no supersymmetric  vacuum  in the
weak coupling regime, a SUSY preserving  vacuum may exist in the domain of
strong coupling.  For more details  see  Secs.\ \ref{sec61} and \ref{sec7}.  

\section{$\!\!\!\!\!\!$ Fermion-Boson Degeneracy: $\!$ Magic Backgrounds}
\label{sec4}

\renewcommand{\theequation}{3.\arabic{equation}}
\setcounter{equation}{0}

If supersymmetry is unbroken, i.e. 
\begin{equation}
Q_\alpha|0\rangle=0\,, \quad \bar Q_{\dot\alpha}|0\rangle=0\,,
\end{equation}
then, from the supersymmetry algebra~(\ref{susyalgebra}), it follows that the
vacuum energy vanishes. Moreover, all states at nonzero energy are degenerate:
each boson state is accompanied by a fermion counterpartner. These statements
are exact. It is instructive, however, to trace how they manifest themselves in 
perturbation theory.

In the previous sections we have constructed the classical vacua in various models.
The corresponding field configurations have zero energy and, hence,
are invariant
under the SUSY transformation.  
The vanishing of the vacuum energy persists when (perturbative)
quantum effects 
are switched on. This  miracle of supersymmetry, which was discovered by
the founding fathers,  is  due to the cancellation between the  bosonic
and  fermionic degrees of freedom.  Indeed, at one loop the vacuum energy is the
sum of the frequencies of the zero-point oscillations  for each mode of the 
theory,
\begin{equation}
{ E}_{\,\rm vac} = \frac{1}{2}\sum_{i} \left\{ \omega^{\rm 
boson}_{i}-
\omega^{\rm ferm}_{i}\right\}\, .
\label{ffve}
\end{equation}
This expression implies that the spectrum is discretized. For instance,
the system is put in a three-dimensional finite box with the periodic boundary
conditions. The boson and fermion terms enter with the opposite signs.
Equation (\ref{ffve})  is  general, this  formula is not specific for supersymmetry. 

The special feature of supersymmetry is the fermion-boson degeneracy of the
excitations over the vacuum: for all nonvanishing frequencies $ \omega^{\rm 
boson}_{i}=\omega^{\rm ferm}_{i}$. At one-loop level the frequencies are those of
free particles
$
\omega_{i}= (m^2 +{\vec k_i}^2)^{1/2}
$,
where ${\vec k_i}$ is the discretized three-momentum. The cancellation of 
quantum corrections to the vacuum energy continues to take place  in two loops,
three loops, and so on; in fact, it  persists to any finite order in perturbation theory.

A nonvanishing result for ${E}_{\rm vac}$ can appear only nonperturbatively.
From the general arguments above it follows that this can happen only provided
that the  set of the classical vacua contain both, the  boson and fermion
configurations. In four-dimensional theories this  means the presence of {\em
massless fermions} at the
classical level (note that we discuss now perturbation theory).  Only if there is an
appropriate massless fermion ``prefabricated" in the spectrum, can it assume the
role of Goldstino. 

The calculation of the vacuum energy is equivalent to the calculation 
of loops in the ``empty space".  When we speak of the instanton 
calculations, they are carried out in the background instanton field.
The presence of a given background field, generally speaking, breaks 
supersymmetry, the supercharges are not conserved anymore, and 
the degeneracy between the eigenfrequencies $\omega^{\rm 
boson}_{i}$ and $\omega^{\rm ferm}_{i}$ is gone. 
Gone with it is the cancellation of the quantum corrections. 
In some special background fields, however, a part of 
supersymmetry is preserved. Some supercharges are broken while 
some others remain conserved. This is sufficient for 
the cancellation of loops, order by order.  In the beginning of the section we
referred to such backgrounds as magic.

The simplest example of the background preserving a part of supersymmetry
can be found in the Wess-Zumino model.  Let us consider the background
superfields of the form
\begin{eqnarray}
Q_i (x_L,\theta)&=& a_i +b_i^\alpha \theta_\alpha + c_i
\theta^2\,,\nonumber\\[0.1cm]
\bar Q_i (x_R,\bar\theta)&=&0\,,
\label{confchir}
\end{eqnarray}
where $a_i,b_i^\alpha,c_i$ are constants. It is clear that this configuration is
invariant under the action of the supercharges $\bar Q_{\dot\alpha}$ but is {\em
not} invariant with respect to the $ Q_\alpha$-\-generated  transformations.  The
residual  invariance is sufficient to ensure the boson-fermion degeneracy and,
hence, the cancellation of the  quantum corrections.

Substituting the
background field (\ref{confchir}) into  the classical action one sees that the kinetic
term vanishes while the superpotential term survives and is expressed via
the constants
$a_i,b_i^\alpha,c_i$ (of course, the resulting action is proportional to $L^3 T$).
Correspondingly, nothing can be said about the kinetic term (and, as we know, it
{\em is} renormalized perturbatively). At the same time, 
the absence of perturbative quantum corrections 
implies that the superpotential ${\cal W}$ is
not renormalized. This is an alternative way of  proving
 the non-renormalization
theorem~\cite{GRS}. 

Two comments are in order here. First, the configuration~(\ref{confchir}) implies
that the background is not invariant under the complex conjugation; we treat the
background fields $Q_i$ and $\bar Q_i$ as independent. Such a treatment presents
no problem in perturbation theory.  Second, if some fields are massless one should
be careful: certain infrared singular $D$ terms could arise in loops. They are not
forbidden by the general argument above. However, in the continuous limit, $L\to
\infty$, they are indistinguishable from $F$ terms~\cite{Louis}.

In the gauge theories  any self-dual (or anti-self-dual field) gluon field preserves
one 
half of SUSY. The instanton field is a particularly important example of such
configurations. Another example is given by spatially constant
 self-\-dual fields
(to\-rons~\cite{torHooft}). In the self-dual backgrounds the fermion-boson
cancellation takes place. This phenomenon generalizes the fermion-boson
cancellation in the static backgrounds.  At one loop this property was first noted
in~\cite{DADV}. The theorem extending the cancellation to all loops was 
established in \cite{NSVZdop}. 

Thus,  finding the instanton contribution to certain quantities becomes
a purely classical problem, much simpler than a general instanton 
analysis and calculations in non-supersymmetric gauge theories.  Below we will 
explain in more detail how it works. It is instructive, however, to 
start from a non-gauge theory where a similar phenomenon --
conservation of one half of supersymmetry and the subsequent 
cancellation of the loop corrections -- takes place. 
 This will allow us to have a closer look at the basic ingredients of 
this  phenomenon peeled off of  conceptually irrelevant technicalities, such 
as the analytic continuation to the Euclidean space, gauge freedom  
and so on. We will deal with all these issues in due course,
but for now it seems reasonable to defer them. 

\subsection{A prototype -- supersymmetric  domain walls}
\label{sec42}

Let us  return to the minimal Wess-Zumino model 
discussed in Sec.\ \ref{sec31}, see Eqs.\  (\ref{lagrwz}), (\ref{spot}). 
As was mentioned, the model has two degenerate vacua~(\ref{vacua}).  Field
configurations interpolating between  two degenerate vacua are called the {\em
domain walls}.  They have the following properties: (i) the corresponding solutions
are  static and depend only on one spatial coordinate; (ii) they are
topologically stable and  indestructible -- once a wall is created  it cannot
disappear.  Assume for  definiteness that the wall lies in the $xy$ plane. Then the
wall  solution $\phi_{\rm w}$ will depend only on $z$. Since the wall 
extends indefinitely in the $xy$ plane, its energy $ E_{\rm w}$ is 
infinite. However, the wall tension ${\cal E}_{\rm w}$  (the energy per unit area 
${\cal E}_{\rm w} = E_{\rm w}/{A}$)  is finite,
in principle measurable, and has a clear-cut physical meaning.

The wall solution of the classical equations of motion is  known from the ancient
times,
\begin{equation}
\phi_{\rm w} = \frac{m}{\lambda}\tanh (|m|z)\, .
\label{wallsol}
\end{equation}
Note that the parameters $m$ and $\lambda$ are not assumed to be real.
A remarkable feature of this solution is that it preserves one half of
supersymmetry. Indeed, the SUSY transformations~(\ref{susytr}) generate the
following transformation of fields,
\begin{equation}
\delta \phi=\sqrt{2} \varepsilon \psi\,,\qquad
\delta \psi^{\alpha} = 
\sqrt{2}\left [{\varepsilon}^{\alpha} F + 
i\,\partial_{\mu}\phi\,
({\sigma}^{\mu})^{\alpha\dot\alpha}\,\bar\varepsilon_{\dot\alpha}\right]\, .
\label{love}
\end{equation}  
The domain wall we consider is purely bosonic, $\psi=0$. Moreover,
\begin{equation}
 F  =- \left.\frac{\partial\bar{\cal
W}}{\partial\bar\phi}\right|_{\bar
\phi=\phi_{\rm w}^*}=-e^{-i\eta}\,\partial_z\phi_{\rm w} (z)\,,
\label{lineom}
\end{equation}
where 
\begin{equation}
\eta=\arg \frac{m^3}{\lambda^2}\,.
\label{phaseW}
\end{equation}

The relation~(\ref{lineom}) means that the domain wall actually satisfies the first
order differential equation, which is by far a stronger constraint than the classical
equations of motion. Due to this feature 
\begin{equation}
\delta \psi_\alpha\propto {\varepsilon}_{\alpha}+i\,e^{i\eta}\,
({\sigma}^{z})_{\alpha\dot\alpha}\,\bar\varepsilon^{\dot\alpha}
\end{equation}
vanishes provided that
\begin{equation}
\varepsilon_{\alpha}=-i\,e^{i\eta}\,
({\sigma}^{z})_{\alpha\dot\alpha}\,\bar\varepsilon^{\dot\alpha}\,.
\end{equation}
This condition singles out two supertransformations (out of four) which do not act
on the domain wall (alternatively it is often said  that they act trivially).

Now, let us calculate the wall tension at the classical level. To this end we rewrite
the expression for the tension as
\begin{equation}
{\cal E} =  \int^{+\infty}_{-\infty} {\rm d} z \left[ 
\partial_z\bar\phi \,\partial_z \phi   +\bar F F \right]
\equiv  \int^{+\infty}_{-\infty} {\rm d} z \left\{ \left[e^{-i\eta}\,
 \partial_z{\cal W} +{\rm H.c.}\right] +\left| \,\partial_z \phi +e^{i\eta}\,F\,\right|^2 
\right\}\,,
\label{clwen}
\end{equation}
where $F=-\partial\bar{\cal W}/\partial \bar \phi$ and it is implied that $\phi$
 depends only on $z$. This form makes it clear why the domain wall satisfies the
first order differential equation~(\ref{lineom}). As a result the wall tension ${\cal
E}_{\rm w}$ coincides with the modulus of the topological charge
${\cal Z}$ 
\begin{equation}
{\cal E}_{\rm w}=\left|{\cal Z}\right|\,,
\end{equation}
where ${\cal Z}$ is defined as
\begin{equation}
{\cal Z}=2 \left\{ {\cal W}(\phi (z=\infty)) -  {\cal W}(\phi (z=-\infty)) \right\}
=\frac{8\,m^3}{3\,\lambda^2}
\,.
\label{cctls}
\end{equation}
Note that the phase of ${\cal Z}$ coincides with $\eta$ introduced in Eq.\
(\ref{phaseW}). 

Such states are called BPS or {\em BPS-saturated}.  BPS stands
for Bogomolny, Prasad and Sommerfield.  The works of these authors
\cite{BPS} have nothing to do with supersymmetry;
they considered the Abrikosov vortices and monopoles in the non-supersymmetric
models and observed that in certain limits these objects satisfy the first order
equations and their masses coincide with the topological charges.
In the context of supersymmetry we see that the BPS saturation is equivalent 
to the residual supersymmetry. 

How come that we got a nonvanishing energy for the state which is annihilated by
some supercharges? This is because the superalgebra~(\ref{susyalgebra}) gets
modified,
\begin{equation}
\left\{Q_\alpha\,,Q_\beta\right\}=-4\,\Sigma_{\alpha\beta}\,\bar {\cal
Z}\,,\qquad
\left\{\bar Q_{\dot\alpha}\,,\bar
Q_{\dot\beta}\right\}=-4\,\bar\Sigma_{\dot\alpha\dot\beta}\, {\cal Z}\,,
\label{cccdw}
\end{equation}
where 
\begin{equation}
\Sigma_{\alpha\beta}=-\frac 12\int {\rm d} x_{[\mu} {\rm d} x_{\nu ]}\,
(\sigma^\mu)_{\alpha\dot\alpha} (\bar \sigma^\nu)_{\beta}^{\dot\alpha}
\end{equation}
is the wall area tensor. All other commutation relations remain intact.
In the context of the modified SUSY algebra,  the topological charge ${\cal Z}$
bears the name of the {\em central} charge$\,$\footnote{For a recent discussion
of the general theory of the tensorial central charges in various superalgebras
see~\cite{SFMP}.}. 

The connection between the BPS saturation and the central extension of the
superalgebra was revealed long ago  by  Olive  and Witten \cite{WO},  shortly after 
the advent of supersymmetry.  In centrally extended superalgebras
the fact that a state is annihilated by a supercharge does not imply the vanishing
of the energy of the state -- instead its mass is equal to the central charge of the
state.

To derive this relation  let us consider the representations of the centrally
extended superalgebra. In the problem of the domain walls  we are interested not
in a generic representation but, rather, in a special one where one half of the
supercharges annihilate all states.  The existence of such
supercharges was  demonstrated above at the classical level. The covariant
expressions for the residual  supercharges $\tilde Q_\alpha$ are
\begin{equation}
\tilde Q_\alpha= e^{i\eta/2}\,Q_\alpha-\frac{2}{A}\,
e^{-i\eta/2}\,\Sigma_{\alpha\beta}\, n^\beta_{\dot\alpha} \,\bar Q^{\dot\alpha}
\,,
\end{equation}
where 
\begin{equation}
n_{\alpha\dot\alpha}= \frac {P_{\alpha\dot\alpha}}{{\cal E}_{\rm w} A}
\end{equation}
is the unit vector proportional to the wall four-momentum
$P_{\alpha\dot\alpha}$; it has only the time component in the rest frame.
The subalgebra of these residual supercharges in the rest frame is
\begin{equation}
\left\{\tilde Q_\alpha\,,\tilde Q_\beta\right\}=8\,\Sigma_{\alpha\beta}\left\{{\cal
E}_{\rm w} - |{\cal Z}|\right\}\,.
\end{equation}
We will refer to this subalgebra as {\em stationary}. 

The existence of the stationary subalgebra immediately
proves that the wall tension ${\cal E}_{\rm w}$ is equal to the central charge ${\cal
Z}$.  Indeed, $\tilde Q|{\rm wall}\rangle =0$ implies
that ${\cal E}_{\rm w} - |{\cal Z}|=0$. This equality is valid both to any order in
perturbation theory and nonperturbatively.

An interesting question is the supermultiplet structure of the domain walls,
i.e. the representations of the centrally extended algebra.  The minimal
representation in the sector where $\tilde Q_\alpha$ is realized trivially is
two-dimensional~\cite{Intri}. 

 From the non-renormalization theorem for the
superpotential~\cite{GRS}  we additionally infer that the central
charge ${\cal Z}$ is not renormalized. Thus, the result
\begin{equation}
{\cal E}_{\rm w}=\frac{8}{3}\left|\frac{m^3}{\lambda^2}\right|
\label{clewt}
\end{equation}
for the wall tension is {\em exact}~\cite{Dvali1}. 

The wall tension ${\cal E}_{\rm w}$ is a physical parameter and, as such, should be
expressible in terms of the physical (renormalized) parameters $m_{\rm ren}$
and $\lambda_{\rm ren}$. One can easily verify that this is compatible with the
statement of nonrenormalization of  ${\cal E}_{\rm w}$. Indeed, 
$$
 m = Z\, m_{\rm ren}\,\qquad \lambda = Z^{3/2}\lambda_{\rm ren}\, ,
$$
where $Z$ is the $Z$ factor coming from the kinetic term. Consequently,
$$
\frac{m^3}{\lambda^2} = \frac{m^3_{\rm ren}}{\lambda^2_{\rm ren}}\, . 
$$ 
Thus, the absence of the quantum corrections to Eq.\ (\ref{clewt}), the
renormalizability of the theory, and the  non-renormalization 
theorem for superpotentials -- all these three elements are 
intertwined with each other. In fact, every two elements taken separately 
imply the third one. 

What lessons have  we  drawn from the example of the domain walls? 
In the centrally extended SUSY algebras the exact relation $E_{\rm vac}=0$
is replaced by the exact relation ${\cal E}_{\rm w}- |{\cal Z}|=0$. Although
this statement is valid both perturbatively and nonperturbatively,
it is  very instructive to visualize it as an explicit cancellation between bosonic and
fermionic modes in perturbation theory -- we will do the 
exercise  in Sec.\ \ref{sec424},
where we consider 1+1 dimensional models with minimal supersymmetry.

The nonrenormalization of the central charge is not a general feature. 
In the domain wall problem it is due to the extended supersymmetry in
the effective 1+1 dimensional theory to which the minimal
four-dimensional Wess-Zumino model reduces, in a sense.
In the two-dimensional models with minimal
supersymmetry (Sec.~\ref{sec424})  the central charge gets quantum
corrections. 

\subsection{Superpotential and anomaly}
\label{secSAA}

The domain walls make the superpotential observable. 
More exactly, the central
charge is related, on the one hand, to the wall tension, and on the other hand,
to the jump of the superpotential in passing from one spatial infinity
to the other, through the wall. In the Wess-Zumino model the central charge
is proportional to the tree-level superpotential, see Eq.\ (\ref{cctls}).
A natural question one should  ask is whether the central charge exists
in ${\cal N}= 1$ SUSY gauge theories, and if yes, what replaces
the tree-level superpotential in Eq.\ (\ref{cctls}). We will see that
the answer to this question is important for  understanding of  the
nonperturbative gauge dynamics.

A hint of the existence of the central extension can be obtained
from the analysis of SUSY gluodynamics (Sec.\ \ref{secEL}). In this theory there
are $T_G$ distinct supersymmetric vacua. Hence, there should exist
field configurations interpolating between them, the domain walls.
We know  already that the domain walls go
hand in hand with occurrence of the  central charge 
in ${\cal N}= 1$ SUSY.  

The actual calculation of the  central charge in the gauge theories 
requires a careful treatment of the anomalies: besides the tree-level
superpotential, ${\cal Z}$ contains an anomalous term~\cite{Dvali2}.
We will not dwell on details of the derivation here -- they will lead us far
astray, and, after all, we need only the final result. It is not difficult
to  show
that the anomaly in the central charge is not a  new one
-- it is related~\cite{CS}
to the supermultiplet of anomalies (\ref{geom}). Equation (\ref{cccdw})
is still valid, being general; for the central charge ${\cal Z}$,
instead of ${\cal Z} = 2\Delta ({\cal W})$, one gets
\begin{eqnarray}
{\cal Z} &\!\!= &\!\!\frac{2}{3} 
\Delta\left\{
\left[ 3{\cal W } - \sum_f Q_f\, \frac{\partial{\cal W }}{\partial Q_f}
\right] \right.
 \nonumber\\[0.2cm]
&\!\!- &\!\!
\left.
\left[ \frac{3T_G- \sum_f T(R_f)}{16\pi^2}\,{\rm Tr}\,W^2 + 
\frac{1}{8}\sum_f\gamma_f 
\bar{D}^2 (\bar{Q}_{\!f}\, e^{V} Q_f) \right]
\right\}_{\theta=0}\, .
\label{achgt}
\end{eqnarray}
We hasten to make a few comments concerning this relation. 
The first term in the second line is of purely quantum origin: it presents
the gauge anomaly in the central charge. The second term in the second line is
a total superderivative. Therefore, it vanishes after averaging over any 
supersymmetric vacuum state. Hence, it can be safely omitted.
The first line presents the classical result. At the classical level
$ Q_f(\partial{\cal W }/\partial Q_f)$ is a total superderivative too
(cf. Eq.\ (\ref{ka1})). If we discard it for a short while
(forgetting about the quantum effects), we return to ${\cal Z} = 2\Delta ({\cal
W})$, the formula obtained in the Wess-Zumino model. At the quantum level
$ Q_f(\partial{\cal W }/\partial Q_f)$ ceases to be a total superderivative
because of the Konishi anomaly. 
It is still  convenient to eliminate $ Q_f(\partial{\cal W }/\partial Q_f)$
in favor of Tr$W^2$ by virtue of the Konishi relation (\ref{ka1}). In this
way one arrives at
\begin{equation}
{\cal Z} = 2
\Delta\left\{ {\cal W} - \frac{T_G- \sum_f T(R_f)}{16\pi^2}\,{\rm Tr}\,W^2
\right\}_{\theta=0}\, .
\label{achgf}
\end{equation}
We see that the superpotential ${\cal W}$ is amended by the anomaly;
in the operator form
\begin{equation}
 {\cal W} \longrightarrow  {\cal W} - \frac{T_G- \sum_f T(R_f)}{16\pi^2}\,{\rm
Tr}\,W^2\,.
\label{ofasp}
\end{equation}
The anomaly may or may not materialize as a nonperturbative 
correction in the
low-energy effective superpotential. For instance, in the SU(2) SQCD
with one flavor the anomalous term  Tr$\,\lambda\lambda$ in Eq.\ (\ref{ofasp})
materializes at low energies as a $\Phi^{-2}$ term, see Sec.\ \ref{sec471}.
On the other hand, in the two-flavor model (Sec.\ \ref{sec472}), where
$T_G- \sum_f T(R_f) = 0$, the anomalous contribution  vanishes. This explains
why in the two-flavor model the superpotential is not generated beyond the
tree level even after the mass terms of the matter fields are switched on.

In the next section we will see how  the anomaly
modifies the operator of the central charge in a
much simpler setting of a  nongauge
theory. We will consider a two-dimensional analog of the Wess-Zumino model,
with a minimal supersymmetry. The model is instructive in two aspects:
it very transparently demonstrates the occurrence of the anomaly
in the central charge and the boson-fermion cancellations in perturbation
theory. 

\subsection{Digression:  solitons in two-dimensional theories with 
the minimal SUSY}
\label{sec424}

As was mentioned in Sec.\ \ref{sec42}, in some aspects  the problem of the
domain wall is
obviously  two-dimensional$\,$\footnote{The  presentation in this section is based
on Ref.\ \cite{SVV}.}. The remnant of the original four-dimensional formulation is
the extended  supersymmetry of the emerging  two-dimensional model. Indeed,
four supercharges imply ${\cal N}=2$ supersymmetry in 1+1. Considering
two-dimensional  models on their own right, we can descent to  the minimal 
${\cal N}=1$ supersymmetry, with two supercharges, by deforming 
the ${\cal N}=2$
model obtained as a dimensional reduction of the
four-dimensional Wess-Zumino model.

To perform the reduction to two dimensions 
it is sufficient to assume that all fields
are independent of $x$ and $y$, and depend on $t$ and $z$ only. In two
dimensions the Lagrangian~(\ref{wzcomp}) can be presented in the following
form$\,$\footnote{To distinguish between the superpotentials in four and two
dimensions the latter is denoted by $W$ as opposed to the calligraphic ${\cal W}$
in four dimensions.}:
\begin{equation}
{\cal L}=\frac{1}{2}\left\{ \partial _\mu\varphi_i
\,\partial^\mu\varphi_i
+i \bar \psi_i \gamma^\mu \partial_\mu \psi_i + f_i f_i +2 f_i
\,\frac{\partial
{ W}}{\partial \varphi_i} - \frac{\partial^2 {W}}{\partial
\varphi_i \partial
\varphi_j}\, \bar
\psi_i
\psi_j \right\}
\;.
\label{n2lag}
\end{equation}
We introduced real fields $\varphi_i$, $\psi_i$ and $f_i$, where
$i=1,2$.  Up to
normalization, they are just the  real and imaginary parts  of the original
fields, e.g,
$\varphi=(\varphi_1+i \varphi_2)/\sqrt{2}$. Summation over
$i$ is implied.  The fermionic field $\psi_i$ is a two-component Majorana spinor,
a convenient representation for the two-dimensional 
$\gamma$ matrices  is
$$
\gamma^0 = \sigma_2\, , \,\, \gamma^1 = i\sigma_3\, .
$$
We will stick to it in this section.
The superpotential $ W=2\, {\rm Re} {\cal
W} $ is a function of two variables
$\varphi_1$ and $\varphi_2$,
\begin{equation}
{W}(\varphi_1, \varphi_2) = \sqrt{2}\left[\frac{m^2}{\lambda} \,\varphi_1 -
\frac{\lambda}{6} \,\varphi_1^3 +\frac{\lambda}{2}\,\varphi_1\varphi_2^2\right]
\,,
\label{arkmi}
\end{equation}
where $m$ and $\lambda$ are now assumed to be real.

The presence  of extended supersymmetry is reflected in the harmonicity
of this  superpotential,
\begin{equation}
\frac{\partial^2 {W}}{\partial \varphi_i \partial
\varphi_i}=0\qquad \mbox{for ${\cal N}=2$}
\,.
\end{equation}
To break ${\cal N}=2$ down to ${\cal N}=1 $  we consider  a more general case
of nonharmonic functions  ${W}\,(\varphi_1, \varphi_2) $,
\begin{equation}
{ W}\,(\varphi_1, \varphi_2) =\sqrt{2}\left[ \frac{m^2}{\lambda} \,\varphi_1
-
\frac{\lambda}{6} \,\varphi_1^3 +\frac{\lambda}{2}\,\varphi_1\varphi_2^2
+\frac{\mu}{\sqrt{2}}\,\varphi_2^2\right]\,,
\label{nonharm}
\end{equation}
where at $\mu \neq 0$ the ${\cal N}=2$ SUSY is broken down to the minimal ${\cal
N}=1$.

The two supercharges $Q_\alpha$ of the model are
\begin{equation}
Q_\alpha=\int \! {\rm d}z\, J^{\,0}_\alpha\, ,\qquad
J^\mu = \sum_{i=1,2} \left[(\partial_\nu \varphi_i )\gamma^\nu
\gamma^\mu\psi_i - i f_i\gamma^\mu\psi_i\right]\, .
\label{cosc}
\end{equation}
The original Wess-Zumino model contained two more supercharges transforming 
$\varphi_1\to \psi_2$ and $\varphi_2\to \psi_1$. This invariance is broken at 
$\mu\neq 0$.

The canonical commutation relations imply that
\begin{equation}
\left\{J_{\alpha}^\mu, \bar Q_\beta \right\}
=2\,(\!\gamma^\nu)_{\alpha\beta} \,\vartheta^{\mu}_{\nu} +
2i\,(\!\gamma^5)_{\alpha\beta}\, \zeta^\mu
\;,
\label{celoc}
\end{equation}
where $\vartheta^{\mu}_{\nu}$  is the energy-momentum tensor
and $\zeta^\mu$ is the conserved topological current,
\begin{equation}
\zeta^\mu=\epsilon^{\mu\nu} \partial_\nu { W}~.
\label{topcurr}
\end{equation}
The notation
$\gamma^5 = \gamma^0\gamma^1= -\sigma_1$ is used.
Integrating the $\mu=0$ component of \mbox{Eq.\ (\ref{celoc})}  over
space gives
the SUSY algebra:
\begin{equation}
\{{Q}_\alpha, \bar Q_\beta\}
=2\,(\!\gamma^\mu)_{\alpha\beta} \,P_\mu +
2i\,(\!\gamma^5)_{\alpha\beta}\,
 {\cal Z}\;.
\label{ce}
\end{equation}
Here $P_\mu=\int\! {\rm d} z
\,\vartheta^{0}_{\mu}$ are operators of the total energy and
momentum, and ${\cal Z}$ is the central charge,
\begin{equation}
{\cal Z} = \int \! {\rm d}z \,\zeta^0= \int \! {\rm d}z\, \partial_z
{ W }(\phi) =
{ W}[\phi (z=\infty )]-
 { W}[\phi (z=-\infty )]\, ,
\label{central}
\end{equation}
which coincides with the topological one.

The classical kink solution for $\varphi_1$ is the same, up to normalization,  as in 
Eq.\ (\ref{wallsol}), while  the second field $\varphi_2$  vanishes,
\begin{equation}
\varphi_1=\varphi_{\rm kink}=\frac{m\sqrt{2}}{\lambda}\tanh mz\,,
\quad
\varphi_2=0~.
\label{N2sol}
\end{equation}
The solution is annihilated by $Q_2$, the corresponding supersymmetry is
preserved in the kink background. The action of $Q_1$ produces the fermion zero
mode -- the fermion kink. The classical mass of the kink is
\begin{equation}
M={\cal Z}= \frac{8}{3}\,\frac{m^3}{\lambda^2}\,.
\label{2dcsm}
\end{equation}

Now let us move from the
classical to the quantum level and study quantum corrections
to the kink mass. The issue of quantum corrections to the soliton mass in
two-dimensional models with ${\cal N} =1$ supersymmetry has a long and
dramatic history. As was noted in \mbox{Ref.\ \cite{WO}},  in the models with
central extensions topologically stable solitons can be BPS saturated. The mass
of the BPS saturated solitons must be  equal to the central charge. The simplest
case of the model with one real scalar field and one
two-component real spinor (such models are often called the supersymmetric
Ginzburg-Landau models) was considered in~\cite{DADV}.  It was
argued~\cite{DADV} that, due to a residual supersymmetry, the mass of the soliton
calculated at the classical level remains intact at  the one-loop level. A few years
later it was noted
\cite{KR}  that the non-renormalization theorem~\cite{DADV} cannot possibly be
correct, since the classical soliton mass is proportional to $m^3/\lambda^2$,
 and the physical mass of the scalar field gets
a logarithmically infinite renormalization (there is no ultraviolet logarithm in  the
correction to $\lambda$). Since the soliton mass is an observable physical
parameter, it must stay finite in the limit of the infinite  ultraviolet cut off. This
implies, in turn, that the quantum corrections cannot vanish -- they ``dress" $m$
in the classical expression (\ref{2dcsm}), converting  the bare mass parameter
into the renormalized one.  The one-loop renormalization of the soliton mass
was first calculated in \cite{KR}. 

Since then a number of one-loop calculations were carried out
\cite{IM,AHV,JFS,SR,AU,yama,CM1,RN,Jaffe}. The results reported
and the  conclusion of saturation/non-saturation oscillated with time.
Although all authors clearly agree that the
logarithmically divergent term corresponds to the renormalization of
$m$, the finite term comes out differently,  varying from work to work.
The resolution of the paradox came recently~\cite{SVV}: the mass stays equal to
the central charge to all orders, but the latter gets modified by a quantum
anomaly.  We will illustrate this assertion below.

The exact equality of the mass and the central charge follows from the SUSY
algebra~(\ref{ce}). Moreover, a similar relation is true for the densities
(i.e. local rather than integrated quantities),
\begin{equation}
\langle {\rm sol}\,| {\cal H}(x) - \zeta^{\,0} (x)|\,{\rm
sol}\rangle=0\,,
\label{locenergy}
\end{equation}
where ${\cal H}(x)=\vartheta_{0}^{0}(x)$ is the
 Hamiltonian density. It follows from
Eq.\ (\ref{celoc}) with $\mu=0,\,\alpha=2,\,\beta=1$ and the residual
supersymmetry $Q_2|{\rm sol}\rangle =0$.

Let us show how the fermion-boson cancellation in ${\cal H} - \zeta^{\,0}$
manifests itself  in the language of modes. In the one-loop approximation we
need a quadratic expansion of the Hamiltonian around the soliton,
\begin{eqnarray}
\left[{\cal H}-\zeta^{\,0}\right]_{\rm quad}
&=&\frac{1}{2}\left\{ \dot \chi_1^2
+[P_1\,\chi_1]^2
-i\,\eta_1\, P_1^\dagger\,\xi_1 +i\, \xi_1\, P_1\,\eta_1 \right.\nonumber\\[0.1cm]
 &&+\left.\dot \chi_2^2
+[P_2^\dagger\,\chi_2]^2
+i\,\eta_2\, P_2\,\xi_2
-i\, \xi_2\,P_2^\dagger\,\eta_2
\right\}\,,
\label{H22}
\end{eqnarray}
where the following notation is used:
\begin{equation}
\chi_1=\varphi_1-\varphi_{\rm kink}\,,\quad \chi_2=\varphi_2\,, \quad
\left(
\begin{array}{l}
\xi_i
\\
\eta_i
\end{array}\right)
=\psi_i
\;,
\end{equation}
and  the differential operators $P_{1,2}$ are defined as
\begin{equation}
P_1=\partial_z +\frac{\partial f_1}{\partial \varphi_1}
\,,\qquad P_2=\partial_z -\frac{\partial f_2}{\partial \varphi_2}
\,.
\end{equation}
Here  the derivatives of $f_i= -\partial W/\partial\varphi_i $ are evaluated 
at the kink solution,
\begin{equation}
\frac{\partial f_1}{\partial \varphi_1}=\sqrt{2}\,\lambda\, \varphi_{\rm kink}\,,
\qquad \frac{\partial f_2}{\partial \varphi_2}=-\sqrt{2}\,\lambda \,\varphi_{\rm
kink} -2\mu
\,.
\end{equation}
From the expression~(\ref{H22})  it is easy to see that a convenient basis for the
mode expansion of $\chi_1$ and $\eta_1$ is provided by the eigenfunctions
$v_n (z)$ of the Hermitian differential operator $ P_1^\dagger P_1$, 
\begin{equation}
P_1^\dagger P_1\,v_n (z)=\omega_n^2\, v_n (z)\,.
\label{eigenA}
\end{equation}
As for $\xi_1$ it is the operator $ P_1 P_1^\dagger$ that defines the
mode composition,
\begin{equation}
P_1 P_1^\dagger\,\tilde v_n (z)=\omega_n^2\, \tilde v_n (z)\,.
\label{eigenB}
\end{equation}
Similarly, the mode decomposition for $\xi_2$ runs in the eigenmodes $w_n$ of
the operator 
$P_2^\dagger P_2$ while the mode decomposition  for
$\chi_2,\, \eta_2$ runs in the eigenmodes $\tilde w_n$ of the operator $P_2
P_2^\dagger$,
\begin{equation}
P_2^\dagger P_2\,w_n (z)=\nu_n^2\, w_n (z)\,, \qquad P_2 P_2^\dagger \,\tilde
w_n (z)=\nu_n^2\, \tilde w_n (z)\,.
\label{eigenA2}
\end{equation}

To discretize the spectrum of the modes let us put the system in a large
box, i.e. impose the boundary conditions at $z=\pm L/2$ (at the end $L\to
\infty$). It is convenient to choose  the boundary conditions in a form which is
compatible with the residual supersymmetry implemented by the action of
$Q_2$. The suitable boundary conditions are
\begin{equation}
\begin{array}{ll}
\left. \left(\partial_z\varphi_1 +f_1 \right) \right|_{z=\pm L/2}=0\,,
&~~~~\left. \varphi_2\right|_{z=\pm  L/2}=0\,,\\[0.2cm]
 \left. \left(\partial_z +\partial f_1/\partial
\varphi_1\right) \eta_1\right|_{z=\pm  L/2}=0\,, &~~~~\left. 
\eta_2\right|_{z=\pm  L/2}=0\,,\\[0.2cm]
  \left. \xi_1\right|_{z=\pm  L/2}=0\,,&~~~~\left. \left(\partial_z-\partial
f_2/\partial \varphi_2
\right)\xi_2
\right|_{z=\pm L/2}=0
\,.
\end{array}
\label{boundary}
\end{equation}
It is easy to verify the invariance of these conditions under the 
transformations generated by $Q_2$. They are also consistent with the
classical solutions, both for the flat vacuum and for the kink. In
particular, the soliton solution $\varphi_{\rm kink}$  satisfies 
$\partial_z\varphi_{\rm kink} + f_1=0$ everywhere, and the  boundary
conditions~(\ref{boundary}) do not  deform it --  we can indeed place the kink
in the box.

 The boundary conditions for the modes in the box follow from the linear
expansion of \mbox{Eq.\ (\ref{boundary})},
\begin{equation}
\begin{array}{ll}
\left. P_1\,
v_n\right|_{z=\pm L/2}=0 \,, &~~~~\left. P_2\, w_n\right|_{z=\pm
L/2}=0\,,\\[0.2cm]
\left. \tilde v_n\right|_{z=\pm L/2}=0\,, &~~~~\left. \tilde w_n\right|_{z=\pm
L/2}=0\,.
\end{array}
\label{boundary1}
\end{equation}

With these boundary conditions all eigenvalues of the operators $P_i^\dagger P_i$
and $P_i P_i^\dagger$ are the same, with the exception of the zero modes. 
The operators $P_i P_i^\dagger$ have no zero modes, while $P_i^\dagger P_i$
do 
have. Moreover, for nonzero modes the eigenfunctions $v_n$ and $\tilde v_n$ 
(and
$w_n$ and $\tilde w_n$) are algebraically related
\begin{eqnarray}
&&\tilde v_n=\frac{1}{\omega_n} P_1 \,v_n\,, \qquad
v_n=\frac{1}{\omega_n} P^\dagger_1 \,\tilde v_n\,,\nonumber\\[0.2cm]
&&\tilde w_n=\frac{1}{\nu_n} P_2 \,w_n\,, \qquad
w_n=\frac{1}{\nu_n} P^\dagger_2 \,\tilde w_n
\,.
\label{relmodes}
\end{eqnarray}

The expansion in the eigenmodes has the form,
\begin{eqnarray}
&&\chi_1 (x) =\sum_{n\neq 0} \chi_{1n} (t) \,v_n (z)\,, \quad
\eta_1 (x)=\sum_{n\neq 0} \eta_{1n} (t)\,v_n (z)\,, 
\quad \xi_1 (x)=\sum_{n\neq 0} \xi_{1n} (t)\,\tilde v_n (z)\,,\nonumber \\[0.2cm]
&&\chi_2 (x) =\sum_{n\neq 0} \chi_{2n} (t) \,\tilde w_n (z)\,, \quad
\eta_2 (x)=\sum_{n\neq 0} \eta_{2n} (t)\,\tilde w_n (z)\,, 
\quad \xi_2 (x)=\sum_{n\neq 0} \xi_{2n} (t)\,w_n (z)\,.\nonumber \\[-0.2cm]
&&~~~~
\label{expansion}
\end{eqnarray}

Note that the summation above does not include the zero modes $v_0$ and
$w_0$. The mode $v_0\propto \partial_z \varphi_{\rm kink}$ is due to the
translational  invariance of the problem at hand, it reflects the possibility of
shifting the kink center. We fix the center to be nailed at the 
origin. That is why this
mode should not be included in the mode decomposition. The same mode $v_0$
in
$\eta_1$ corresponds to the fermion partner of the bosonic kink, it is generated
by the action of $Q_1$. The zero mode $w_0$ in $\xi_2$ coincides with $v_0$ at
$\mu=0$. Setting $\mu=0$ we ensure extended supersymmetry, so that the
bosonic kink has two fermionic zero modes. At $\mu\neq 0$ the extended SUSY is
gone but the zero mode in $\xi_2$ survives,  due to the Jackiw-Rebbi index
theorem~\cite{Jackiw}. It is curious to note that at $\mu< m$ this mode sits on
the kink while at $\mu>m$ it runs away to the boundary of the box.

The coefficients $\chi_{in}$, $\eta_{in}$ and $\xi_{in}$ are time-dependent
operators. Their equal time commutation relations are determined by the
canonical commutators,
\begin{equation}
[\chi_{im},\dot \chi_{kn}]=i\delta_{ik}\delta_{mn}\,,\quad
\{\eta_{im},\eta_{kn}\}=\delta_{ik}\delta_{mn}\,,\quad
\{\xi_{im},\xi_{kn}\}=\delta_{ik}\delta_{mn}\,.
\end{equation}
The anticommutators of $\eta$ and $\xi$ vanish.
Thus, the mode decomposition reduces dynamics of the system under
consideration to quantum mechanics of an infinite set of supersymmetric
harmonic oscillators. The ground state of the quantum soliton corresponds to
setting each oscillator from the set to the ground state.

Constructing the creation and annihilation operators in the standard way
we find the following nonvanishing expectations values of the bilinears
built from the operators $\chi_{in}$, $\eta_{in}$ and $\xi_{in}$ in  the ground
state:
\begin{eqnarray}
&&\langle \dot \chi_{1n}^2\rangle_{\rm sol}=\frac{\omega_n}{2}\,,\quad
\langle  \chi_{1n}^2\rangle_{\rm sol}=\frac{1}{2\omega_n}\,,\quad
\langle \eta_{1n} \,\xi_{1n}\rangle_{\rm sol}=\frac{i}{2}\,;\nonumber\\[0.2cm]
&& \langle \dot \chi_{2n}^2\rangle_{\rm sol}=\frac{\nu_n}{2}\,,\quad
\langle  \chi_{2n}^2\rangle_{\rm sol}=\frac{1}{2\nu_n}\,,\quad
\langle \eta_{2n} \,\xi_{2n}\rangle_{\rm sol}=\frac{i}{2}\,.
\label{expect}
\end{eqnarray}
The expectation values of other bilinears obviously vanish.
Combining \mbox{Eqs.\ (\ref{H22})}, (\ref{expansion}) and
(\ref{expect}) we  get
\begin{eqnarray}
&&\langle {\rm sol}\,| \left[{\cal H}(x) - \zeta^{\,0}\right]_{\rm
quad}|\,{\rm sol}\rangle\nonumber\\[0.2cm]
&&=\sum_{n \neq 0} \frac{\omega_n}{4}\left\{v_n^2 +\tilde v_n^2 -
v_n^2 -\tilde v_n^2 \right\} 
+\sum_{n \neq 0} \frac{\nu_n}{4}\left\{\tilde w_n^2+ w_n^2 
 -\tilde w_n^2 -w_n^2\right\} 
\equiv 0\, .~~~~~~~~
\label{H3}
\end{eqnarray}
The eight terms  in \mbox{Eq.\ (\ref{H3})}  are in
one-to-one correspondence with the eight terms in \mbox{Eq.\ (\ref{H22})},
the terms with the plus sign come from bosons while those with the minus sign
from fermions. In  proving the vanishing of the right-hand side we did
{\em not} perform  integrations by parts --
this is essential to guarantee that
no surface terms arise from such integrations. The vanishing of the right-hand
side of (\ref{H22}) demonstrates explicitly the residual supersymmetry (i.e. the
conservation of $Q_2$) at work. Note that the sums over the modes are
quadratically divergent in the ultraviolet if considered separately for bosons and
fermions, therefore a regularization is required.
 Any regularization preserving the
residual supersymmetry will maintain the above cancellation.

Equation~(\ref{locenergy}) must be considered as a local version of BPS
saturation (i.e. conservation of a residual supersymmetry). The
nonrenormalization of the expectation value of ${\cal H} -\zeta^{\,0}$
over the soliton state is the main lesson we draw; a similar result will be
exploited later in the instanton calculations.

The equality between the kink mass and the central charge renders 
the calculation of the quantum corrections to the mass a simple task reducing it to 
calculating $\langle W \rangle$  in the flat vacuum, see Eq.\ (\ref{central}).
In one loop
\begin{equation}
\left\langle  { W} \right\rangle_0 = { W}_0
+\frac{1}{2}\,\frac{\partial^2 {W}_0 }{\partial \varphi_1^2}\, \left\langle \chi_1^2
\right\rangle_{\,0}+ \frac{1}{2}\,\frac{\partial^2 {W}_0 }{\partial \varphi_2^2}\,
\left\langle \chi_2^2\right\rangle_{\,0}\,,
\label{wcorr}
\end{equation}
where the subscript 0 marks the flat vacua 
\begin{equation}
\varphi_{1{\rm vac}}=\pm \frac{m\sqrt{2}}{\lambda}\,,\quad \varphi_{2{\rm
vac}}=0\,,
\end{equation}
at $z=\pm \infty$. The average quadratic fluctuation is given by a simple tadpole
graph,
\begin{equation}
\left\langle\,\chi_i^2\,\right\rangle_{\,0} = \int
\frac{{\rm d}^2 p}{(2\pi )^2}\,\,
\frac{i}{p^2 -({\partial^2 {W}_0 }/{\partial
\varphi_i^2})^2} =\frac{1}{4\pi}\ln \frac{M_{\rm uv}^2}{({\partial^2 {W}_0
}/{\partial \varphi_i^2})^2}
\,,
\label{5}
\end{equation}
where $M_{\rm uv}$ is the ultraviolet cut-off, it drops out in ${\cal Z}$.
The result has the form
\begin{equation}
M\!=\!{\cal Z}\!=\!\left[ {W}+\frac{{\partial^2 {W}
}/{\partial \varphi_1^2}}{4\pi}\,\ln\left|
\frac{{\partial^2 {W}
}/{\partial \varphi_2^2}}{{\partial^2 {W}
}/{\partial \varphi_1^2}}\right| \right]_{\{ \varphi\}=
\{ \varphi_{\rm vac} (z=+\infty)\}} \!\!\!- \left[ (z=+\infty)\to 
(z=-\infty)\right]\,.
\end{equation}
In this form the result is valid for any superpotential $W$ with the
property
\begin{equation}
\left.\sum_{i=1,2} \frac{\partial^2 {W} }{\partial
\varphi_i^2}\right|_{\{ \varphi\}=
\{ \varphi_{\rm vac} (z=+\infty)\}}=
\left.\sum_{i=1,2} \frac{\partial^2 {W} }{\partial
\varphi_i^2}\right|_{\{ \varphi\}=
\{ \varphi_{\rm vac} (z=-\infty)\}}\,.
\end{equation}
In the particular case of Eq.\ (\ref{nonharm}) the explicit expression is
\begin{equation}
M= \frac{8}{3}\,\frac{m^3}{\lambda^2}- \frac{m}{2\pi}\ln
\left|\frac{\mu^2-m^2}{m^2}\right| - \frac{\mu}{2\pi}\ln
\left|\frac{\mu+m}{\mu-m} \right|\,.
\end{equation}

At $\mu=0$ we are back to extended supersymmetry. The one-loop correction
vanishes as  was expected. This also explains the finiteness of the correction
to the soliton mass at $\mu\neq 0$. Indeed,
 at virtual momenta much larger than $\mu$ the extended SUSY is
recovered. In the opposite limiting case $\mu\gg m$ the fields
$\varphi_2$ and $\psi_2$  become very heavy and can viewed as the ultraviolet
regulators for the light fields. In this limit
\begin{equation}
M= \frac{8}{3}\,\frac{m^3}{\lambda^2}- \frac{m}{2\pi}\left[\ln
\frac{\mu^2}{m^2} +2\right]\,.
\label{masskink}
\end{equation}
This formula has a clear interpretation in the theory with one supermultiplet
$\{\varphi_1, \psi_1\}$ of light fields$\,$\footnote{Although we say that the
interpretation is transparent, quite remarkably, it was not discovered until
recently.}. Namely, the definition~(\ref{central}) of the central charge
${\cal Z}$ should be modified by the substitution
\begin{equation}
W(\varphi_1)\to W(\varphi_1) +\frac{W^{\prime\prime}(\varphi_1)}{4\pi}\,,
\label{qa2dim}
\end{equation}
where
\begin{equation}
{W}(\varphi_1) = \sqrt{2}\left[\frac{m^2}{\lambda} \,\varphi_1 -
\frac{\lambda}{6} \,\varphi_1^3 \right]
\,.
\label{wphi1}
\end{equation}
The logarithmic term in Eq.\ (\ref{masskink}) is the quantum correction to the
expectation value of ${W}(\varphi_1)$ over the flat vacuum. The nonlogarithmic 
term in the square brackets presents the quantum anomaly
$W^{\prime\prime}/{4\pi}$ in the central charge. This anomaly is  a superpartner
to the anomalies in the trace of the energy-momentum tensor
$\vartheta^\mu_\mu$ and in $\gamma^\mu J_\mu$, see Ref.\ \cite{SVV} for
details. It is instructive to compare Eq.\ (\ref{qa2dim}) with Eq.~(\ref{ofasp}).

Summarizing, although the two-dimensional model we have considered is 
no more than a
methodical example, it allows one  to draw two important parallels with the
gauge theories. First, the presence of the residual supersymmetry guarantees
the fermion-boson cancellations in certain quantities, both in perturbation
theory and nonperturbatively. Second, some nonrenormalization theorems are
modified by anomalies. Here the effect showed up in the superpotential at one
loop. In the gauge theories similar phenomena manifest themselves
in the effective low-energy superpotentials
nonperturbatively.

\subsection{Instanton tunneling transitions in gauge theories -- 
continuation to  the Euclidean space}\label{sec43}

The soliton examples above teach us that there exist  certain coordinate 
dependent background fields in which one half of supersymmetry
is preserved. In such magic backgrounds the quantum corrections 
vanish, order by order. Instantons in four-\-dimensional gauge 
theories belong to this class too. Physical interpretation of solitons and instantons
is very different, however. Solitons are extended objects that have a
particle-like interpretation in the Minkowski formulation of the theory. 
Instantons, on the other hand, are related to the tunneling amplitudes connecting
the vacuum state to itself. In the gauge theories this is the main source of
nonperturbative physics  shaping the vacuum structure. 

In the semiclassical treatment of the tunneling transitions instantons present the
extremal trajectories (classical solutions) in the {\em imaginary} time.  Thus, the
analytical continuation to the imaginary time becomes necessary. The theory in
the imaginary time often can be formulated as a
field theory in the Euclidean space. This is the standard starting point of the
instanton practitioners. However, the Euclidean formulation does not exist in the
${\cal N}  =1$ SUSY theories  because they contain the Weyl (or Majorana) 
fermions.   The easy way to see that this is the case is  as follows.  One 
observes  that it is impossible to find four real four-by-four matrices
with the algebra $\{\gamma_\mu\gamma_\nu\}=\delta_{\mu\nu}$
necessary for constructing the Euclidean version of the theory with 
the Majorana spinors. Only the theories with the extended 
superalgebras, ${\cal N} =2$ or 4, where all spinor fields can be written in the
Dirac form,   admit  the Euclidean   formulation \cite{Zumino}.  

The main statement of the present section is that by no means the Euclidean  
formulation of the theory is  necessary for the imaginary time analysis.
We start with  the original (Minkowski) formulation of 
the theory and show how it defines the imaginary time continuation.
To this end let
us turn to the basics of the functional-integral representation 
\cite{Bere}. It starts from a quantum Hamiltonian $\hat H(\hat p_i,\hat \phi_i,
\hat\psi_k,\hat{\bar\psi_k})$ which is an operator function of 
 the bosonic coordinates
$\hat\phi_i$ and their  conjugate momenta $\hat p_i$, and the fermionic
variables
$\hat\psi_k$ and $\hat{\bar\psi_k}$. The hat marks the operators. Then one
can represent the evolution  operator $\exp(-i\hat H T)$ as the functional
integral
\begin{eqnarray}
&&\langle \phi_{\rm out},  \psi| \exp(-i\hat HT) |\phi_{\rm in}, {\bar\psi}\rangle
=
\int\prod_{i,t} {\rm d}\phi_i(t){\rm d} p_i(t) \prod_{k,t} {\rm d}\psi_k(t) {\rm
d}\bar\psi_k(t)\nonumber\\[0.2cm]
&&~~~\times\exp
\left\{ i\int_0^T \!{\rm d}t \left[ \sum \left(p_j\dot{\phi}_j
+i\bar\psi_k\dot\psi_k\right) - H(p(t),
\phi(t),\psi(t),\bar\psi(t)) \right]\right\}
\,,
\label{repfi}
\end{eqnarray}
where $H$ is the symbol of the operator $\hat H$ corresponding to a certain
operator ordering, and the integration runs over all trajectories
with the given boundary conditions.  The variables $\phi$, $p$ are the
$c$-numbers while $\psi$, $\bar\psi$ are the anticommuting Grassmann
numbers. 

Now, the Euclidean continuation reduces to the rotation of the time 
parameter, $T\to -i\tau$. Instead of $\exp(-i\hat HT)$ we consider $\exp(-\hat
H\tau)$,
\begin{eqnarray}
&&\langle \phi_{\rm out},  \psi| \exp(-\hat H\tau) |\phi_{\rm in},
{\bar\psi}\rangle =
\int\prod_{j,t} {\rm d}\phi_j(t){\rm d} p_j(t) \prod_{k,t} {\rm d}\psi_k(t) {\rm
d}\bar\psi_k(t)\nonumber\\[0.2cm]
&&\times\exp
\left\{ -\int_0^\tau \!{\rm d}t \left[ -i \sum \left(p_j\dot{\phi}_j
+i\bar\psi_k\dot\psi_k\right) + H(p(t),
\phi(t),\psi(t)\bar\psi(t)) \right]
\right\}
\,.
\label{repfie}
\end{eqnarray}
No redefinition of fields is made, the integration runs over the same variables
as in Eq.\ (\ref{repfi}). The Hamiltonian symbol is also the same.
For the bosonic variables one can perform the integration over
$p_j$ (assuming the quadratic dependence of the Hamiltonian on $p$). One then 
arrives at the standard functional integral over
$\phi$ with the exponent
\begin{equation}
- \int_0^\tau \!{\rm d}t \left[  \sum \dot{\phi}_j^2
 + H_{\rm bos}(i\dot\phi(t),\phi(t)) \right]=-\int \!{\rm d}^4 x{\cal L}_{\rm Eucl}
\,. 
\end{equation}
Here ${\cal L}_{\rm Eucl}$ is an invariant Euclidean Lagrangian, it is obtained
from the original Minkowski Lagrangian by the  substitution 
\begin{equation}
t=-ix_4\,.
\end{equation}

As for the fermion part we leave it as is: there is no way to cast it in the
explicitly invariant Euclidean form. The problem is due to the fact that the
fermionic integration runs over the holomorphic variables,  and the operation of
involution (i.e. complex conjugation) which relates $\psi$ and $\bar\psi$ has no 
Euclidean analog.  In spite of this the representation~(\ref{repfie}) is fully suitable 
for the semiclassical analysis of the tunneling transitions.
With this representation in hands,
we can find the extremal trajectories (both,
bosonic and fermionic),  solving  the classical equations of motion. 
To illustrate the procedure we will consider below the BPST instanton and the
gluino zero modes in  SUSY gluodynamics.

\subsubsection{Instanton solution in the spinor notation}
\label{issn}

Here we develop the spinorial formalism in application to  the instantons.
All conventional vectorial indices are replaced by spinorial. It is particularly
convenient in SUSY theories where bosons and fermions are related. An
additional bonus is that there is no need to introduce the 't Hooft symbols.

The spinor notation introduced in Sec.\ \ref{sec30} is based on
SU(2)$_L\times$SU(2)$_R$ algebra of the Lorentz group (the undotted and
dotted
indices, correspondingly). In the Minkowski space the SU(2) subalgebras are
related by the complex conjugation (involution). In particular, this allows one to
define the notion of a real vector,
$(A_{\alpha\dot\beta})^*=A_{\beta\dot\alpha}$. As it was mentioned above, the
property of involution is lost after the continuation to imaginary time.  For the
same reason the notion of the Majorana spinors is also lost.

Consider the simplest non-Abelian gauge theory -- supersymmetric SU(2)
gluodynamics.  The Lagrangian is given in Eq.\ (\ref{ld}).  
As was  explained above, the classical equations are the same as in the
Minkowski space with the substitution $t=-ix_4$ (no substitution is made for the
fields). In particular, the duality equation has the form
\begin{equation}
\bar G_{\dot \alpha\dot\beta} \equiv-\frac 12
G_{\mu\nu}
(\bar\sigma^{\mu})_{\dot\alpha\gamma}(\sigma^{\nu})^\gamma_{\dot\beta}
 \equiv  ({E}^a +i\,{B}^a)({\bar \sigma}^a)_{\dot\alpha\dot\beta}=0
\, .
\label{dopone}
\end{equation}
Here we introduce  two triplets of matrices 
\begin{equation}
(\sigma^a)_\alpha^\beta=(\tau^a)_{\alpha\beta}\,, \qquad
(\bar \sigma^a)^{\dot \alpha}_{\dot\beta}=(\tau^a)_{\dot\alpha\dot\beta}
\,,\qquad a=1,2,3\,,
\end{equation}
which represent the generators of SU(2)$_L$ and  SU(2)$_R$  subalgebras of the
Lorentz group. This expression shows that $\bar G_{\dot \alpha\dot\beta}$ is the  (0, 1)
representation of this group, while  $G_{\alpha\beta}$ defined similarly is 
the (1, 0) representation.

The four-potential which is the solution of Eq.\ (\ref{dopone}) is
\begin{equation}
A_{\beta\dot\beta}^{\{\alpha\gamma\}} = - 2i \,\frac{1}{x^2+\rho^2}
\left( \delta^\alpha_\beta x^\gamma_{\dot\beta} + 
\delta^\gamma_\beta 
x^\alpha_{\dot\beta} \right)\, .
\label{instspinn}
\end{equation}
Where is the familiar color index $a=1,2,3$?  It is traded in for two spinorial
indices
$\{\alpha\gamma\}$,
\begin{equation}
 A^{\{\alpha\gamma\}}\equiv {A}^a \,
({i\tau_2\tau^a})_{\alpha\gamma}\, .
\label{trade}
\end{equation}
The symmetric in $\alpha$, $\gamma$ tensor $A^{\{\alpha\gamma\}}$ is the
adjoint representation of the color SU(2). The instanton presents a
hedgehog configuration which is invariant under the simultaneous  rotations
in the SU(2)$_{\rm color}$  and SU(2)$_L$ spaces, this invariance is explicit in
Eq.\ (\ref{instspinn}). We will do the trade-in (\ref{trade})
frequently. The braces will remind us  that this symmetric pair of
spinorial indices is  connected with the color index $a$. 

All the definitions above are obviously taken from the Minkowski 
space. The Euclidean aspect of the problem reveals itself only in the fact 
that $x_0$ is purely imaginary.  As a concession to the Euclidean nature 
of the instantons we will consistently imply that
\begin{equation}
x^2 \equiv -x_\mu x^\mu = \vec{x}^2 -x_0^2 = \vec{x}^2 +x_4^2\, .
\label{defx2}
\end{equation}
The minus sign in Eq.\ (\ref{defx2}) is by no means necessary; it is 
obviously a compromise, which turns out to be rather  convenient, 
though. 

We can check that the field configuration (\ref{instspinn}) 
reduces to the standard anti-instanton of Belavin {\em et al.} \cite{BPST}. Indeed,
\begin{equation}
A^{ a}_\mu = \frac{1}{4} 
A_{\beta\dot\beta}^{\alpha\gamma}(-i \tau^a \tau_2)_{\gamma\alpha}(
\bar\sigma_\mu )^{\dot\beta\beta}=
\left\{ \begin{array}{ll}
2i\,{x^a} (x^2 +\rho^2 )^{-1}\qquad\qquad\qquad~~\mbox{at}\,\,\, \mu = 0\,  ,
\\[0.2cm]
2\, (\epsilon^{amj}x^j - \delta^{am} x_4)\, (x^2 +\rho^2 )^{-1}\,\,\, 
\mbox{at}\,\,\, \mu = m\, .
\end{array}
\right. 
\label{winst}
\end{equation}
This can be seen to be the standard anti-instanton solution (in the 
non-singular gauge), provided 
one takes into account the fact that 
$$
{A}^a_0 = i A^a_4\, .
$$
Let us stress that it is $A_\mu$ with the {\em lower vectorial index} which is
related to the standard Euclidean solution, for further details see Ref.\ \cite{ABC}. 
The  time component of $A_{\mu}^{a}$ in Eq.\ (\ref{winst}) is purely  imaginary. 
This is alright -- in fact, $A_0$ is not the integration variable  in the canonical
representation~(\ref{repfie}). The spatial components
$A_m^a$ are real.

From Eq.\ (\ref{instspinn}) it is not difficult 
to get the anti-instanton gluon field strength tensor
\begin{eqnarray}
G_{\alpha\beta}^{\{\gamma\delta\}} &\equiv&\!\!-\frac 12
G_{\mu\nu}^{\{\gamma\delta\}}
(\sigma^{\mu})_{\alpha\dot\gamma}(\bar\sigma^{\nu})^{\dot\gamma}_{\beta}
 \equiv  ({E}^j
-i\,{B}^j)^{\{\gamma\delta\}}({ \sigma}^j)_{\alpha\beta}\nonumber\\[0.2cm]
&=&\!\!
8i \,\left(\delta^\gamma_\alpha
\delta^\delta_\beta + \delta^\delta_\alpha
\delta^\gamma_\beta
\right)\, \frac{\rho^2}{(x^2+\rho^2)^2}\, . 
\label{doptwo}
\end{eqnarray}
This expression  implies that
\begin{equation}
E^a_n = 4 i\, \delta^a_n\,\frac{\rho^2}{ (x^2 + \rho^2 )^{2}}\, , \qquad
B^a_n= -4\, \delta^a_n\,\frac{\rho^2}{ (x^2 + \rho^2 )^{2}}\, .
\end{equation}

This completes the construction of the anti-instanton. As for the instanton
it presents the solution of the constraint $G_{\alpha\beta}=0$, it can be obtained by
the replacement of all dotted indices by undotted and {\em vice versa}. 

The advantages of the approach presented here become fully 
apparent when the fermion fields are included. Below we  briefly 
discuss the impact of the fermion fields in SU(2) supersymmetric 
gluodynamics.

The supersymmetry transformations in SUSY gluodynamics take the 
form
\begin{equation}
\delta\lambda^a_\alpha = G_{\alpha\beta}^a\, \varepsilon^\beta\, , 
\qquad
\delta\bar\lambda^a_{\dot\alpha} = \bar
G_{\dot\alpha\dot\beta}^a\,\bar{\varepsilon}^{\dot\beta}\, .
\label{sstSSGL}
\end{equation}
Since in the anti-instanton background $\bar G_{\dot\alpha\dot\beta} 
=0$, the supertransformations with the dotted parameter 
$\bar{\varepsilon}^{\dot\beta}$ do not act on the background field.
Thus, one half  of SUSY is preserved, much in the same way it occurs
in the problem of the domain walls, although  in the instanton 
problem it is a different combination of the supercharges that does not act on 
the classical  solution. 

On the other hand, the supertransformations with the undotted 
parameter $\varepsilon^\beta$ do act nontrivially. 
When applied to the gluon background field, they create two 
fermion  zero modes,
\begin{equation}
\lambda_{\alpha(\beta)}^{\{\gamma\delta\}} \propto
G_{\alpha\beta}^{\{\gamma\delta\}} \propto
\left(\delta^\gamma_\alpha
\delta^\delta_\beta + \delta^\delta_\alpha
\delta^\gamma_\beta
\right)\, \frac{\rho^2}{(x^2+\rho^2)^2}\, ,
\label{sszm}
\end{equation}
the subscript $(\beta)=1,2$ performs numeration of  the zero modes.

These two zero modes are built basing on  supersymmetry, hence they  are
called {\em supersymmetric}.  Somewhat less obvious is the existence of 
two extra zero modes. They are related to the superconformal transformations
(see Sec.\ \ref{sec44} for more explanations) and  are called {\em
superconformal}. The superconformal  transformations 
have the same form as in Eq.\ (\ref{sstSSGL}), with the parameter $\varepsilon$
substituted by a linear function of the coordinates $x_\mu$
\begin{equation}
\varepsilon^\alpha\to x^{\alpha}_{\dot\gamma}\,\bar\beta^{\dot\gamma}\,.
\end{equation}
In this way we get
\begin{equation}
\lambda_{\alpha(\dot\gamma)}^{\{\gamma\delta\}} \propto
G_{\alpha\beta}^{\{\gamma\delta\}}\, x^{\beta}_{\dot\gamma}\propto
\left(\delta^\gamma_\alpha
x^\delta_{\dot\gamma} + \delta^\delta_\alpha
x^\gamma_{\dot\gamma}
\right)\, \frac{\rho^2}{(x^2+\rho^2)^2}\, ,
\label{sczm}
\end{equation}
where the subscript $(\dot \gamma) =1,2$ enumerates two
modes.

Thus, we constructed four zero modes, in full accord with the index 
theorem following from the chiral anomaly~(\ref{3anom}). It is instructive to verify
that they  satisfy the Dirac equation ${\cal
D}_{\alpha\dot\alpha}\lambda^{\alpha}=0$. For  the supersymmetric zero
modes (\ref{sszm}) this equation  reduces to the equation ${\cal D}^\mu
G_{\mu\nu}=0$ for the instanton field. As for the superconformal modes
(\ref{sczm}) the additional term containing
$\partial^{\alpha}_{\dot\alpha} x^\beta_{\dot\gamma}\propto
\epsilon^{\alpha\beta}\epsilon_{\dot\alpha\dot\gamma}$ vanishes upon
contraction with $G_{\alpha\beta}$.

All four zero modes are chiral (left-handed). There are no 
right-handed zero modes on the anti-instanton, i.e. the equation ${\cal
D}_{\alpha\dot\alpha}\bar\lambda^{\dot\alpha}=0$ has no solution.
This is another manifestation of the loss of involution, the operator 
${\cal D}_{\alpha\dot\alpha}$ ceases to be Hermitian.

We use the anti-instanton field as a reference point throughout the review.
In the instanton field the roles of $\lambda$ and $\bar\lambda$ interchange,
together with the dotted and undotted indices.

This concludes our explanatory remarks regarding the analytic 
continuation necessary in developing instanton calculus in ${\cal N} 
=1$ SUSY gauge theories. We hope that the basics of the formalism are 
now  clear.  Some further aspects related to the proper introduction of the 
collective coordinates and the cancellation of non-zero modes are 
elucidated in the following sections. 

\subsection{Instanton calculus in  supersymmetry}
\label{sec44}

In this section we discuss the basic elements of instanton calculus in 
supersymmetric gauge theories -- ``ABC of superinstantons".
These elements are: collective coordinates (instanton moduli) both for the gauge
and matter fields, the instanton measure in the moduli space, the cancellation of the
quantum corrections. In the recent years instanton calculus evolved in a rather
contrived formalism,  especially in the multi-instanton
problems~\cite{DKMS}.  Our task is limited to basics. We will focus on the
one-instanton calculations in various models. The
presentation in  this section essentially follows Refs.\ \cite{NSVZ3,Yung}.

\subsubsection{Collective coordinates}
\label{ssecCC}

The instanton solution~(\ref{instspinn}) has only one collective 
coordinate, the instanton size $\rho$. In fact, the classical BPST 
instanton depends on eight collective coordinates: the
instanton size $\rho$, its center $(x_0)_\mu$, and three angles that describe the
orientation of the instanton in one of the SU(2) 
subgroups of the Lorentz group (or, equivalently, in the SU(2) color space). 
If the gauge group is larger than SU(2),
there are additional coordinates describing the embedding of the instantonic 
SU(2) ``corner" in the full gauge group $G$, we will treat them separately.

The procedure allowing one to introduce all  these eight coordinates is very well
known (see \cite{ABC});  here our focus is mainly on the Grassmann collective
coordinates and on the way supersymmetry acts in the space of 
the collective coordinates.

The general strategy is as follows. One starts from finding all symmetries 
of classical field equations. These symmetries form some group ${\cal G}$.
The next step is to consider a particular classical solution (instanton). 
This solution defines  a {\em stationary group} ${\cal H}$ of transformations
-- i.e. those that  
 act trivially,  do not change the original solution at all. 
It is evident that ${\cal H}$ is a subgroup of ${\cal G}$.
The space of the collective coordinates is determined by the quotient ${\cal
G}/{\cal H}$. Construction of this quotient is a convenient way of introducing the
collective coordinates.  

An example of the transformation belonging to the stationary subgroup ${\cal H}$
for the anti-instanton~(\ref{instspinn}) is  the SU(2)$_R$ subgroup of the
Lorentz group. An example of the transformations acting nontrivially is given
by 
 the four-dimensional translations which are a part of the group ${\cal G}$.

An important comment is in order here. In  SUSY gluodynamics the construction
sketched above generates the full one-instanton moduli space. However, in the
multi-instanton problem, or in the presence of matter, some extra moduli
appear which are not tractable via the classical symmetries. An example
of this type is the 't Hooft zero mode for the matter fermions. Even in such
situations  supersymmetry acts on these extra moduli in a certain way, and we
will study  the issue below.

Following the program outlined above let us start from identifying the symmetry 
group ${\cal G}$ of the classical equations in SUSY gluodynamics.
The obvious symmetry is the Poincar\'e invariance extended to SUSY by inclusion 
of 
the supercharges $Q_\alpha$, $\bar Q_{\dot \alpha}$. The Poincar\'e 
group includes translations $P_{\alpha\dot\alpha}$, and the Lorentz rotations
$M_{\alpha\beta}$, $\bar M_{\dot\alpha\dot\beta}$. Additionally the fermions
bring in the chiral rotation $\Pi$. 

In fact, the classical Lagrangian (\ref{ld})  has a 
wider symmetry -- the superconformal group (a  pedagogical discussion
of the superconformal group can be found in \cite{sconfrev}).  The additional
generators are dilatation $D$, special conformal transformations
$K_{\alpha\dot\alpha}$, and superconformal transformations $S_{\alpha}$ and
$\bar{S}_{\dot\alpha}$. 

Thus, the superconformal algebra includes sixteen bosonic generators 
and eight fermionic.  They all are of the geometric nature -- they can be realized as 
transformations of the coordinates in the superspace. Correspondingly,
the generators  are presented as the differential operators acting in the superspace,
in particular,
\begin{eqnarray}
&&P_{\alpha\dot\alpha}=i\partial_{\alpha\dot\alpha}\,,\qquad
\bar M_{\dot\alpha\dot\beta}=- \frac 12 x^\gamma_{\{
\dot\alpha}\partial_{\gamma\dot\beta\}}-\bar\theta_{\{
\dot\alpha}\frac{\partial}{\partial\bar\theta^{
\dot\beta\}}}\,,\nonumber\\[0.2cm]
&&
D=\frac{i}{2}\left[
x^{\alpha\dot\alpha}\partial_{\alpha\dot\alpha}+\theta^{
\alpha}\frac{\partial}{\partial\theta^{
\alpha}}+\bar\theta^{
\dot\alpha}\frac{\partial}{\partial\bar\theta^{
\dot\alpha}}\right]\,, \qquad \Pi=\theta^{
\alpha}\frac{\partial}{\partial\theta^{
\alpha}}-\bar\theta^{
\dot\alpha}\frac{\partial}{\partial\bar\theta^{
\dot\alpha}}\,,\nonumber\\[0.2cm]
&&
 Q_\alpha=-i\frac{\partial}{
\partial
\theta^\alpha}  +\bar{\theta}^{\dot\alpha}\partial_{\alpha\dot\alpha}\,, \qquad
\bar Q_{\dot\alpha}=i\frac{\partial}{
\partial
\bar\theta^{\dot\alpha}} 
-{\theta}^{\alpha}\partial_{\alpha\dot\alpha}\,, \nonumber\\[0.2cm]
&&
S_{\alpha}= -(x_R)_{\alpha\dot\alpha} \bar Q^{\dot\alpha} -2\theta^2
D_\alpha \,, \qquad 
\bar S_{\dot\alpha}=-(x_L)_{\alpha\dot\alpha} Q^\alpha+ 2\bar\theta^2 \bar
D_{\dot\alpha}\,.
\label{difform}
\end{eqnarray}
Here the symmetrization in $\dot\alpha$, $\dot\beta$ is marked by the braces.
The generators as written above act  on the superspace coordinates. In
application to the fields the generators should be supplemented by some extra
terms (e.g. the spin term in $\bar M$, the conformal weight in $D$).

The differential realization~(\ref{difform}) allows one to establish the full set of the
(anti)\-commutation relations in the superconformal group. The  set  can be
found$\,$\footnote{Warning: our normalization of some generators differs from that
in Ref.\ \cite{sconfrev}.} in~\cite{sconfrev}. What we will need for the
supersymmetry transformations of the collective coordinates is the commutators of
the supercharges with all generators,
\begin{eqnarray}
&&\{ Q_\alpha\,,\bar{Q}_{\dot\beta} \}= 2 P_{\alpha\dot\beta} \, ,\,\,\,
\{ Q_\alpha\,,\bar{S}_{\dot\beta} \} =0 \, ,\,\,\,
\{ \bar{Q}_{\dot\beta}\,,
\bar{S}_{\dot\alpha}\}=-4i\bar M_{\dot\alpha\dot\beta}
+2D\,\epsilon_{\dot\alpha\dot\beta}+3i\,\Pi \,
\epsilon_{\dot\alpha\dot\beta}\, ,\nonumber\\[0.2cm]
&&[ Q_\alpha\,,D] =\frac{i}{2}\,Q_\alpha\, ,\,\,\,[ \bar Q_{\dot\alpha}\,,D]
=\frac{i}{2}\,\bar Q_{\dot\alpha}\, ,\,,\quad [ Q_\alpha\,,\Pi] =Q_\alpha\,,\quad [\bar Q_{\dot\alpha}\,,\Pi] =-\bar
Q_{\dot\alpha}\nonumber\\[0.2cm]
&& [ Q_\alpha\,,M_{\beta\gamma}]
=-\frac 12
\,(Q_\beta\,\epsilon_{\alpha\gamma}+Q_\gamma\,\epsilon_{\alpha\beta})
\,,\quad [\bar Q_{\dot\alpha}\,,M_{\beta\gamma}]=0\,,
\nonumber\\[0.2cm]
&&[ Q_\alpha\,,K_{\beta\dot\beta}]=2i\,\epsilon_{\alpha\beta}\,\bar S_{\dot \beta}
\,.
\label{sccrel}
\end{eqnarray}

Now, what is the stationary group ${\cal H}$ for the anti-instanton
solution~(\ref{instspinn})? 
This bosonic solution is obviously invariant under the chiral transformation
$\Pi$ which acts only on fermions. Besides, the transformations
$K_{\alpha\dot\alpha}+(\rho^2/2)P_{\alpha\dot\alpha}$ does not act  on this
solution. The subtlety to be taken into account is that this and other similar
statements are valid modulo the gauge transformations. A simple way to verify
that $K_{\alpha\dot\alpha}+(\rho^2/2)P_{\alpha\dot\alpha}$ does not act is to
apply it to the gauge invariant objects like Tr$\,G_{\alpha\beta}
G_{\gamma\delta}$. Another possibility is to observe that the conformal
transformation is a combination of translation and inversion. Under the inversion
the instanton  in the regular gauge becomes the very same instanton in the
singular gauge. 

Unraveling the gauge transformations is particularly  important for the
instanton orientations. At first glance, it seems that  neither SU(2)$_R$ nor
SU(2)$_L$ Lorentz rotations act on the instanton solution. The
expression~(\ref{doptwo}) for the gluon field strength tensor contains  no dotted
indices, which explains the first part of the statement, while the  SU(2)$_L$
rotations of $G_{\alpha\beta}$ can be compensated by those in the gauge group.
This conclusion is misleading, however. In Sec.\
\ref{secORIENT} we will show that the instanton orientations are coupled to 
the SU(2)$_R$ Lorentz rotations, i.e.\ to $\bar M_{\dot\alpha\dot\beta}$
generators, while the SU(2)$_L$ rotations are indeed 
compensated by the gauge
transformations.

Thus, we count eight bosonic generators of the stationary group ${\cal H}$.
It contains also four fermionic generators $\bar Q_{\dot \alpha}$ and $S_\alpha$.
It is easy to check  that these twelve generators indeed form  a graded algebra.
To help the reader we collected the generators of ${\cal G}$ and ${\cal H}$
in Table~\ref{tabgen}.
\begin{table}
\begin{center}
\begin{tabular}{|c|c|c|}
\hline
~ & ~  & ~   \\[-0.1cm]
Group  & bosonic &  fermionic  \\[0.2cm]\hline
\vspace*{-0.2cm}
~ & ~  & ~  \\
${\cal G}$ &$ P_{\alpha\dot\alpha}\,,\, M_{\alpha\beta}\,,\, \bar
M_{\dot\alpha\dot\beta}\,,\,D\,,\,\Pi\,,\,K_{\alpha\dot\alpha}$&
$Q_\alpha\,, \,\bar Q_{\dot \alpha}\,,\, S_\alpha\,, \,\bar S_{\dot \alpha}$  
\\[0.2cm]\hline
\vspace*{-0.2cm}
~ & ~  & ~  \\ 
${\cal H}$ &$ 
\,\Pi\,,\, K_{\alpha\dot\alpha}+(\rho^2/2)P_{\alpha\dot\alpha}\,, \,
M_{\alpha\beta}$&
$\bar Q_{\dot \alpha}\,,\, S_\alpha$  
\\[0.2cm]\hline
\end{tabular}
\caption{The generators of the classical symmetry group ${\cal G}$ and the
stationary subgroup ${\cal H}$.}\label{tabgen}
\end{center}
\end{table}

Now we are ready to introduce the set of the collective coordinates  (the
instanton moduli) parametrizing the quotient ${\cal G}/{\cal H}$.  To this end 
let us start from the purely bosonic anti-instanton solution~(\ref{instspinn}) of
the size $\rho=1$ and centered at the origin, and apply to it \cite{NSVZ4}
a generalized shift operator ${\cal V}(x_0,\rho,\bar\omega, \theta_0,
\bar\beta )$.
\begin{eqnarray}
\Phi(x,\theta,\bar \theta;\, x_0 ,\rho,\bar\omega, \theta_0, \bar\beta ) \!\!&=&\!\!
{\cal V}(x_0 ,\rho,\bar\omega, \theta_0, \bar\beta  )\,
\Phi_0(x,\theta,\bar \theta)\,,\nonumber\\[0.2cm]
{\cal V}(x_0 ,\rho,\bar\omega, \theta_0, \bar\beta )
\!\!&=&\!\! e^{iPx_0}e^{-iQ \theta_0}e^{-i\bar{S}\bar\beta}e^{i\bar M \bar\omega}
e^{iD\ln\rho}\,\,  ,
\label{genshift}
\end{eqnarray}
where $\Phi_0(x,\theta,\bar \theta)$ is a 
superfield constructed from the
original bosonic solution~(\ref{instspinn}). Moreover,
$P_{\alpha\dot\alpha},Q_\alpha, \bar{S}_{\dot \alpha},\bar M_{\alpha\beta},D$
are the generators in the differential form~(\ref{difform}) (plus
nonderivative terms related to the conformal weights and spins of the fields) . The
differential representation appears since we deal with the classical fields, in the
operator language the action of the operators at hand would  correspond to the
standard commutators, e.g.
$[P_{\alpha\dot\alpha},\Phi]=i\partial_{\alpha\dot\alpha}\Phi$.

To illustrate how the generalized shift operator ${\cal V}$ acts we apply it to  the
superfield Tr$\,W^2$,
\begin{equation}
{\rm Tr} \,(W^\alpha W_\alpha )_0= \theta^2\, 
\frac{96}{(x^2 
+1)^4}= {\theta^2}\,  \frac{96}{(x^2_L +1)^4}\, . 
\label{w2bpst}
\end{equation}
Applying ${\cal V}$ to this expression one gets
\begin{equation}
{\rm Tr} \,(W^\alpha W_\alpha )={\cal V}(x_0 ,\rho,\bar\omega, \theta_0,
\bar\beta )\, 
\frac{96\,\theta^2}{(x^2_L +1)^4}
 =  
\frac{96\,\tilde\theta^2\,\rho^4}{[(x_L-x_0)^2 +\rho^2]^4}\, ,
\label{w2shifted}
\end{equation}
where 
\begin{equation}
\tilde{\theta}_\alpha = (\theta -\theta_0)_\alpha +(x_L-
x_0)_{\alpha\dot\alpha}\bar{\beta}^{\dot\alpha}\, . 
\label{tildeth}
\end{equation}
In deriving this expression we used the representation~(\ref{difform}) for the
generators. Note that the generators $\bar M$ act trivially on the Lorentz scalar
$W^2$. As for the dilatation $D$, the nonderivative term should be added to account
for the nonvanishing dimension of $W^2$, equal to 3.
The value of Tr$W^2$ 
depends on the variables $x_L$ and $\theta$ and on the moduli
$x_0$, $\rho$, $\theta_0$ and $\bar\beta$. It does not depend on $\bar\omega$
because we consider the Lorentz and color singlet.

Of course, the most detailed information is contained in the matrix superfield $V$
which is the supergeneralization of the gauge four-potential. Applying the
generalized shift operator ${\cal V}$ to $V_0$,
\begin{equation}
V_0^{\{\alpha\gamma\}} =4i
 \,\frac{1}{x^2+1}
\left(\theta^\alpha x^\gamma_{\dot\beta}\bar\theta^{\dot\beta} + 
\theta^\gamma x^\alpha_{\dot\beta}\bar\theta^{\dot\beta}
 \right)\, ,
\end{equation}
we obtain a generic instanton configuration which depends on all collective
coordinates. One should keep in mind that, in distinction with Tr$W^2$, the
superfield $V^{\{\alpha\gamma\}}$ is not a gauge-invariant object. Therefore,
the application of ${\cal V}$ should be supplemented by a subsequent gauge
transformation
\begin{equation}
e^V\to e^{i\bar\Lambda}\,e^Ve^{-i\Lambda}\,,
\end{equation}
where the chiral superfield $\Lambda$ must be  chosen in such a way that the
original gauge is maintained.

\subsubsection{The symmetry transformations of the moduli}
\label{ssecstm}

Once all relevant collective coordinates are introduced,
it is natural to pose a question of how the classical symmetry group acts on them.
Although the  complete set of the superconformal transformations of the 
instanton moduli
could be readily  found, we focus on the exact symmetries -- the Poincar\'e
group plus supersymmetry.  Only the exact symmetries are preserved by the
instanton measure, and we will use them for its reconstruction. 

The following consideration
shows how to find the transformation laws for the collective coordinates. Assume
we are interested in translations, $x\to x+a$. The operator generating the
translation is
$\exp (iPa)$. Let us apply it to the configuration 
$\Phi(x,\theta,\bar \theta;\, x_0 ,\rho, \bar\omega, \theta_0, \bar\beta )$, see Eq.\
(\ref{genshift}),
\begin{eqnarray}
e^{iPa}\Phi(x,\theta,\bar \theta;\, x_0 ,\rho,\bar\omega, \theta_0, \bar\beta
)\!\!&=&\!\! e^{iPa} e^{iPx_0}e^{-iQ
\theta_0}e^{-i\bar{S}\bar\beta}e^{i\bar M \bar\omega}
e^{iD\ln\rho}\,\Phi_0(x,\theta,\bar
\theta)\nonumber\\[0.1cm]
\!\!&=&\!\! \Phi(x,\theta,\bar \theta;\, x_0+a
,\rho,\bar\omega, \theta_0, \bar\beta )\,.
\end{eqnarray}
 Thus, we obviously get the original configuration with $x_0$ replaced by $x_0 +
a$ with no change in other collective coordinates.
Alternatively, one can say that the interval $x-x_0$ is an invariant of 
the translations; the  instanton field configuration
must not depend on $x$ and $x_0$ separately, but only on the 
invariant  combinations. 

Passing to supersymmetry, the transformation generated by $\exp
(-iQ\varepsilon)$ is the simplest to deal with,
\begin{equation}
\theta_0\to\theta_0+\varepsilon  \,.
\end{equation}
Other moduli stay intact.  

For the supertranslations with the parameter 
$\bar\varepsilon$, we do the same: act by $\exp (\!-i \bar{Q}\bar\varepsilon )$
on   the configuration $\Phi$,
\begin{equation}
e^{-i\bar{Q}\bar\varepsilon }\Phi(x,\theta,\bar \theta;\, x_0 ,\rho,\bar\omega,
\theta_0, \bar\beta )=e^{-i \bar{Q}\bar\varepsilon } e^{iPx_0}e^{-iQ
\theta_0}e^{-i\bar{S}\bar\beta}e^{i\bar M \bar\omega}
e^{iD\ln\rho}\,\Phi_0(x,\theta,\bar
\theta)\,.
\end{equation}
Our goal is to put $\exp (-i \bar{Q}\bar\varepsilon )$ to the rightmost position --
when $\exp (-i \bar{Q}\bar\varepsilon )$ acts on the original anti-instanton
solution $\Phi_0(x,\theta,\bar \theta)$ it reduces to unity. In the process we get
various commutators listed in Eq.\ (\ref{sccrel}). For instance, the first nontrivial
commutator we encounter is $[\bar{Q}\bar\varepsilon,\,Q \theta_0]$. This
commutator produces $P$ which effectively shifts $x_0$ by $-4i 
\theta_0\bar\varepsilon$. Proceeding further in this way we arrive at  
the following results~\cite{NSVZ4} for the SUSY transformations of the moduli:
\begin{eqnarray}
&&\delta(x_0)_{\alpha\dot\alpha} = -4i 
(\theta_0)_\alpha\bar\varepsilon_{\dot\alpha}\, , \quad
\delta\rho^2 = - 4i(\bar\varepsilon\bar\beta )\rho^2
\,,
\nonumber\\[0.2cm]
&&\delta (\theta_0)_\alpha = \varepsilon_\alpha\, , \quad
\delta\bar\beta_{\dot\alpha} = - 4 i\bar\beta_{\dot\alpha}\,
(\bar\varepsilon\bar\beta 
)\,,
\nonumber\\[0.2cm]
&&\delta\Omega^{\dot \alpha}_{\dot\beta}=4i\left[\bar\varepsilon^{\dot\alpha}\,
\bar\beta_{\dot\gamma}+\frac 1 2 \delta^{\dot\alpha}_{\dot\gamma}\,
(\bar\varepsilon \bar\beta)\right] \Omega^{\dot\gamma}_{\dot\beta}
\,,
\label{strlaws}
\end{eqnarray}
where we have introduced the rotation matrix $\Omega$ defined as
\begin{equation}
 \Omega^{\dot\alpha}_{\dot\beta}=\exp\left[ -i
\bar\omega^{\dot\alpha}_{\dot\beta}\right]\,.
\label{orientM}
\end{equation}
This definition of the rotation matrix $\Omega$ corresponds to the rotation of
spin-1/2 objects. 

Once the transformation laws for the
instanton moduli are established one can construct invariant combinations of these
moduli.  It is easy to verify that such invariants are
\begin{equation}
\frac{\bar\beta}{\rho^2}\,,\qquad {\bar\beta}^2\,F(\rho)\,,
\label{invglue}
\end{equation}
where $F(\rho)$ is an arbitrary function of $\rho$.

{\em A priori}, one could have expected that the above invariants could appear
in the quantum corrections to the instanton measure. In fact, the transformation
properties of the collective coordinates under the chiral U(1) preclude this from
happening. The chiral charges of all fields are given  in Sec.\ \ref{sec34}. In terms
of the collective coordinates it implies that the  chiral charge of $\theta_0$
and $\bar\beta$ is unity while that of $x_0$ and $\rho$ is zero. This means that
the invariants~(\ref{invglue}) are chiral nonsinglets and cannot appear in the
corrections to the measure. 

The chiral U(1) symmetry is anomalous in SUSY gluodynamics. It has a
non-anomalous discrete subgroup $Z_4$, however (see Sec.\ \ref{secEL}).
This subgroup is sufficient to forbid the presence of the invariants~(\ref{invglue})
nonperturbatively. 

A different type of invariants are those built from 
the superspace coordinates and the instanton moduli.
An example from the non-SUSY instanton is the interval $x-x_0$ which is the
invariant of translations. We elevate the notion to superspace.

The first invariant of this type evidently is
\begin{equation}
(\theta-\theta_0)_\alpha
\label{firstinv}
\end{equation}
Furthermore,  $x_L -x_0$ does not change under
translations and under a part of supertransformations generated by $Q_\alpha$.
It does change, however, under the $\bar Q_{\dot\alpha}$ transformations.
Using Eqs.\ (\ref{susytr}) and (\ref{strlaws}) one can built a combination of 
$\theta-\theta_0$ and $x_L -x_0$ that is invariant,
\begin{equation}
\frac{\tilde{\theta}_\alpha}{\rho^2} =\frac{1}{\rho^2}\left[ (\theta
-\theta_0)_\alpha +(x_L-
x_0)_{\alpha\dot\alpha}\bar{\beta}^{\dot\alpha}\right]\, . 
\label{tildeinv}
\end{equation}
The superfield Tr$\,W^2$ given in Eq.\ (\ref{w2shifted}) can be used as a check.
It can be presented as
\begin{equation}
\mbox{Tr}\,W^2=\frac{\tilde{\theta}^2}{\rho^4} \,
F\left(\frac{(x_L-x_0)^2}{\rho^2}\right)
\,.
\label{winv}
\end{equation}
Although the first factor is invariant, the ratio $(x_L-x_0)^2/\rho^2$ is not.
Its variation, however, is proportional to $\tilde{\theta}$; therefore the
product~(\ref{winv}) is invariant (the factor $\tilde{\theta}^2$ acts as
$\delta(\tilde{\theta})$).

\subsubsection{The measure in the moduli space}
\label{ssecmeasure}

Now, that the proper collective coordinates are introduced,
we come to the important ingredient of superinstanton calculus -- the {\em
instanton measure},  or the formula of  integration in the space of the collective
coordinates.  The general procedure of obtaining the measure is well known,
it is based on the path integral representation in the canonical form~(\ref{repfi}).
In terms of the mode expansion this representation reduces to the integration
over the coefficients of the mode expansion. The integration measure splits in
two factors: integration over the zero and nonzero mode coefficients. 
It is just the zero modes' coefficients that are related to moduli.

We follow the route pioneered by 't~Hooft~\cite{Hooft1}. In one-loop
approximation the functional integral, say, over the  scalar field can be written as
\begin{equation}
 \left[\frac{{\rm det} \,(L_2 +M_{\rm PV}^2)}{{\rm det}\, L_2}\right]^{1/2}
\end{equation}
where $L_2$ is the differential operator appearing in the expansion of the
Lagrangian near the given background in the quadratic approximation,
$L_2=-{\cal D}^2$. The numerator is due to the ultraviolet regularization.
Following \cite{Hooft1} we use the Pauli-Villars regularization -- there
is no  alternative in the instanton calculations.  The mass term of the regulator
fields is  $M_{\rm PV}$. Each given  eigenmode of $L_2$ with the eigenvalue
$\epsilon^2$ contributes $M_{\rm PV}/\epsilon$. For the scalar field there are no
zero eigenvalues. However, for the vector and spinor fields
the  zero modes do exist,
the set of the zero modes corresponds to that of moduli, generically to be denoted
in this section 
as $\eta_i$. For the bosonic zero modes the factor $1/\epsilon$
(which, of course,  explodes at $\epsilon\to 0$)  is replaced by the
integration  over the corresponding collective coordinate ${\rm d}\eta^b$, up to a
normalization factor.  Similarly, for the fermion zero mode $\sqrt{\epsilon}\to
{\rm d}\eta^f$, see the discussion below.

 The zero modes can be obtained by differentiating the field
$\Phi(x,\theta,\bar\theta;\eta)$ over the collective coordinates $\eta_i$ at the
generic point in the  space of the instanton moduli. In the instanton problem
$\{\eta_i\}=\{x_0,\rho,\bar\omega,\theta_0,\bar\beta\}$. The derivatives $\partial
\Phi/\partial \eta_i$ differ from the corresponding zero modes by the
normalization. It is just these normalization factors that determine the measure:
\begin{equation}
{\rm d} \mu= e^{-8\pi^2/g^2}\,\prod_i {\rm d}\eta_i^b\,\frac{M_{\rm
PV}}{\sqrt{2\pi}}\,
\left\|\frac{\partial
\Phi(\eta)}{\partial\eta_i^b}\right\|\;
\prod_k
{\rm d}\eta_k^f \, \frac{1}{\sqrt{M_{\rm PV}}}\, \left\|\frac{\partial
\Phi(\eta)}{\partial \eta_i^f}\right\|^{-1}\,,
\label{genmeas}
\end{equation}
where the
norm $\|\Phi\|$ is defined as the square root of the integral over $|\Phi|^2$. 
The superscripts $b$ and $f$ mark the bosonic and fermionic collective
coordinates, respectively.
Note that we
have also included $\exp (-{\cal S})$ in the measure (the instanton action ${\cal
S}=8\pi^2/g^2$). In the expression above it is implied that the zero modes are
orthogonal.  If this is not  the case, which often  happens
in practice,  the measure is given by a more general formula
\begin{equation}
{\rm d} \mu= e^{-8\pi^2/g^2}\,(M_{\rm PV})^{n_b-\frac{1}{2}
n_f}\,(2\pi)^{-n_b/2}\,
\prod_i
{\rm d}\eta_i \, \left\{ \mbox{Ber}\, \left\langle\frac{\partial
\Phi(\eta)}{\partial \eta_j}\bigg|\frac{\partial
\Phi(\eta)}{\partial \eta_k}\right\rangle\right\}^{\frac{1}{2}}\,,
\label{genmeas1}
\end{equation}
where Ber stands for the Berezinian (superdeterminant).
The normalization of the fields is fixed by the requirement that their kinetic
terms are canonic.

We pose here to make a remark  regarding the fermion part of the measure.
The fermion part of the Lagrangian is $i\lambda^{\alpha}
{\cal D}_{\alpha\dot\alpha}\bar\lambda^{\dot\alpha}$. For the mode
expansion of the field $\lambda^{\alpha}$  it is convenient to use the Hermitian
operator 
 \begin{equation}
(L_2)^\alpha_\beta=-{\cal D}^{\alpha\dot\alpha}{\cal D}_{\beta\dot\alpha}\,,
\qquad L_2\,\lambda=\epsilon^2 \lambda
\,.
\label{quadop}
\end{equation}
The operator determining the $\bar\lambda$ modes is
\begin{equation}
(\tilde L_2)^{\dot\alpha}_{\dot\beta}=-{\cal D}^{\alpha\dot\alpha}{\cal
D}_{\alpha\dot\beta}\,,
\qquad \tilde L_2\,\bar\lambda=\epsilon^2 \bar\lambda
\,.
\end{equation}
The operators $(L_2)^\alpha_\beta$ and $(\tilde L_2)^{\dot\alpha}_{\dot\beta}$
are not identical -- we have already encountered with a similar situation
in the analysis of the solitons, Sec.\ \ref{sec424}.

In the anti-instanton background the operator $L_2$ has four zero modes,
discussed above, while $\tilde L_2$ has none. As for the nonzero modes, they are 
degenerate and related as follows:
\begin{equation}
\bar\lambda^{\dot\alpha}=\frac{i}{\epsilon}\,{\cal D}^{\dot\alpha}_\alpha\,
\lambda^\alpha
\,.
\end{equation}
The parallel with the discussion of the fermion modes in the
two-dimensional soliton problem in  Sec.\  \ref{sec424} is quite transparent. 

Taking into account the relations above, we find that the  modes with the given
$\epsilon$ contribute the term $\epsilon\int \!{\rm d}^4 x \,\lambda^2$ into the
fermion part of the action. For the given mode
$\lambda^2=\epsilon^{\alpha\beta}\lambda_\beta\lambda_\alpha$ vanishes,
literally speaking. However, there are two modes, $\lambda_{(1)}$ and
$\lambda_{(2)}$, for each given $\epsilon$ and it is, actually, the product 
$\lambda_{(1)}\lambda_{(2)}$ that enters. This consideration provides us with the
definition of the norm matrix for the fermion zero modes,
\begin{equation}
\int \!{\rm d}^4 x \,\lambda_{(i)}\lambda_{(j)}\,,
\end{equation}
which should be used in calculating the Berezinian.

The norm factors depend on
$\eta_i$, generally speaking; Eq.\ (\ref{genmeas1}) gives the measure at any point
of the
instanton  moduli space. Thus, the relation~(\ref{genmeas1}) conceptually solves
the problem of the construction of the measure.  

In practice, the measure comes out simple in certain points on the moduli space.
For instance, instanton calculus always starts from the purely bosonic instanton.
Then, to reconstruct the measure everywhere on the
instanton  moduli space one can apply
the exact symmetries of the theory. By exact we mean those symmetries which
are preserved at the quantum level -- the Poincar\'e symmetries plus 
supersymmetry, in the case at hand, rather than the full superconformal group.
As we will see, this is sufficient  to get the full measure in supersymmetric
gluodynamics but not in theories with matter.

In non-SUSY  gluodynamics the measure  was found in~\cite{Hooft1}.
Let us briefly remind 't~Hooft's construction, then we will add the 
 fermion part specific to SUSY gluodynamics. 

\vspace*{0.2cm}
\noindent
{\em Translations}

\vspace*{0.1cm}

\noindent
The translational zero modes are obtained by differentiating the instanton field
$A_\nu/g$ over $(x_0)_\mu$ where $\mu$ performs the numeration of the
modes: there are four of them.  The factor $1/g$ reflects the transition to the
canonically normalized field,
 a requirement mentioned after Eq.\ (\ref{genmeas1}).
Up to the sign,  differentiation over $(x_0)_\mu$  is the same as differentiation
over $x_\mu$. The field
$a^{(\mu)}_\nu=g^{-1}\partial_\mu A_\nu$ obtained in this way does not satisfy
the gauge condition ${\cal D}^\nu a_\nu=0$. Therefore, it must be supplemented by
a gauge transformation, $\delta a_\nu=g^{-1}{\cal D}_\nu \varphi$. In the case at
hand  the gauge function $\varphi^{(\mu)}=-A_\mu$. As a result, the translational
zero modes take the form
\begin{equation}
a^{(\mu)}_\nu=g^{-1}\left(\partial_\mu A_\nu -{\cal D}_\nu
A_\mu\right)=g^{-1}G_{\mu\nu}\,.
\end{equation}
Note that now the gauge condition is satisfied. The norm of each translational
mode is obviously $\sqrt{8\pi^2/g^2}$.

\vspace*{0.2cm}
\noindent
{\em Dilatation}

\vspace*{0.1cm}

\noindent
The dilatational zero mode is 
\begin{equation}
a_\nu=\frac{1}{g}\,\frac{\partial A_\mu}{\partial
\rho}=\frac{1}{g\rho}\,G_{\nu\mu}\, x^\mu\,,
\qquad
\|a_\nu\|=\frac{4\pi}{g}\,.
\end{equation}
The gauge condition is not broken by the differentiation over $\rho$.

\vspace*{0.2cm}
\noindent
{\em Orientations}

\vspace*{0.1cm}

\noindent
The orientation zero modes look as a particular gauge transformation of
$A_\nu$~\cite{Hooft1},
\begin{equation}
(a_\nu)^{\alpha}_\beta=g^{-1}\left({\cal
D}_\nu\,
\Lambda\right)^{\alpha}_\beta
\,,
\label{misha}
\end{equation}
where the spinor notation for color is used and the gauge function $\Lambda$ has
the form
\begin{equation}
\Lambda^{\alpha}_\beta=\left(U\,\bar\omega\, U^T\right)^{\alpha}_\beta=
U^{\alpha}_{\dot\alpha}\,U_{\beta}^{\dot\beta}\,
\bar\omega^{\dot\alpha}_{\dot\beta}\,,
\label{arksix}
\end{equation}
where
\begin{equation}
U_{\dot{\alpha}}^\alpha  = \frac{x_{\dot{\alpha}}^\alpha}{\sqrt{x^2+\rho^2}}\, ,
\label{umatrix}
\end{equation}
and $\bar\omega^{\dot\alpha}_{\dot\beta}$ are three orientation parameters.
It is easy to check that Eqs.\ (\ref{misha}), (\ref{arksix}) do indeed produce 
the normalized zero modes, satisfying the condition ${\cal D}^\nu a_\nu=0$.
The gauge function~(\ref{arksix}) presents special gauge transformations which 
are absent in the topologically trivial sector. 

The procedure that led to the occurrence of $\bar\omega^{\dot\alpha}_{\dot\beta}$
as the orientation collective coordinates is described above, probably, too sketchy.
We will return to the issue of the geometrical meaning of these coordinates  in 
Sec.\ \ref{secORIENT},  after we introduce the matter fields in the fundamental
representation.

Note that the matrix $U$ satisfies the equation
\begin{equation}
{\cal D}^2 U_{\dot{\alpha}}=0\,,
\label{arkfour}
\end{equation}
where the undotted index of $U$ is understood as the color index. 
Correspondingly, ${\cal D}$ in Eq.\ (\ref{arkfour}) acts as the covariant
derivative in the fundamental representation.
 Equation  (\ref{arkfour}) will be exploited
below, in considering the matter fields in the fundamental representation.
Note also that 
\begin{equation}
{\cal D}^2\Lambda=0\, .
\label{arkfive}
\end{equation}

This construction -- making a string built from several
 matrices $U$ -- can be extended to the arbitrary representation
of SU(2).  The representation with spin $j$ is obtained by multiplying $2j$
matrices $U$ in a manner analogous to that exhibited in Eq.\ (\ref{arksix}).

Calculating ${\cal D}_\nu\Lambda$ explicitly, we get the following
expression for the orientation modes and their norm:
\begin{equation}
a_{\beta\dot\beta}^{\{\alpha\gamma\}}=\frac{1}{4g}\,
G_{\beta\sigma}^{\{\alpha\gamma\}}\,x^{\sigma\dot\sigma}\,
\bar\omega_{\dot\sigma\dot\beta}\,,\qquad
\left\|\frac{\partial a_\nu^a}{\partial\bar \omega^b}\right\| =\frac{2\pi\rho}{g}\,.
\end{equation}

\vspace*{0.2cm}
\noindent
{\em Supersymmetric modes}

\vspace*{0.1cm}

\noindent
We started discussing these modes  in Sec.\ \ref{issn},
\begin{equation}
\lambda_{\alpha(\beta)}^{\{\gamma\delta\}} = 
g^{-1} G_{\alpha\beta}^{\{\gamma\delta\}}\,, \qquad 
\left\langle\lambda_{(1)}\,\bigg| \,\lambda_{(2)}\right\rangle
=\frac{32\pi^2}{g^2}\,.
\end{equation}
Up to a numerical matrix, the supersymmetric modes coincide with the
translational ones.  There are four translational modes and  two
supersymmetric. The factor two, the ratio of the numbers of the
bosonic to fermionic modes,  reflects the
difference in the number of the spin components. This is, of course, a natural
consequence of supersymmetry.

\vspace*{0.2cm}
\noindent
{\em Superconformal modes}

\vspace*{0.1cm}

\noindent
These modes were also briefly  discussed in Sec.\ \ref{issn},
\begin{equation}
\lambda_{\alpha(\dot\beta)}^{\{\gamma\delta\}} = 
g^{-1} x^\beta_{\dot \beta} G_{\alpha\beta}^{\{\gamma\delta\}}\,, \qquad 
\left\langle\lambda_{(\dot 1)}\,\bigg| \,\lambda_{(\dot 2)}\right\rangle
=\frac{64\pi^2\rho^2}{g^2}\,.
\end{equation}
The superconformal modes have the same $x\,G$ form as the orientational and
dilatational modes. Again we have four bosonic and two fermionic modes.

\vspace*{0.3cm}

The relevant normalization factors, as well as the
accompanying factors from the regulator fields for all  modes, are collected in
Table~\ref{tabmodes}. 
\begin{table}
\begin{center}
\begin{tabular}{|l|l|}
\hline
~ &  ~   \\[-0.1cm]
Boson modes & Fermion modes\\ 
~ &  ~   \\[-0.1cm]
\hline 
~ &  ~   \\[-0.1cm]
$4\,{\rm T}\,\,\to {\cal S}^2 (2\pi )^{-2} M_{PV}^4 {\rm d}^4x_0$ & $ 
2\,{\rm 
SS}\,\,\to {\cal S}^{-1} (4M_{PV})^{-1} {\rm d}^2\theta_0$  \\
~ &  ~   \\[-0.1cm]
  \hline 
~ &  ~   \\[-0.1cm]
$1\,{\rm D}\,\,\to  {\cal S}^{1/2} (\pi )^{-1/2} M_{PV} {\rm d}\rho $ & $
2\,{\rm SC}\,\,\to {\cal S}^{-1} (8M_{PV})^{-1} \rho^{-2}\,{\rm d}^2\bar\beta$  \\
 ~ &  ~   \\[-0.1cm]
\hline
~ &  ~   \\[-0.1cm]
$3\,{\rm GCR}\,\,\to {\cal S}^{3/2} (\pi )^{1/2}  M_{PV}^3\rho^3$ & $ 2\,{\rm 
MF}\,\,\to (M_{PV})^{-1} (8\pi^2|v|^2\rho^2)^{-1}{\rm d}^2\bar\theta_0$ \\
~ &  ~   \\
\hline
\end{tabular}
\end{center}
\caption{The contribution of the zero modes to the instanton 
measure.  The notation is as follows: 4~T stands for the four translational modes, 
1~D  one  dilatational mode, 3~GCR  three modes associated with the orientations
(the group volume is included), 2~SS  two supersymmetric gluino modes, 2  SC 
two  superconformal gluino modes,  2~MF  two matter fermion  zero  modes, ${\cal
S}\equiv 8\pi^2 / g^2$.}\label{tabmodes}
\end{table}
Assembling all factors  together we get the measure for a specific point in the
moduli space: near the original bosonic anti-instanton solution~(\ref{instspinn}),
\begin{equation}
d\mu_0 = \frac{1}{256\pi^2}\,e^{ -8\pi^2/{g^2}}\, (M_{\rm PV})^6\, \left(
\frac{8\pi^2}{g^2}\right)^2 
\frac{{\rm d}^3 \bar\omega}{8\pi^2}\, {\rm d}^4 x_0 \, 
{\rm d}^2\theta_0 \, {\rm d}\rho^2 \,{\rm d}^2\bar\beta \, . 
\label{fimsg}
\end{equation}

How this measure transforms under the  exact symmetries? First, let us check the
SUSY transformations~(\ref{strlaws}). They imply that ${\rm d}^4 x_0$ and ${\rm
d}^2\theta_0$ are invariant. As for the last two differentials,
\begin{equation}
{\rm d}\rho^2 \to {\rm d}\rho^2\, [1-4i (\bar\varepsilon\bar\beta )]\, , \qquad
{\rm d}^2\bar\beta \to {\rm d}^2\bar\beta \,[1+ 4i (\bar\varepsilon\bar\beta )]
\,, 
\end{equation}
so that the product is invariant too. 

The only noninvariance of the measure~(\ref{fimsg}) is that of ${\rm d}^3
\bar\omega$ under the SU(2)$_R$ Lorentz rotation generated by
$\bar M_{\dot\alpha\dot\beta}$. It is clear that for the generic instanton
orientation $\bar\omega$ the differential
${\rm d}^3\bar \omega$ is replaced by the SU(2) group measure ${\rm
d}^3\Omega_{\rm SU(2)}={\rm d}^3\bar \omega\sqrt{G}$ where $G$ is the
determinant  of the Killing metric on the group SU(2) and the matrix $\Omega$
defined in Eq.\ (\ref{orientM}) is a general element of the group. In
fact, this determinant is a part of the Berezinian in the general
expression~(\ref{genmeas1}). The SU(2) group is compact: the integral over all
orientations yields the volume of the group$\,$\footnote{ Actually, the group of the
instanton orientations is O(3)=SU(2)/$Z_2$ rather than SU(2). This distinction is
unimportant for the algebra, it is important, however, for the group volume.} which
is equal to
$8\pi^2$. Performing this integration we arrive at the final result for the instanton
measure in SU(2) SUSY gluodynamics:
\begin{equation}
{\rm d}\mu_{\,{\rm SU}(2)} = \frac{1}{256\pi^2}\, e^{ -8\pi^2/{g^2}}\,(M_{\rm
PV})^6\, \left(
\frac{8\pi^2}{g^2}
\right)^2 
  {\rm d}^4 x_0 \, 
{\rm d}^2\theta_0 \, {\rm d}\rho^2 \,{\rm d}^2\bar\beta \, . 
\label{fimsg1} 
\end{equation}

Note that the regulator mass $M_{\rm PV}$ can be viewed as a complex parameter.
It appeared from the regularization of the operator~(\ref{quadop}) which has a
certain chirality.

\subsubsection{Including matter: SQCD with one flavor}
\label{ssecmatter} 

Now we extend the analysis of the previous  sections to include matter. A
particular model  to be considered  is SU(2) SQCD with one flavor, see Sec.\
\ref{sec312}. 

In the Higgs phase the instanton configuration is  an approximate solution. A
manifestation of this fact is the  $\rho$ dependence of the classical
action~\cite{Hooft1}. The solution  becomes exact  in the limit $\rho\to 0$.
For future applications only this limit is of importance, as we will see later.
A new feature of the theories with matter is the occurrence of extra fermionic
zero modes in the matter sector, which  gives rise to additional 
collective coordinates. Supersymmetry provides a geometrical meaning to these
collective coordinates.

As above, we  start from a bosonic field configuration  and apply supersymmetry
to build the full instanton  orbit. In this way we find a realization of SUSY in the
instanton moduli space. 

We have already learned that  SQCD with one flavor  classically  has  a 
one-dimen\-sional
$D$-flat direction, 
\begin{equation}
(\phi^\alpha_f)_{\rm vac} = v\,\delta^\alpha_f\,,\qquad 
(\bar\phi^{\,\alpha}_f)_{\rm vac} = \bar v\,\delta^\alpha_f
\label{flatvac}
\end{equation}
where $v$ is an arbitrary complex parameter, the vacuum 
expectation  value of the squark fields. Here $\alpha$ is the color 
index  while $f$  is  the subflavor index, $\alpha , f = 1,2$. The color 
and flavor indices get  entangled, even in the topologically trivial 
sector, although in  a  rather  trivial manner.

What changes occur in the instanton background?  The equation for the scalar field
$\phi^\alpha_f$ becomes
\begin{equation}
{\cal D}_\mu^2 \phi_f = 0\,, \qquad {\cal D}_\mu=\partial_\mu-i A_\mu^a
\tau^a/2
\,.
\end{equation}
Its solution in the anti-instanton background~(\ref{instspinn}) has the form
\begin{equation}
\phi_{\dot{f}}^\alpha  = v \, U_{\dot{f}}^\alpha=
v \,\frac{x_{\dot{f}}^\alpha}{\sqrt{x^2+\rho^2}}\, .
\label{scsol}
\end{equation}
Asymptotically, at $x\to\infty$, 
\begin{equation}
\phi_{\dot{f}}^\alpha\to \tilde U_{\dot{f}}^\alpha v\,,\quad A_\mu\to i
\tilde U\partial_\mu \tilde U^\dagger\,,\quad
\tilde U_{\dot{f}}^\alpha=\frac{x_{\dot{f}}^\alpha}{\sqrt{x^2}}\,,
\end{equation}
i.e. the configuration is gauge equivalent to the flat vacuum~(\ref{flatvac}).
Note that the equation for the field $\bar \phi$ is the same. With
the boundary conditions~(\ref{flatvac}) the solution is
\begin{equation}
\bar\phi_{\dot{f}}^\alpha  =\bar v\, U_{\dot{f}}^\alpha=\bar v
\,\frac{x_{\dot{f}}^\alpha}{\sqrt{x^2+\rho^2}}\, .
\label{scsolbar}
\end{equation}

To generate the full instanton orbit, with all collective coordinates 
switched  on, we again apply all generators of the superconformal group to
the field configuration $\Phi_0$ which presents now a set of the  superfields 
$V_0$, $Q_0$ and $\bar Q_0$. The bosonic components are given in Eqs.\
(\ref{instspinn}),
 (\ref{scsol}) and (\ref{scsolbar}), the fermionic ones vanish.
The superconformal group is still the symmetry group of the classical equations. 
Unlike SUSY gluodynamics now, at $v\neq 0$, all generators act nontrivially.
At first glance one might suspect that one needs to introduce 16+8 collective
coordinates. 

In fact, a part of the generators act nontrivially already in the flat vacuum with 
$v\neq 0$. For instance, the action of $\exp(i\Pi\alpha)$ changes the phase of $v$.
Since we want to consider the theory with the given vacuum state such
transformation should be excluded from the set generating the instanton collective
coordinates. This situation is rather general, for a more detailed  discussion see
Sec.\ \ref{secORIENT} and  Ref.\ \cite{AVMS}.

As a result, the only new collective coordinates to be added are conjugated
to $\bar Q_{\dot\alpha}$.  The differential operators $\bar Q_{\dot\alpha}$,
defined$\,$\footnote{The supercharges and the matter superfields
are denoted by one and the same letter $Q$. We hope that this unfortunate
coincidence will cause no confusion. The indices help to figure out what is meant
in the given context. For the supercharges we usually indicate the spinorial indices
(the Greek letters from the beginning of the alphabet). The matter superfields
carry the flavor indices (the Latin letters). Moreover, $Q_0$ and $\bar Q_0$
(with the subscript 0) is the starting purely bosonic
configuration of the matter superfields.}
in Eq.\ (\ref{diffq}), annihilate $V_0$ (modulo a supergauge
transformation) and $\bar Q_0$. It acts non-trivially on $Q_0$ producing  the 
't~Hooft zero modes of the matter fermions,
\begin{equation}
\bar Q^{\dot\alpha}\, (Q_0)^\alpha_{\dot
f}=-2\theta^\beta\left(\frac{\partial}{\partial
x_L}-i A\right)_{\beta}^{\dot\alpha}\left[ v
\, U_{\dot{f}}^\alpha(x_L)\right]
 =
4\,\delta^{\dot\alpha}_{\dot f} \,\theta^\alpha  v 
\,
\frac{\rho^2}{(x_L^2 + \rho^2)^{3/2}}\, .
\label{thhofts}
\end{equation}
We remind that the superscript of $Q_0$ is the color index, the subscript is
subflavor, and they got entangled with the Lorentz spinor index of the supercharge.
Note that only the left-handed matter fermion fields have zero modes, similarly
to gluino. We see how the 't~Hooft zero modes get a geometrical interpretation
through supersymmetry. It is natural to call the corresponding  fermionic
coordinates $(\bar\theta_0)_{\dot\alpha}$. The supersymmetry transformations
shift it by $\bar\varepsilon $.

In order to determine the action of SUSY in the expanded moduli space let us
write down the generalized shift operator,
\begin{equation}
{\cal V}(x_0 ,\theta_0, \bar\beta,  \bar\zeta,\bar\omega,\rho  )
=e^{iPx_0}e^{-iQ\theta_0}e^{-i\bar{S}\bar\beta}
e^{-i\bar Q\bar \zeta}e^{i\bar M \bar\omega}e^{iD\ln\rho}\,.
\end{equation}
Here new Grassmann coordinates $\bar \zeta^{\dot\alpha}$ conjugated to $\bar
Q_{\dot\alpha}$ are introduced.  Repeating the
procedure of Sec.\ \ref{ssecstm} in the presence of
$\bar \zeta$ we obtain the SUSY transformations of the moduli. They are the
same as in Eq.\ (\ref{strlaws}) plus the transformations of $\bar \zeta$,
\begin{equation}
\delta \bar\zeta_{\dot\alpha}=
\bar\varepsilon_{\dot\alpha} -4 i \bar\beta_{\dot\alpha}\,
(\bar\zeta\bar\varepsilon )\, .
\label{thetatr}
\end{equation}
In the linear order in the fermionic coordinates the SUSY  transformation of
$\bar\zeta$ is the same as that of $\bar\theta$ but the former contains nonlinear
terms. The combination which transforms linearly, exactly as $\bar \theta$, is
\begin{equation}
(\bar\theta_0)^{\dot\alpha}=\bar\zeta^{\dot\alpha}[1-4i (\bar \beta
\bar\zeta)]\,,\qquad 
\delta(\bar\theta_0)^{\dot\alpha}=\bar\varepsilon^{\dot\alpha}\,. 
\label{thetazero}
\end{equation}
The variable $\bar\theta_0$ joins the set $\{x_0,\theta_0\}$ describing the
superinstanton center$\,$\footnote{In the original work~\cite{NSVZ3} the notation
$\bar\theta_0$ was used for what is called here $\bar\zeta$. The
combination~(\ref{thetazero}) was not introduced, it first appeared in Ref.\
\cite{Yung}.  }.

A more straightforward way to introduce the collective coordinate $\bar\theta_0$
is to use a different ordering in the shift operator ${\cal V}$,
\begin{equation}
{\cal V}(x_0 ,\theta_0, \bar\theta_0,\bar\beta_{\rm inv},\bar\omega_{\rm
inv},\rho_{\rm inv}
 ) =e^{iPx_0}e^{-iQ\theta_0}e^{-i\bar Q\bar
\theta_0}e^{-i\bar{S}\bar\beta_{\rm inv}} e^{i\bar M \bar\omega_{\rm
inv}}e^{iD\ln\!\rho_{\rm inv}}\;.
\label{invcoor}
\end{equation}
Needless to say that this reshuffling changes the definition of the other
collective coordinates. With the ordering~(\ref{invcoor}) it is clear that $x_0$,
$\theta_0$, and $\bar\theta_0$ transform as $x_L$, $\theta$ and $\bar\theta$,
respectively, while  the other moduli are the invariants of the SUSY
transformations. For this reason we marked them by the subscript inv.  Certainly
we can find the relation between the two sets of the collective coordinates,
\begin{eqnarray}
&&\bar\beta_{\rm inv}=\bar\beta\,[1 + 4 i \,(\bar\beta
\bar\zeta )]=\frac{\bar\beta}{1 - 4 i \,(\bar\beta
\bar\theta_0 )}\,,\nonumber\\[0.2cm]
&&\rho^2_{\rm inv}= \rho^2 \,[1 + 4 i \,(\bar\beta 
\bar\zeta )]=\frac{\rho^2}{1 - 4 i \,(\bar\beta
\bar\theta_0 )}\,,\nonumber\\[0.2cm]
&&\left[\Omega_{\rm inv}\right]^{\dot\alpha}_{\dot\beta}\equiv \left[ e^{-i
\bar\omega_{\rm inv}}\right]^{\dot\alpha}_{\dot\beta}=\exp\left\{-4i
\left[\bar\zeta^{\dot\alpha}\,
\bar\beta_{\dot\gamma}+\frac 1 2 \delta^{\dot\alpha}_{\dot\gamma}\,
(\bar\zeta \bar\beta) \right]\right\}\Omega^{\dot\gamma}_{\dot\beta}
\,,
\label{rhoinv}
\end{eqnarray}
It is worth emphasizing
 that all these SUSY invariants, built from 
the instanton moduli, are due to introduction of the coordinate $\bar\zeta$
conjugated to $\bar Q$.

Let us recall that in the theory with matter there is the non-anomalous
$R$ symmetry, see Sec.\ \ref{sec34}. We did not introduce the corresponding
collective coordinate because it is not new compared to the moduli of the flat vacua.
Nevertheless, it is instructive to consider the $R$ charges of the collective
coordinates. We collected these charges in Table~\ref{tabrcharge}.
\begin{table}
\begin{center}
\begin{tabular}{|c|c|c|c|c|c|c|}
\hline
~ &   ~ &  ~ & ~ & ~ & ~ & ~\\[-0.2cm]
Coordinates & $\theta_0$  & $\bar\beta$ & $\eta$ & $\bar\theta_0$ & $x_0$ &
$\rho$\\  ~ &   ~ &  ~ & ~ & ~ & ~ & ~\\[-0.2cm]
\hline 
~ &   ~ &  ~ & ~ & ~ & ~ & ~\\[-0.1cm]
$R$ charges & 1 & 1 & 1 & $-1$ & 0 & 0 \\[0.1cm]
\hline 
\end{tabular}
\end{center}
\caption{The $R$ charges of the instanton collective coordinates.}\label{tabrcharge}
\end{table}

From this Table it is seen that the only invariant with the vanishing
$R$ charge is $\rho^2_{\rm inv}$. This fact has  a drastic impact.  In SUSY
gluodynamics no combination of moduli was invariant under SUSY and U(1)
simultaneously.  This fact was used, in particular, in constructing the instanton
measure; the expression for the measure comes out unambiguous.
In the
theory with matter, corrections to the instanton measure proportional to powers of
$|v|^2\rho^2_{\rm inv}$ can emerge, generally speaking.
Actually, they do emerge, although all terms beyond the leading
$|v|^2\rho^2_{\rm inv}$ term are accompanied by powers of the
coupling constant $g^2$. 

Let us now pass to the invariants constructed from the coordinates
in the superspace and the moduli. 
Since the set $\{x_0,\theta_0,\bar\theta_0\}$ transforms the same way as
the superspace coordinate $\{x_L,\theta,\bar\theta\}$ such invariants are the
same as those built from two points in the superspace, namely
\begin{equation}
z_{\alpha\dot\alpha}=  (x_L-x_0)_{\alpha\dot\alpha} + 
4i\,(\theta-\theta_0)_\alpha\, (\bar\theta_0)_{\dot\alpha}\,,\quad
\theta-\theta_0\,,\quad
\bar \theta-\bar \theta_0
\,.
\label{inv2}
\end{equation}
All other invariants can be obtained by combining the sets of Eqs.\ (\ref{inv2})
and (\ref{rhoinv}). For instance, the invariant combination $\tilde{x}^2/\rho^2$
where
\begin{equation}
\tilde{x}_{\alpha\dot\alpha} = (x_L - x_0 )_{\alpha\dot\alpha} 
+ 4 i \,\tilde\theta_\alpha\bar\zeta_{\dot\alpha}\, ,
\label{tildex}
\end{equation}
which frequently appears in applications can be rewritten in such a form,
\begin{equation}
\frac{\tilde{x}^2}{\rho^2}=\frac{z^2}{\rho_{\rm inv}^2}
\,.
\end{equation}

One can exploit these invariants to immediately generate various superfields
with the collective coordinate switched on, starting  from the original bosonic
anti-instanton configuration. For example~\cite{NSVZ3},
\begin{eqnarray}
{\rm Tr}\, (W^\alpha W_\alpha )& 
{\longrightarrow} & 96\,\tilde\theta^2\, \frac{\rho^4}{(\tilde{x}^2 
+\rho^2)^4}\, , \nonumber\\[0.2cm]
Q^{\alpha\dot{f}}Q_{\alpha\dot{f}}& 
{\longrightarrow} &  
2 \,\frac{v^2\tilde{x}^2}{\tilde{x}^2+\rho^2}\, ,\nonumber\\[0.2cm]
\bar{Q}^{\dot\alpha{f}}\bar{Q}_{\dot{\alpha} f}& 
{\longrightarrow} & 
2 \,\frac{\bar{v}^2z^2}{z^2+\rho^2_{\rm inv}}\, .
\label{instglsuf}
\end{eqnarray}
The difference between $\tilde{x}$ and $x_L-x_0$ is unimportant
in Tr$\,W^2$ because of the factor  
$\tilde\theta^2$.
Thus, the superfield Tr$\,W^2$ remains intact: the matter fields do not 
alter the result for Tr$\,W^2$ obtained in SUSY gluodynamics.  The difference 
between 
$\tilde{x}$ and $x_L-x_0$ is very important, however, in the 
superfield  $Q^2$. Indeed, putting $\theta_0 =\bar\beta = 0$ and 
expanding Eq.\ (\ref{instglsuf}) in $\bar\theta_0$ we recover, in the 
linear  approximation, the same 't Hooft zero modes as in Eq.\ (\ref{thhofts})
\begin{equation}
\psi^{\alpha\dot{f}}_\gamma = 2\sqrt{2} i v 
(\bar\theta_0)^{\dot{f}}\delta^\alpha_\gamma \, \frac{\rho^2}{((x-
x_0)^2 + \rho^2)^{3/2}}\, .
\label{thhofts1}
\end{equation}

Note that the superfield $\bar{Q}^{\dot\alpha{f}}\bar{Q}_{\dot{\alpha} f}$
contains a fermion component if $\theta_0\neq 0$. What is the meaning
of this fermion field? (We keep in mind that the
  Dirac  equation  for $\bar\psi$ has no zero modes.) The origin of this fermion
field is the Yukawa interaction $(\psi\lambda )\bar\phi$ generating a source term
in the classical  equation for $\bar\psi$,
namely, ${\cal D}_{\alpha\dot\alpha}\bar\psi^{\dot\alpha} \propto
\lambda_\alpha \bar\phi$. 

\subsubsection{Orientation collective coordinates as the Lorentz  
SU(2)$_R$  rotations}
\label{secORIENT}

In this section we focus on the orientation collective coordinates 
$\bar\omega^{\dot\alpha}_{\dot\beta}$, in an attempt to explain  their origin in the
most transparent manner. The presentation below is adapted from
Ref.\ \cite{AVMS}.  The main technical problem with the introduction of the
orientations is the necessity of untangling them from the nonphysical gauge
degrees of freedom. Introduction of matter  is the most straightforward way to
make this untangling transparent. 

To begin with, let us define a gauge invariant vector field $W_\mu$ 
\begin{equation}
(W_\mu)^{\dot f\dot g}=\frac{i}{|v|^2}\left[\bar \phi^{\dot f} {\cal D}_\mu
\phi^{\dot g} - ({\cal D}_\mu\bar \phi^{\dot f})
\phi^{\dot g}\right]\, ,
\label{arktwelve}
\end{equation}
where $\dot f , \, \dot g$ are the SU(2) (sub)flavor indices, $\phi_{\dot g}$ is the
lowest component of the superfield $Q_{\dot g}$, the color
indices are suppressed.  In the flat vacuum~(\ref{dflat2}) the field $W_\mu$
coincides with the gauge field $A_\mu$ (in the unitary gauge).

What are the symmetries of the flat vacuum?  They obviously include
the Lorentz SU(2)$_L \times$SU(2)$_R$. Besides, the vacuum is invariant under the
flavor SU(2) rotations. Indeed, although $\phi^\alpha_{\dot f}\propto 
\delta^\alpha_{\dot f}$ is not invariant under the multiplication by the unitary 
matrix $S^{\dot f}_{\dot g}$, this noninvariance is compensated by the rotation in
the gauge SU(2). Another way to see this is to observe that the only modulus field
$\phi^\alpha_{\dot f}\phi_\alpha^{\dot f}$ in the model at hand  is flavor singlet.

For the instanton configuration, see Eq.\ (\ref{instspinn}) for $A_\mu$ and 
Eq.\ (\ref{scsol}) for $\phi$, the field  $W_{\alpha\dot\alpha}$  reduces to
\begin{equation}
(W_{\alpha\dot\alpha}^{\rm inst})^{\dot f\dot g}=2i\frac{\rho^2}{(x^2+\rho^2)^2}
\left( x^{\dot g}_\alpha\delta^{\dot  f }_{\dot\alpha} + x^{\dot f
}_\alpha\delta^{\dot  g}_{\dot\alpha}\right)\,. 
\label{arkeleven}
\end{equation}

The next task is to examine the impact of SU(2)$_L
\times$SU(2)$_R\times$SU(2)$_{\rm flavor}$ rotations on the 
$W_\mu^{\rm inst}$. It is immediately seen that Eq.\ (\ref{arkeleven}) 
remains intact under the action of SU(2)$_L$.  It is also invariant under the
simultaneous rotations in SU(2)$_R$ and SU(2)$_{\rm flavor}$. Thus, only one 
SU(2) acts on $W_\mu^{\rm inst}$ nontrivially. We can choose it to be the
SU(2)$_R$ subgroup of the Lorentz group. This explains why we introduced
the orientation coordinates through $\bar M \bar \omega$.

Note that the scalar fields play an auxiliary role in the construction presented,
they allow one to introduce a relative orientation. At the end one can consider the
limit $v\to 0$ (the unbroken phase). 

Another comment refers to  higher groups. Extra orientation coordinates
describe the orientation of the instanton SU(2) within the given gauge group.
Considering the theory in the Higgs regime allows one to make analysis again
in the gauge invariant manner. The crucial difference, however, is that the extra
orientations, unlike three SU(2) ones, are not related to the exact symmetries
of the theory
in the Higgs phase.  Generally speaking, the classical action becomes dependent on
the extra orientations, see  examples in Sec.\ \ref{gmc}.

\subsubsection{The instanton measure in the one-flavor model}
\label{imofm}

The approximate nature of the instanton configuration
at $\rho v \neq 0  $ implies that the classical action is $\rho$-dependent.  From 
't~Hooft's calculation~\cite{Hooft1} it is well known that in the limit $\rho v \to
0 $ the action becomes
\begin{equation}
\frac{8\pi^2}{g^2} \longrightarrow \frac{8\pi^2}{g^2} + 
4\pi^2|v|^2\rho^2\, .
\label{arkseven}
\end{equation}
The coefficient of $|v|^2\rho^2$ is twice larger than in 't~Hooft's case because
there are two scalar (squark) fields in the model at 
hand, as compared to one scalar doublet in 't~Hooft's calculation. 
Let us remind that  the $|v|^2\rho^2$ term (which is often referred to in the
literature as the 't~Hooft term)
 is entirely due to a surface contribution
in the action,
\begin{equation}
\int \!{\cal D}_\mu \bar\phi{\cal D}_\mu \phi\, {\rm d}^4 x =
- \int\! \bar\phi{\cal D}^2 \phi {\rm d}^4 x + \int\! {\rm d}\Omega_\mu
\partial^\mu \left( \bar\phi{\cal D}_\mu \phi
\right) {\rm d}^4 x =\int\! {\rm d}\Omega_\mu\,
\partial^\mu \left( \bar\phi{\cal D}_\mu \phi
\right)  {\rm d}^4 x \, .
\label{surface}
\end{equation}

Since the 't~Hooft term is saturated on the large sphere a question
immediately comes to one's mind as to a possible ambiguity in its calculation.
Indeed, what would happen if  from the very beginning one started from  the
bosonic Lagrangian with the kinetic term $- \bar\phi{\cal D}^2 \phi$
rather than ${\cal D}_\mu \bar\phi{\cal D}_\mu \phi$? Or, alternatively, one could
start from an arbitrary linear combination of these two kinetic terms.
In fact, such a linear combination naturally appears in supersymmetric
theories from $\int {\rm d}^4\theta \bar Q e^V Q$.  These questions are fully
legitimate. In Sec.\ \ref{secSURF} we demonstrate that the result quoted in
Eq.\ (\ref{arkseven}) is unambiguous and correct, it can be substantiated
by a dedicated analysis.  

The term  $4\pi^2|v|^2\rho^2$ is obtained for the purely bosonic field configuration.
For non-vanishing fermion fields an additional contribution to the action comes
from the Yukawa term $(\psi\lambda )\bar\phi$.  We could have calculated this
term by substituting the classical field $\phi$ and the zero modes for $\psi$ and
$\lambda$.  However, it is much easier to do the job by using the SUSY invariance
of the action. Since $\rho^2_{\rm inv}$ (see Eq.\ (\ref{rhoinv})) is the only
appropriate 
invariant which could be constructed from the moduli,  the action at
$\bar\theta_0\neq 0$ and 
$\bar\beta\neq 0$ becomes
\begin{equation}
\frac{8\pi^2}{g^2} + 
4\pi^2|v|^2\rho^2_{\rm inv}\,  .
\label{thooftrhoinv}
\end{equation}

To obtain the full instanton measure we proceed the same way as in Sec.\
\ref{ssecmeasure}.  Besides the classical action, the only change is due to
the additional integration over ${\rm d}^2 \bar\theta_0$. From the general 
formula~(\ref{genmeas1}) we infer that it brings in an extra power of $M_{\rm
PV}^{-1}$ and a normalization factor which could be read off from the
expression~(\ref{thhofts1}). Overall, the extra integration takes the form (see
Table \ref{tabmodes}),
\begin{equation}
\frac{1}{M_{PV}}\, \frac{1}{8\pi^2v^2\rho^2}\,{\rm d}^2\bar\zeta=
\frac{1}{M_{PV}}\, \frac{1}{8\pi^2v^2\rho_{\rm inv}^2}\,{\rm d}^2\bar\theta_0
\,.
\end{equation}
Note, that the SUSY transformations (\ref{strlaws}) and (\ref{thetatr}) leave
this combination invariant. Note also, that the 't Hooft zero modes are 
chiral, it is $1/v^2$ that appears, rather than $1/|v|^2$. The instanton measure
``remembers" of  the phase of the vacuum expectation value of the scalar field. As
we will see shortly, this is extremely important for recovering  the proper chiral
properties  of  the instanton-induced superpotentials. 

Combining the $d^2\bar\theta_0$ integration with the previous 
result one gets
\begin{equation}
{\rm d}\mu_{\rm one-fl}= \frac{1}{2^{11}\pi^4 v^2 }\, M_{\rm PV}^5 \left( 
\frac{8\pi^2}{g^2}
\right)^2 \!\exp\left( -\frac{8\pi^2}{g^2}-4\pi^2|v|^2\rho^2_{\rm inv}
\right) \frac{ {\rm d}\rho^2}{\rho^2}\,
 {\rm d}^4 x_0\, {\rm d}^2\theta_0 \,
 {\rm d}^2\bar\beta_{\rm inv}\,  {\rm d}^2\bar\theta_0 \, .
\label{imesqcd}
\end{equation}
This measure is explicitly invariant under the SUSY transformations. Indeed, ${\rm
d}\rho^2/\rho^2$ reduces to ${\rm d}\rho^2_{\rm inv}/\rho^2_{\rm inv}$ (up to
a subtlety at the singular point $\rho^2=0$ to be discussed later).

Let us remind that the expression~(\ref{imesqcd}) is obtained under the
assumption that the parameter $\rho^2|v|^2\ll1$ and accounts for the zero and first
order terms in the expansion of the action in this parameter. Summing up the
higher orders leads to some function of $\rho^2_{\rm inv}|v|^2$ in the exponent. 

\subsubsection{Verification of the 't Hooft term}
\label{secSURF}

In the previous section we mentioned the ambiguity in the 't~Hooft term
due to its surface nature. The surface terms call for the careful consideration of
the boundary conditions. Instead, we suggest an alternative route via
the scattering amplitude technique~\cite{SVZscat}.  Calculation of the scattering
amplitude takes care of the correct boundary conditions automatically.

As a simple example let us consider the non-supersymmetric SU(2) model with one
Higgs doublet $\phi^\alpha$.  Our task is to demonstrate that the instanton-induced
effective interaction of the $\phi$ field is
\begin{equation}
\Delta {\cal L}= \int {\rm d}\mu
\exp\left\{-2\pi^2\rho^2\left[\bar\phi(x)\phi(x)-|v|^2\right]\right\}
\,,
\label{scatL}
\end{equation}
where ${\rm d}\mu$ is the instanton measure of the model, it includes, in
particular, the factor 
$\exp (-2\pi^2\rho^2|v|^2)$.

We want to compare two alternative calculations of one and the same amplitude --
one based on the instanton calculus, and another following from the  effective
Lagrangian~(\ref{scatL}). Let us start from the emission of one physical Higgs by a
given instanton with the fixed collective coordinates. The 
interpolating field $\sigma$ for the physical Higgs
 can be defined as
\begin{equation}
\sigma(x)=\frac{1}{\sqrt{2}\,|v|}\left[\bar\phi(x)\phi(x)-|v|^2\right]
\,.
\end{equation}
The Lagrangian~(\ref{scatL}) implies that the emission amplitude $A$ is equal to 
\begin{equation}
A=-2\sqrt{2}\,\pi^2 \rho^2 |v|
\,.
\end{equation}

On the hand let us calculate  the expectation value of $\sigma(x)$ in the instanton
background. In the leading (classical) approximation,
\begin{equation}
\left\langle\sigma(x)\right\rangle_{\rm inst}
=\frac{1}{\sqrt{2}\,|v|}\left[\bar\phi_{\rm inst}(x)\phi_{\rm inst}(x)-|v|^2\right]
=- \frac{|v|}{\sqrt{2}}\, \frac{\rho^2}{ x^2+\rho^2} 
\,.
\end{equation}
Considering $x\gg \rho$ we arrive at
\begin{equation}
\left\langle\sigma(x)\right\rangle_{\rm inst} \to 
-2\sqrt{2}\,\pi^2 \rho^2 |v|\cdot \frac{1}{4\pi^2x^2}
\,.
\end{equation}
The first factor is the emission amplitude $A$, while the second factor is the free
particle propagator.  

Thus, the effective Lagrangian~(\ref{scatL}) is verified in the linear in $\sigma$
order.  To verify the exponentiation it is sufficient to show the factorization of the
amplitude for the emission of  arbitrary number of the  $\sigma$ particles. In the
classical approximation this factorization is obvious.

\subsubsection{The instanton measure: general gauge group}
\label{imgc}

By the general case we mean generalization 
of the SU(2) model with one flavor
to arbitrary gauge groups $G$ and 
arbitrary matter sector. Passing to higher groups we still consider the same SU(2)
instanton of Belavin, Polyakov, Schwarz and Tyupkin. What we need to do is to
specify its orientation in the group
$G$. Correspondingly, new collective  bosonic coordinates emerge. To introduce
them  we start from identifying a stationary subgroup $H$ of the group $G$ which
does not act on the given instanton solution. The generators of $H$ do not give rise
to new moduli. Therefore, the number of additional orientations is $d_G -d_H -3$
where we subtracted 3 to take into account three SU(2) orientation already
considered (the corresponding moduli
are $\bar\omega^{\dot\alpha\dot\beta}$). Here $d_G$ and  $d_H$ stand for  the
number of the generators of the groups $G$ and $H$, respectively,
i.e. dimensions of the groups. Note, that for any
group
$G$ the following relation takes place:
\begin{equation}
d_G -d_H -3=4T_G -8\,,
\end{equation}
where the dual Coxeter number $T_G$ is defined after Eq.\ (\ref{3anom}).

The color generators can be classified with respect to the instanton SU(2) as
follows: $d_H$ singlets of SU(2), one triplet, and $(d_G -d_H -3)/2$ doublets.
The singlets produce no collective coordinates, the triplet was already accounted
for, what should be  added is the collective coordinates due to the doublets. 
The additional orientational modes can be found along the same line of reasoning
as in Sec.\ \ref{ssecmeasure} (see  in Ref.\ \cite{Bernard} where the issue is
discussed in detail),
\begin{equation}
a_\nu^\alpha=g^{-1}\left[{\cal D}_\nu\,
U_{\dot\beta}\,\bar\omega^{\dot\beta}\right]^\alpha\,,\qquad 
\left\|\frac{\partial a_\nu^\alpha}{\partial \bar\omega^{\dot\beta}}\right\|
=\frac{\sqrt{2}\,\pi\rho}{g}\,.
\end{equation}
Here the matrix $U^\alpha_{\dot\alpha}$ satisfying the equation ${\cal D}^2 U=0$
is given in Eq.\ (\ref{umatrix}). We remind that in the case at hand both
$a_\nu^\alpha$ and $\bar\omega^{\dot\beta}$ are SU(2) doublets. As a result, the 
$d_G -d_H -3$ extra orientations  bring in the following extra factor in the
measure
\begin{equation}
\left[\frac{\sqrt{\pi}}{g}\, \rho\, M_{\rm PV}\right]^{4T_G
-8}\frac{\Omega_G}{8\pi^2\,\Omega_H}\,,
\end{equation}
where $\Omega_G$ and $\Omega_H$ are volumes of the groups $G$ and
$H$, respectively.  The factor $8\pi^2$ in the denominator is due to fact that 
$\Omega(\mbox{SU}(2)/Z_2)=8\pi^2$ is already included in the measure. 

The gauge group SU($N$)  is of most practical importance. In this case the
groups $G$ and $H$ are
\begin{equation}
G=\frac{\mbox{SU}(N)}{Z_N}\,,\qquad 
H=\frac{\mbox{SU}(N-2)}{Z_{N-2}}\times \frac{\mbox{U}(1)}{Z_N}\,,
\end{equation}
with dimensions $d_G= N^2 -1$ and $d_H=(N-2)^2$.
The ratio of the volumes is~\cite{Bernard}
\begin{equation}
\frac{\Omega_G}{\Omega_H}=\frac{2^{4N-5} \pi^{2N-2}}{(N-1)! \,(N-2)!}\,.
\label{omn}
\end{equation}
We accounted for the fact that the center of the group does not act in the adjoint
representation. This is important in the calculation of the volume of the group.

In full analogy with  the SU(2) model, the bosonic orientation modes have gluino
counterpartners of the same form,
\begin{equation}
\lambda^\alpha_\beta=g^{-1}\left[{\cal
D}_\beta^{\dot\beta}\,U_{\dot\beta}\right]^\alpha=
\frac{4}{g}\,\frac{\delta^\alpha_\beta\,\rho^2}{(x^2+\rho^2)^{3/2}}\,,\qquad
\left\|\lambda^\alpha_\beta\right\|
=\frac{4\,\pi\rho}{g}\,.
\label{scolor}
\end{equation}
It  is not difficult to verify directly that it is a zero mode, indeed. There are $2N-4$
such zero modes, the corresponding collective coordinates will be
denoted$\,$\footnote{These fermionic coordinates are marked by bar to emphasize
that they are partners to the color orientations carrying the dotted indices. Note
that the origin of $\bar\beta$ is similar.} by
$\bar\xi_i$. By normalizing this mode in the way similar to the orientation
modes, in essence, we gave a geometrical interpretation to the coordinates
$\bar\xi_i$. Note that the U(1) charge of $\bar\xi_i$ is the same as that of
$\theta_0$ and $\bar\beta$, i.e. equal to one.

Assembling all pieces, we arrive at the final expression for the instanton measure
in SU($N$)  SUSY gluodynamics,
\begin{equation}
{\rm d}\mu_{\,{\rm SU}(N)} = \frac{\pi^{2N-6}\,e^{ -8\pi^2/g^2}}{2^{8}\,(N-1)!\,
(N-2)!} \, (M_{\rm PV})^{3N}
\left( 
\frac{8\pi^2}{g^2} \right)^{N}\! \! {\rm d}\rho^2\,
 {\rm d}^4 x_0\, {\rm d}^2\theta_0 \,
 {\rm d}^2\bar\beta\,  \prod_{i=1}^{2N-4}\rho\,{\rm d}\bar\xi_i \, .
\label{sunqcd}
\end{equation}
It is useful to present also the measure for the  arbitrary gauge group $G$,
\begin{equation}
{\rm d}\mu_{G} =
\frac{e^{-8\pi^2/g^2}}{2^{\,4T_G+3}\,\pi^4}\,\frac{\Omega_G}{\Omega_H}\,
 ( M_{\rm PV})^{\,3T_G}
\left( 
\frac{8\pi^2}{g^2} \right)^{T_G}\! {\rm d}\rho^2\,
 {\rm d}^4 x_0\, {\rm d}^2\theta_0 \,
 {\rm d}^2\bar\beta\,  \prod_{i=1}^{2T_G-4}\!\rho\,{\rm d}\bar\xi_i \, .
\label{genqcd}
\end{equation}

\subsubsection{The instanton measure: general matter content }
\label{gmc}

In SU(2) SQCD we have discussed the matter in the fundamental representation.
We found that for each doublet superfield there appeared one fermionic zero
mode, i.e. one fermionic collective coordinate. We related this mode to the scalar 
component of the same superfield in the Higgs phase (the solution of the equation
${\cal D}^2 \phi=0$ with a nontrivial asymptotic behavior). This procedure 
can be easily generalized to arbitrary representation $R(j)$ of SU(2) where $j$ is
the color spin. In this case the solution for the scalar field in the anti-instanton
background can be written as
\begin{equation}
\phi^{\alpha_1,\dots,\alpha_{2j}}= v \,U^{\alpha_1}_{\dot\alpha_1}\, \cdots\, 
 U^{\alpha_{2j}}_{\dot\alpha_{2j}}\,c^{\dot\alpha_1\cdots\dot\alpha_{2j}}
\,,\quad
\bar \phi^{\alpha_1,\dots,\alpha_{2j}}= \bar v \,U^{\alpha_1}_{\dot\alpha_1}\,
\cdots\, U^{\alpha_{2j}}_{\dot\alpha_{2j}}\,
\bar c^{\dot\alpha_1\cdots\dot\alpha_{2j}}
\,,
\label{arkeight}
\end{equation}
where  $c^{\dot\alpha_1\cdots\dot\alpha_{2j}}$ is a symmetric tensor,
$\bar c= c^*$, normalized to unity 
\begin{equation}
c^{\dot\alpha_1\cdots\dot\alpha_{2j}} \, \bar
c_{\dot\alpha_1\cdots\dot\alpha_{2j}}=1
\,.
\end{equation}
The particular form of $c$ is determined by the choice of the vacuum configuration
from the vacuum manifold.

The total number of the matter fermion zero modes is equal to 
\begin{equation}
2T(j)=\frac{2}{3}\, j\,(j+1)(2j+1)\,,
\end{equation}
where $T(j)$ is the dual Coxeter number for the representation $R(j)$.
Correspondingly, $2T(j)$ fermionic coordinates $\bar\eta_i$ must be  introduced.
The extra factor $\prod {\rm d}\bar \eta_i$ appears in the measure. Besides,  if
$v\neq 0$, i.e. we are in the  Higgs phase, the classical action gets modified.  Our
task is to establish the form of the $|v|^2 \rho^2$ term in the action. 

The modification of the bosonic part is pretty obvious,
\begin{equation}
 \Delta S= 4\pi^2\, j\, |v|^2\, \rho^2\,.
\label{matact}
\end{equation}
This is a generalization of the 't~Hooft term.
It worth stressing that this term in the action  is proportional to $j$. This is a
topological feature, due to the fact that $\Delta S$ is expressible as a surface
integral, see Eq.\ (\ref{surface}). Since $\phi\propto U^{2j}$ the surface integral is
proportional to $j$.

As previously, the impact of the fermion zero modes  reduces to the
replacement of
$\rho^2$ in Eq.\ (\ref{matact}) by $\rho^2_{\rm inv}$. What remains to be clarified
is the definition of $\bar \theta_0$ in terms of $\bar\eta_i$.
Out  of all $2T(j)$ fermion zero modes only $2j$ zero modes of the form 
\begin{equation}
\psi^{\alpha_1,\dots,\alpha_{2j}}_\gamma=\frac{i}{4\pi\rho\sqrt{2j}}\,
{\cal D}_\gamma^{\dot \alpha_1}\left[ U^{\alpha_1}_{\{\dot\alpha_1}\,
\cdots\,  U^{\alpha_{2j}}_{\dot\alpha_{2j}\}}\right] \bar\eta^{\dot
\alpha_2\cdots\dot\alpha_{2j}}
\label{twoj}
\end{equation}
are involved in the problem at hand. Here $\bar\eta^{\dot
\alpha_2\cdots\dot\alpha_{2j}}$ present $2j$ fermionic collective coordinates.
The explicit  form of the fermion zero modes  displayed in Eq.~(\ref{twoj}) shows
that these particular modes can be understood as the SUSY transformation of the
$2j+1$ bosonic solutions of the   equation ${\cal D}^2 \phi=0$. The latter  have the
form~(\ref{arkeight}), with the  fixed coefficients $c$ replaced  by an arbitrary
symmetric  tensor.

The parameter $\bar \theta_0$ is introduced through the SUSY transformation
of the bosonic configuration~(\ref{arkeight}) with the parameter 
$\bar \varepsilon=\bar \theta_0$. Performing this transformation and comparing
the result with Eq.\ (\ref{twoj}) we find
\begin{equation}
(\bar \theta_0)_{\dot \alpha_1}=\frac{1}{4\pi\rho v\sqrt{j}} \,
\bar c_{\dot\alpha_1\cdots\dot\alpha_{2j}}
\bar\eta^{\dot \alpha_2\cdots\dot\alpha_{2j}}
\end{equation}

If the matter sector contains several irreducible representations, then $\Delta S$
is the sum over all representations, each one enters with its own $v$ and
$c$. In the simplest case of two doublets considered in Sec.\ \ref{imofm}
the $D$-flatness implies that $v$'s are the same, and the color orientations given
by $c$'s are opposite. Then the summation over two doublets
 returns us to Eq.\ (\ref{thooftrhoinv}), with $\rho^2_{\rm inv}$
defined in Eq.\ (\ref{rhoinv}). 

The resulting  measure is 
\begin{equation}
{\rm d}\mu= {\rm d}\mu_{\,{\rm SU}(2)} \,\prod_{R(j)} \left\{
(M_{\rm PV})^{-T(j)}\,e^{-4\pi^2\, j\, |v_j|^2\, \rho^2_{\rm inv}}\,\prod_{i=1}^{2T(j)}
{\rm d}\,
\bar\eta_i\right\}\,,
\label{sumatt}
\end{equation}
where ${\rm d}\mu_{\,{\rm SU}(2)}$ is given in Eq.\ (\ref{fimsg1}). 

What changes in passing to higher groups? Not much. One should decompose the
matter representation in the group $G$ with respect to the instanton SU(2)
``corner".
The SU(2) singlet fields can be dropped out, other fields contribute according to 
Eq.\ (\ref{sumatt}) where $v$ should be replaced by its SU(2) projection
$v_{\,{\rm SU}(2)}$.  Therefore, the $\rho^2$ terms in the action certainly depend
on the orientation of the instanton SU(2) subgroup within the group $G$.  The
measure differential with respect to this orientation  is
\begin{equation}
{\rm d}\mu={\rm d}\mu_G\,\frac{{\rm d} \Omega_{G/H}}{\Omega_{G/H}}
\,(M_{\rm PV})^{-T(R)}\prod_{R(j)} \left\{
e^{-4\pi^2\, j\, |v_{\,{\rm SU}(2)}|^2\, \rho^2_{\rm inv}}\,\prod_{i=1}^{2T(j)}
{\rm d}\,
\bar\eta_i\right\}\,.
\label{measmat}
\end{equation}
Here an obvious relation $T(R)=\sum T(j)$ is used.

\subsection{Cancellation of quantum corrections to the measure }
\label{sec45}

So far, our analysis of the instanton measure was in essence  classical. 
Strictly speaking, it would be  better to call it semiclassical. Indeed,
let us not forget that  the calculation
of the pre-exponent is related to the one-loop corrections.  In our case the
pre-exponent is given by the integral over the collective coordinates. In
non-supersymmetric theories the  pre-exponent is not exhausted by this
integration -- non-zero modes contribute as well. Here we will show that the
non-zero modes cancel out in SUSY theories. Moreover, in the unbroken phase the
cancellation of the non-zero modes persists to any order in perturbation theory
and even beyond, i.e. nonperturbatively. Thus, we obtain the extension of the
non-renormalization theorem~\cite{GRS} to the instanton background. The
specific feature of this background, responsible for the  extension, is preservation 
of one half of SUSY. Note that in the Higgs phase the statement of cancellation  is
also valid  in the zeroth and first order in the parameter $\rho^2 |v|^2$.

In the first loop the cancellation is pretty obvious.  Indeed, in SUSY gluodynamics
the differential operator $L_2$ defining the mode expansion has one and the same
form, see Eq.\ (\ref{quadop}), for both the gluon and gluino fields,
\begin{eqnarray}
 -{\cal D}^{\alpha\dot\alpha}{\cal D}_{\beta\dot\alpha}\,
a_n^{\beta\dot\gamma}\!\!&=&\!\! \omega_n^2\,
a_n^{\alpha\dot\gamma}\,,\nonumber\\[0.2cm]
 -{\cal D}^{\alpha\dot\alpha}{\cal D}_{\beta\dot\alpha}\,
\lambda_n^{\beta}\!\!&=&\!\! \omega_n^2 \,\lambda_n^{\alpha}\,.
\end{eqnarray}
The residual supersymmetry  (generated by $\bar Q_{\dot \alpha}$) is reflected in
$L_2$ in the absence  of free dotted indices. 
Therefore, if the  boundary conditions respect the residual supersymmetry --
which we assume they do --
the eigenvalues and eigenfunctions are the same for $a^{\alpha\, \dot 1}$,
$a^{\alpha\, \dot 2}$, and $\lambda^{\alpha}$. For the field $\bar
\lambda^{\dot\alpha}$ the relevant operator is $-{\cal D}^{\alpha\dot\alpha}{\cal
D}_{\alpha\dot\beta}=-\delta^{\dot\alpha}_{\dot\beta}\,
{\cal D}^{\alpha\dot\gamma}{\cal D}_{\alpha\dot\gamma}/2$,
\begin{equation}
-{\cal D}^{\alpha\dot\gamma}{\cal D}_{\alpha\dot\gamma}
\,\bar\lambda_n^{\dot\alpha}= \omega_n^2\,\bar\lambda_n^{\dot\alpha}\,.
\end{equation}
This equation shows$\,$\footnote{The equality ${\cal D}^{\alpha\dot\alpha}{\cal
D}_{\alpha\dot\beta}=(1/2)\, \delta^{\dot\alpha}_{\dot\beta}\,
{\cal D}^{\alpha\dot\gamma}{\cal D}_{\alpha\dot\gamma}$ 
exploits the fact that
$\bar{G}_{\dot\alpha\dot\beta}=0$ for the anti-instanton.}
that the modes of
$\bar\lambda$  coincide with those of the scalar field $\phi$ in the same
representation of the gauge group,
\begin{equation}
-{\cal D}^{\alpha\dot\gamma}{\cal D}_{\alpha\dot\gamma}
\,\phi_n= \omega_n^2\,\phi_n\,.
\end{equation}
Moreover, {\em all} nonzero modes are expressible in terms of $\phi_n$ (This nice
feature was noted in Ref.\ \cite{BCCL}). This is quite evident for 
$\bar\lambda^{\dot 1}$ and $\bar\lambda^{\dot 2}$. As for the non-zero modes of $a$ and
$\lambda$ they are
\begin{equation}
a_n^{\alpha\dot 1(\dot\beta)}=a_n^{\alpha\dot 2(\dot\beta)}=\frac{1}{\omega_n}
{\cal D}^{\alpha\dot\beta}\, \phi_n\,,\qquad 
\lambda_n^{\alpha(\dot\beta)}=\frac{1}{\omega_n} {\cal D}^{\alpha\dot\beta}\,
\phi_n\,.
\label{modrel}
\end{equation}
Thus, the integration over $a$ produces $1/\omega_n^4$ for each given eigenvalue.
The integration over $\lambda$ and $\bar\lambda$ produces  $\omega_n^2$.
The balance is restored by the contribution of the scalar ghosts which provides 
the remaining $\omega_n^2$. 

The same cancellation is extended to the matter sector. In every
supermultiplet each mode of the scalar field $\phi$ is accompanied by two modes
in $\psi^\alpha$ and  $\bar \psi^{\dot \alpha}$, see \mbox{Eq.\ (\ref{modrel})}.
Correspondingly, one gets $\omega_n^2/\omega_n^2$ for each eigenvalue.

From the one-loop consideration it is clear that the cancellation is due to the
boson-fermion pairing enforced  by the residual supersymmetry of the instanton
background. The very same supersymmetry guarantees the cancellation in higher
loops.  First of all, on general symmetry grounds, corrections, if present, could not
be  functions of the collective coordinates: it was shown previously that no
appropriate invariants exist. Therefore, the only possibility left is a purely
numerical series in powers of $g^2$.

In fact, even such series does not appear.  Indeed, let us consider the two-loop
supergraph in the instanton background (Fig.\ 1). This graph has two vertices. 
Its contribution is the integral over the  supercoordinates of both vertices,
$\{ x,\theta ,\bar\theta\}$ and $\{ x',\theta' ,\bar\theta'\}$, respectively. Let us
integrate  over the supercoordinates of the  second vertex  and over the coordinates
$x$ and $\theta$ (but not
$\bar\theta$!) of the first vertex.  Then the graph can be presented as  the integral
$
\int  {\rm d}^2 \bar\theta\, F(\bar\theta)
$.
The function
$F$ is invariant under the simultaneous SUSY transformations of $\bar\theta$ and
the instanton collective coordinates. As it was shown in Sec.\ \ref{ssecstm}, in SUSY
gluodynamics there are no invariants  containing
$\bar\theta$. Therefore, the function $ F(\bar\theta)$ can be only a constant,
and then the  integration over $\bar\theta$ yields zero~\cite{NSVZdop}. 

\begin{figure}[h]    
\vskip3mm
\epsfxsize=8.5cm
 \centerline{\epsfbox{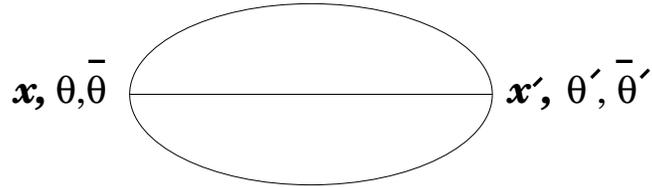}}
 \caption{A typical two-loop supergraph. The solid lines denote
the propagators of the quantum superfields in the (anti)instanton background.
We rely only on the most general features of the supersymmetric
 background field technique.  For a
 pedagogical introduction to supergraphs and supersymmetric
 background field technique  the reader is referred  to Ref.\ \cite{SSOT}.  }
\end{figure}

The proof above is a version of the arguments based on the residual
supersymmetry.  Indeed, no invariant can be built of $\bar\theta$ because
there is no collective coordinate $\bar\theta_0$. The absence of $\bar\theta_0$
is, in turn, the consequence of the residual supersymmetry.  Introduction of 
matter in the Higgs phase changes the situation. At $v\neq 0$ no 
residual supersymmetry
survives. In terms of the collective coordinates this is reflected in the emergence
of $\bar\theta_0$. Correspondingly, the  function $ F(\bar\theta)$ becomes
a function of the invariant $\bar \theta-\bar \theta_0$ (see Eq.\ (\ref{inv2})), and
the integral does not vanish.

Therefore, in the theories with matter, in the Higgs phase, the instanton does get
corrections. However, these corrections vanish~\cite{rarensvz} in the limit
$|v|^2\rho^2\to 0$. Technically, the invariant above containing $\bar\theta$
disappears at small $v$ because $\bar\theta_0$ is proportional to $1/v$.

Summarizing, the instanton measure gets no quantum corrections in SUSY
gluodynamics and in the unbroken phase in the presence of  matter. In the
Higgs phase the corrections start from the terms
$g^2|v|^2\rho^2$.

One important comment is in order here regarding the consideration above. Our
proof  assumes that there exists a supersymmetric ultraviolet regularization of the
theory.  At one-loop level the Pauli-Villars regulators do the job. In higher loops
the regularization is achieved  by  a combination of the Pauli-Villars regulators with
higher derivatives terms. We do not use this regularization explicitly; rather, we
rely on the theorem of its existence. That is all we need. As for the
 infrared
regularization, it is provided by the instanton  field itself. Indeed, at fixed collective
coordinates all eigenvalues are nonvanishing. The zero modes should not be
counted when the  collective coordinates are fixed.

\section{Sample Applications}
\label{secSA}
\renewcommand{\theequation}{4.\arabic{equation}}
\setcounter{equation}{0}

The stage is set, and we are ready to apply the formalism outlined 
above  in concrete problems that arise in SUSY gauge theories.
In this section we start discussing  applications of instanton calculus which
are of practical interest. First, we will derive the NSVZ $\beta$  function. 
Then in  SU(2) SQCD with one  flavor  we will obtain  an
instanton-generated  superpotential~\cite{ADS2}. In the theory with two
flavors we will calculate the  quantum deformation of the moduli space. Both
effects  are due  to  the one-instanton contribution. The two-flavor SU(2) model  is
a major component of the  ITIY  mechanism \cite{IT} on which we will  dwell in
Sec.\ \ref{sec63}.

\subsection{ Novikov-Shifman-Vainshtein-Zakharov 
$\beta$  function}\label{sec46}

The exact results for the instanton measure obtained above,
in conjunction with the renormalizability, can be converted into 
exact relations for the $\beta$ functions.

\subsubsection{Exact $\beta$ function in supersymmetric gluodynamics}
\label{ebfisg}

Consider first supersymmetric gluodynamics. The gauge group $G$ 
can be arbitrary. The expression for the instanton measure is given in 
Eq.\ (\ref{genqcd}). What are the input theoretical parameters in this expression?
There are two such parameters: the bare coupling constant $g$ and the regulator
mass $M_{\rm PV}$. Although the instanton measure is a theoretical construction 
it will be directly related, as we will see, to the physically observable quantities.
The renormalizability of the theory implies that the latter, as well as the former,
depends only on a special combination of $g$ and $M_{\rm PV}$: the ultraviolet 
cutoff  $M_{\rm PV}$ must  conspire with the bare coupling $g$ to make the
instanton  measure expressible in terms of the renormalized coupling $g_{\rm ren}
(\rho )$. This means that $g$ should be understood as a function $g(M_{\rm PV})$
such that the combination entering the instanton measure does not depend on
$M_{\rm PV}$,
\begin{equation}
(M_{\rm PV})^{3T_G} 
\left(\frac{1}{g^2(M_{\rm PV})}\right)^{T_G}\!
\exp \left(-\frac{8\pi^2}{g^2(M_{\rm PV})}\right)={\rm const}\,.
\label{efbf}
\end{equation}
The dimensionful constant on the right-had side is related to  the physical scale
parameter $\Lambda$ defined in the standard perturbative schemes, see Sec.\
\ref{pertL} for a more detailed discussion.

The independence of the left-hand side on $M_{\rm PV}$ gives the exact answer
for the running coupling (in the Pauli-Villars scheme)
$$
\alpha (\mu) = \frac{g^2(\mu)}{4\pi}\,.
$$
The result can be formulated, of course, in the form of the exact $\beta$
function.
Taking the logarithm and differentiating with respect to
$\ln M_{\rm PV}$, we arrive at
\begin{equation}
\beta (\alpha ) \equiv \frac{{\rm d}\,\alpha(M_{\rm PV})}{{\rm d}\ln M_{\rm
PV}}= -\frac{3T_G\,\alpha^2}{2\pi}\left(1-\frac{T_G\,\alpha}{2\pi} 
\right)^{-
1}\, .
\label{ebfpg}
\end{equation}

We pause  here to make a remark regarding  the complexified structure of the
objects considered. In the derivation above it was assumed that both the gauge
coupling
$g$ and the  Pauli-Villars regulator mass $M_{\rm PV}$ are real.
As we know, see Eq.\ (\ref{itheta}), the gauge coupling is complexified,
$8\pi^2/g^2\to8\pi^2/g^2-i\vartheta$. In the instanton measure the regulator
mass $M_{\rm PV}$ and the fermion collective coordinates are complex too.
Therefore,  it is instructive to  study the phase dependence related to the
(anomalous) U(1) transformation of the fields. Under these transformations every
fermion collective coordinate is multiplied by $\exp(-i\alpha)$. Correspondingly,
the instanton measure~(\ref{genqcd}) is multiplied by $\exp(2i\alpha T_G)$. This
is an explicit manifestation of the chiral anomaly. The phase factor can be absorbed
into the phase of the regulator mass $M_{\rm PV}$, it  is the regulators mass terms
which break the U(1). Thus, the chiral properties of the measure are consistent with
the anomaly.  

Alternatively, instead of rotating $M_{\rm PV}$,  one can  shift the vacuum angle 
$\vartheta\to \vartheta +2\alpha T_G$ in the complexified exponent,
$\exp(-8\pi^2/g^2+i\vartheta)$, 
 in the measure. The subtlety is that the factors $8\pi^2/g^2$ in the
{\em pre-exponent} are not shifted. In fact, it is Re($8\pi^2/g^2$) that enters in the
pre-exponent. This is  is the so-called holomorphic anomaly~\cite{AVMS1}. 
Have we used $(8\pi^2/g^2-i\vartheta)^{2T_G}$ in the pre-exponent we would
obtain a $\vartheta$ dependence of the $\beta$ function starting from the fourth
loop. This obviously cannot happen in perturbation theory. 

The holomorphic anomaly is related to supersymmetric regularization of the
higher loops. As we mentioned it is done through higher derivative $D$ terms.
The corresponding ultraviolet regulator is {\em not} chiral, unlike the Pauli-Villars
regulators used in the first loop. In our derivation of the $\beta$ function we tacitly
assumed that the absolute values of all regulator masses are the same.

\subsubsection{Theories with matter: $\beta$ function via anomalous
dimensions}
\label{twmbfvad}

Now, let us introduce the matter fields in arbitrary representation $R$.
This representation can be reducible, $R=\sum R_i$. Besides the gauge interaction,
the matter fields  can have arbitrary  (self)interactions, i.e. an arbitrary
renormalizable  superpotential is allowed. The possible superpotential does not
explicitly show up in our final formula, Eq.\ (\ref{nsvzbetaf}). It is hidden
in the anomalous dimensions which certainly  do depend on the
presence/absence of the superpotential.

The instanton measure  is given in Eq.\ (\ref{measmat}). 
It assumes that the kinetic terms of the matter are normalized canonically,
i.e.  the matter part of the Lagrangian is
\begin{equation}
{\cal L}_{\rm matter}= \frac{1}{4}\sum_i \int\!{\rm d}^2\theta{\rm d}^2\bar\theta
\,\bar Q_i e^V Q_i +\frac{1}{2}\left\{\int\!{\rm d}^2\theta\,{\cal W}(Q_i)
+{\rm H.c.}\right\}
\,.
\end{equation}
As we know the superpotential is not renormalized but the kinetic terms are.
Therefore, it is more convenient to allow for arbitrary $Z$ factors in the bare
Lagrangian,
\begin{equation}
{\cal L}_{\rm matter}= \frac{1}{4}\sum_i Z_i \int\!{\rm d}^2\theta{\rm
d}^2\,\bar\theta
\bar Q_i e^V Q_i +\frac{1}{2}\left\{\int\!{\rm d}^2\theta\,{\cal W}(Q_i)
+{\rm H.c.}\right\}
\,.
\label{zmatt}
\end{equation}
These  $Z$ factors are bare ones, normalized at the ultraviolet cut off, $Z_i( M_{\rm
PV})$. They can be fixed by the condition that the  $Z$ factors become unity in the
infrared.  As a result the measure~(\ref{measmat}) acquires the extra factor
\begin{equation}
\prod_i \left( Z_i\right)^{-T(R_i)}
\end{equation}
This makes the formulas symmetric with respect to the $Z$ factors 
of the gauge fields $Z_g=1/g^2$. The integration  in the measure over each
collective coordinate  is accompanied by $Z^{\mp1/2}$ (plus for bosonic,  minus for
fermionic moduli). 

Note, that in the instanton calculations the infrared cut off is provided by the
instanton size $\rho$.  Thus, we can choose  $Z(\rho)=1$ at the
infrared scale, adjusting $Z_i( M_{\rm PV})$. This is  similar to the gauge
coupling: we treat $g(\rho )$ as a physical fixed coupling allowing the bare
coupling $g$ to ``float".

Then the generalization of Eq.\ (\ref{efbf}) is
\begin{equation}
(M_{\rm PV})^{3T_G-\sum T(R_i)} \left(\frac{1}{g^2}\right)^{T_G}\!
\exp \left(-\frac{8\pi^2}{g^2}\right) \, \prod_i \left( Z_i\right)^{-T(R_i)}
={\rm const}
\,.
\label{efbfm}
\end{equation}
The right-hand side is independent of
$M_{\rm PV}$; the factor
$(M_{\rm PV})^{3T_G-\sum T(R_i)}$ on the left hand side must be compensated by
an implicit
$M_{\rm PV}$-dependence of $g$ and $Z_i$. Differentiating
over $\ln M_{\rm PV}$ one gets  the $\beta$ function.

In distinction with the pure gauge case Eq.\ (\ref{efbfm}) does not fix 
the running of the gauge coupling  {\em per se}; rather, it expresses the running of
the gauge coupling via the anomalous
dimensions of the matter fields,
\begin{equation}
\gamma_i \equiv - \frac{{\rm d} \ln Z_i}{{\rm d} \ln M_{\rm PV}}\, .
\label{defgamma}
\end{equation}
Taking the logarithm and  differentiating over $\ln M_{\rm PV}$
we arrive at
\begin{equation}
\beta (\alpha) \equiv \frac{{\rm d}\,\alpha\,(M_{\rm PV})}{{\rm d}\ln M_{\rm
PV}}= -\frac{\alpha^2}{2\pi}\left[3\,T_G -\sum_i T(R_i)(1-\gamma_i )
\right]\left(1-\frac{T_G\,\alpha}{2\pi} \right)^{-1}
\, .
\label{nsvzbetaf}
\end{equation}
This is the NSVZ $\beta$ function. 

It is worth noting that one can readily 
derive it in perturbation theory too, from a {\em one-loop} 
calculation, using some general properties of supersymmetry. A pedagogical 
discussion of the one-loop perturbative calculation leading to 
Eq.\ (\ref{nsvzbetaf}) and  related ideas are presented in Ref.\ \cite{little}. 
The relation between the NSVZ $\beta$ function and those obtained in other 
schemes (for instance, the standard dimensional reduction, and so on)
was the subject of a special investigation \cite{Jack}.

It is interesting to examine how the general formula~(\ref{nsvzbetaf}) works
in some particular cases. 
Let us start from the theories with the extended SUSY, ${\cal N}=2$.
They can be presented as ${\cal N}=1$ theories containing one matter
field in the adjoint representation which enters the same extended
supermultiplet as the gluon field. Therefore, its $Z$ factor $Z=1/g^2$ and
$\gamma=\beta/\alpha$. In addition, we allow for some number of the
matter hypermultiplets in the arbitrary color representations (let us remind that
every hypermultiplet consists of two ${\cal N}=1$ chiral superfields).  
The ${\cal N}=2$ SUSY leads to $Z=1$ for all hypermultiplets.
Indeed, for ${\cal N}=2$ the K\"ahler potential is in-one-to-one
correspondence with the superpotential. The latter is not renormalized
perturbatively owing to ${\cal N}=1$ SUSY.  Hence,   the K\"ahler potential for the 
hypermultiplets is not renormalized too, implying that  $Z=1$.

 Taking into
account these facts we derive from Eq.\ (\ref{nsvzbetaf}) the following gauge
coupling 
$\beta$ function:
\begin{equation}
\beta_{\,{\cal N}=2} (\alpha) = -\frac{\alpha^2}{2\pi}\left[2\,T_G -
\sum_i T(R_i)
\right]
\, .
\label{ntwobetaf}
\end{equation}
Here the summation runs over the ${\cal N}=2$ matter hypermultiplets.
This result proves that the $\beta$ function is one-loop in  ${\cal N}=2$ theories.

We can now make  one step further passing to ${\cal N}=4$. In terms of ${\cal
N}=2$ this theory corresponds to one matter hypermultiplet in the adjoint
representation.  Substituting $\sum T(R_i)=2\,T_G$ in  Eq.\ (\ref{ntwobetaf})
produces the vanishing  $\beta$ function. Thus, the ${\cal N}=4$ theory  is finite.

In fact, Eq.\ (\ref{ntwobetaf}) shows that the class of finite theories is much wider.
Any ${\cal N}=2$ theory with the matter hypermultiplets satisfying the 
condition $2\,T_G - \sum_i T(R_i)=0$ is finite. An example is provided  by 
$T_G$  hypermultiplets in the fundamental representation.

The NSVZ $\beta$ function allows one to find finite theories even in the class of
${\cal N}=1$.  The simplest example  was
suggested in  Ref.\ \cite{PWJM}, further developments are presented
in~\cite{HSU}. The general idea is to have the matter sector such that the conditions
$3\,T_G - \sum_i T(R_i)=0$ and $\gamma_i=0$ are met
simultaneously. For instance~\cite{PWJM},
consider the SU(3) gauge model with nine triplets $Q^i$ and nine antitriplets
$\tilde Q_i$ and the superpotential
\begin{equation}
{\cal W}=h\left(Q^1 Q^2 Q^3 +Q^4 Q^5 Q^6 +Q^7 Q^8 Q^9 +
\tilde Q_1 \tilde Q_2 \tilde Q_3 +\tilde Q_4 \tilde Q_5 \tilde Q_6 +\tilde Q_7 \tilde
Q_8 \tilde Q_9\right)\,,
\end{equation}
where an implicit contraction of the color indices by virtue of $\epsilon_{ijk}$
is implied.
The flavor symmetry of the model ensures that there is only one $Z$ factor for all
matter fields. Since the condition $3\,T_G - \sum_i T(R_i)=0$ is satisfied, the
finiteness is guaranteed provided that the anomalous dimension $\gamma$
vanishes. At small $g$ and $h$ the anomalous dimension $\gamma (g, h)$ is
determined by  a simple one-loop calculation,
\begin{equation}
\gamma (g, h)=-\frac{g^2}{3\pi^2} + \frac{|h|^2}{4\pi^2}\,.
\end{equation}
This shows that the condition $\gamma (g, h)=0$ has a solution, at least in the
vicinity of small couplings.

\subsubsection{NSVZ $\beta$ function and Wilsonean action}
\label{Wilsact}
In the instanton derivation of the $\beta$ function presented above the zero
modes play a crucial role. If they were absent the $\beta$ function would vanish.
The impact of zero modes is two-fold: first, they determine the power of $M_{\rm
PV}$, i.e. the one-loop coefficient $b $ of the $\beta$ function, 
\begin{equation}
b=n_b-\frac 1 2 n_f\,,
\end{equation}
where $n_b$ and $n_f$ are the numbers of bosonic and fermionic zero modes,
respectively. Second, the higher loop coefficients in  the $\beta$ function
are completely determined by the renormalization of the zero modes. 

In terms of the perturbation theory the latter implies that the first loop is factored
out from higher loops.  It is instructive to trace  this phenomenon in perturbation
theory {\em per se}. Let us examine the two-loop graph of Fig.\ 1. The graph
refers to the covariant background field technique~\cite{SSOT}. The difference with
the instanton calculation is that the gauge background field is assumed to be weak,
the gauge coupling renormalization is determined by the quadratic in the external
field term $W^\alpha W_\alpha$. Performing the integration over the primed
supercoordinate $\{ x',\theta' ,\bar\theta'\}$ we find that the effective action
generated by this graph takes the form 
\begin{equation}
\Delta S= \int {\rm d}^4 x {\rm d}^2 \theta {\rm d}^2 \bar \theta {\cal F}[W]
\,,
\end{equation}
where the gauge invariant functional ${\cal F}$ depends on $W_\alpha$ and
$\bar W_{\dot\alpha}$. If  there were no infrared singularities ${\cal F}$ would be
expandable  in powers of $W$, and the quadratic terms $W^2$, $\bar W^2$ would
obviously drop out. This would mean the vanishing of the second loop in the
$\beta$ function, which certainly cannot be true. The loophole is in the infrared
singularities,   ${\cal F}$ is {nonlocal}. The nonlocality is of the type
\begin{equation}
{\cal F}\propto W\,\frac{D^2}{\partial^2}\,W
\,,
\end{equation}
which leads to the local expression $W^2$ after the integration over ${\rm d}^2
\bar \theta$.
Note, that the proof of the nonrenormalization theorem in the instanton background
 in Sec.\ \ref{sec45} does not suffer from this problem -- the instanton field itself
provides the infrared regularization.

Thus, we see that all higher loops penetrate in the $\beta$ function through the
infrared, the first loop is factored out. In the Wilsonean action, where the  infrared
effects are excluded by construction, the gauge coupling is renormalized at one loop
only~\cite{anomaly}. Higher loops are absorbed into matrix elements and
produce the relation between the Wilsonean and the standard gauge 
coupling$\,$\footnote{The Wilsonean coupling is often referred to as the holomorphic
coupling in the current literature.}.

\subsubsection{Making contact with the perturbative definition of $\Lambda$}
\label{pertL}
The expressions~(\ref{efbf}) and (\ref{efbfm})  establish combinations of the bare
Lagrangian parameters and $M_{\rm PV}$ which are cut-off independent.
Then it is natural to relate these combinations to the physical scale parameter
$\Lambda$ of the type used in perturbative QCD.  

The standard convention used by the QCD
practitioners~\cite{PDG} is
\begin{equation}
\Lambda_{\rm pt}^b =\mu^b\left(\frac{16\pi^2}{b \, g^2(\mu)} \right)^{b_1/b}
\! \exp\left( - \frac{8\pi^2}{g^2 (\mu)}  
\right)\, ,
\label{Lptdefin}
\end{equation}
where $b$ and $b_1$ are the first and and the second coefficients in the $\beta$
function and the third and higher loops are neglected. 

Let us start from SUSY gluodynamics.  Then,
\begin{equation}
b=3T_G\,,\qquad b_1= 3T_G^2\,,
\end{equation}
and the expression for $\Lambda$ takes the form
\begin{equation}
\Lambda_{G} =M_{\rm PV}\left(\frac{16\pi^2}{3T_G \, g^2} \right)^{1/3}
\! \exp\left( - \frac{8\pi^2}{3T_G \,g^2 }  
\right)\, ,
\label{Ltg}
\end{equation}
where we substituted $\mu$ by $M_{\rm PV}$ and $g^2(\mu)$ by the bare
coupling  $g^2$. In distinction with the general QCD case the expression~(\ref{Ltg})
is exact rather than two-loop. This can be seen from comparison with the exact
relation~(\ref{efbf}). Let us stress that, due to factorization, the coupling constant
and the group factors enter in the combination $T_G \,g^2$ for any gauge group.
This scaling is well-known in the large $N$ limit of SU($N$). In the supersymmetric
theory factorization makes the scaling exact.
Later on we will calculate the gluino condensate and express it in terms of
this $\Lambda_{G}$.

In the theories with matter the situation becomes more complex.
In this case the gauge coupling constant is not the only parameter defining the
renormalization procedure. The parameters of the 
superpotential (masses and  the Yukawa couplings) are also
involved in the renormalization procedure. They are renormalized through the
$Z$ factors of the matter fields. 
Thus, besides the running gauge couplings one has to take into account the running
masses and the Yukawa couplings. 

Unlike SUSY gluodynamics where the general formula (\ref{Lptdefin}) was in fact
exact, in the theories with matter it looses its exact nature.  
 For this reason, it is inconvenient to use Eq.\  (\ref{Lptdefin})
as a starting definition. The
exact renormalization group invariant combination is displayed in Eq.\
(\ref{efbfm}).  We will invoke it to introduce the scale parameter, 
\begin{equation}
\Lambda^b= (M_{\rm PV})^b
\left(\frac{16\pi^2}{b g^2}\right)^{T_G}\!
\exp \left(-\frac{8\pi^2}{g^2}\right) \, \prod_i
Z_i^{-T(R_i)} 
\,,
\label{efbfm2}
\end{equation}
where 
\begin{equation}
b=3T_G-\sum T(R_i)\,.
\end{equation}

This definition assumes that $Z_i=1$ in the infrared.
Certainly, this $\Lambda$ can be related to $\Lambda_{\rm pt}$, Eq.\
(\ref{Lptdefin}), at the two-loop level. For example,  
one may consider the theory with purely gauge interactions, i.e. the superpotential
reduces to the mass terms. 
For the two-loop comparison, it is sufficient to know $Z_i$ in the one-loop
approximation,
\begin{equation}
Z_i=\left[\frac{g^2}{g^2(v)}\right]^{-2C_2(R_i)/b}
\,,
\end{equation}
where $C_2$ is the quadratic Casimir
\begin{equation}
C_2(R_i)=T(R_i)\,\frac{{\rm dim}(G)}{{\rm dim}(R_i)}\,,
\end{equation}
and dim stands for dimension of the representation. It is implied that the theory is
fully Higgsed and $v$ is the scale of all moduli. Comparing with the
definition~(\ref{Lptdefin}) we find
\begin{equation}
\Lambda=\Lambda_{\rm pt} \left[\frac{16\pi^2}{bg^2(v)}\right]^{2\sum
C_2(R_i)T(R_i)/b^2}
\,.
\end{equation}

\subsubsection{Perturbative versus nonperturbative $\beta$ functions}
\label{pvnbf}

The NSVZ $\beta$ function derived above was shown to be exact in perturbation
theory. A natural question arises about nonperturbative effects in the
$\beta$ function. In some cases it is known for a long time that such effects are
present. The most famous example is ${\cal N}=2$ gluodynamics where Seiberg
found~\cite{seib88} the  one-instanton exponential term in the gauge
coupling.
 Later on the full answer
containing all exponential terms was obtained by Seiberg and Witten~\cite{SEIW}. 

Although our consideration does not include the nonperturbative
corrections to the $\beta$ function we present here symmetry arguments
which prompt us  in which theories the NSVZ $\beta$ function is
nonperturbatively exact and in which cases  nonperturbative
corrections are possible. The symmetry we keep in mind is the
$R$ symmetry. 

One can  formulate a general {\em theorem}. Consider a generic point on the
moduli space of vacua.   Assume that this point corresponds  either to the Higgs
phase or to the Abelian Coulomb phase, in the weak coupling regime.  Then, if no
combination of moduli (respecting the flavor symmetry of theory) has the
vanishing $R$ charge, then the NSVZ $\beta$ function is nonperturbatively exact.

The proof is quite straightforward.  Under the assumption above, we deal 
essentially with one (flavor symmetric) modulus $M$ of a nonzero $R$ charge. If
two or more  such moduli existed one could always organize an $R$ neutral
combination. Then,  for each given chiral quantity the $R$ symmetry uniquely
fixes its dependence on the modulus. This dependence is of the  form
$ \Lambda^k/M^n$ where $\Lambda$ is the scale parameter of the theory.
Since $\Lambda\propto \exp[-8\pi^2/(3T_G -\sum T(R_i))g^2]$ no iteration of the
exponential terms occur.  We will further elaborate this
topic in a slightly different, although related context, in Sec.\ \ref{sec48}.

An example is provided by SQCD with the gauge group SU($N$) and the number of
flavors $N_f=N-1$. The appropriate chiral quantity to consider is the superpotential
${\cal W}$ which depends on
\begin{equation}
M={\rm det}\left\{ Q^f \tilde Q_g\right\}\,.
\end{equation}
The $R$ charge of $M$ is equal to $-2$ which fixes ${\cal W}$ to be ${\cal
W}=\Lambda^{2N+1}/M$.

On other hand, in the ${\cal N}=2$ gluodynamics with the gauge group SU(2) 
the only modulus Tr\,$\Phi^2$ has the vanishing $R$ charge -- thus, the series in
$\Lambda^4/({\rm Tr}\,\Phi^2)^2$ is not ruled out, and, sure enough,  it actually 
emerges~\cite{SEIW}.

Note, however, that these nonperturbative effects in the gauge coupling
physically are due  to monopoles which are heavy in weak coupling.
For this reason the nonperturbative effects do not contribute to the
running of the coupling --
the running is governed by the one-loop $\beta$ function.
The situation is similar to the effect of the heavy quark in the  QCD
gauge coupling, The heavy quark does not
contribute to the running below the heavy quark threshold.  

\subsection{Gluino condensate in SUSY gluodynamics}
\label{secglucon}

As a first example of the nonperturbative phenomenon let us consider
the calculation of the gluino condensate in SU(2) gluodynamics~\cite{NSVZ1}.
To this end we consider the two-point function 
\begin{equation}
\langle 0| T \left\{{\rm Tr}\,W^2(x_L,\theta), {\rm
Tr}\,W^2(x^\prime_L,\theta^\prime)\right\}|0\rangle\,.
\label{wwcor}
\end{equation}
Supersymmetry fixes the superspace coordinate dependence of this correlator to be
\begin{equation}
{\rm const} +(\theta-\theta^\prime)^2 F(x_L -x^\prime_L)\,.
\end{equation}
We see that the correlator of the lowest components ${\rm Tr}\,\lambda^2$ can
only be constant, which, due clusterization, implies
\begin{equation}
\Pi=\langle 0| T \left\{{\rm Tr}\,\lambda^2(x_L,\theta), {\rm
Tr}\,\lambda^2(x^\prime_L,\theta^\prime)\right\}|0\rangle
=\langle 0|{\rm Tr}\,\lambda^2|0\rangle^2\,.
\end{equation}

It is obvious that $\Pi$ vanishes in  perturbation theory. The one-instanton
contribution to $\Pi$ does not vanish, however. This is readily seen
from the balance of the zero modes. The calculation
is quite straightforward,
\begin{equation}
\Pi=\int {\rm d}\mu_{{\rm SU}(2)}\, {\rm Tr}\,W^2_{\rm inst}(x_L,\theta=0) \, 
{\rm Tr}\,W_{\rm inst}^2(x^\prime_L,\theta^\prime=0)\,,
\end{equation}
where the instanton measure ${\rm d}\mu_{{\rm SU}(2)}$ is displayed in 
Eq.\ (\ref{fimsg1}) and ${\rm Tr}\,W^2_{\rm inst}$ is given in Eq.\
(\ref{w2shifted}). Integrating over the fermionic coordinates $\theta_0$
and $\bar\beta$ we arrive at 
\begin{equation}
(x-x^\prime)^2\,\int\! {\rm d}\rho^2{\rm d}^4 x_0 \, \frac{\rho^8
}{\left[(x-x_0)^2+\rho^2\right]^4\left[(x^\prime-x_0)^2+\rho^2\right]^4}\,.
\label{invtop}
\end{equation}
The integral over $x_0$ and $\rho$ is well-defined. On dimensional grounds it is
proportional $1/(x-x^\prime)^2$, the integral over $\rho^2$ is totally saturated
at $\rho^2\sim (x-x^\prime)^2$. Thus, the expression~(\ref{invtop}) is just a
number, $\pi^2/45$. 

Collecting all numericals, we get 
\begin{equation}
\langle 0|{\rm Tr}\,\lambda^2|0\rangle^2= \frac{2^{10}\pi^4}{5}\,
e^{-8\pi^2/g^2}\,\frac{M_{\rm PV}^6}{g^4}=\frac{144}{5}\,\Lambda_G^6
\,,
\label{glucon}
\end{equation}
where the scale parameter $\Lambda_G$ is defined in Eq.\ (\ref{Ltg}).
The gluino condensate, which is obviously a nonperturbative effect, is expressed
here in terms of the scale parameter $\Lambda_G$ introduced through a standard
perturbation theory formula. Note a relatively large ($\sim$30) numerical coefficient
in Eq.\ (\ref{glucon}). Note also that the gluino condensate comes out
double-valued. This is in agreement with the existence of two bosonic vacua
in the theory. Two vacua correspond to $I_W=2$ and to the spontaneous breaking 
$Z_4\to Z_2$, see Sec.\  \ref{secEL}.  

What is the theoretical status of this  derivation performed in the strong coupling
regime? Formally, one can start from the correlator at $(x-x^\prime)^2\ll
\Lambda_G^{-2}$ where the coupling is small and the semiclassical analysis should
by reliable. Inside the instanton calculation of the correlator $\Pi$  we
see no corrections, either perturbative, or nonperturbative. 

On the other hand, there are still unresolved issues. One may ask, for instance,
how the result~(\ref{glucon}) is compatible with the fact that the one-instanton
calculation of $\langle 0|{\rm Tr}\,\lambda^2|0\rangle$ yields zero. It looks as an
apparent contradiction with the cluster decomposition. The answer to this question
is not yet clear. A possible explanation was suggested in Ref.\ \cite{Amati}:
it was argued that the instanton calculation refers to an average over the two
bosonic vacua of the theory. This averaging makes $\langle 0|{\rm
Tr}\,\lambda^2|0\rangle$ to vanish while the square is the same for both vacua.

Unfortunately, it is not the end of the story. One can calculate the same gluino
condensate starting from the one-flavor model (Sec.\ \ref{sec471}) in the weak
coupling regime. Proceeding from the small to large quark mass and using the
holomorphic dependence of $\langle 0|{\rm Tr}\,\lambda^2|0\rangle$ on $m$
we return back to strong coupling. The gluino condensate found in such a way, see
Eq.\ (\ref{lamwc}),  contains an extra factor $\sqrt{5/4}$ compared to the
condensate following from Eq.\ (\ref{glucon}).  
Within the hypothesis\cite{Amati}
about averaging over vacua the discrepancy can be interpreted as a signal
of the
existence of an extra chirally symmetric vacuum (see Sec.~\ref{sec7}). 

There is, however,  one more relevant test: the ${\cal N}=2$ theory softly
broken to
${\cal N}=1$ by a small mass of the matter fields. The Seiberg-Witten
solution\cite{SEIW}  then gives the value of gluino condensate 
which is larger than what follows from the correlator
calculation, again by the same factor $\sqrt{5/4}$ (for details
see\cite{RVto}). There is no extra
vacuum in this problem, which shows that the vacuum averaging hypothesis
does not work in the case at hand. The problem of instanton calculations
in strong coupling remains unsolved.

The correlator~(\ref{wwcor}) is the simplest in its class. One can extend
the approach to include more chiral operators:  $n$-point functions of
$W^2$ and/or chiral operators constructed from the matter fields~\cite{RoVe}.
All correlators of this type are of the topological nature -- this feature was revealed
in the most transparent way in the topological field theories constructed later by
Witten~\cite{wtop}.

\subsection{One-flavor model}
\label{sec471}

The classical structure of  SQCD with the gauge group SU(2)
and one flavor was discussed
in Sec.\ \ref{sec312}. The model has one modulus 
\begin{equation}
\Phi=\sqrt{ Q_\alpha^f  Q^\alpha_f/2}\,.
\end{equation}
In the absence of the superpotential all vacua with different $\Phi$ are
degenerate. The degeneracy is  not lifted to any finite order of perturbation
theory. As shown below it is lifted nonperturbatively~\cite{ADS2}
 by an instanton
generated superpotential ${\cal W}(\Phi)$.

Far away from the origin of the valley, when $|\Phi|\gg \Lambda$, the
gauge SU(2) is spontaneously broken, the theory is in the Higgs 
regime, and the gauge bosons are  heavy. In addition, the gauge coupling 
is small, so that the quasiclassical treatment is reliable. 
At weak coupling the leading nonperturbative contribution is due to 
instantons. Thus, our task is to find the instanton-induced effects.

The exact $R$ invariance of the model is sufficient to establish 
the functional form of the effective superpotential ${\cal W}(\Phi)$,
\begin{equation}
{\cal W}(\Phi) \propto \frac{\Lambda^5_{\rm one-fl}}{\Phi^2} \,,
\label{arkatwo}
\end{equation}
where the power of $\Phi$ is determined by its $R$ charge ($R_\Phi=-1$)
and the power of $\Lambda$ is fixed on  dimensional grounds.
Here we introduced the notation
\begin{equation}
\Lambda^5_{\rm one-fl}=\frac{e^{-8\pi^2/g^2}}{Z g^4}\,(M_{\rm PV})^{5}
\,,
\label{lambda1}
\end{equation}
which coincides with the general definition~(\ref{efbfm2}) up to a numerical factor.

To see that one instanton  does induce this superpotential, 
we consider the instanton transition in the background  field $\Phi(x_L,\theta)$
weakly depending on the superspace coordinates. To this end one generalizes the
result~(\ref{imesqcd}), which assumes $\Phi=v$ at distances much larger than
$\rho$, to a variable superfield $\Phi$,
\begin{equation}
{\rm d}\mu = \frac{1}{2^{5}} \,\frac{\Lambda^5_{\rm one-fl}}{\Phi^2
(x_0,\theta_0) }
\,
\exp\left(-4\pi^2\,\bar\Phi \Phi \,\rho^2_{\rm inv}\right) 
\frac{ {\rm d}\rho^2}{\rho^2}\,
 {\rm d}^4 x_0\, {\rm d}^2\theta_0 \,
 {\rm d}^2\bar\beta\,  {\rm d}^2\bar\theta_0 \, .
\label{imesqcd1}
\end{equation}

There exist many alternative ways to  verify
that this generalization is indeed correct. For instance,   one could  calculate the
propagator of the quantum part of
$\Phi = v + \Phi_{\rm qu}$ using the constant background
$\Phi=v$ in the measure, see Sec.\ \ref{secSURF} for more details.

The effective superpotential is obtained by integrating over $\rho$, $\bar\beta$
and $\bar\theta_0$. Since these variables enter the measure only through 
$\rho^2_{\rm inv}$, at first glance the integral seems to be vanishing.
Indeed, changing the variable $\rho^2$ to $\rho^2_{\rm inv}$  makes the
integrand independent of $\bar\beta$ and $\bar\theta_0$. This is not the case,
however. 
The loophole is due to the singularity at $\rho^2_{\rm inv}=0$. To resolve the
singularity let us integrate first over the fermionic variables. For an arbitrary
function $F(\rho^2_{\rm inv})$ the integration takes the form
\begin{equation}
\int \frac{ {\rm d}\rho^2}{\rho^2} \,{\rm d}^2\bar\beta\,  {\rm d}^2\bar\theta_0
\,F(\rho^2[1+4i\bar\beta\bar\theta_0])=\int \frac{ {\rm d}\rho^2}{\rho^2} \,16
\rho^4 F^{\prime\prime}(\rho^2)= 16\, F(\rho^2=0)\,.
\end{equation}
The integration over $\rho^2$ was performed by integrating by parts twice.
It was assumed that $ F(\rho^2\to\infty)=0$. 
It is seen that the result depends only on the zero-size instanton. In other words,
\begin{equation}
\frac{ {\rm d}\rho^2}{\rho^2} \,{\rm d}^2\bar\beta\,  {\rm d}^2\bar\theta_0
\,F(\rho^2_{\rm inv})= 16\, {\rm d}\rho^2_{\rm inv}\,\delta(\rho^2_{\rm
inv})F(\rho^2_{\rm inv})\,.
\end{equation}

The instanton generated superpotential is$\,$\footnote{A dedicated analysis of 
various one-instanton effects has been  carried out 
recently~\cite{FiPo}, with the aim of collecting numerical factors that had been
usually ignored in the previous works. Equation (3.1) in~\cite{FiPo}
coincides with Eq.\ (\ref{effw}), provided we restore the factor $g^{-4}Z^{-1}$
omitted (deliberately) in Ref.\ \cite{FiPo}. }
\begin{equation}
{\cal W}_{\rm inst}(\Phi) =\frac{\Lambda^5_{\rm one-fl}}{\Phi^2
}\, .
\label{effw}
\end{equation}
The result presented in Eq.\ (\ref{effw}) bears
a topological nature: it does not depend on the particular form of the  integrand
$F(\rho^2_{\rm inv})$ since the integral  is determined by $\rho^2 = 0$. The
integrand is given by the exponent only at small $\rho^2$. No matter how it
behaves as a function of
$\rho^2$, the formula for the superpotential
 is the same, provided that the
integration over
$\rho^2$ is convergent
at large $\rho^2$. We advertised this assertion -- the saturation at $\rho^2 =
0$ -- more than once previously. Technically,  the saturation at $\rho^2 = 0$
makes the calculation self-consistent
(remember, at $\rho^2 = 0$ the instanton solution becomes exact in the
Higgs phase) and explains why the result  (\ref{effw}) 
gets no perturbative corrections in higher orders.

We see
that in the model at hand the instanton does generate a superpotential which lifts
the vacuum degeneracy.  The result is exact both perturbatively and
nonperturbatively. 

In the absence of the tree-level superpotential the induced superpotential leads to
a run-away vacuum -- the lowest energy state is achieved at the infinite value of
$\Phi$. One can stabilize the theory by adding  the mass term $m \Phi^2$ in the
classical superpotential. The total superpotential then  takes the form
\begin{equation}
{\cal W}(\Phi) =m \Phi^2+{\cal W}_{\rm inst}(\Phi)\, .
\label{effwtot}
\end{equation}
One can trace the origin of the second term to the anomaly in Eq.\ (\ref{ofasp})
in the original full theory (i.e. the theory before integrating out the gauge fields). 

Minimizing energy we get two supersymmetric vacua at
\begin{equation}
\langle \Phi^2\rangle = \pm \left[\frac{\Lambda^5_{\rm one-fl}}{m }\right]^{1/2}
\,.
\end{equation}
To obtain the gluino condensate one can use the Konishi relation~(\ref{ka1})
which in the case at hand implies
\begin{equation}
\langle {\rm Tr}\,\lambda^2\rangle= 16\pi^2 m \langle \Phi^2\rangle=
\pm 16\pi^2\left[m\Lambda^5_{\rm one-fl}
\right]^{1/2}=\pm 16\pi^2\left[\frac{m\, e^{-8\pi^2/g^2}}{ Z g^4}\,(M_{\rm PV})^{5}
\right]^{1/2}\,.
\label{lamwc}
\end{equation}
With our convention~(\ref{zmatt}) the bare quark mass $m_{\rm bare}$ is $m/Z$, 
therefore the gluino condensate dependence on $m_{\rm bare}$ is holomorphic.
In fact, the square root dependence on  $m_{\rm bare}$ is an exact 
statement~\cite{SVMO}. This allows one to pass to large $m_{\rm bare}$ where 
the matter could be viewed as one of the regulators. Setting  $m_{\rm
bare}=M_{\rm PV}$ we return to supersymmetric gluodynamics. Comparing the
square of $\langle {\rm Tr}\,\lambda^2\rangle$ from Eq.\ (\ref{lamwc}) with Eq.\
(\ref{glucon}) we find~\cite{NSVZ3} a mismatch factor $4/5$ (Eq.\ (\ref{lamwc})
yields a larger result).

In addition to the holomorphic dependence on $m_{\rm bare}$, the gluino
condensate depends holomorphically on the regulator mass $M_{\rm}$. 
As for the gauge coupling, $1/g^2$ in the exponent can be complexified according to
Eq.\ (\ref{itheta}), but in the pre-exponential factor it is Re$g^{-2}$ that enters.
This is the holomorphic anomaly~\cite{AVMS1} -- the dependence on the Wilsonean
coupling constant $g_W$ of SUSY gluodynamics
\begin{equation}
\frac{1}{g_W^2}=\frac{1}{g^2}-\frac{1}{4\pi^2}\ln \left[{\rm Re}\, \frac{1}{g^2}
\right]
\end{equation}
is holomorphic.

Let us stress the statement of the holomorphic dependence refers to the bare
parameters. For instance, if one expresses the very same gluino condensate 
in terms perturbative scale $\Lambda_{\rm pt}$ and the physical mass 
of the modulus field $m_\Phi=2m$ one would get at the two-loop level
\begin{equation}
\langle {\rm Tr}\,\lambda^2\rangle^2 \propto m_\Phi \, \Lambda_{\rm pt}^5 
\left[\ln  \frac{\Lambda_{\rm pt}}{m_\Phi }\right]^{3/10}
\,.
\end{equation}
Not only the holomorphy in $m$ is lost -- this ugly expression is only approximate,
higher loops result in further logarithmic corrections.

\subsection{Two-flavor model}
\label{sec472}
The two-flavor model presents a new phenomenon: although the superpotential is
not generated, some points of the classical moduli space become inaccessible.
In other words, the geometry of the moduli space is changed~\cite{Nati1}. 

Compared to the previous section we add one extra flavor, i.e. now
we deal with four  chiral superfields $Q_f$ where $f = 1,2,3,4$ is the subflavor
index.  The $D$-flat direction is parametrized by six chiral invariants,
\begin{equation}
M_{fg} = -M_{gf}=Q^{\alpha}_fQ^{\beta}_g\,\epsilon_{\alpha\beta}\,.
\end{equation}
The matrix $M_{fg}$ is antisymmetric in 
$f,g$ and is  subject to one classical constraint,
\begin{equation}
{\rm Pf} (M) \equiv \frac 1 2 \,\epsilon^{fgpq}M_{fg}M_{pq} = 0\, .
\label{pfaf}
\end{equation}
The combination on the right-hand side  is called {\em Pfaffian}.
The flavor rotations allow one to render the matrix $M_{fg}$ in the form where
$M_{12}$ and $M_{34}$ are the only nonvanishing elements.
Then the constraint~(\ref{pfaf}) means that the classical moduli space
${\cal M}$ consists of  two manifolds ${\cal M}_1$ and ${\cal M}_2$,
\begin{equation}
{\cal M}_1=\{M_{34}=0\,, ~M_{12}\mbox{~arbitrary}\}\,,\qquad
{\cal M}_2=\{M_{12}=0\,,~M_{34}\mbox{~arbitrary}\}\,.
\end{equation}

Our task is to find a superpotential on each branch. In the absence of 
tree-level superpotential ${\cal W}_{\rm tree}$
the  effective superpotential  is not generated, unlike the one-flavor model.
Let us introduce a classical mass term which can be chosen as 
\begin{equation}
{\cal W}_{\rm tree}=m^{12}M_{12} +m^{34}M_{34}\,,
\label{clw}
\end{equation}
without loss of generality.
The theory is defined as a limit when the parameters $m^{fg}$ go to zero.
The instanton-generated effects should be calculated separately on two manifolds
${\cal M}_1$ and ${\cal M}_2$. 

For the manifold ${\cal M}_1$ the matter field configuration appropriate for the
instanton calculation is
\begin{eqnarray}
&& Q^\alpha_{1,2}(x_L,\theta;\bar\theta_0)=
v\left\{\frac{(x_L)^\alpha_{1,2}}{\sqrt{x_L^2+\rho^2}}+4(\bar\theta_0)_{1,2}
\theta^\alpha\frac{\rho^2}{(x_L^2+\rho^2)^{3/2}}\right\}\,,\nonumber\\[0.2cm]
&& \bar Q^\alpha_{1,2}(x_R,\theta)=
\bar v\,\frac{(x_R)^\alpha_{1,2}}{\sqrt{x_R^2+\rho^2}}\,,\nonumber\\[0.2cm]
&& Q^\alpha_{3,4}(x_L,\theta;\bar\eta_{3,4})=
\frac{\sqrt{2}}{\pi}\,\bar\eta_{3,4}\,
\theta^\alpha\frac{\rho}{(x_L^2+\rho^2)^{3/2}}\,,\quad 
\bar Q^\alpha_{3,4}=0\,.
\label{config}
\end{eqnarray}
The fields $Q^\alpha_{1,2}$ contain the boson component, corresponding to 
$M_{12}=v^2$ and the fermion zero modes.  The fields $Q^\alpha_{3,4}$ contain
only the zero fermion modes. Substituting this configuration into the classical action
we arrive at the following expression for the instanton measure:
\begin{equation}
{\rm d}\mu={\rm d}\mu_{\rm one-fl}\,\exp\left(- m^{34}\,\bar\eta_3\bar\eta_4\right) \frac{{\rm
d}\bar\eta_3{\rm d}\bar\eta_4}{ M_{\rm PV}Z}\,,
\label{twofl}
\end{equation}
where ${\rm d}\mu_{\rm one-fl}$ is given in Eq.\ (\ref{imesqcd}).
This can be compared with general formula~(\ref{sumatt}). In the general formula
we put $v_j=v_{3,4}=0$ but accounted for the effect of the mass term
$m^{34}M_{34}$. The mass term $m^{12}M_{12}$ is neglected -- we are looking for
the effects which do not vanish in the limit $m^{12}\to 0$.

The  difference with the one-flavor case is factorized in Eq.\ (\ref{twofl}).  The
integration over $\bar\eta$ can be carried out. It yields
\begin{equation}
{\rm d}\mu={\rm d}\mu_{\rm one-fl}\,\frac{m^{34}}{M_{\rm
PV} Z}\,.
\label{twofl1}
\end{equation}
Combining this with Eq.\ (\ref{effw}) we get for the total effective potential 
on the manifold ${\cal M}_1$ the following expression:
\begin{equation}
{\cal W}=m^{12}M_{12}+\frac{m^{34}\,\Lambda^4_{\rm
two-fl}}{M_{12}}\,.
\label{twotot}
\end{equation}
Here 
\begin{equation}
\Lambda^4_{\rm two-fl}=\frac{\Lambda^5_{\rm one-fl}}{M_{\rm PV}Z}=
\frac{e^{-8\pi^2/g^2}}{Z^2 g^4}\,(M_{\rm PV})^{4}\,.
\end{equation}
In fact, Eq.\ (\ref{twotot}) is exact, both perturbatively and
nonperturbatively. This can be proven by analyzing the $R$ charges.

The superpotential~(\ref{twotot}) fixes the vacuum value of the modulus $M_{12}$,
\begin{equation}
\langle 
M_{12}\rangle_{\rm vac}=\pm\,\sqrt{\frac{m^{34}}{m^{12}}}\,\,\Lambda^2_{\rm
two-fl}\,.
\end{equation}
The modulus $M_{34}$ classically was zero. Nonperturbative
effects shift its vacuum value  from zero. It can found by virtue of the
nonanomalous Konishi relation (see Eq.\ (\ref{ka1}))
\begin{equation}
\bar D^2 (\bar Q^1 e^V Q_1 -\bar Q^3 e^V
Q_3)=4\,\left(m^{12}M_{12}-m^{34}M_{34}\right)\,.
\end{equation}
The vacuum average of the left-hand side vanishes implying
\begin{equation}
\langle M_{34}\rangle_{\rm vac}=\frac{m^{12}}{m^{34}}\,\langle 
M_{12}\rangle_{\rm vac}=\pm\,\sqrt{\frac{m^{12}}{m^{34}}}\,\,\Lambda^2_{\rm
two-fl}\,.
\label{konets}
\end{equation}

The remarkable phenomenon we encounter here is the change of the geometry 
of the moduli space. Classically, the product $M_{12}\,M_{34}$ vanishes while 
at the nonperturbative level $M_{12}\,M_{34}=\Lambda^4_{\rm two-fl}$.
Invariantly, this can be written as
\begin{equation}
{\rm Pf} (M_{fg})=\Lambda^4_{\rm two-fl}\,.
\label{constr}
\end{equation}

The analysis we performed referred to the manifold ${\cal M}_1$
under the assumption that $m^{12}\ll m^{34}$. This assumption was crucial to
ensure the weak coupling regime. Note that we could not perform the weak
coupling analysis on ${\cal M}_2$ if $m^{12}\ll m^{34}$. However, once
established in the weak coupling, the constraint~(\ref{constr}) remains valid
everywhere on the moduli space including the domain of the strong coupling. 
Note that the origin of the moduli space which was the singular point of
intersection of ${\cal M}_1$ and ${\cal M}_2$ disappears from the quantum
moduli space~(\ref{constr}) whose metric is smooth everywhere.

Note also that the vacuum average of the superpotential~(\ref{twotot}) can be
rewritten as 
\begin{equation}
\left\langle\, {\cal W}\,\right\rangle_{\rm vac}=\left\langle
m^{12}M_{12}+m^{34}M_{34}\right\rangle_{\rm vac}\,,
\end{equation}
which is exactly equal to the vacuum average of the original
superpotential~(\ref{clw}). This is in distinction with the one-flavor model
and can be understood given the expression for the central charge discussed in 
Sec.\ \ref{secSAA}. In the two-flavor model the anomaly vanishes, see 
Eq.\ (\ref{ofasp}), which forces the superpotential in the low-energy theory of the
moduli to coincide with the tree-level superpotential of the original full theory.

\subsection{How effective is the effective Lagrangian?}
\label{secHEEL}

The notion of the effective Lagrangian was used above to determine the vacuum
structure in the moduli space. It was assumed that the theory is in the Higgs phase
and the only light fields are the moduli. All heavy fields,  massive gauge bosons
and gauginos, were integrated out. As a result we get an effective Lagrangian
for the light degrees of freedom in the form of expansion in their momenta. In
particular, the instanton-generated superpotential, see e.g. Eq.\  (\ref{effw}), gives
rise to terms of the zeroth order in momentum. 

Such terms obviously describe the amplitudes at momenta below $m_W$.
The question is what happens above $m_W$. The point is that the scale $m_W$
is not a relevant parameter in the instanton calculations. Indeed, the integral over
instanton size is saturated at $\rho\sim v^{-1}$ which is 
parametrically smaller than $m_W^{-1}$ (remember, we are in the weak coupling
regime, $g^2$ is a small parameter). Moreover, the instanton-generated scattering
amplitudes do not depend~\cite{Ring,MLVV} on the particle momenta at all (on
mass shell). This means that the scale $v\sim m_W/g$ is also irrelevant.
Technically, this  is reflected in the possibility of rewriting  the integral over $\rho$
in such a form that only the
$\rho=0$ contribution survive.

As a result the cross sections of the instanton-generated processes, e.g. the baryon 
 number violation in the  Standard Model, 
grow with the energy and reach the unitary limit~\cite{BNVrev} at energies $\sim
v/g$. This scale is  the only one  relevant physically  for the nonperturbative
effects. At this scale new physics comes into play -- production of
sphalerons~\cite{sphaleron} or monopoles.  Here we enter the
strong coupling regime even at small $g^2$: multi-instanton effects become as
important as one-instanton, their summation is needed~\cite{ZMSV}.

In the interval of energies  $gv\lsim E \ll v/g$ one can limit oneself to  one
instanton;   the heavy fields should be kept, they  should not be integrated out,
however. Our task is constructing an  effective Lagrangian (or action)
which includes both, the chiral superfields and those of gluons/gluinos. 
Thus,  the problem we face is  a supergeneralization of the baryon number
violation at high   energies (but not superhigh, though, we stay much below the 
sphaleron energy).

In non-supersymmetric models the one-instanton effective action is known
for a long time. In the SU(2) model with one scalar and one Dirac field in the
fundamental representation it has the form,
\begin{equation}
{S}_{\rm eff}=\int {\rm d}\mu \exp\left\{-2\pi^2\rho^2\, (\bar \phi \phi
-|v|^2) -2\pi\rho\, \bar \eta_{\dot\alpha}
\bar\psi^{\dot\alpha}
+4i\pi^2\,\frac{\rho^2}{g^2}\,\Omega^{\dot\gamma}_{\dot\alpha}
\,\Omega^{\dot\delta}_{\dot\beta} \, \bar
G^{\{\dot\alpha\dot\beta\}}_{\dot\gamma\dot\delta}\right\}
\,,
\label{nonsusy}
\end{equation}
where the instanton measure ${\rm d}\mu$ should be modified to include the
nonzero modes and  $\Omega$ is the instanton orientation matrix defined in Eq.\
(\ref{orientM}). The fields in the exponent present the interpolating fields for the
particles which are scattered in the instanton background. In order to get the
scattering amplitude with the given number of external particles one expands the
exponent to the appropriate power of the appropriate field.

Let us generalize ${S}_{\rm eff}$ to SQCD with one flavor. This can be readily
done by applying the collective coordinate technique we have presented,
\begin{eqnarray}
S_{\rm one-fl}&\!\!=&\!\!\int\! {\rm d}\mu \,
\frac{2 v^2}{Q^f_\alpha Q_f^\alpha}\,\exp\left\{-2\pi^2\rho^2_{\rm inv}\,
 (\bar Q^f e^V Q_f -2|v|^2) \right. \nonumber\\[0.2cm]
&\!\!+&\!\!\left. 4\pi^2\,\frac{\rho^2_{\rm
inv}}{g^2}\,\Omega^{\dot\gamma}_{\dot\alpha}
\, \Omega^{\dot\delta}_{\dot\beta} \left[i \bar \nabla_{\dot\gamma}\bar
W^{\{\dot\alpha\dot\beta\}}_{\dot\delta}+8(\bar\beta_{\rm
inv})_{\dot\gamma} \bar W^{\{\dot\alpha\dot\beta\}}_{\dot\delta}\right]\right\}
\,,
\label{yung}
\end{eqnarray}
where $\rho^2_{\rm inv}$ and $\bar\beta_{\rm inv}$ are defined in Eq.\
(\ref{rhoinv}), and  the background-field-covariant spinor derivative
$\bar \nabla_{\dot\gamma}$ is defined as 
\begin{equation}
\bar \nabla_{\dot\gamma}\bar W= e^{V} \left[\bar  D_{\dot\gamma}\left(
e^{-V}\bar W e^{V}\right) \right] e^{-V}\,.
\label{covsupder}
\end{equation}
This result can be extracted from Ref.\  \cite{Yung}, where the ${\cal N}=2$
case was considered, by dropping off irrelevant for  ${\cal N}=1$ terms.

A few comments are in order about the origin of various terms in Eq.\ (\ref{yung}).
The first term in the exponent is a straightforward generalization of the first two
terms in Eq.\ (\ref{nonsusy}), SUSY combines interactions of squarks and quarks.
The term $\bar \nabla_{\dot\gamma}\bar
W^{\{\dot\alpha\dot\beta\}}_{\dot\delta}$ generalizes the last term in  Eq.\
(\ref{nonsusy}), it includes also a part of the  gluino interaction (related to the
supersymmetric zero modes). Finally, the last term presents the remaining part 
of the
gluino interactions (due to the superconformal zero modes). Indeed, if in the
non-supersymmetric case we  considered the matter in the adjoint
representation we would have four zero modes instead of two. The
supersymmetric modes would generate the fermion part in $\bar
\nabla_{\dot\gamma}\bar W^{\{\dot\alpha\dot\beta\}}_{\dot\delta}$, while the
superconformal ones the fermion part in $(\bar\beta_{\rm inv})_{\dot\gamma}
\bar W^{\{\dot\alpha\dot\beta\}}_{\dot\delta}$.

The measure ${\rm d}\mu$ contains integrations over $x_0$, $\theta_0$ and 
$\bar\theta_0$, as well as over $\rho^2_{\rm inv}$ and $\bar\beta_{\rm inv}$
(note  the absence of the nonzero modes in distinction with the
non-supersymmetric case). The external fields $Q$,  $\bar Q$,  and $\bar W$ are
taken at the point
$x_0, \theta_0,\bar\theta_0$ in the superspace. The integrations over
$\rho^2_{\rm inv}$ and $\bar\beta_{\rm inv}$ can be done explicitly.
Let us start with the integration over $\bar\beta_{\rm inv}$. If it were not for the
last term in the exponent, explicitly proportional to $\bar\beta_{\rm inv}$, the
result of the $\bar\beta_{\rm inv}$ integration would be naively zero. In fact,
as was discussed previously,  the integral 
 does not vanish due to the singularity at
$\rho^2_{\rm inv}=0$. The contribution from  $\rho^2_{\rm inv}=0$ coincides with
Eq.\ (\ref{effw}), because the exponent vanishes at  $\rho^2_{\rm inv}=0$. This
contribution is an $F$ term. 
The only part which comes from the domain $\rho^2_{\rm inv}\neq 0$
is due to the last term in Eq.\ (\ref{covsupder}).
At $\rho^2_{\rm inv}\neq 0$ we need to expand in 
$(\bar\beta_{\rm
inv})_{\dot\gamma} \bar W^{\{\dot\alpha\dot\beta\}}_{\dot\delta}$ and keep
the term of the second order. Performing then integrations over $\bar\beta_{\rm
inv}$ and $\rho^2_{\rm inv}$ we get the $D$ term. Overall,
\newpage
\begin{eqnarray}
&&S_{\rm one-fl}=\Lambda^5_{\rm one-fl}
\left\{\int\! {\rm d}^4 x \,{\rm d}^2
\theta \,\,
\frac{1}{Q^f_\alpha Q_f^\alpha} \right.\nonumber\\[0.2cm]
&&\left.+8\int\! {\rm d}^4 x \,{\rm d}^2 \theta \,{\rm d}^2 \bar
\theta\,\frac{{\rm d}^3\Omega}{8\pi^2}
\,\frac{1}{Q^f_\alpha Q_f^\alpha} \,
\frac{g^{-4}\,\Omega^{\dot\delta}_{\dot\beta}\,\bar
W^{\{\dot\beta\dot\alpha\}}_{\dot\delta}\,\bar
W_{\{\dot\alpha\dot\gamma\}\dot\rho}\,\Omega^{\dot\gamma\dot\rho}}
{\left[\bar Q^f e^V Q -2i g^{-2}\Omega^{\dot\gamma}_{\dot\alpha}
\, \Omega^{\dot\delta}_{\dot\beta}\bar \nabla_{\dot\gamma}\bar
W^{\{\dot\alpha\dot\beta\}}_{\dot\delta}\right]^2 }\right\}
\,.
\label{yung1}
\end{eqnarray}

This instanton-generated action contains the superpotential term (the first line),
which describes the scattering of the matter fields,  and the $D$ term (the second
line) which  involves both the matter and the gauge fields. It contains two extra
derivatives (or two extra fermions). The expression we obtained sums up all
tree-level  instanton-generated interactions. At the quantum level  the
superpotential term stays intact, while the $D$ term gets corrected.  

\subsection{When one instanton is not enough?}
\label{sec48}

The vast majority of calculations in supersymmetric gauge theories 
that 
are classified as ``reliable" are based on the strategy ascending to the 
1980's: the matter sector is arranged in such a way that the
non-Abelian gauge group is spontaneously broken, and the running
of the gauge coupling in the infrared is aborted at the scale
correlated with the expectation values of the moduli fields.
Perturbative physics plays no dynamical role; nonperturbatively 
induced superpotentials (if they are  actually 
induced) are  saturated by a single instanton.
This is a typical situation in all the cases when the $R$ charges of the
moduli involved are nonvanishing (e.g. Sec.\ \ref{sec471}). However, there is an 
important  class of models where the multi-instanton contributions are 
instrumental in shaping dynamics of the flat directions.
These models are peripheral  in the range of questions related to 
supersymmetry breaking -- they are important in other aspects
of nonperturbative SUSY gauge dynamics. Therefore, there are no 
reasons why we should  dwell on them in this review. Nevertheless, 
such models deserve a brief discussion from the point of view of the 
instanton formalism. We will outline some general features and 
explain  why one-, two-, three-, ... , $n$-instanton contributions are 
equally important.  

The pattern of the gauge symmetry breaking in the models at hand 
is different -- the non-Abelian gauge group is spontaneously broken 
down  to an Abelian U(1) (sometimes, one deals with several U(1)'s; for 
simplicity we will limit ourselves to a single U(1)).  The theory is said to 
be in  the {\em Coulomb phase} (more exactly, Abelian 
 Coulomb phase). This pattern is 
sufficient for preventing  the gauge coupling constant
from growing  in the infrared. 
The moduli fields are assumed to be large. One more requirement is the 
vanishing of the $R$ charges of at least some  moduli fields.

The most famous example of this type is the SU(2) model with the 
extended supersymmetry, ${\cal N}=2$, solved by Seiberg and 
Witten
\cite{SEIW}. In terms of ${\cal N}=1$ superfields one can view this 
theory as SUSY gluodynamics plus one chiral matter superfield 
$\Phi^a$
in the adjoint representation of SU(2).  The vacuum valley is 
parametrized by one complex invariant, $\Phi^2\equiv \Phi^a \Phi^a$; the gauge 
SU(2)
is obviously broken down to U(1). Thus, a photon field and all its  
${\cal 
N}=2$ superpartners remain massless, while all other states acquire 
masses (there are some exceptional points on the 
valley where some extra states become  massless, though).

The conserved anomaly-free $R$ current in this model has the form
\begin{equation}
\lambda^a_\alpha\bar{\lambda}^a_{\dot\alpha} -
\psi^a_\alpha\bar{\psi}^a_{\dot\alpha} 
\end{equation}
where $\psi$ is the matter fermion. In fact, this current is vector 
rather 
than axial. Thus, the $R$ charge of $\psi$'s is $-1$ while that
of the superfield $\Phi^a$ is zero. The $R$ charge of $\Phi^a \Phi^a$
vanishes too, which implies that no superpotential can be 
generated$\,$\footnote{If one adds a mass term to the matter field
a superpotential is generated. It is proportional to the mass parameter.}.
Does this mean that there are no nonperturbative effects in the 
low-energy limit, when all massive states
are integrated out? The answer is negative.
Nonperturbative effects show up in the kinetic term of
the photon field  which takes the form
\begin{equation}
{\cal L }_\gamma = \frac{1}{8 }  \int { \rm d}^2\theta \, f(\Phi^2 )\,W^2  + { \rm
H.c.}
\label{arkten}
\end{equation}
where $f$ is an analytic function of $\Phi^2 $.  The ${\cal N}=2$ SUSY relates this 
term with $D$ terms of the moduli field $\Phi$. 
In perturbation theory 
\begin{equation}
f= \frac{1}{g^2} - \frac{1}{4\pi^2}\ln\frac{M_{\rm PV}^2}{2\Phi^2}=
-\frac{1}{4\pi^2}\ln\frac{\Lambda^2}{2\Phi^2}\, ,
\label{Mone}
\end{equation}
only one loop contributes.
Nonperturbatively, an infinite series of the type
\begin{equation}
\sum_n C_n \left( \frac{\Lambda^4}{\Phi^4}\right)^n
\label{Mthree}
\end{equation}
is generated on the right-hand side, so that
\begin{equation}
f= -\frac{1}{4\pi^2}\ln\frac{\Lambda^2}{2\Phi^2}
+ \sum_n C_n
\left(
\frac{\Lambda^4}{\Phi^4}\right)^n\, .
\label{Monark}
\end{equation}
Since $\Phi^2$ has zero $R$ 
charge, 
all  $n$ are allowed. The factor $\Lambda^4$ comes from the 
instanton 
measure (the first -- and the only -- coefficient in the $\beta $ 
function 
is four), so the index $n$ actually counts the number of instantons. 
The one-instanton term in this series was found a decade ago 
\cite{seib88}. 
Instead of summing up the infinite series the exact answer was found~\cite{SEIW}
by applying holomorphy and duality. Later on it was subject to scrutiny of direct 
multi-instanton calculations (e.g.~\cite{DKM,DKMS}). The Seiberg-Witten result
passed all  tests. 

In view of the importance of the issue of  one instanton vs. many,
we would like to exhibit pictorially the difference between the instantons
in distinct models. As a representative of the first class of models  (the
one-instanton
 saturation) we will consider
SU(2) SQCD with one flavor, see Sec.\ \ref{sec471}. The counterpart from the second
class (the multi-instanton
 saturation) is the SU(2) ${\cal N}=2$ theory.

The instanton contribution in SQCD is depicted symbolically in Fig.\ 2.
The solid lines attached to the anti-instanton ($\bar I$) indicate the fermion zero
modes. The dashed  lines marked by  the crosses stand for the squark condensate in
the vacuum,  $\langle\bar\phi \rangle =\bar v$.
The anti-instanton has four gluino and two quark (left-handed)
 zero modes. Two gluino zero modes and two quark modes are paired 
by virtue 
of
$\langle\bar\phi \rangle$. There is no way to neutralize  the remaining two,
although it is possible
to transform the modes of $\lambda$ into those of $\psi$. 
We see that one instanton inevitably produces two fermion zero modes.
That is why it generates the interaction vertex of the type$\,$\footnote{All
 powers in Eq.\
(\ref{arkaone}) ``miraculously" fit each other. The fourth power of $\bar v$
is fixed by the number of the zero modes, while the eighth power of $\rho_{\rm
char}$ follows from dimensional arguments.  After all, Eq.\ (\ref{arkaone}) contains
only $v$, while $\bar v$ cancels. The cancellation of $\bar v$ ensures the proper
analytic structure of the (anti)instanton-induced interaction.}
\begin{equation}
\Lambda^5 \bar v^4 \rho_{\rm char}^8 \, \psi\psi 
\label{arkaone}
\end{equation}
 where $\rho_{\rm char}$ is a
characteristic value of the instanton size, 
$\rho_{\rm char}^2
\sim 1/(v\bar v)$. This interaction is exactly what one needs for
the generation of the superpotential ${\cal W} \propto \Lambda^5 /\Phi^2$
(remember, $\int{\rm d}^2\theta \,\Phi^{-2} \propto \psi\psi\, v^{-4}$).
 On the other hand,
the two-instanton configuration will have four fermion zero modes 
non-neutralized,
three instantons six zero modes and so on. It is obvious that
neither of them contribute to the superpotential. 

\begin{figure}[h]   
\epsfxsize=9cm
\centerline{\epsfbox{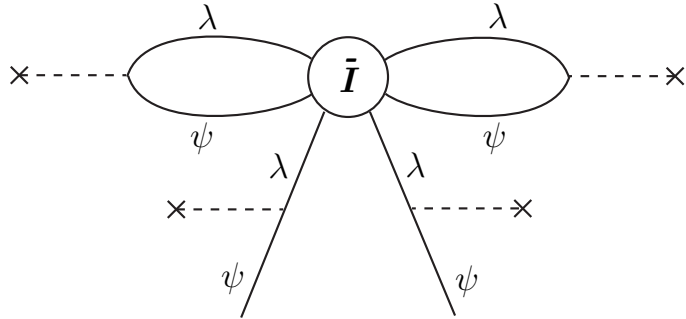}}
 \caption{One-instanton
 contribution in the SU(2) model with one flavor.}
\end{figure}

In the ${\cal N}=2$ theory with no matter hypermultiplets the instanton has four
gluino and four quark zero modes attached to it (Fig.\ 3). One can neutralize  them
all 
by virtue of the insertion of $\langle\bar\Phi\rangle$. That is why in the model at
hand two, three, etc. instantons manifest themselves essentially
in the same way  as one.
None  contributes to the superpotential. Rather, they generate (\ref{arkten}) --  the
integral $\int{\rm d}^2\theta \, W^2 f(\Phi^2)$ does not vanish on purely 
bosonic fields. 

\begin{figure}[h]  
\epsfxsize=8cm
\centerline{\epsfbox{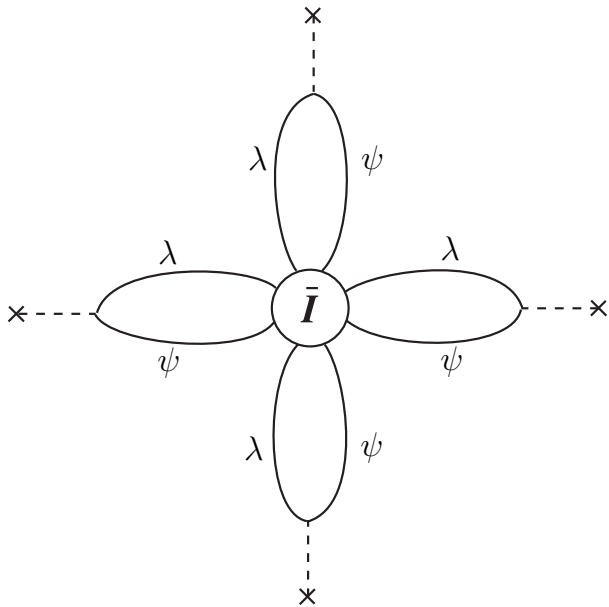}}
\caption{One-instanton contribution
in the ${\cal N}=2$ theory (with no matter hypermultiplets).}
\end{figure}

The issue of one instanton vs. many was briefly touched
in Sec.\ \ref{pvnbf},  in connection with the NSVZ $\beta$ function.
In the models where the one-instanton results are exact,
this $\beta$ function is exact not only perturbatively, 
but at the nonperturbative
level too. On the other hand, in those models where the
summation of the multi-instantons is needed,  the NSVZ $\beta$ function
 ceases to be exact at the nonperturbative
level. The function $f$ in Eq.\ (\ref{Monark}) can be 
considered as a generalized (inverse) coupling  constant. 
Just this function $f(\Phi^2)$ was the central object of the Seiberg-Witten analysis 
based on holomorphy and duality. This novel approach provides  a  wealth of
breakthrough  insights regarding the nonperturbative SUSY gauge
dynamics~\cite{SEIW}.  This interesting topic is outside the scope of this review.

The only question to be raised in connection with instanton calculus
is whether one can construct ${\cal N}=1$ rather than ${\cal N}=2$ 
theories with  the similar properties -- the Coulomb phase,
the vanishing $R$ charge of the relevant moduli, and an infinite 
series for the generalized coupling constant saturated by multi-instantons.

A systematic search for  ${\cal N}=1$ theories with the simple gauge 
group and  the Coulomb phase everywhere on the moduli space was 
carried out in Ref.\ \cite{CSASKI}.  The necessary condition on the 
matter 
sector was shown to be
\begin{equation}
\sum_i T(R_i) = T_G\, ,
\label{Mtwo}
\end{equation}
where the sum runs over all matter fields.
Note that the basic ${\cal N}=2$ model considered by Seiberg and 
Witten does obey this condition, automatically, since the matter field is in 
the adjoint
representation. The number of models satisfying Eq.\ (\ref{Mtwo}) is 
limited. One has to examine all of them in order to determine whether
the pattern of the gauge symmetry breaking is indeed such that
one is left with the unbroken U(1) irrespective of the values of the 
moduli. A general solution turns out to be unique --
the SO($N$) models with $N-2$ vector matter fields (exceptional 
solutions exist also for SU(6) and Sp(6)).  The conserved $R$ current 
in  this model has the form
$$
\lambda^a_\alpha\bar{\lambda}^a_{\dot\alpha} - \sum_{i=1}^{N-2}
\psi^i_\alpha\bar{\psi}^i_{\dot\alpha}\, , 
 $$
so the $R$ charge of $\psi's$ is $-1$, and the $R$ charge of the matter 
superfields vanishes. The generalized inverse coupling constant has the form
similar  to  (\ref{Monark}). The series can be summed up to produce a 
hyperelliptic  curve, much in the same way as in the 
Seiberg-Witten solution of the ${\cal N}=2$ model.

\section{Spontaneous Supersymmetry Breaking in Weak Coupling}
\label{sec5}

\renewcommand{\theequation}{5.\arabic{equation}}
\setcounter{equation}{0}

In this and  subsequent sections we will review  what is
known today regarding the dynamical supersymmetry breaking.
The material covered in the previous  part can be viewed an extended 
introduction  to this critically  important topic, which, in a sense, is 
the 
{\em raison  d'etre} of multiple SUSY-based theoretical constructions. 
The spontaneous SUSY breaking can be realized by virtue of  
tree-level  mechanisms discovered in the early 1970's and described 
in 
text-books, and by virtue of nonperturbative mechanisms which are 
divided, in  turn, in two classes -- weak coupling and strong coupling 
mechanisms. There is no clear-cut boundary between the classes 
since
some of the mechanisms under discussion in the current literature 
combine  elements inherent to  both classes. Some of the 
nonperturbative mechanisms were discovered in the 1980's, others
present a relatively recent development,  dating back to
Seiberg's results of 1993/94. Finally, certain ideas are still
controversial, and 
not all implications for the dynamical SUSY breaking are  clear at the 
moment. The most notable example of this type is the chirally 
symmetric vacuum in SUSY gluodynamics \cite{kovner2}.
If this vacuum indeed exists  (we hasten to add that the final 
confirmation is still pending) then this would have a drastic impact 
on 
many conclusions regarding the dynamical SUSY breaking. Theories 
with 
no supersymmetric  vacuum  of the ``old" type could develop one, of 
the 
Kovner-Shifman type, in  the domain of strong coupling. This is 
especially transparent, for instance, in  the one-generation SU(5) 
model 
(Sec.\ \ref{sec61}). Since the situation with the  chirally symmetric 
vacuum is not yet settled,  we will discuss the known mechanisms 
of 
the dynamical SUSY  breaking forgetting for a while about its 
possible 
existence. Comments on the possible role of the Kovner-Shifman 
vacuum
are collected at the very end (Sec.\ \ref{sec7}); a few marginal 
remarks 
are scattered in Secs.\ \ref{sec55} and \ref{sec61}.

Many models exhibiting the  dynamical  SUSY breaking
reduce in the low-energy limit to an effective theory of light degrees of freedom
in which one of the known tree-level mechanisms acts. Therefore,
to begin with, we briefly review them.

\subsection{Traditional tree mechanisms (brief look at the old guide 
book)}\label{sec51}

Before submerging into nonperturbative dynamics
it seems instructive to outline the conventional tree-level 
mechanisms leading to the spontaneous SUSY breaking.
Our task here is to refresh  memory
and to emphasize common features and distinctions with the 
dynamical mechanisms.

Two such mechanisms are familiar for a long time.
 The first one, usually referred
to  as the O'Raifeartaigh, or $F$ term mechanism \cite{ORAI}, 
works due to a ``conflict of interests" between $F$ terms
of various matter fields belonging to the matter sector.
The necessary and sufficient condition for the existence of the SUSY 
vacuum is the vanishing of {\em all} $F$ terms. 
In the O'Raifeartaigh approach, the superpotential is arranged in such 
a way that it is impossible to make all $F$ terms vanish 
simultaneously.
One needs  minimally three  matter fields in order to realize the 
phenomenon in the renormalized models with the polynomial 
superpotentials. With one or two matter fields and 
the polynomial superpotential the supersymmetric vacuum solution 
is always possible. With three superfields and  the generic 
superpotential, supersymmetric  solution exists too; it ceases to exist 
for some degenerate superpotentials.

Consider the superpotential
\begin{equation}
{\cal W} (\Phi_1, \Phi_2, \Phi_3) =
\lambda_1\Phi_1 (\Phi_3^2-M^2) + \mu \Phi_2 \Phi_3\, .
\label{raisp}
\end{equation}
Then
\begin{equation}
\bar{F}_i= -\frac{\partial{\cal W}}{\partial\Phi_i}
=\left\{
\begin{array}{l}
\lambda_1(\phi^2_3 - M^2)\, , \,\,\, ~~~~~~i=1\, ,\\[0.1cm]
\mu\phi_3\, ,  \qquad ~~~~~~~~~~~~~i=2\, ,\\[0.1cm]
2\lambda_1\phi_1\phi_3 +\mu\phi_2\, , \,\,\,\, ~~~i=3\, .
\end{array}\right.
\label{raifc}
\end{equation}
The vanishing of the second line implies that $\phi_3 =0$,
then the first line cannot vanish. There is no solution with
$F_1=F_2=F_3=0$ -- supersymmetry is spontaneously broken.

What is the actual minimal energy configuration?  It  depends on the ratio
$\lambda_1 M/\mu$. For instance, at  $M^2 < \mu^2 /(2\lambda_1^2)$, 
 the minimum of the scalar potential occurs at $\phi_2=\phi_3=0$.
The value of $\phi_1$ can be arbitrary: an indefinite equilibrium 
takes place at the tree level. (The loop corrections to the K\"ahler potential lift
this degeneracy and lock the vacuum at $\phi_1=0$.) Then
$F_2=F_3=0$, and the vacuum energy density is, obviously,
${\cal E} = |F_1|^2 = \lambda_1^2 M^4$.

Since $F_1\neq 0$ the corresponding fermion is Goldstino,
$m_{\psi_1} =0$. It is not difficult to calculate the masses
of other particles. Assume  that the vacuum expectation
value (VEV) of the
field 
$\phi_1$ vanishes. Then the fluctuations of $\phi_1$ remain 
massless (and degenerate with $\psi_1$). The Weyl field
$\psi_2$ and the quanta of  $\phi_2$ are also degenerate; their 
common mass is $\mu$. At the same time, the fields from $\Phi_3$
split: the Weyl spinor $\psi_3$ has the mass $\mu$, while 
\begin{equation}
m_{a}^2 =\mu^2 -2\lambda_1^2 M^2\, , \,\,\, m_{b}^2 =\mu^2 
+2\lambda_1^2 M^2\, ,
\end{equation}
where
$$
\phi_3 \equiv \frac{1}{\sqrt{2}}(a + ib )\, .
$$
Note that in spite of the splitting
\begin{equation}
m_{a}^2 +m_{b}^2 -2m_{\psi_3}^2 = 0\, ,
\label{strpa}
\end{equation}
as if there is no SUSY breaking. Equation (\ref{strpa}) is a particular
example of the general supertrace relation \cite{FGP}
\begin{equation}
\mbox{Str}\,{\cal M}^2 \equiv \sum_J (-1)^{2J}(2J+1)m_J^2 = 0\, ,
\label{strge}
\end{equation}
where Str stands for supertrace, ${\cal M}^2$ is the mass squared 
matrix of the real fields in the supermultiplet; the subscript $J$ 
marks the spin of the particle. Equation (\ref{strge}) is 
valid only at the tree level -- quantum corrections arising due to 
SUSY breaking do modify it. It  also gets modified in the theories
where (a part of) SUSY breaking occurs due to the Fayet-Iliopoulos 
mechanism, speaking of which we must turn to a rather narrow 
subclass of theories that possess a U(1) gauge symmetry.

The Fayet-Iliopoulos mechanism \cite{FIL}, which is also called
the $D$ term SUSY breaking, applies  in the models
where the gauge sector includes an U(1) subgroup. The simplest and 
most transparent example is SQED (Sec.\ \ref{sec311}).

In order to launch the spontaneous SUSY breaking a (Lorentz scalar)
field transforming nontrivially under supersymmetry must acquire a 
VEV. In the O'Raifear\-taigh mechanism this role was played by an $F$ 
component of a chiral superfield. The $D$ component of the
vector superfields is also non-invariant under SUSY transformations. 
If  it develops a non-vanishing VEV, supersymmetry is 
spontaneously broken too.   In order to make $D$ develop a VEV, a 
$\xi$ term, which is called the Fayet-Iliopoulos term, must be added in the action,
see Eq.\ (\ref{xiterm}).

Although the Fayet-Iliopoulos term literally does not exist in the  non-Abelian 
gauge theories an analog of the phenomenon does exist: the kinetic part of the
action $\bar Q e^V Q$ determines the form of the $D$ term in the Higgs phase.
The conflict between the requirements of vanishing of the $D$ and $F$ terms
may lead the spontaneous SUSY breaking.

Returning to SQED,  Eq.\ (\ref{polpot}) shows that with the massive 
matter ($m\neq 0$)
the zero vacuum energy is not attainable. Indeed, the mass term 
requires $S$ and $T$ to vanish in the vacuum,
while the $D$ term in the potential requires
$|T|^2= |S|^2 +\xi$. If $\xi\neq 0$ both conditions cannot be met 
simultaneously.

If $\xi > m^2/e^2$ the minimal energy is achieved at
$$
S^+S = 0\, , \,\,\, T^+T = \xi - \frac{m^2}{e^2}\, .
$$
The minimal energy
$${\cal E} = m^2\left( \xi - \frac{m^2}{2e^2}\right)
$$
is positive, so that SUSY is spontaneously broken. 
In addition, the gauge U(1) is broken too. The phase of $T$ is eaten 
up in 
the Higgs mechanism, the photon becomes massive,
$m_\gamma = e\sqrt{\xi -(m^2/e^2)}$.
A linear combination of the photino and $\psi_t$ is the Goldstino; it 
remains massless. Another linear combination, as well as the scalar 
and 
spinor fields from $S$ are massive.

If $0<\xi < m^2/e^2$ the scalar fields develop no VEVs, the vacuum 
configuration corresponds to 
$$
S^+S = T^+T =0\, , 
$$
while the vacuum energy is ${\cal E} =e^2\xi^2/2$. The gauge U(1)
remains unbroken: the photon is massless, while the photino assumes 
the role the Goldstino. The fermion part of the matter sector does not 
feel the broken supersymmetry (at the tree level),
$$
m_{\psi_s} = m_{\psi_t} =m \, ,
$$
while the boson part does
$$
m_{s}^2 = m^2 +e^2\xi\, , \,\,\,  m_{t}^2 = m^2 - e^2\xi\, .
$$

\subsection{Dynamical mechanisms: preliminaries}
\label{sec52}

In the majority of examples discussed so far the gauge symmetry is 
typically  spontaneously broken \cite{ADS2,ADS3,NSVZ2} (it would 
be 
more exact  to say that it  is realized in the Higgs regime). Instantons 
are  instrumental in this phenomenon, which provides a number of 
evident 
advantages; the weak coupling regime and calculability of the 
nonperturbative  effects are  the most important ones. The phenomenon 
is  quite universal. In  almost  any model with matter there are 
classically 
flat directions, vacuum valleys, and together with them there arises a 
potential  possibility for destabilization of the unbroken vacuum at 
the  origin (vanishing values of the moduli). Roughly speaking, 
the  theory is pushed away from the origin by instantons. 

Our main subject here is the dynamical SUSY breaking. 
Among a vast variety of
models with the spontaneously broken gauge symmetry we will 
choose those  where SUSY is broken too. This will provide us with calculable 
scenarios  of dynamical SUSY breaking. In a few instances it was 
argued  that SUSY is broken in the strong coupling regime, where direct 
calculations  are impossible. One then relies on indirect arguments. 
We will  discuss  such scenarios too. Our strategy is pragmatic: starting 
from  simpler  scenarios we will be moving towards  more complicated 
ones.

The very same aspect -- weak vs. strong coupling SUSY  breaking -- can be viewed
from a slightly different angle. Let us discuss a  generic  hierarchy of scales
inherent to a typical supersymmetric theory with  matter. First, there exists an
intrinsic  scale $\Lambda_g$ of the 
underlying gauge theory$\,$\footnote{In weak coupling the relevant scale is $g|v|$
rather than  $\Lambda_g$. }.
The scale of the dynamical SUSY breaking 
$\Lambda_{\rm dsb}$ can be of order $\Lambda_g$ or much lower 
than 
$\Lambda_g$. In the latter case when one descends below 
$\Lambda_g$
all gauge degrees of freedom can be integrated out, and one is left 
with the dynamics of the would-be moduli. This effective low-energy 
dynamics can be described by  generalized Wess-Zumino models
which can be studied relatively easily.
SUSY is dynamically broken if a generalized O'Raifeartaigh 
mechanism 
takes place.

The most common scenario of this type is as follows. Assume that in 
a  model with  chiral matter (so that Witten's index vanishes)
there are classically flat directions in the limit of vanishing  Yukawa 
terms. If instantons generate a repulsive superpotential, if the model 
can be completely stabilized by the Yukawa terms, so that 
the exits from all valleys are blocked (the run-away vacua totally 
excluded), if all Yukawa couplings are small (dimensionless couplings 
$\ll 1$, dimensional $\ll \Lambda_g$) -- if 
all  these ``ifs" are satisfied -- then dynamical SUSY breaking will 
most 
certainly occur in the weak coupling regime. 

There is an alternative version of the  weak coupling 
scenario. In some models (e.g. Sec.\ \ref{sec55})
the original gauge group is only partly broken, and the 
superpotential 
for the moduli is generated by the gluino condensation in the 
unbroken 
subgroup, rather than through the one-instanton mechanism. 
These models are still calculable because (i)
 so is the gluino condensate, and (ii) the SUSY breaking is due to 
dynamics of the moduli fields at a  low-energy scale $\Lambda_{\rm 
dsb}\ll \Lambda_g$, where all (strongly interacting) degrees of 
freedom 
associated with the unbroken subgroup can be integrated out. 

On the other hand, if  $\Lambda_{\rm dsb}\sim \Lambda_g$,
relevant dynamics is that of the strong coupling regime.
Here one can hope, at best, to establish the very fact of
SUSY breaking using global symmetry arguments and/or 
't Hooft matching \cite{tHooft2}.

Two criteria are known in the literature \cite{ADS3,ARV}, each of 
which  guarantees spontaneous SUSY breaking.

\vspace{0.2cm}

\noindent {\em Criterion 1}. Suppose that the Yukawa terms in the 
superpotential introduced for the sake of stabilization at large values 
of 
moduli do not contain some matter superfield $Q$, or a linear 
combination 
of several superfields. In this case the gluino condensate is an order 
parameter -- a non-zero VEV of the gluino density, $\langle 
\mbox{Tr}\,\lambda^2 \rangle \neq 0$, implies the spontaneous SUSY breaking.
This statement is quite transparent. Assume some superfield $Q$ enter in 
the original action only through its kinetic term
$\left. \bar Q e^V Q\right|_D$. In this case the anomalous Konishi relation
\cite{Konishi} takes the form (Sec.\ \ref{sec34})
\begin{equation}
\bar{D}^2(\bar Q e^V Q)\propto \,{\rm Tr}\,W^2\, .
\end{equation}
Then the vacuum expectation value of Tr$\,W^2$ is equivalent to 
the vacuum expectation value of the operator $\bar{D}^2(\bar Q e^V 
Q) $. Taking the  spinor derivative is equivalent to the (anti)commutation 
with the supercharge. Hence in the theories with the  unbroken SUSY 
no VEV's of full superderivatives can develop. Thus, for $\langle 
\mbox{Tr}\,\lambda^2 \rangle \neq 0$ we observe a contradiction.

\vspace{0.2cm}

\noindent {\em Criterion 2}. Suppose that in a theory under  
consideration the vacuum valleys are completely absent, i.e. they are 
non-existent from the very beginning (as in the SU(5) theory with 
one quintet and one (anti)decuplet), or 
the vacuum degeneracy is fully lifted by the tree-level 
superpotential. 
If in such a theory some {\em exact continuous global } symmetry
is spontaneously broken, so is SUSY.

A sketch of the proof \cite{31} is as follows. If a continuous global 
invariance is spontaneously broken, there is a massless Goldstone 
boson, 
call it $\pi$. Suppose, SUSY is unbroken. Then $\pi$ must be 
accompanied by massless superpartners, in particular,
a scalar particle $\sigma$ with spin 0. Since the field $\pi$, being the 
Goldstone boson, appears in the Lagrangian with a zero potential, the 
potential for $\sigma$ must vanish too, and, as a consequence, the 
vacuum expectation value $\langle\sigma\rangle$ is not fixed.
In other words, $\sigma + i \pi$
is a modulus, which can be varied arbitrarily, corresponding to a 
continuously degenerate vacuum manifold$\,$\footnote{The argument above
shows that the global symmetries in the superpotential are always complexified}. 
This contradicts,  however,  the  initial assumption of no flat directions. The only
possibility of getting  rid  of the contradiction is to conclude that SUSY is
spontaneously broken. 

\vspace{0.2cm}

The above two criteria are not completely independent.  In fact, if some superfield
does not enter into the classical  superpotential  (Criterion 1), then there exists an
axial $R$ current -- a linear  combination of  the matter current and gluino current
-- which is strictly conserved.  Further, the operator Tr$\,\lambda^2$ is obviously
noninvariant with respect to the transformations  generated  by this current.
Therefore, the gluino condensation, $\langle \mbox{Tr} \,\lambda^2\rangle \neq
0$, automatically implies spontaneous breaking  of  the corresponding axial
symmetry.

\subsection{Dynamical SUSY breaking in the {\em 3-2}  model}
\label{sec53}

The simplest model where dynamical SUSY breaking takes place is the  {\em 3-2}
model of Affleck, Dine and Seiberg~\cite{ADS3}.  This chiral model  was discussed
at length in Sec.\ \ref{sec3-2}  where a description of the classical moduli space (its
complex dimension is three)  was given. The K\"ahler potential on the moduli space
was calculated.

\begin{figure}[h]    
\vskip3mm
 \centerline{\epsfbox{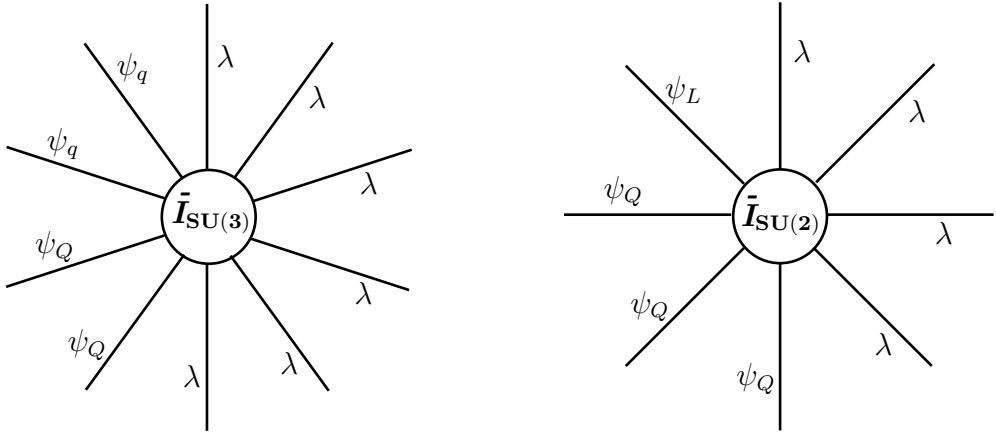}}
 \caption{The {\em 3-2} model: the zero mode structure
 in the field of the anti-instanton in the gauge groups SU(3) and SU(2),
 respectively. $\lambda$ denotes the SU(3) gluinos in the first case and
SU(2) gluinos in the second. }
\end{figure}

Two chiral invariants, $I_1$ and $I_2$, where $I_{1,2}$ were defined  in Eq.\
(\ref{cuinv}), are cubic; a combination thereof  can be used as a (renormalizable)
tree-level superpotential.  Its purpose is to block the valleys and spare us of the
run-away vacua. 
Note that  the kinetic part possesses a global flavor SU(2) symmetry:
$\bar{u}$ and $\bar{d}$ fields can be freely  rotated into one  another. In addition
to this global SU(2), the theory has two strictly conserved  U(1) currents. One of
them is that of the hypercharge ($=2\times$the  mean electric charge in the weak
SU(2) multiplet). In SM this  current would be coupled to the U(1) gauge boson, but
since we  remove the latter from the  model, the corresponding symmetry is  just a 
global U(1) invariance. The second U(1) is the  anomaly-free $R$ current involving
gluinos. 

The fastest way of establishing the U(1) charges is through instantons
themselves. The  instanton in the SU(3) gauge group has six gluino zero modes, two
$\psi_Q$ and two $\psi_q$. (A concise mnemonical formula  is  $\lambda^6 \,
\psi_Q^2 \, \psi_q^2\,$, see Fig.\ 4. We will consistently use similar formulas
below).  The instanton in SU(2) has four  gluino zero modes,
three $\psi_Q$ and one $\psi_L$, i.e.  $\lambda^4\,  \psi_Q^3 \, \psi_L$
for a shorthand. It is convenient to choose the conserved anomaly-free Konishi
current in the form
$$ 
\frac{1}{3} \bar Q^f Q_f - \frac{1}{3}\bar q^{\bar f} q_{\bar f} - \bar L^fL_f\, ,
  $$
where the summation over the  SU(3) indices in the first two terms is implicit.
As for the $R$ current, we conveniently define it in such a way that
$R(L) = 0$.  Then, $R(\psi_L) = -1$ and $R(\psi_Q) = -1$ too
(the latter is immediately seen from the  $\lambda^4 \psi_Q^3 \psi_L$
structure of the SU(2) instanton). Hence, $R(Q) = 0$. From the
$\lambda^6 \psi_Q^2 \psi_q^2$ structure of the SU(3) instanton we conclude
that $R(\psi_q) = -2$ which entails, in turn, $R(q) = -1$. The Konishi and $R$
charges of the superfields and 
 invariants $I_{\bar f}$ and $J$ are summarized in
Table~\ref{KR23}

By an appropriate global SU(2) rotation one can always arrange
that the tree-level superpotential has the form
\begin{equation}
{\cal W}_{\rm tree} = hQ^{\alpha f }\bar d_\alpha L^g\varepsilon_{fg} = 
hI_2\, 
,
\label{tlsp32}
\end{equation}
so that $I_1$ does not appear in the superpotential. This means, in  turn,  that
$\bar u_\alpha $ does not appear in ${\cal W}$; we will make  use  of this fact
later.

\begin{table}
\begin{center}
\begin{tabular}{|c|c|c|c|c|c|}
\hline
~ & ~  & ~ & ~& ~& ~  \\[-0.1cm]
Field & $Q$ & $q$ & $L$ & $I_{\bar f}$ &  $J$ \\[0.2cm]\hline
\vspace*{-0.2cm}
~ & ~  & ~& ~& ~& ~  \\
U(1)$_K$ &1/3&
$-1/3$&$-1$&$-1$&0
\\[0.2cm]\hline
\vspace*{-0.2cm}
~ & ~  & ~  & ~& ~& ~\\ 
U(1)$_R$ & 0&
$-1$&0&$-1$&$-2$
\\[0.2cm]\hline
\end{tabular}
\caption{The U(1) charges of the superfields $Q$, $q$, $L$ and  invariants $I_{\bar
f}$ and
$J$ defined in Eqs.\ (\ref{cuinv}), (\ref{quainv}) in the {\em 3-2} model.}
\label{KR23}
\end{center}
\end{table}

As in MSSM it will be assumed that the SU(3) gauge coupling $g_3$ is 
much larger than the SU(2) gauge coupling $g_2$, and, in addition,
\begin{equation}
h\ll g_2 \ll g_3. 
\end{equation} 
The smallness of $h$ will automatically ensure that
$\Lambda_{\rm dsb} \ll \Lambda_{3,2}$ where $\Lambda_{3,2}$ are the 
scale parameters of SU(3) and SU(2), respectively.

A generic point on the moduli space 
corresponds to complete breaking of all gauge symmetries; the 
theory is 
totally Higgsed. It is not difficult to check that the tree-level 
superpotential
(\ref{tlsp32}) locks all exits from the valleys --  escape to large 
values  
of the moduli  is impossible since there the energy is positive rather 
than vanishing. (This exercise is left for the reader.)

Now we switch on nonperturbative effects taking into account only 
instantons in the gauge SU(3) and neglecting the SU(2) instantons
whose impact is much weaker because $g_2\ll g_3$.
What is the general constraint on the instanton-induced superpotential?
First, it cannot depend on $I_{\bar f}$ since $I_{\bar f}$ is not singlet with 
respect
to global SU(2), and it is impossible to built any singlet from
$I_{\bar f}$. On the other hand, $J$ is singlet and its Konishi charge vanishes.
Thus, it is a suitable candidate. Examining the $R$ charge of $J$ we conclude
that the only possible 
 superpotential  has the form
\begin{equation}
{\cal W}_{\rm inst} =\frac{2(\Lambda_3)^7}{J}\, .
\label{instsp32}
\end{equation}
The factor 2 is singled out for convenience; the power of 
$\Lambda_3$ is established on dimensional grounds, it exactly matches
the first coefficient of the $\beta$ function in the SU(3) model with two triplets
and two anti-triplets.

Is the superpotential (\ref{instsp32}) actually generated
by the SU(3) instanton?  To answer this question
it is not necessary to perform the calculation anew -- the problem can be readily
reduced to that of our reference model, SU(2) SQCD with one flavor
(Sec.\ \ref{sec471}). Indeed, let us assume that $|\tau_1| \gg |\tau_2|$
in Eq.\ (\ref{arktwo}). Then the pattern of breaking of SU(3) is step-wise:
first one breaks SU(3)$\to$SU(2) at a high scale, and then at a much lower scale
SU(2) is broken down completely.  Below the higher scale
the fields $Q$ and $q$ with the SU(2) index $f=\bar f =1$
disappear --
they are eaten up by five SU(3) gauge bosons$\,$\footnote{The gauge bosons
from the coset $G/H$, where $G$ is the original gauge group and $H$ is
the unbroken subgroup, are sometimes referred to as {\em elephants}
in the Russian literature. This emphasizes the fact that they become
heavy while the gauge bosons $\in H$ remain light.}
which become very heavy and, in their turn,
 also disappear from the light
spectrum. We are left with  the SU(2) theory with
one flavor,
$Q^{\alpha 2}\,  ,\,\,  q_{\alpha\bar 2}\, , \,\,\, \alpha = 2,3$. 
In this corner of the moduli space the invariant $J$ becomes
$J\to \tau_1^2\, Q^{\alpha 2}\, q_{\alpha\bar 2}$ where the summation over
$\alpha$ runs over $\alpha=2,3$. Then, confronting Eq.\ (\ref{instsp32})
with Eq.\ (\ref{effw}) we immediately conclude that the superpotential
(\ref{instsp32}) is indeed generated, the relation between the scale parameter
of SU(2) and $\Lambda_3$ being $\Lambda_3= (\tau_1^2\,\Lambda^5_{\rm
one-fl})^{1/7}$.

The instanton-induced superpotential pushes the scalar fields
away  from the  origin  and launches the spontaneous breaking 
of  SUSY in the weak coupling regime.
To prove the dynamical SUSY breaking quickly, we can use Criterion 1 
(Sec.\ \ref{sec52}), since $\bar{u}$ does not appear in ${\cal W}_{\rm tree}$. 
From the  $\lambda^6\,  \psi_Q^2\,  \psi_q^2$ structure of the
zero modes it is perfectly clear
 that the condensate of Tr$\,\lambda^2$ actually 
develops.  (Two gluino zero modes are neutralized by the operator
 Tr$\,\lambda^2$ whose vacuum expectation value
we calculate. The remaining four gluino zero modes are paired
with $\psi_q$ and $\psi_Q$, as in Fig.\ 3.). The run-away 
vacuum is  impossible. 
Since Tr$\,\lambda^2\neq 0$, supersymmetry is broken.

To obtain a more 
detailed  information on the particle spectrum it is necessary
to investigate the  effective low-energy 
Wess-Zumino model for the moduli. One can either use the explicit 
parametrization of the moduli fields given in Eq.\ (\ref{arktwo}), with the canonic
kinetic terms$\,$\footnote{More exactly,
the parametrization  (\ref{arktwo}) must be generalized
to include the flavor rotations associated with the global SU(2)
in $\{\bar u, \bar d\}$.}, or, instead, deal 
directly  with the moduli $I_{\bar f}$, $\, J$. In the latter case the
superpotential  term is 
trivial, but the kinetic term acquires the K\"{a}hler potential ${\cal K}$. 
Both 
methods 
have their advantages and disadvantages; the second approach is 
more 
common, however.  One of the reasons is that it  automatically takes care of the
global symmetry on the moduli space.

The total  superpotential is
\begin{equation}
{\cal W} = {\cal W}_{\rm tree}+ {\cal W}_{\rm inst}\, . 
\end{equation}
The subsequent procedure is standard: one 
 minimizes the scalar potential $U$  obtained from the given superpotential and the
K\"ahler potential presented in \mbox{Eq.\ (\ref{kkt})},
\begin{equation}
U =\left[ G^{-1}\right]_{i\bar j}
\frac{\partial {\cal W}}{\partial \phi_i }\,\frac{\partial \bar{\cal W}}{\partial
\bar\phi_{\bar j} } 
\end{equation}
where 
 \begin{equation}
G^{i\bar j}=\frac{\partial^2\,{\cal K}}{\partial \phi_i \partial \bar \phi_{\bar j}}
\label{arkmetric}
\end{equation}
is the K\"ahler metric,
 $G^{-1}$ is the inverse matrix, and $\phi_{1,2,3}$ are the lowest components
of the moduli $I_{\bar f}$ and $J$.

Unlike the superpotential, which is exact both perturbatively and 
nonperturbatively, the  K\"ahler metric is renormalized in higher loops.
However, if the solution for the vacuum lies at large values of the 
moduli --
which is the case, as we will see {\em a posteriori}, provided $h\ll 1$
-- then all gauge bosons are heavy, the gauge coupling is small, and 
the  corrections to ${\cal K}$ are negligible. 

Minimizing $U$ one gets~\cite{ADS3}:
\begin{equation}
(\tau_1)_{\rm vac} \approx 1.29\,\frac{\Lambda_3}{h^{1/7}}\, , \,\,\,
(\tau_2)_{\rm vac} \approx  1.25\,\frac{\Lambda_3}{h^{1/7}}\, , \,\,\,
{\cal E} \approx  3.59\,h^{10/7}\,\Lambda_3^4\, .
\end{equation}
In the limit $h\to 0$ the parameters $\tau_1$ and $\tau_2$ tend to infinity, as 
was 
expected. This justifies the assertion of the weak coupling regime.
The masses of the gauge bosons are $M\sim g\Lambda_3h^{-1/7} 
\to\infty$.

The (former) massless moduli either remain massless or become light 
particles. One Weyl fermion, Goldstino, is exactly massless. The role of 
Goldstino  belongs to the electron, as can be seen either directly from 
the effective low-energy Lagrangian or indirectly, from 
the 't Hooft matching.

The relevant anomalous triangle is that of three hypercharges. 
Remember, the  hypercharge is not gauged in the model at hand, so 
that this 
is  an ``external" anomaly eligible for  the 't Hooft matching. The 
anomalous 
triangle does not vanish due to the fact that the positron was 
removed 
from the matter sector in passing from SM to the {\em 3-2} model 
under 
consideration.
In addition, we get a neutral fermion with mass $\approx 11.3 
h^{6/7} \Lambda_3$. 

Among the scalar partners there is one strictly 
massless 
Goldstone boson, corresponding to the spontaneous breaking of the 
$R$  symmetry, and one charged and three neutral scalars with masses 
$\sim h^{6/7}\Lambda_3$.

The {\em 3-2} model presents  a 
standard pattern of the dynamical SUSY breaking
in the weak coupling regime. Analyses of other models
with the same dynamical behavior go in parallel. 

\subsection{SU(5) model with two generations}
\label{sec54}

The analysis of this model at the classical level was carried out
in Sec.\ \ref{su52} 
where an explicit parametrization of the $D$-flat directions was presented
and the K\"ahler potential on the moduli space built.
A few additional remarks are necessary in order to complete the 
treatment of the model. 
First of all, in accordance with the general strategy, the
exits from the valleys must be locked by a tree-level superpotential 
term.  If one limits oneself to renormalizable theories,
this term can only contain the  $M$ invariants cubic in the
chiral superfields (see Eq.\ (\ref{MBinv})). The chiral invariants of the $B$ type are
quartic  and, 
hence,  if added, would ruin renormalizability.
Of course, if one considers the model as an effective low-energy 
approximation, one is free to add the quartic invariants too. This has 
no  qualitative impact on the dynamical picture, however, and we will 
not  do that.

The most generic form of the tree-level cubic superpotential is
\begin{equation}
{\cal W}_{\rm tree} = \lambda M_1\, .
\label{su5ts}
\end{equation}
It is always possible to rotate the fields to
 eliminate the $M_2$ term by using
the global  SU(2)$_X$ invariance of the kinetic term. The superpotential 
(\ref{su5ts}) breaks SU(2)$_X$; the global SU(2)$_V$
remains intact. Two anomaly-free axial U(1) symmetries
also survive as exact symmetries of the model (see below).
The coupling constant $\lambda$ is assumed to be small, $\lambda 
\ll 1$,
to ensure calculability. Away from the origin of the moduli space 
 the theory is in the Higgs regime. Generically, the gauge
group SU(5) is completely broken. If we are far away from the origin, the gauge
bosons are very  heavy in the scale $\Lambda$, and the theory is weakly coupled. 
Small $\lambda$ will eventually lead to a vacuum located far away 
from the origin. 

It is not difficult to check, using the 
explicit  parametrization (\ref{MM6}), that the tree-level 
term  (\ref{su5ts}) locks all exits, so that the run-away vacuum is 
impossible. The next  question  one must address is whether 
instanton generates a superpotential pushing the theory away 
from  the  origin.

Again, as in {\em 3-2} model, we first examine
what can be said of the  superpotential on general grounds. 
The instanton-induced superpotential term
(in the limit \mbox{$\lambda\to 0$}) must involve both quintets and both 
(anti)decuplets, since there is one zero mode in each quintet and three 
zero  modes in each decuplet (in addition to ten gluino zero modes, of 
course). 
Moreover,  the above numbers of the zero modes imply that
each $V$ superfield must be accompanied 
in the superpotential by three $X$ superfields.
Second, the  instanton-induced superpotential term
must respect the global SU(2)$_X\times$SU(2)$_V$ invariance.
This leaves us with a unique choice
\begin{equation}
{\cal W}_{\rm inst} = \mbox{const}\, \times\frac{ 
(\Lambda_5)^{11}}{ B_{\bar gf}B_{\bar g'f'}\epsilon^{\bar g\bar
g'}\epsilon^{ff'}}\, ,
\label{su5ins}
\end{equation}
where the power of $\Lambda$ in the numerator is established
from dimensional counting, and the invariants $B_{\bar gf}$ are
defined in Eq.\ (\ref{MBinv}). The expression in the denominator
is nothing but the invariant $I_2$ defined in Eq.\ (\ref{miful}).

Equation (\ref{su5ins}) goes through  additional checks.
First, the power of $\Lambda$, eleven, matches the first coefficient 
of 
the $\beta$ function $3T_G - \sum_i T(R_i)$. We remind
that$\,$\footnote{Incidentally, instanton calculus is the easiest and fastest way of
calculating the Dynkin index, if you do not have handy an appropriate text book 
where they all are tabulated, of course. The procedure is as follows.  Assume, a
group $G$ and a representation $R$ of this group are   given.
Then one must pick up an SU(2) subgroup of $G$ and decompose $R$ 
with respect to this SU(2). 
For each irreducible SU(2) multiplet of spin $j$
the index $T = j(j+1)(2j+1)/3$. Hence, the number of zero modes
in the SU(2) instanton background is $(2/3)j(j+1)(2j+1)$.
In this way one readily establishes the total number of the zero 
modes 
for the given representation $R$. This is nothing but the Dynkin 
index.
 The value of  $T(R)$ is one  half of this number.
For instance, in SU(5) a good choice of SU(2) would be weak isospin. 
Each quintet has one weak isospin doublet; the remaining elements 
are singlets. Each doublet has one zero mode. As a result, $T(V) = 
1/2$. Moreover,  each decuplet has three weak isospin doublets 
while the remaining  elements are singlets. Hence, $T(X) = 3/2$. }
$T(V) = 1/2$ and 
$T(X) = 3/2$. In SU(5) with two generations
$3T_G - \sum_i T(R_i) = 11$.

The second test of Eq.\ (\ref{su5ins}) comes from 
counting the U(1) charges. As was mentioned, the
model has two conserved
anomaly-free U(1) currents.  
The Konishi current requires each $V$ to be accompanied by three $X$'s.
We already know this.
As for  the $R$ charges, it is convenient to define
the conserved $R$ current in such a way that $R(\psi_X)=0$. Then,
the $R$ charges of other fields are unambiguously fixed$\,$\footnote{
Again, the easiest way of establishing
the form of the anomaly-free $R$ current is instanton calculus {\em 
per  se}.  The zero mode formula is $\lambda^{10}\, \psi_{X_1}^3\,
\psi_{V_1}\, \psi_{X_2}^3\, \psi_{V_2}$.
 Hence, the anomaly-free $R$ current 
is obtained provided that
the ratio of the $R$ charges is $R(\psi_{V})/R(\lambda )=
-5$.}, they are collected in  Table~\ref{tab2genR}.

\begin{table}
\begin{center}
\begin{tabular}{|c|c|c|c|c|c|}\hline
~ & ~  & ~ & ~& ~& ~  \\[-0.1cm]
 Field & ~ $\psi_X$ ~ & $\psi_V$ ~ &~~ $X$ ~~ &
~~$V$~~ &~~$B$~~
\\[0.2cm]\hline 
\vspace*{-0.2cm}
~ & ~  & ~& ~& ~& ~  \\
 $ R\,\, \mbox{ charge}$  &  ~$ 0$~  &$-5$ & ~$1 $~& $-4 $
&$-1$ \\ [0.2cm]
\hline
\end{tabular}
\caption{The $R$ charges in the SU(5) model with two 
generations.}\label{tab2genR}
\end{center}
\end{table}

With the given $R$ charges of the $X,V$ superfields
the $R$ charge of 
${\cal W}_{\rm inst} $ in Eq.~(\ref{su5ins}) is two, in full
accord with the exact $R$ invariance (in the limit $\lambda\to 0$).

What remains to be done is to verify, by a straightforward  
 one-instanton calculation,  that the dimensionless numerical 
constant in 
Eq.\ (\ref{su5ins}) does not vanish. There are no reasons for it to 
vanish,  and it does not. Technically, one can exploit the very same trick
we described in connection with the {\em 3-2} model. Choose
a corner of the moduli space corresponding to a two-stage breaking of the
gauge group, 
SU(5) $\to$ SU(2) $\to$ nothing.  If the scale of the first breaking is 
much 
higher than that of the second breaking,
the elephant gauge bosons (i.e. those  belonging to SU(5)/SU(2)) 
become very heavy and decouple, and the problem  again reduces to the 
SU(2) model with one flavor,  Sec.\ \ref{sec471}. We urge the
 reader to go through
this calculation, and find the value of the numerical constant,
this is a good exercise.

The resulting dynamical picture in the SU(5) model is very close
to that in the {\em 3-2} model of Sec.\ \ref{sec53}. The 
instanton-induced 
superpotential pushes the theory away from the origin of the valley. 
The tree-level superpotential does not allow the vacuum to run 
away. 
An equilibrium is achieved at large values of the chiral invariants, 
where the theory is in the weakly coupled (calculable) regime. 

It is not difficult to check that the gluino condensate develops, much 
in the same way as in the {\em 3-2} model. Since the superfield $X_2$ 
does not participate in the tree-level superpotential, Criterion 1 of 
Sec.\  \ref{sec52} tells us that supersymmetry must be spontaneously 
broken. The SU(5) model under consideration is singled out from the 
zoo 
of  others  by its special feature: this  is the only one in the class of 
models with the simple gauge  group,  
$T_G >\sum_i T(R_i)$, and  purely chiral matter,  which leads to the 
spontaneous breaking of SUSY  in the weak coupling regime 
\cite{DMS}. 

If one wants to know the light  particle content of the theory in the
SUSY breaking vacuum, one  must work out an effective low-energy 
description. Again, this could be done either in terms of the explicit 
parametrization or by analyzing the effective Lagrangian for  the 
moduli. The K\"ahler potential on the moduli space is presented in
Eq.\ (\ref{kaposu5}). 

Analysis of the vacuum and light excitations 
was carried out in Ref.\ \cite{Veldhuis}. In the vacuum the values of the
moduli scale as $v\sim \lambda^{-1/11} \Lambda$, while the vacuum energy
density scales as ${\cal E}_{\rm vac} \sim \lambda^{18/11} \Lambda^4$.
The model has four massless Goldstone bosons corresponding to the
spontaneously broken global symmetries. Note that
with the tree-level superpotential switched on, the global flavor symmetry is
SU(2)$_X\times$U(1)$^2$. It is spontaneously broken down to
U(1).  In addition, there are eight bosons with masses proportional to
$\lambda v \sim \lambda^{10/11}\Lambda$. 

\subsection{ The {\em 4-1} model: SUSY breaking through gluino 
condensation}\label{sec55}

This model was treated in Refs.\ \cite{DNNS,PT}. The field content of the 
matter sector is easy to memorize: one starts from the SU(5) model 
with one generation and pretends that SU(5) is broken down to
SU(4)$\times$U(1).  Then
$$
X_{\alpha\beta}\to A_{\alpha\beta} + Q_\alpha\, , \,\,\, 
V^\beta \to \bar Q^\beta + S \, .
$$
In other words, we deal with $6+4+\bar{4} + 1$. As we will see 
shortly, 
with this matter set the
theory is not fully Higgsed, an 
unbroken SU(2) subgroup survives, it is in
the  strong  coupling regime. Nevertheless, 
the light fields surviving at  energies below $\Lambda_2$ 
are weakly coupled, their dynamics is described by a Wess-Zumino model
 which is
fully calculable and ensures the spontaneous SUSY breaking.

As a warm-up  exercise let us turn off the U(1) coupling
and focus on the SU(4) theory (our intention is to take into account 
the U(1) later). In the SU(4) theory there are three moduli,
\begin{equation}
M= Q\bar Q\, , \,\,\, P = \mbox{Pf}(A)\, , \,\,\, \mbox{and} \,\,\, 
S\, . 
\end{equation}
We collected in Table \ref{tab41} the $R$ charges of 
the elementary fields and the moduli.

\begin{table}
\begin{center}
\begin{tabular}{|c|c|c|c|c|c|}\hline
~ & ~  & ~ & ~& ~& ~  \\[-0.1cm]
 $ A $  & ~ $Q$~ & ~ $\bar Q$~ &~$S$ ~ &~
$M$ 
~ &
~$P$~ \\ [0.2cm]\hline 
\vspace*{-0.2cm}
~ & ~  & ~& ~& ~& ~  \\
2 & $ -3$  & $ -1$  &4 & $-4$ & 4  \\ [0.2cm]
\hline
\end{tabular}
\caption{The $R$ charges in the {\em 4-1} model.}\label{tab41}
\end{center}
\end{table}

A generic point from the vacuum valley corresponds to the  $\mbox{SU(4)} \to
\mbox{SU(2)}$ pattern of the gauge symmetry breaking.
Thus, in the low-energy limit one deals with
SU(2) SUSY gluodynamics, supplemented by three gauge-singlet 
moduli fields. Below the scale of SU(2) theory only the moduli 
survive; 
a nonperturbative superpotential for them is generated
through  the gluino condensation,
\begin{equation}
{\cal W} \sim (\Lambda_2)^3 \sim \frac{(\Lambda_4)^5}{\sqrt{MP}}\, ,
\label{supxyz}
\end{equation}
where $\Lambda_{2,4}$ are the scale parameters
of SU(2) and SU(4), respectively.

Now it is time to turn on the U(1) coupling. Then the $D$-flatness 
conditions with respect to U(1) eliminate one moduli field out of 
three.
The remaining moduli are
$$
I_1 = MP\, \,\, \mbox{and}\,\,\, I_2 = MS\, .
$$
The superpotential (\ref{supxyz}) does not depend on $I_2$;
it pushes $I_1$ towards infinity. To lock the valleys one can 
introduce a tree-level superpotential, much in the same way as in 
the 
{\em 3-2} model. An appropriate choice of the superpotential is
\begin{equation}
{\cal W}_{\rm tree} = h I_2 \equiv h SM\, .
\end{equation}
It is not difficult to show \cite{DNNS,PT} that adding $h SM$ one 
locks all $D$-flat directions,
SU(4) and U(1), simultaneously, eliminating the possibility of the 
run-away behavior.

The resulting low-energy theory presents a close parallel to the 
{\em 3-2} model  and can be analyzed in the weak coupling regime.
Basically, the only difference is the origin of the nonperturbative 
superpotential. In the {\em 3-2} model it was provided by 
instanton, here by the gluino condensation. The gluino condensate is 
known exactly
\cite{SVMO}. 
The procedure of the vacuum state searches is straightforward, 
albeit numerically tedious, as in the {\em 3-2} model itself. 
We will not go into details here referring the reader to the original 
publications. The conclusion is that supersymmetry is dynamically 
broken. A non-supersymmetric vacuum is found at
$$
v\sim \Lambda_4 h^{-1/5}\gg \Lambda_4
$$
where $v$ is the scale of the typical expectation value of the matter 
fields,
so that all corrections (say, to the K\"{a}hler function)
turn out to be small. The vacuum energy density scales as
$$
{\cal E}_{\rm vac} \sim h^{6/5} (\Lambda_4)^4 \, .
$$

The {\em 4-1} model is only the simplest representative of its class.
It can be generalized in various directions. The basic idea --
a nonperturbative superpotential through the gluino condensation
in an  unbroken subgroup,  plus
a cleverly chosen tree-level superpotential -- persists.
In particular, {\em 2k-1} models ($k>2$) were considered in 
\cite{DNNS,PT}. 

\subsection{A few words on other calculable models}
\label{sec56}

Using the strategy outlined above one can construct more
complex calculable models with the spontaneously broken SUSY.
The models based on products of unitary groups, 
SU($N$)$\times$SU($M$),  are listed in Table \ref{tabNM}.

\begin{table}
\begin{tabular}{|p{2.0cm}|p{4.0cm}|p{3.7cm}|p{3.5cm}|}
\hline
~ & ~  & ~ & ~  \\[-0.1cm]
Model & {~~~~~~~~~~\em N\,-2}& {~~~~~~\em N\,-(N$-$1)}  &{~~~~~~\em N-(N$-$2)}\\
[0.2cm]
\hline 
\vspace*{-0.2cm}
~ & ~  & ~& ~ \\
Matter &\mbox{$\{ N,2\} +2\,\{ \bar N,1\}$} \mbox{$+\{
1,2\}$}&\mbox{$\{N,N-1\}$}
 \mbox{$+(N-1)\,\{\bar 
N,1\}$} \mbox{$+N\{1,\overline{N-1}\}$}\}&\mbox{$\{
N,N-2\}$} \mbox{$+(N-2)\,\{ \bar  N,1\}$} \mbox{$+N\,\{1,\overline{N-2}\}$}\\ 
[0.2cm]
\hline
\vspace*{-0.2cm}
~ & ~  & ~& ~ \\
{Comments, references} &$N$ odd, at $N>5$ quartic  superpotential 
required, \cite{DNNS} &Non-renormalizable ${\cal W}_{\rm tree}$, 
\cite{PST}&Via duality relat\-ed to \mbox{$ N$-{\em
2}} mo\-dels, \cite{PST}\\[0.2cm] 
\hline
\end{tabular}
\caption{The SU($N$)$\times$SU($M$) models, usually referred to as 
the {\em 
N-M} models.}\label{tabNM}
\end{table}

Simplectic and some other subgroups were considered too (e.g.
\cite{DNNS,KIST,CSS}). It is hardly necessary to  dwell on  these 
examples  since, with a single exception,  they do not go beyond the 
range of  ideas that are 
already familiar. The interested reader may consult  the original 
publications. 

The exception mentioned above is the use of Seiberg's duality
\cite{Nati2}. Assume that we have a dual pair such that the ``electric" 
theory is weakly coupled  while the ``magnetic" theory is strongly 
coupled. Assume that one can establish that under a certain choice of 
parameters the ``electric" theory spontaneously breaks 
supersymmetry.
Then, SUSY must be broken in the ``magnetic" theory too, and one 
learns
something about this noncalculable breaking from consideration
in the weak coupling regime. The information is not too abundant, 
though, since the full mapping between the ``electric" and  
``magnetic"  theories is not known. The best-known problem belonging 
to this  class is the {\em N-(N--2)} model analyzed along these lines in 
\cite{PST}. Its dual strongly coupled partner is the {\em N-2} model. 

Another  issue worth mentioning is a 
device, dating back to  the 1980's \cite{SVMO}, allowing  one to 
convert  
``non-calculable" models into calculable. The main idea 
is  as follows. Given a model where SUSY is broken in the strong 
coupling 
regime, one  introduces in this model, additionally, 
non-chiral matter (i.e. a set of matter fields with mass terms that are 
assumed to be small).  Such an expansion definitely does not change  
Witten's index. The extra matter fields may result in the emergence 
of 
the $D$ flat  directions. If they do, and  if a repulsive 
 superpotential is generated, supersymmetry may be spontaneously 
broken in the weak coupling regime.  The best known example of 
this  type  \cite{PT,Mura} is the SU(5) theory with
one anti-decuplet, one  quintet plus one or two extra
$\{ {\bf 5} +{\bf\bar  5}\}$.
One may call it a non-minimal one-generation SU(5) model
(non-minimal, because of the  extra nonchiral matter).
If the mass term of the extra flavor is small, the corresponding 
dynamics turns out quite similar to  that of the {\em 3-2} model: the 
theory is fully Higgsed, instanton does generate a  superpotential 
term 
repulsing the theory from the origin,  the exits from the 
valleys are locked. When the mass terms are small, the vacuum lies
far away from the origin. Everything is calculable.

When the mass terms of the non-chiral matter fields
increase, the vacuum moves towards the origin of the valley,
 and the coupling constant becomes stronger. Eventually, when all 
mass  terms become of order $\Lambda_5$, the  calculability is 
lost. (One still can argue,  following  an indirect line of reasoning, 
based 
on   Criterion 2 in Sec.\ \ref{sec52} and holomorphy, that SUSY stays 
broken. But this is a different story.) 

One last remark in conclusion of this mini-survey.
A class of promising  calculable models based on the {\em novel} 
idea -- the ITIY nonchiral mechanism of the dynamical SUSY breaking --
was engineered by Dimopoulos {\em et al.} \cite{MDDGR}.  The ITIY 
model {\em per se} is noncalculable, and is considered in 
Sec.\ \ref{sec6}. 
The modification necessary in order to make it calculable,
sometimes called the plateau mechanism, logically belongs to the 
current  
section, but it would be hardly possible to consider it here,
prior to the discussion of the ITIY model. Therefore, we postpone 
a ``get acquainted with the plateau mechanism" part till 
Sec.\ \ref{sec63}. 

\section{Spontaneous Supersymmetry Breaking in\\ Strong Coupling}
\label{sec6}

\renewcommand{\theequation}{6.\arabic{equation}}
\setcounter{equation}{0}

\subsection{SU(5) model with one generation}
\label{sec61}

This is the simplest  model
where the dynamical SUSY breaking may occur, as was understood 
shortly after the search began \cite{ADS1,MV}.
In fact, this is the  simplest simple-gauge-group model
with the nonchiral matter.  It is truly strongly coupled, the only 
relevant parameter is $\Lambda$, no small parameters exist,
and if SUSY is broken, the scale of the breaking has to be $\sim 
\Lambda$. 

We are already familiar with many features of this model.
It has no flat directions, no superpotential is possible, and 
it possesses two conserved U(1) currents -- the Konishi 
current and the $R$ current.
The corresponding charges of the matter fields are
\begin{equation}
Q_K(X)=-1, \,\,\, Q_K(V) =3\,;\qquad  R(X)= 1\,,\,\,\,
R(V)=-9\,,
\label{su5rchar}
\end{equation}
where our convention regarding
 the $R$ charge of $\psi_X$ is $R(\psi_X)=0$, the same as in the SU(5) model with
two generations. The zero mode formula is $\lambda^{10}\,\psi_X^3\,\psi_V$.

Both criteria of the dynamical SUSY breaking formulated in
Sec.\ \ref{sec52} are applicable. Here is how it works.

\vspace{0.2cm}

\noindent
{\em (i) Gluino condensation} 

\vspace{0.1cm}

\noindent
Following \cite{MV} (see also Sec.\ \ref{secglucon}) one can consider the
correlation function
\begin{equation}
\Pi (x,y,z) = \langle T(\lambda^2 (x) , \lambda^2 (y), \sigma (z) 
)\rangle\, ,
\label{mfcf}
\end{equation}
where 
\begin{equation}
\sigma =\left.
\epsilon^{\alpha\beta\gamma\delta\rho} 
X_{\alpha\beta}X_{\gamma\delta} \, (V^\kappa 
X_{\kappa\chi}\lambda^\chi_\psi\lambda^\psi_\rho )\, \right|_{\theta =0}\, , 
\end{equation}
and the Greek letters are used for the SU(5) gauge indices
(the Lorentz indices of the gluino field are suppressed). 

\begin{figure}[h]   
\vskip3mm
\epsfxsize=8.5cm
\centerline{\epsfbox{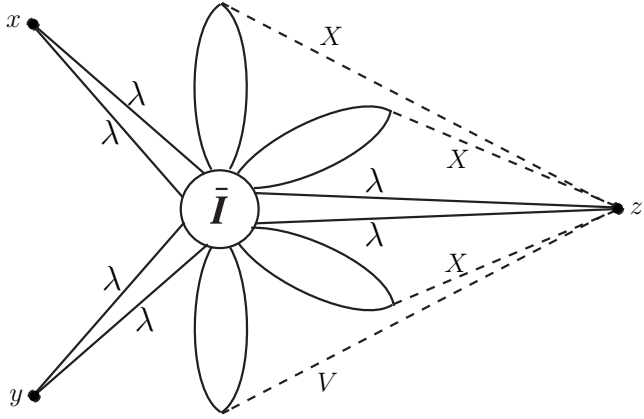}}
\caption{The instanton saturation of the
correlation function (\ref{mfcf}) in the one-generation SU(5) model.}
 \end{figure}

All  operators in the correlation function $\Pi (x,y,z)$
are the lowest components of the chiral superfields. Then  
supersymmetry tells us that, if this correlation function
does not vanish, it can only be an $x,y,z$ independent constant
\cite{NSVZ1}. It goes without saying that
$\Pi (x,y,z)$ vanishes in perturbation theory. The one-instanton 
contribution is non-vanishing, however, (see Fig.\ 5)
and does, indeed, produce $\Pi\propto (\Lambda_5)^{13}$ (The exponent 13 
coincides with the first coefficient of the $\beta$ function.)
If $x,y,z\ll \Lambda^{-1}$ one expects that the one-instanton 
contribution saturates $\Pi (x,y,z)$, so that the constant obtained in 
this way is reliable. If so, one can pass to the limit
$x,y,z\to\infty$ 
and use 
the property of clusterization at large $x,y,z$ to
argue that the gluino condensate develops$\,$\footnote{The solution 
with 
$\langle \lambda^2\rangle = 0$ and $\langle\sigma\rangle 
\to\infty$ is ruled out due to the absence of the flat directions 
(vacuum valleys).}, 
$\langle \lambda^2\rangle \neq 0$.  Since the superpotential 
is absent,
the gluino condensate is the order parameter for SUSY breaking.
One concludes that supersymmetry is spontaneously broken 
\cite{MV}.

\vspace{0.2cm}

\noindent
{\em (ii) Breaking of global symmetries}

\vspace{0.1cm}

\noindent
An alternative line of reasoning \cite{ADS1} is based
on Criterion 2. Assume that neither of the two U(1)  symmetries of 
the 
model (see Eq.\ (\ref{su5rchar})) is spontaneously broken. Then one has to match 
six 't~Hooft anomalous triangles \cite{tHooft2}:
\begin{eqnarray}
&&
\mbox{Tr}\, Q_R = -26, ~~~\,\,\,  \mbox{Tr}\,Q_K=5, ~~~~~\qquad
 \mbox{Tr} \,Q_K^3=125, \nonumber\\[0.1cm\,]
&&
 \mbox{Tr}\, Q_R^3 = -4976, \,\,\,  \mbox{Tr} 
\,Q_RQ_K^2 = -450, \,\,\,  \mbox{Tr}\, Q_R^2Q_K = 1500\, . 
\label{teqa}
\end{eqnarray}
Solutions of these matching conditions do exist \cite{ADS1}
but they look rather awful. For instance, a minimal solution
requires five massless composite Weyl fermions with the following 
assignment
of the $\{Q_K, Q_R\}$ charges:
$$
(-5,26), \,\,\, (5,-20), \,\,\, (5,-24), \,\,\, (0,1) , \,\,\, (0,-9)\, . 
$$
Since the solution is so complicated it is then natural to conclude 
(natural but not {\em mandatory}) that at least one of the axial 
symmetries is 
spontaneously broken, which, according to Criterion 2, would entail 
the spontaneous SUSY breaking. The spectrum of the theory must 
include at least  several composite massless particles: the Goldstino, 
and one  massless boson for each broken axial 
symmetry.
 
\vspace{0.2cm}

\noindent
{\em (iii) Adding nonchiral matter}

\vspace{0.1cm}

\noindent
Finally, the third alternative route providing  an additional 
argument that SUSY is broken
follows the strategy of \cite{SVMO} -- adding nonchiral matter with a 
small mass term $m$ makes the theory more manageable by 
Higgsing (a part of) the gauge group and creating a theory whose 
dynamical behavior is simpler than that of the original one. The 
pattern 
of SUSY 
breaking/conservation can be  established in this auxiliary model;
then one can return back to the original model by sending
$m\to\infty$. If no phase transition in $m$ happens {\em en route}
one gets an idea of the vacuum structure of the original model.

This strategy was applied to the SU(5) model in \cite{Mura} and
\cite{PT}, where one and two additional  flavors, respectively, were 
introduced. For instance, with one extra flavor $(Q, \bar Q)$ one 
arrives at a four-dimensional
moduli space parametrized by four chiral invariants $XX\bar Q$, $Q 
\bar Q$, $XVQ$, and $V \bar Q$.
A generic point from the vacuum valley corresponds to
the breaking of the gauge SU(5) down to SU(2), with four 
color-singlet  moduli.
The low-energy theory for these moduli represents a generalized 
Wess-Zumino model. A nonperturbative superpotential is generated
through the gluino condensation in the SU(2) theory. If the most general 
renormalizable tree-level superpotential is added, one arrives at
a typical  O'Raifeartaigh model of the spontaneous SUSY breaking 
\cite{Mura}. Certainly, the limit $m\to\infty$  that returns
us to the original theory and the 
strong coupling regime, cannot be achieved in a rigorous manner.
The very fact of the SUSY breaking is a rigorous statement, however.
Indeed, the gluino condensate $\langle\lambda\lambda\rangle$
depends on $m$ linearly$\,$\footnote{The proof that 
$\langle\lambda\lambda\rangle\propto m$ is suggested to the reader as an
exercise.}.  Therefore, if $\langle\lambda\lambda\rangle\neq 0$ at
small $m$, it cannot vanish at large $m$. 

Concluding this section let us note that the 
S0(10) model with a single spinor (i.e.  {\bf 16}-plet) representation  is 
very similar to the one-generation SU(5).Ê
It is both chiral -- no mass term is possible -- and free from the 
internal anomalies. The model was treated along the same lines as the
one-generation  SU(5)
model in the 1980's \cite{ADS3}. Calculable deformations obtained by 
supplementing the original model by  one nonchiral matter 
superfield in  the representation  {\bf 10} were considered by Murayama
\cite{Mura}. 

 \subsection{The Intriligator-Seiberg-Shenker model}
\label{sec62}

The mechanism suggested by these authors \cite{ISS} is the SU(2) 
gauge theory with a single matter superfield $Q$
in the representation 3/2, i.e. $Q_{\alpha\beta\gamma}$, symmetric 
in  all three indices ($\alpha ,\beta ,\gamma = 1,2$). This theory is 
chiral  since no mass term is possible. Indeed, it is easy to see that the only 
quadratic invariant one could write
$Q_{\alpha\beta\gamma}Q^{\alpha\beta\gamma}$ vanishes 
identically. Obviously, there are no cubic invariants. The first (and the only)
nontrivial invariant is quartic,
\begin{equation}
u = 
Q_{\alpha\beta\gamma}Q^{\alpha\beta\rho}
Q^{\tilde\alpha\tilde\beta
\gamma}Q_{\tilde\alpha\tilde\beta\rho}\, .
\label{quin}
\end{equation}
Thus, a one-dimensional vacuum valley exists; it is parametrized by 
the modulus 
$u$. At the origin of the valley, $u = 0$, the classical theory  exhibits 
massless fields (the gauge bosons and gauginos). At $u\neq 0$ the 
gauge  SU(2) is fully broken, the theory is fully Higgsed. If $u\gg\Lambda$, 
the  gauge bosons are heavy, and the theory is weakly coupled. 

The model possesses a single anomaly-free $R$ current$\,$\footnote{The
 specific form of the $R$ current depends on the group 
factors. The expression in Eq.\ (\ref{clint}) takes into account that
 $T(3/2) = 5$, and the zero mode formula is
$\lambda^4 \psi_Q^{10}$.}, with the following 
$R$ charges: 
\begin{equation}
R(\psi_Q ) = -\frac{2}{5} \, , \quad R(Q) = \frac{3}{5}\, .
\label{clint}
\end{equation}
 The $R$ charge of $u$ is 12/5 while that of $\psi_u$ is 7/5.
The $R$ charge conservation is spontaneously broken at
$u\neq 0$.

Following \cite{ISS} let us assume that a (nonrenormalizable) 
tree-level superpotential is added, ${\cal W} = u/M_0$, where $M_0$ is some 
ultraviolet parameter, $M_0\gg\Lambda$. This superpotential should 
be considered as a low-energy limit of some fundamental theory 
defined at  the scale
$M_0$. What exactly this fundamental theory is need not concern us 
here. The tree-level superpotential is treated in the first nontrivial
order in $\Lambda/M_0$. It is not iterated --
the iterations cannot be considered without the knowledge
of the fundamental theory because of the nonrenormalizable nature 
of ${\cal W} = u/M_0$. Higher order effects in $\Lambda/M_0$
are discarded.

Classically the effective Lagrangian for the moduli field takes the 
form
\begin{equation}
{\cal L} = \frac{1}{4}\int d^4\theta (\bar u u)^{1/4} +
\left\{ \frac{1}{2}\int d^2\theta \frac{u}{M_0} +\mbox{H.c.}\right\} 
\, .
\label{naiv}
\end{equation}
It has a unique supersymmetric vacuum state at $u=0$. The 
tree-level 
superpotential pushes the theory towards the origin of the valley. 
The existence of the  supersymmetric vacuum at $u=0$ is entirely
due to the fact that the kinetic term for $u$ in Eq.\ (\ref{naiv}) is 
singular 
at $u=0$.

While the kinetic term $(\bar u u)^{1/4} $
is definitely correct at large $u$, at small $u$ the singularity could be 
smoothed out  provided that the theory at small $u$ is in the 
confining 
phase.
Then, generally speaking, there are no reasons to expect that
there are massless physical states, other than those described by the 
moduli field $u$, and, if so, one can expect the K\"{a}hler function to 
be 
regular at the origin,
$$
{\cal K} = (\bar u u)^{1/4}\, \longrightarrow\, {\cal K} = \Lambda^{-
6}(\bar u u)+ 
...\,,\>\qquad (|u|\ll \Lambda )\, .
$$
Let us accept this assumption as a working hypothesis. Then the 
would-be  supersymmetric vacuum at $u=0$ disappears. Indeed,
at small $u$
\begin{equation}
F_u = \frac{\Lambda^6}{M_0}\,\,\, \mbox{and} \,\,\, {\cal E} =   
\frac{\Lambda^6}{M_0^2}\, .
\end{equation}

The vacuum energy density is very small being measured in its 
natural 
units, ${\cal E}/\Lambda^4\ll 1$, but is definitely non-zero. Since at large 
$u$
the valley is locked and there is no supersymmetric solution either, 
one
arrives at the conclusion of the spontaneous supersymmetry
breaking in the model under consideration.

What are the arguments in favor of the hypothesis of the
confining regime at small $u$ and no massless fields other than $u$
itself?

In essence, there is only one argument \cite{ISS}: the 't Hooft 
matching 
at $u=0$.
At the origin of the valley the $R$ charge is conserved provided the 
small tree-level superpotential is switched off. 
There are two 't Hooft triangles: Tr$\,Q_R$ and Tr$\,Q_R^3$.
If the theory is in the confining regime, with no massless composites,
both triangles must be matched by the contribution of $\psi_u$.
And they do match! Indeed, it is easy to see that
$$
\mbox{Tr}\, Q_R = \frac{7}{5}\,,\qquad \mbox{Tr}\, Q_R^3 
= 
\left(\frac{7}{5}\right)^3\, .
$$
Since the $R$ charge of $\psi_u$ is 7/5, the matching is automatic.

Is this argument sufficiently conclusive in order to say that
the theory is in the confining regime, and the 
Intriligator-Seiberg-Shenker mechanism works? Perhaps, not.
In fact, one can give a rather strong argument pointing in the 
opposite 
direction.

Although the model at hand is asymptotically free, it is barely so.
The first coefficient of the $\beta$ function is abnormally small,
$\beta_0 = 6 - 5 =1$. The second coefficient of the $\beta$ function
is of a normal size and positive. So it is very likely that
the $\beta$ function has a zero at a small value of $\alpha$,
and the theory is conformal and relatively weakly coupled in the 
infrared, much in the same way as QCD with
$N_f$ close to $11N_c/3$ \cite{MB,BZ} or 
SQCD with $N_f$ slightly lower than $3N_c$, i.e. near the right edge 
of 
the conformal window \cite{Nati2}. The only distinction is that 
in the latter case, by sending $N_c\to \infty$, we can approach to the
edge of the conformal window arbitrarily closely, so that the infrared 
fixed point of the $\beta$ function occurs at an arbitrarily small 
value
of $\alpha$, and higher order effects are parametrically suppressed.
In the Intriligator-Seiberg-Shenker model one has to rely on the
numerical suppression of the higher order effects in $\gamma$.
In order to give the reader the idea of the degree of suppression let 
us 
do this simple numerical exercise. 
The numerator of the NSVZ $\beta$ function in the model at hand 
is proportional
to
\begin{equation}
3T_G - T(3/2)(1-\gamma (\alpha )) = 1 + 5\gamma (\alpha )\, , 
\end{equation}
where $\gamma (\alpha )$ is the anomalous dimension of the
field $Q_{\alpha\beta\gamma}$. The infrared fixed point occurs
at $\alpha_*$ such that
\begin{equation}
\gamma (\alpha_* ) = - \frac{1}{5}\, . 
\label{exone}
\end{equation}
In the leading order
\begin{equation}
\gamma (\alpha ) = - C_2(R) \frac{\alpha}{\pi}\, ,
\label{extwo}
\end{equation}
where 
$$
C_2 (R) = T(R) \frac{\mbox{dim (adj) }}{\mbox{dim}(R)} = 
\frac{15}{4}
\,\,\, \mbox{for}\,\,\, R=3/2\, . 
$$
The coefficient $C_2$ is rather large by itself; what is important, this
numerical enhancement does not seem to propagate to  higher
order coefficients in $\gamma$. Combining Eqs.\ (\ref{exone}) and 
(\ref{extwo}) one arrives at$\,$\footnote{For comparison: in QCD with 
fifteen massless flavors  $\alpha_* /\pi = 1/22$.} 
\begin{equation}
\frac{\alpha_* }{\pi} = \frac{4}{75}\, .
\end{equation}
It seems rather unlikely that,  with this small value of the critical 
coupling constant,  the higher-order corrections are so abnormally 
large 
that
they eliminate the infrared fixed point altogether. 

What if the infrared fixed point exists in the weak coupling regime?
Then, the  infrared limit is conformal (the non-Abelian Coulomb 
phase).
Needless to say that the 't Hooft matching becomes uninformative:
if the theory is in the non-Abelian Coulomb phase,
the 't Hooft triangles are trivially saturated by the
unconfined gluino and quark fields themselves. Moreover,
the singularity in the K\"{a}hler function need not be smoothed out, 
and 
the conclusion of the dynamical SUSY  breaking 
becomes unsubstantiated. 
It is natural then to find  a supersymmetric vacuum at $u=0$.

Everything seems self-consistent,
except a single question: why then the two 't~Hooft triangles
of the model, Tr$\,Q_R$ and Tr$\,Q_R^3$,  are  successfully saturated by 
the 
composite $\psi_u$? Can this matching be a mere coincidence?

We cannot be sure. It is worth adding, though, that at least one 
example
of such a coincidental matching 
was found in a rather similar model
\cite{BCI}. The model consists of the SO($N$) gauge sector,
plus a single matter (chiral) superfield in the (two-index symmetric 
traceless) tensor representation of SO($N$). The moduli fields of this 
model are known to saturate the 't Hooft triangles
corresponding to all unbroken global axial symmetries at the origin 
of 
the moduli space. One would then naturally expect 
to find the confining phase at the origin.  At the same time, in Ref.\ \cite{BCI}  it
was argued  that the origin of the moduli space 
 belongs to the non-Abelian Coulomb phase rather than to the 
confining 
phase, and the above matching by the moduli fermions is 
coincidental.
The essence of the argument is beyond the scope of this review.
The interested reader is referred to \cite{BCI}.

 In summary, the Intriligator-Seiberg-Shenker model
was designed to provide a new mechanism of the dynamical SUSY 
breaking. It remains to be seen whether it actually does the job.
Even if the original SU(2) model is in the infrared-conformal phase
and preserves supersymmetry 
one can try to construct other, more complicated models based on the 
same idea which, hopefully, do not fall in the conformal window.
A promising candidate  of this type is an SU(7) model \cite{CSSe}
with two anti-symmetric tensors,  six (anti)fundamentals
and an appropriate tree-level superpotential, believed to be $s$ 
confining. 

\subsection{The Intriligator-Thomas-Izawa-Yanagida model}
\label{sec63}

The model is nonchiral, Witten's index is two. Nevertheless, 
supersymmetry is dynamically broken, in a manner  briefly 
discussed in 
Sec.\ \ref{sec36}.  Here we explain in detail how the ITIY  mechanism 
works. 

The model \cite{IT}  is a close relative of the SU(2) model considered 
in Sec.\  \ref{sec472}. The matter chiral superfields $Q^{\alpha}_ f$ 
are 
four  color doublets ($\alpha = 1,2$ and $f= 1,2,3,4$ are color and 
subflavor  indices).  In addition to the ``quark'' superfields 
$Q^{\alpha}_ f$,  six color-singlet chiral
superfields $S^{fg}=-S^{gf}$ are introduced.  Their 
 interaction  with  $Q_{\alpha f}$ is due to the superpotential, 
\begin{equation}
{\cal W} = \frac{h}{2} \, S^{fg} \,Q^{\alpha}_ f \, 
Q^{\beta}_g\,\epsilon_{\alpha
\beta}\;.
\label{spit}
\end{equation}
The theory is globally invariant under the SU(4) rotations of the 
subflavors. It also has a conserved $R$ charge: $Q^{\alpha}_ f$ is
neutral while the $R$ charge of $S^{fg}$ is two.

Let us start from the case  $h=0$ when the
superpotential (\ref{spit}) is switched off and the singlet fields 
$S^{fg}$ 
are  decoupled. This case is well studied~\cite{Nati1}, see also 
Sec.~\ref{sec472}. The classical moduli space is  spanned by the gauge 
invariants
\begin{equation}
M_{fg} = -M_{gf}=Q^{\alpha}_fQ^{\beta}_g\,\epsilon_{\alpha\beta}
\end{equation}
The matrix $M_{fg}$
is antisymmetric in 
$f,g$, its six elements are  subject to one classical constraint (\ref{pfaf}).
Nonperturbative quantum corrections change the geometry of 
the moduli space, the constraint 
${\rm Pf} (M)=0$ is replaced by ${\rm Pf} (M)=\Lambda^4_{\rm two-fl}$,
so that the origin is excluded.

Let us now switch on the interactions (\ref{spit}). It is clear that the
consideration of Sec.\ \ref{sec472}  remains valid, with the substitution
$
m^{fg} \to h S^{fg} 
$ (the lowest component of $S^{fg}$ is implied).
Thus one gets,
\begin{equation}
\left\langle M_{fg}\right\rangle =
-\frac{1}{16\pi^2 h} \,\left[\frac{1}{S}\right]_{fg}  
\left\langle \mbox{Tr}\,W^2\right\rangle\;; ~~~~~
\left\langle \mbox{Tr}\,W^2\right\rangle= {\rm const}\, 
h\,\sqrt{{\rm
Pf}(S)} \,\Lambda^2_{\rm two-fl}\;.
\label{anomS}
\end{equation}
Now the  fields $S^{fg}$ are dynamical; to determine  the vacuum
state one has to take into consideration their $F$ terms, 
\begin{equation}
 \left\langle F_{S^{fg}}\right\rangle= \left\langle 
\frac{\partial}{\partial
S^{fg}}\,{\cal W}\right\rangle =h \left\langle M_{fg}\right\rangle\;.
\end{equation}
Since  $\left\langle M_{fg}\right\rangle$ does not vanish, 
$\left\langle 
M_{fg}\right\rangle \sim h \Lambda^2$, this 
leads to a clear
contradiction with the presumed supersymmetry of the vacuum 
state.

To get a clearer picture of the phenomenon (and, in particular, to 
answer 
the question how SUSY can be broken in the theory where 
Witten's index does not vanish,  $I_W=2$)
let us add a small mass term to the fields $S$.
To keep our presentation as transparent as possible
we will assume that
$$ 
S^{12} = S^{34} \equiv S , \,\,\, \mbox{all other components vanish.}
$$
In fact, for any general set of $S^{fg}$ one can always perform a global 
SU(4) rotation to eliminate  all components other than $S^{12}$ and $S^{34}$.

The part of the superpotential containing $S$ consists of the
tree-level term $mS^2$,  plus the nonperturbative term due to the 
gluino condensation$\,$\footnote{The ambiguity in the sign of the 
constant  in Eq.\ (\ref{xyz}) is in one-to-one correspondence 
with the $\pm$ sign in Eq.\ (\ref{konets}).},
\begin{equation}
{\cal W}_S = \frac{1}{2}\, m\, S^2 \pm \mbox{const}\, h\,S\,
\Lambda^2_{\rm two-fl}\, .
\label{xyz}
\end{equation}
We find two supersymmetric vacua at
\begin{equation}
S = \pm \mbox{const}\, h \, m^{-1} \Lambda^2_{\rm two-fl}\, , 
\end{equation}
in full accord with Witten's index. 
In the limit of small $m$, however, these supersymmetric vacua
lie very far in the space of fields.
Let us see what happens at a finite distance in the space of fields.

If $|S|\ll h\,  m^{-1} \Lambda^2_{\rm two-fl}$, the vacuum energy density
${\cal E}$ is just a constant, 
\begin{equation}
{\cal E} = |F_S|^2 = \mbox{const}\, h^2\Lambda^4\, .
\label{xyt}
\end{equation}
We get a plateau, or indefinite equilibrium.

The equilibrium is destroyed by the perturbative corrections to the 
K\"{a}hler function \cite{Shir}. Assume that $h^{-1} \Lambda \ll |S|\ll 
h \, m^{-1} \Lambda^2$. Then the renormalization of the kinetic term is 
weak; at one loop it is given by the following $Z$ factor
\begin{equation}
Z = 1 +C\, \frac{h^2}{16\pi^2}\ln \frac{M_{\rm UV}}{h \, |S|}\, ,
\end{equation}
where $C$ is a {\em positive} constant and $M_{\rm UV}$ is the ultraviolet 
cut off. Inclusion of this correction tilts the plateau,
\begin{equation}
{\cal E} = |F_S|^2\, Z^{-1} = \mbox{const}^2\, h^2\, \Lambda^4
\left( 1 +C\, \frac{h^2}{16\pi^2}\ln \frac{M_{\rm UV}}{h \, |S|}\right)^{-1}\, ,
\label{konets2}
\end{equation}
making the theory slide towards smaller values of $|S|$.
A minimum (or minima) is achieved somewhere at $h\, |S| \lsim \Lambda$.
At $h\, |S| \sim \Lambda$ Eq.\ (\ref{konets2})  becomes inapplicable, of course.
Perturbation theory becomes useless for the calculation of
the K\"ahler potential.  One can show~\cite{CLP}, however, that at 
 $ |S| \lsim \Lambda$
 noncalculable terms in the K\"ahler potential are
suppressed by powers of $h$ and are negligible
provided that $h\ll 1$. One gets a stable non-supersymmetric
vacuum at $\langle S \rangle =0$. Besides the Goldstino, at
$\langle S \rangle =0$ the theory exhibits five massless Goldstone bosons
corresponding to the spontaneous breaking
of the global SU(4) down to Sp(4).  Of course, the analysis
of Ref.\ \cite{CLP} has nothing to say about
the possibility of finding a still lower non-supersymmetric vacuum at
$h\, |S| \sim \Lambda$. In this domain the  K\"ahler potential
is  noncalculable. 

 For finite values of $m$ the non-supersymmetric vacua
are   quasistable because of the  tunneling into the 
supersymmetric vacuum at
$\langle S\rangle \propto 1/m$. 
In the limit $m\to 0$ the tunneling probability vanishes, and the 
non-supersymmetric vacuum becomes stable.

An elegant idea of how to make the ITIY mechanism fully calculable 
by stabilizing the
theory on the plateau at sufficiently large values of $S$
was suggested in Ref.\ \cite{MDDGR}.
Assume that a part of the global SU(4) of the ITIY model is gauged.
The original gauge SU(2) -- call it strong -- is thus supplemented
by additional weaker gauge interactions, with the gauge coupling
$\alpha_a$, such that $\Lambda_a\ll \Lambda$ (the subscript $a$ 
means additional). The particular pattern considered in \cite{MDDGR}
was SU(2)$\times$ SU(2) weak gauge group.
Note that the Yukawa coupling $h$ is a free parameter, so that the 
ratio of $\alpha_a /h$ can be arbitrary. 

Then the $Z$ factor of the gauge fields gets a contribution both, from 
the Yukawa interactions and the weak gauge interactions. If 
$\alpha_a \gg
h$ the tilt of the plateau is reversed (since the signs of these
two contributions are opposite), $|S|$ is pushed towards large rather 
than small values. This drive towards large values of $|S|$
eventually stops since the running of $\alpha_a$ makes this coupling 
constant  smaller at 
large $|S|$ while $h$ gets larger. Sooner or later $h$ wins over
$\alpha_a$. The theory can be stabilized at arbitrarily large
values of $|S|$, where the $Z$ factor is calculable to the degree of  
precision   we  want.

As an additional bonus one finds that at large values of
$|S|$ the supersymmetry breaking effects can be as small as we 
want. At first sight this conclusion seems rather paradoxical
since 
${\cal E}_{\rm vac} \sim  h^2\Lambda^4$ and is seemingly $S$ 
independent.
One should not forget, however, that the natural unit of energy is set 
by the gluino condensate, $\langle\lambda^2\rangle \sim 
S\Lambda^2\gg \Lambda^3$.
Being measured in these natural units, the vacuum energy is very 
small,
$$
{\cal E}_{\rm vac} \sim S^{-2}\langle\lambda^2\rangle^2\, , 
$$
and the spitting between the glueball/glueballino masses is of order 
$S^{-2}\langle\lambda^2\rangle$. The ratio of the splitting to the 
masses themselves
tends to zero as $S^{-2}$.  The $S$ superfield acts as a phantom 
axion/dilaton --
asymptotically massless and non-interacting -- 
much in the same way as the invisible axion of QCD \cite{KSVZ}.

Concluding this section let us note that a straightforward
generalization of the ITIY
mechanism based on  SU($N$) SQCD with $N$ quark flavors and
$N^2+2$ gauge singlets was considered in Ref.\ \cite{CLP}.

\section{Impact of the Chirally Symmetric Vacuum}
\label{sec7}

The existence of $T_G$ distinct (but physically equivalent) chirally 
asymmetric  vacuum states in SUSY gluodynamics is a
 well-established 
fact. We have already mentioned in passing, that
a physically inequivalent chirally symmetric state at 
$\langle\lambda^2\rangle =0$, the Kovner-Shifman vacuum,  is not 
ruled out; moreover, certain arguments make one believe
\cite{kovner2} that chirally symmetric regime is attainable,
the $\langle\lambda^2\rangle =0$ phase of SUSY gluodynamics 
exists.
If this is indeed the case, and if this vacuum does not disappear in 
the 
presence of sufficiently light (or massless) matter,
it would lead to  absolutely drastic consequences in many 
mechanisms 
of the dynamical SUSY breaking: they will simply disappear.

Let us first briefly review the arguments in favor of the 
Kovner-Shifman 
vacuum. First, it is present in the Veneziano-Yankielowicz effective 
Lagrangian (see Sec.\ \ref{secEL})
which encodes information on the anomalous Ward identities in 
SUSY 
gluodynamics and is expected to properly reflect its vacuum 
structure. 
Second, the instanton calculations of the correlation function
$\langle \lambda^2(x) \lambda^2(0)\rangle$ at small $x$,
performed in the strong coupling theory and in the weak coupling 
theory (i.e. adding matter in SU(2) SUSY gluodynamics, Higgsing the 
theory, and then returning back by using holomorphy in the matter 
mass parameter) do not match each other (Secs.\ \ref{secglucon}, \ref{sec471}).
The first method yields a smaller value for the gluino condensate 
than 
the second one \cite{NSVZ3}. The mismatch is $\sqrt{4/5}$ for SU(2). 
The mismatch could be explained if one invokes a hypothesis of 
Amati 
{\em et al.} \cite{Amati}, according to which
the strong coupling calculation of the correlation function
$\langle \lambda^2(x) \lambda^2(0)\rangle$ in fact yields a result 
averaged over all vacuum states of the theory. If the 
$\langle\lambda^2\rangle =0$ vacuum exists, it would contaminate
this correlation function, explaining in a natural way
the 4/5 suppression factor compared to the calculation in the weak 
coupling regime. The evidence for the chirally symmetric state
is admittedly circumstantial. 
In particular, as it was discussed in
Sec.~\ref{secglucon}, the hypothesis of Amati {\em et
al.}\cite{Amati} is in contradiction with
the strong coupling solution of the ${\cal N}=2$ gluodynamics softly broken to 
${\cal N}=1$.

The  Kovner-Shifman vacuum must give zero contribution to Witten's 
index since the latter is fully saturated by the chirally symmetric 
vacua.
If so, it is potentially unstable under various deformations. 
For instance, putting the system in a finite-size box
lifts the vacuum energy density from zero \cite{KKS}. 
This vacuum disappears in finite volume$\,$\footnote{It goes without saying 
that massless fermions are mandatory in the  
$\langle\lambda^2\rangle 
=0$ phase of SUSY gluodynamics. }. This instability -- the tendency to 
escape under seemingly ``harmless" deformations --  may explain
why the Kovner-Shifman  vacuum at 
$\langle\mbox{Tr}\lambda^2\rangle =0$ is not seen in Witten's 
$D$-brane construction \cite{EWD} . Perhaps, this is not surprising at 
all. 
Indeed, there is a good deal of extrapolation in this construction, 
against which the chirally asymmetric vacua are stable
(they have no choice since they have to saturate Witten's index)
while the $\langle\mbox{Tr}\lambda^2\rangle =0$ vacuum need not 
be stable and may not  survive the space-time distortions associated 
with the $D$-brane engineering.  Neither it is seen 
in the Seiberg-Witten solution \cite{SEIW} of ${\cal N} = 2$ SUSY
gluodynamics slightly perturbed by a small mass term$\,$\footnote{
The chirally symmetric state $\langle \lambda^2 \rangle= \langle m\Phi^2
\rangle=0$ resembles sphaleron: it realizes 
a saddle point of the energy.}  of the
matter field $m$Tr$\,\Phi^2$,  ($m\ll
\Lambda$). If at all, it can exist only  at large values of $m$, i.e. $m\gg
\Lambda$. 

One may ask a question: ``Are we aware of any  other examples of 
supersymmetric theories where Witten's index vanishes, and the 
supersymmetric vacuum appears/disappears under deformations of 
parameters that do not change the structure of the theory, e. g. 
variations of mass parameters or putting the theory in a box
of size $L$ and considering the theory at large but finite $L$ instead
of the limit $L\to\infty$?"

The answer is yes. Surprisingly, the simplest two-dimensional theory 
with the minimal supersymmetry, ${\cal N} = 1$, is of this type.
The model we keep in mind (to be referred to
as the minimal model) is similar to that described in detail
in Sec.\ \ref{sec424}, but simpler. Assume that instead of two chiral superfields 
we deal with one (i.e. omit $\phi_2$ and $\psi_2$ from Eqs.\ (\ref{n2lag}) and
(\ref{arkmi})). This  cubic  superpotential model has 
Witten's index zero.
This can be seen in many ways. Say, if the mass parameter $m^2$
in the superpotential  is made negative, the equation
$\partial{\cal W}/\partial\phi = 0$ has no real solutions.
Alternatively, one can consider the theory with positive $m^2$ in a 
finite 
box, of length  $L$. If 
$\lambda L\ll 1$,  it is legitimate to  retain only the zero momentum 
modes 
discarding all 
others. Then one gets a quantum-mechanical system
known to have vanishing Witten's index \cite{Witten2}. 

The minimal model  at positive (and large) values of 
$m^2$  has two physically equivalent 
vacua, at $\phi =\pm\sqrt{2} m/\lambda$. The vanishing of  Witten's index 
implies that  if  one of the 
vacua is ``bosonic", the other is ``fermionic".
 
Although Witten's index in the minimal model does vanish,
supersymmetry is {\em not} spontaneously  broken, provided
$m^2$ is large and positive, i.e. $m^2\gg \lambda^2$. The 
supersymmetric vacuum with the zero energy density exists. 
Indeed, one can show that $v^2 =m^2 / \lambda^2$ is the genuine 
dimensionless expansion parameter in the theory at hand (the 
expansion 
runs in powers of $1/v^2$). At large $v^2$ the theory is weakly 
coupled. Since there are no massless fields in the Lagrangian,
(and they cannot appear as bound states in the weak coupling 
regime),
there is no appropriate candidate to play the role of Goldstino, and, 
hence,  SUSY  must be  realized linearly.

At large negative $m^2 $ the field  
$\phi$ does not  develop a vacuum expectation value. SUSY is 
broken.
The $\psi$ field is massless, and is the Goldstino of the 
spontaneously broken SUSY.  In fact, one can argue
that the SUSY breaking takes place at positive but small values
of $m^2 $,  $m^2 \sim \lambda^2$. The classical potential for such 
values 
of parameters has an underdeveloped peak at $\phi = 0$ plus two 
shallow minima at $\phi \sim \pm 1$. The quantum corrections are 
likely to completely smear this structure, 
leaving no physical states at zero energy. The switch  from the 
unbroken to spontaneously broken SUSY occurs as a phase transition 
in 
the mass parameter.  

This example teaches us that it is not unreasonable
to think of the existence of the $\langle\lambda^2\rangle =0$
phase of SUSY gluodynamics, even if
neither the $D$-brane perspective nor the Seiberg-Witten solution of 
${\cal N}=2$ at small $m$ carry indications on such a 
solution. Assuming that it exists, one may ask what are its implications 
on the  SUSY breaking mechanisms we discussed.

The mechanisms that are based on the weak coupling regime are 
least  affected. Formally, it may well happen that, apart from the 
SUSY-breaking vacua at large values of fields, there exists a 
SUSY-conserving vacuum at small  values.
These vacua are well-separated  in the space of fields. The 
SUSY-breaking vacua remain as quasistable.
Under an appropriate choice of relevant parameters the barrier may 
be  arbitrarily large, and if the theory initially finds itself in the
SUSY-breaking vacuum, it will never leak to the SUSY-conserving 
one, for all practical purposes. It is worth noting that the possibility of 
SUSY-conserving vacua, separated by a large distance in the space of 
fields from the SUSY-breaking vacuum, where the theory actually 
resides, was incorporated from the very beginning  in  the plateau 
models of Ref.\ \cite{MDDGR}. The possibility of existence of the 
SUSY-conserving  vacua in the domain of strong coupling  in the {\em  
3-2} model or the two-generation SU(5) model just puts them in the 
same class as the plateau models.

The most affected are the mechanisms based on the gluino 
condensation  in the strong coupling regime, e.g. the one-generation 
SU(5) model or the ITIY model. For them, the Kovner-Shifman vacuum 
may be a fatal blow. The ISS mechanism and its derivatives are neutral.
As far as we see it now, they are independent of this phenomenon, 
and  remain intact.

\section{Concluding Remarks}
The instanton calculus proved to be a powerful tool in analyzing supersymmetric
theories. In this review we focused mainly on formal aspects of the method and
several applications. 

Although the instanton calculus score an impressive list of results,
still some conceptual issues remain unclear. Two points are most disturbing.
First, the result obtained via instantons as a rule are of the topological
nature: they do not depend on specific details. For example, in the one-flavor
model the nonperturbative superpotential is totally determined by the zero-size
instantons. Second, the VEV scale $v$ inherent to the instanton calculations in the
Higgs phase has no direct physical meaning. It is clear that instanton results are an
indirect reflection of physics at the monopole/sphaleron scale $v/g$. 
The whole story reminds a famous Anderson's fairy tail where a soldier cooked a
soup from an axe.

An example where the relation to the monopole scale is established is provided by 
${\cal N}=2$ theories~\cite{SEIW} where one and the same function determines
both the scattering amplitudes in the infrared and the monopole/dyon spectrum.
We believe that a similar relation exists in ${\cal N}=1$ theories. An important task
is to reveal it.

\section*{Acknowledgments}
\addcontentsline{toc}{section}{\numberline{}Acknowledgments}

This review is, in a sense, a remake of the review~\cite{IVS}
we wrote in collaboration with V. Zakharov in 1985. The material
was also presented  by one of the
authors (M.S.)  at the 1985 Bakuriani School on High Energy Physics.
We are grateful to V. Zakharov for valuable insights. 
While working on this review we had  fruitful 
discussions with A. Yung whom we  would like to thank for comments 
and  cooperation.  We are grateful to A.~Faraggi,  V.~Khoze, Jr., M.~Marinov,
M.~Mattis,  A.~Ritz, and T.\ ter Veldhuis for helpful discussions. We are grateful to
X.\ Hou for careful reading of the manuscript.

The work was supported in part by the DOE under  grant 
number DE-FG02-94ER40823.

\vspace{0.4cm}

\subsection*{Recommended Literature}
\addcontentsline{toc}{subsection}{\numberline{}Recommended Literature}

\vspace{0.2cm}

{\bf Books}

\vspace{0.2cm}

\noindent
D. Bailin and A. Love,  {\it Supersymmetric Gauge Field Theory
and String Theory} (IOP Publishing, Bristol, 1994).

\vspace{0.2cm}

\noindent
A brief introduction to supersymmetric instanton calculus is given in

\vspace{0.2cm}

\noindent
{\em Instantons in Gauge Theories}, ed. M. Shifman,
(World Scientific, Singapore, 1994), Chapter VII.

\vspace{0.2cm}

\noindent
A brief survey of those  aspects of supersymmetry which are most 
relevant to the recent developments can be found in 

\vspace{0.2cm}

\noindent
J. Lykken, {\em Introduction to Supersymmetry},
hep-th/9612114.

\vspace{0.3cm}

\noindent
{\bf Reviews on SUSY instantons and instanton-based mechanisms of 
supersymmetrty
breaking}

\vspace{0.2cm}

\noindent
D. Amati, K. Konishi, Y. Meurice, G. Rossi and G. Veneziano,
{\it Phys. Rep.} {\bf 162} (1988) 557.

\vspace{0.2cm}

\noindent
V.V. Khoze, M.P. Mattis and M.J. Slater, 
{\em Nucl. Phys. },  {\bf B536}  (1998) 69.

\vspace{0.2cm}

\noindent
L. Randall,  in Proc.  Int. Workshop {\it Perspectives of Strong 
Coupling Gauge Theories}, Nagoya, Japan,  1996, 
Eds. J. Nishimura and K. Kamawaki (World Scientific, Singapore, 1997), page 258
[hep-ph/9706474].

\vspace{0.2cm}

\noindent
W. Skiba, {\it Mod. Phys. Lett.} {\bf A12} (1997) 737.

\vspace{0.2cm}

\noindent
E. Poppitz, {\it Int. J. Mod. Phys.} {\bf A13} (1998) 3051.

\vspace{0.1cm}

\noindent
S. Thomas, in Proc.  Int. Workshop {\it Perspectives of Strong 
Coupling Gauge Theories}, Nagoya, Japan,  1996, 
Eds. J. Nishimura and K. Kamawaki (World Scientific, Singapore, 1997),
page  272
[hep-th/9801007]. 

\vspace{0.2cm}

\noindent
E. Poppitz and S. Trivedi, {\it Dynamical Supersymmetry Breaking},
 hep-th/9803107 [{\it Ann. Rev. Nucl. Part. Sci.}, to be published]. 


\newpage

\begin{thebibliography}{123}
\addcontentsline{toc}{section}{\numberline{}References}

\bibitem{Witten1}
E.~Witten, {\it Nucl. Phys.} {\bf B185} (1981) 513 [Reprinted in {\em 
Supersymmetry}, Ed.\ S.~Ferrara (North Holland/World Scientific, 
Amsterdam 
-- Singapore, 1987), Vol. 1, page 443].

\bibitem{Witten2}
E. Witten, {\it Nucl. Phys.} {\bf B202} (1982) 253 [Reprinted in {\em 
Supersymmetry}, Ed. S. Ferrara (North Holland/World Scientific, 
Amsterdam 
-- Singapore, 1987), Vol. 1, page 490];
{\em J. High Energy Phys.} 9802:006, 1998 
(see Appendix).

\bibitem{NSVZ1}
V.A. Novikov, M.A. Shifman, A.I. Vainshtein, and V.I. Zakharov {\it 
Nucl. Phys.} {\bf B229} (1983) 407 [Reprinted in {\em 
Supersymmetry}, Ed. S. 
Ferrara (North Holland/World Scientific, Amsterdam -- Singapore, 
1987), Vol. 
1, page 606]. 

\bibitem{RoVe}
G.C. Rossi and G. Veneziano, {\it Phys. Lett.} {\bf B138} (1984) 195
[Reprinted in {\em Supersymmetry}, Ed. S. Ferrara (North 
Holland/World 
Scientific, Amsterdam -- Singapore, 1987), Vol. 1, page 620].

\bibitem{ADS2}
I.  Affleck,  M.  Dine, and N. Seiberg,  {\it Nucl. Phys.} {\bf B241} 
(1984) 493.

\bibitem{ISS}
K. Intriligator, N. Seiberg, and  S.H. Shenker, {\it Phys. Lett.} {\bf 
B342} (1995) 152.

\bibitem{IT}
K. Intriligator and  S. Thomas, {\it Nucl. Phys.} {\bf B473} (1996) 
121;\\
K. Izawa and T. Yanagida, {\it Prog. Theor. Phys.} {\bf 95} (1996) 
829.

\bibitem{BPST}
A.A. Belavin, A.M. Polyakov, A.S. Schwartz, and Yu.S. Tyupkin,  {\it 
Phys. Lett.} {\bf B59} (1975) 85 [Reprinted in {\em Instantons in 
Gauge 
Theories}, Ed. M. Shifman (World Scientific, Singapore, 1994), page 
22].

\bibitem{Hooft1}
G. 't Hooft, {\it  Phys. Rev. } {\bf D14} (1976) 3432; (E) {\it  Phys. Rev. 
} {\bf 
D18} (1978) 2199 [Reprinted in {\em Instantons in Gauge Theories}, 
Ed. M. 
Shifman (World Scientific, Singapore, 1994), page 70; note that in the 
reprinted version the numerical errors summarized in  Erratum 
above 
are  corrected.]

\bibitem{DADV}
A. D'Adda and P. Di Vecchia, {\it Phys. Lett.} {\bf 
B73} (1978) 162 [Reprinted in {\em Instantons in Gauge Theories}, 
Ed. M. 
Shifman (World Scientific, Singapore, 1994), page 293].

\bibitem{NSVZdop}
V.A. Novikov, M.A. Shifman, A.I. Vainshtein, and V.I. Zakharov {\it 
Nucl. Phys.} {\bf B229} (1983) 381;
{\it Phys. Lett.} {\bf B166} (1986) 329. 

\bibitem{NSVZ4}
V.A. Novikov {et al.},  {\it 
Nucl. Phys.} {\bf B229} (1983) 394 [Reprinted in {\em Instantons in 
Gauge 
Theories}, Ed. M. Shifman (World Scientific, Singapore, 1994), page 
298].

\bibitem{NSVZ3}
V.A. Novikov, M.A. Shifman, A.I. Vainshtein, and V.I. Zakharov {\it 
Nucl. Phys.} {\bf B260} (1985) 157 [Reprinted in {\em Instantons in 
Gauge  Theories}, Ed. M. Shifman (World Scientific, Singapore, 1994), 
page 311]. 

\bibitem{RV}
G. Veneziano, {\it Phys. Lett.} {\bf B128} (1983) 199.

\bibitem{BRFS}
T. Banks and E. Rabinovici, 
{\it Nucl. Phys.} {\bf B160} (1979) 349;\\
E. Fradkin and S. Shenker, {\it Phys. Rev.} {\bf D19} (1979) 3682.

\bibitem{MDDGR}
S. Dimopoulos, G. Dvali, G. Guidice, and R. Ratazzi,
{\it Nucl. Phys.} {\bf B510} (1997) 12.

\bibitem{GRM}
R. Arnowitt and P. Nath, 
in Proceedings of Summer School Jorge Andre Swieca: {\it Particles and Fields},
Ed. O.J.P. Eboli and V.O. Rivelles (World Scientific, Singapore, 1994)
[hep-ph/9309277];
G. Guidice and R. Rattazzi, {\em Theories with Gauge Mediated 
Supersymmetry Breaking}, hep-ph/9801271 [{\it Phys. Rep.}, to be published]. 

\bibitem{AHW}
I. Affleck, J. Harvey and E. Witten,
{\it Nucl. Phys.} {\bf B206} (1982) 413.

\bibitem{1} 
V.A. Novikov {et al.},  {\it 
Nucl. Phys.} {\bf B223} (1983) 445.

\bibitem{MV}
Y.  Meurice and  G. Veneziano,  {\it Phys. Lett.} {\bf B141} (1984) 69.

\bibitem{ADS1}
I.  Affleck,  M.  Dine, and N. Seiberg, {\it Phys. Lett.} {\bf B137} 
(1984) 187 [Reprinted in 
{\em 
Supersymmetry}, Ed. S. Ferrara (North Holland/World Scientific, 
Amsterdam 
-- Singapore, 1987), Vol. 1, page 600]; { \it Phys. Rev. 
Lett.} {\bf 52} (1984) 1677.

\bibitem{ADS3}
I.  Affleck,  M.  Dine, and N. Seiberg,  {\it Phys. Lett.} {\bf B140} 
(1984) 59; 
{\it Nucl. Phys.} {\bf B256} (1985) 557. 

\bibitem{NSVZ2}
A.I. Vainshtein, V.I. Zakharov, V.A. Novikov, and M.A. Shifman,
{\it Pis'ma ZhETF} {\bf 39} (1984) 494 [{\it JETP Lett.} {\bf 39} 
(1984) 601].

\bibitem{2}
I.  Affleck,  M.  Dine, and N. Seiberg, { \it Phys. Rev. Lett.} {\bf 52} 
(1984) 1677.

\bibitem{novfu}
J. Fuchs and M. Schmidt,
{\it Z. Phys.} {\bf C30} (1986) 161; J. Fuchs, {\it Nucl. Phys. } {\bf 
B272} 
(1986) 677;\\
V. Novikov, {\it Yad. Fiz.} {\bf 46} (1987) 656; 967 [{\it Sov. J. Nucl. 
Phys.}, {\bf 46} (1987) 366; 554]. 

\bibitem{Yung}
A. Yung,  {\it Nucl. Phys. } {\bf B485} (1997) 38.

\bibitem{DKM}
N. Dorey,  V. Khoze,  and M. Mattis, {\it  Phys. Rev.}
{\bf D54} (1996) 2921; {\bf D54} (1996) 7832; {\it Nucl. Phys.}
{\bf B513} (1998) 681; N. Dorey, T. Hollowood,  V. Khoze,  and M. Mattis, {\it 
Nucl. Phys.} {\bf B519} (1998 ) 470.

\bibitem{DKMS}
V. Khoze, M. Mattis, and  M. Slater, {\it Nucl. Phys.}
{\bf B536} (1998) 69.

\bibitem{seib88}
N. Seiberg, {\it Phys. Lett.} {\bf B206} (1988) 75.

\bibitem{Nati1}
N. Seiberg, {\it Phys. Rev. } {\bf D49} (1994) 6857.

\bibitem{Nati2}
N. Seiberg, {\it Nucl. Phys.} {\bf B435} (1995) 129.

\bibitem{DNNS}
M. Dine, A. Nelson, Y. Nir and Y. Shirman,
{\it Phys. Rev.} {\bf D53} (1996) 2658.

\bibitem{PT}
E. Poppitz and S. Trivedi, {\it Phys. Lett.} {\bf B365} (1996) 125.

\bibitem{BW}
J. Bagger and J. Wess, {\it Supersymmetry and Supergravity},
 (Princeton University Press, 1990).

\bibitem{CS}
 B. Chibisov and  M. Shifman, {\it Phys. Rev.} {\bf  D56} (1997) 7990;
(E) {\bf D58} (1998) 109901.

\bibitem{WZ1}
J. Wess and B. Zumino, {\it Phys. Lett.} {\bf B49} (1974) 52 
[Reprinted in {\it Supersymmetry},
 Ed. S. Ferrara, (North-Holland/World Scientific,  Amsterdam 
-- Singapore, 1987), 
Vol.\ 1, page 77].  

\bibitem{GL}
Yu.A.~Gol'fand and E.P.~Likhtman,
{\it JETP. Lett.} {\bf 13} (1971)  323 [Reprinted in {\it 
Supersymmetry},
 Ed. S. Ferrara, (North-Holland/World Scientific,  Amsterdam 
-- Singapore, 1987), 
Vol.\ 1, page 7]. 

\bibitem{JWBZ}
J. Wess and B. Zumino, {\it Nucl. Phys.} {\bf B70} (1974) 39. 

\bibitem{BDSF}
F. Buccella, J.-P. Derendinger, C. Savoy, and S. Ferrara,
{\it Phys. Lett.} {\bf B115} (1982) 375, and 
in Proc. Europhys. Study Conf. {\it Unification of the Fundamental 
Particle 
Interactions II}, Eds. J. Ellis and S. Ferrara (Plenum Press, New York, 
1983),
page 349. 

\bibitem{TT}
T.R. Taylor, {\it Phys. Lett.} {\bf B125} (1983) 185; {\bf B128} (1983) 
403.

\bibitem{Luty1}
M. Luty and W. Taylor, {\it Phys. Rev. } {\bf D53} (1996) 3399.

\bibitem{HKLR}
N. Hitchin, A. Karlhede, U. Lindstr\"{o}m and M. Ro\v{c}ek,
{\it Commun. Math. Phys.} {\bf 108} (1987) 535.

\bibitem{Amati}
D. Amati, K. Konishi, Y. Meurice, G. Rossi and G. Veneziano,
{\it Phys. Rep.} {\bf 162} (1988) 557.

\bibitem{Witten3}
E. Witten, {\it Phys. Lett.} {\bf B117} (1982) 324 [Reprinted in {\em 
Instantons in 
Gauge 
Theories}, Ed. M. Shifman (World Scientific, Singapore, 1994), page 
230]. 

\bibitem{BPR}
J. Bagger, E. Poppitz and L. Randall, {\it Nucl. Phys.}  {\bf B426} 
(1994) 3.

\bibitem{IVS}
 A. Vainshtein, V. Zakharov, and M. Shifman, {\it Usp. Fiz. Nauk}, {\bf 146} (1985)
683 [{\em Sov. Phys. Uspekhi},  {\bf 28} (1985) 709].

\bibitem{Veldhuis1}
T. ter Veldhuis, {\it Phys. Rev.} {\bf D58} (1998) 015010;
 {\it Unexpected symmetries in classical moduli spaces},  hep-th/9811132.

\bibitem{GRS}
J. Wess and B. Zumino, {\it Phys. Lett.} {\bf B49} (1974) 52;\\
J. Iliopoulos and B. Zumino, {\it Nucl. Phys.} {\bf B76} (1974) 310;\\
P. West, {\it Nucl. Phys.} {\bf B106} (1976) 219;\\
M. Grisaru, M. Ro\v{c}ek, and W. Siegel,
{\it Nucl. Phys. } {\bf B159} (1979) 429.

\bibitem{SSF}
A. Salam and J. Strathdee, {\it Nucl. Phys.} {\bf B87} (1975) 85;\\
P. Fayet, {\it Nucl. Phys.} {\bf B90} (1975) 104.

\bibitem{FEZU}
S. Ferrara and B. Zumino, {\it Nucl. Phys. } { \bf  B87} (1975) 207.

\bibitem{Grisa}
M. Grisaru, in
{\it Recent Developments in Gravitation} (Carg\'{e}se Lectures, 1978),
Eds. M. Levy and S. Deser (Plenum Press, New York, 1979), page 577,
and references therein. 

\bibitem{Konishi}
T.E. Clark, O. Piguet, and K. Sibold, {\it Nucl. Phys.} {\bf B 159} (1979) 
1;\\
K. Konishi, {\it Phys. Lett.} {\bf B135} (1984) 439;\\
K. Konishi and K. Shizuya, {\it Nuov. Cim.} {\bf A90} (1985) 111.

\bibitem{anomaly}
M. Shifman and A. Vainshtein, {\it Nucl. Phys. } {\bf B 277} (1986) 456. 

\bibitem{KSV}
 I. Kogan, M.  Shifman, and A. Vainshtein, {\it Phys. Rev.} {\bf D53}
(1996) 4526.

\bibitem{veneziano}
G. Veneziano and S. Yankielowicz,
{\it Phys. Lett.} {\bf 113B} (1982) 231;\\
T.R. Taylor, G. Veneziano, and S. Yankielowicz, {\it Nucl. Phys.} {\bf 
B218} (1983) 493.

\bibitem{kovner2} 
A. Kovner and M. Shifman,  
{\it Phys. Rev.} {\bf D56} (1997) 2396.

\bibitem{SMILGA}
A. Keurentjes, A. Rosly, and A. Smilga,
{\it Phys. Rev. } {\bf D58} (1998) 081701;
V. Ka\v{c} and A. Smilga, hep-th/9902029.

\bibitem{SVMO}
M. Shifman and A. Vainshtein, {\it Nucl. Phys. } {\bf B296} (1988) 
445;\\
A.~Morozov, M.~Olshanetsky and M.~Shifman, {\it Nucl. Phys. } {\bf 
B304} (1988) 291.

\bibitem{Louis}
L. Dixon, V. Kaplunovsky and J. Louis,
{\it Nucl. Phys.} {\bf  B355} (1991) 649;\\
M. Shifman and A. Vainshtein,
{\it Nucl. Phys.} {\bf B277} (1986) 456;
{\bf B359} (1991) 571;\\
I. Jack, D.R.T. Jones, and P. West, {\it Phys. Lett.}
{\bf B258} (1991) 382. 

\bibitem{torHooft}
G. 't Hooft, {\it Commun. Math. Phys.} {\bf 81} (1981) 267. 

\bibitem{BPS}
E. Bogomol'nyi, {\it Sov. J. Nucl. Phys.} {\bf 24} (1976) 449;\\
M.K. Prasad and C.H. Sommerfield, {\it Phys. Rev. Lett.} {\bf 35}
(1976) 760.

\bibitem{SFMP}
S. Ferrara and  M. Porrati, {\it Phys. Lett.} {\bf B423} (1998) 255. 

\bibitem{WO}
 E. Witten and D. Olive, {\it Phys. Lett. } {\bf B78} (1978) 97.

\bibitem{Intri}
P. Fendley and K. Intriligator, {\it Nucl. Phys.} {\bf B372} (1990) 533.

\bibitem{Dvali1}
 G. Dvali and  M. Shifman, {\it Nucl. Phys.} {\bf B504} (1997) 127.

\bibitem{Dvali2}
 G. Dvali and  M. Shifman, 
{\it Phys. Lett.} {\bf B396} (1997) 64; (E) {\bf B407} (1997) 452;\\
A. Kovner, M. Shifman, and  A. Smilga, {\it Phys. Rev.} {\bf
D56} (1997) 7978. 

\bibitem{SVV}
M. Shifman, A. Vainshtein and M. Voloshin, {\it Phys. Rev.} {\bf D59}
(1999) 045016. 

\bibitem{KR}
R.K. Kaul and R. Rajaraman, {\it Phys. Lett.} {\bf B131} (1983) 357.

\bibitem{IM}
C. Imbimbo amd S. Mukhi, {\it Nucl. Phys.} {\bf B247} (1984) 471.

\bibitem{AHV}
A. D'Adda, R. Horsley,  and P. Di Vecchia, {\it Phys. Lett.} {\bf B76}
(1978) 298;\\ R. Horsley, {\it Nucl. Phys. } {\bf B151} (1979) 399.

\bibitem{JFS}
J.F. Sch\"onfeld, {\it Nucl. Phys. } {\bf B161} (1979) 125.

\bibitem{SR}
S. Rouhani, {\it Nucl. Phys. } {\bf B182} (1981) 462.

\bibitem{AU}
A. Uchiyama, {\it Nucl. Phys. } {\bf B244} (1984) 57;  {\it Prog.
Theoret. Phys.} {\bf 75} (1986) 1214; {\it Nucl. Phys. } {\bf B278}  (1986) 121.

\bibitem{yama}
H. Yamagishi, {\it Phys. Lett. } {\bf B147}
(1984) 425. 

\bibitem{CM1}
A.K. Chatterjee and P. Majumdar,  {\it Phys. Rev.} {\bf D30} (1984)
844;  {\it Phys. Lett.} {\bf B159} (1985) 
37.

\bibitem{RN}
A. Rebhan and P. Nieuwenhuizen, {\it Nucl. Phys. } {\bf B508} (1997) 
449;\\
H. Nastase, M. Stephanov, P. Nieuwenhuizen, and A. Rebhan,
{\it Nucl. Phys.} {\bf B542} (1999) 471 [hep-th/9802074].

\bibitem{Jaffe}
N. Graham and R.L. Jaffe, {\it Nucl. Phys.} {\bf B544} (1999) 432
[hep-th/9808140].

\bibitem{Jackiw}
R. Jackiw and C. Rebbi, {\it Phys. Rev. } {\bf D13} (1976) 3398.

\bibitem{Zumino}
B. Zumino, {\it Phys. Lett.} {\bf B69} (1977) 369. 

\bibitem{Bere}
F. Berezin, {\em The Method of Second Quantization}, (Academic
 Press, New York, 1966). 

\bibitem{ABC}
 A. Vainshtein {\em et al.}, {\it Usp. Fiz. Nauk}, {\bf 136} (1982) 553
[{\em Sov. Phys. Uspekhi},  {\bf 25} (1982) 195  [Reprinted in {\em 
Instantons in Gauge Theories}, 
Ed. M.\ Shifman (World Scientific, Singapore, 1994), page 468].

\bibitem{sconfrev}
P. Fayet and S. Ferrara, {\it Phys. Reports} {\bf 32} (1977) 249.

\bibitem{AVMS}
M. Shifman and A. Vainshtein, {\it Nucl. Phys.} {\bf B362} (1991) 21
 [Reprinted in {\em 
Instantons in Gauge Theories}, 
Ed. M.\ Shifman (World Scientific, Singapore, 1994), page 97]. 

\bibitem{SVZscat}
M. Shifman, A. Vainshtein, and V. Zakharov, {\it Nucl. Phys.} {\bf B163} (1980) 46;
{\bf B165} (1980) 45.

\bibitem{Bernard}
C.~Bernard, {\it Phys. Rev. } {\bf D19} (1979) 3013
 [Reprinted in {\em 
Instantons in Gauge Theories}, 
Ed. M.\ Shifman (World Scientific, Singapore, 1994), page 109].

\bibitem{BCCL}
L.~Brown, R.~Carlitz, D.~Creamer, and C.~Lee,  {\it Phys. Rev. }  {\bf D17} (1978)
1583  [Reprinted in {\em 
Instantons in Gauge Theories}, 
Ed. M.\ Shifman (World Scientific, Singapore, 1994), page 168].

\bibitem{SSOT}
S.J. Gates, M.T. Grisaru, M. Ro\v{c}ek, and W. Siegel,
{\it Superspace, or One Thousand and One Lessons in 
Supersymmetry} (The Benjamin/Cummings, Reading, MA, 1983).

\bibitem{rarensvz}
V. Novikov, M. Shifman, A. Vainshtein, and V.  Zakharov {\it 
Phys. Lett.} {\bf B217} (1989) 103. 

\bibitem{AVMS1}
M. Shifman and A. Vainshtein, {\it Nucl. Phys.} {\bf B359} (1991) 571.

\bibitem{little}
M. Shifman, {\it Int. J. Mod. Phys.} {\bf A11} (1996) 5761. 

\bibitem{Jack}
I. Jack, D.R.T. Jones, and A. Pickering, {\it Phys. Lett.}
{\bf B435} (1998) 61, and references therein. 

\bibitem{PWJM}
A.~Parkes and P.~West, {\it Phys. Lett.} {\bf B138} (1984) 99;\\
D.R.T.~Jones amd L.~Mezincescu, {\it Phys. Lett.} {\bf B138} (1984) 293;\\
S.~Hamidi, J.~Patera and J.~Schwarz, {\it Phys. Lett.} {\bf B141} (1984) 349.

\bibitem{HSU}
R. Leigh and M. Strassler, {\it Nucl. Phys.} {\bf B447} (1995) 95;
C. Lucchesi and  G. Zoupanos, {\it Fortsch. Phys.} {\bf 45} (1997) 129;
A.~Hanany, M.J. Strassler, and A.M. Uranga, {\it J. High Energy Phys.} {\bf 06}
(1998) 011 [hep-th/9803086];
 A.~Hanany and Yang-Hui He, {\it J. High Energy Phys.}
{\bf 9902} (1999) 013
[hep-th/9811183].

\bibitem{PDG}
I.~Hinchliffe, {\it Eur. Phys. J.} {\bf C3} (1998) 81.

\bibitem{SEIW}
N. Seiberg and  E. Witten, {\it Nucl. Phys.} {\bf  B426} (1994) 19, (E) 
{\bf 
B430} (1994) 485; {\bf B431} (1994) 484.

\bibitem{RVto}
A.~Ritz, A.~Vainshtein, in preparation

\bibitem{wtop}
E. Witten, {\it Commun. Math. Phys.} {\bf  117} (1988) 353. 

\bibitem{FiPo}
D. Finnell and  P. Pouliot, {\it Nucl. Phys.} {\bf  B453} (1995) 225.

\bibitem{Ring}
A. Ringwald,
{\it Nucl. Phys. } {\bf {B330}} (1990) 1;\\
O. Espinosa, {\it Nucl. Phys.} {\bf {B343}} (1990) 310;\\
L. McLerran, A. Vainshtein and M. Voloshin,
{\it Phys. Rev.  } {\bf {D42}} (1990) 171.

\bibitem{MLVV}
L. McLerran, A. Vainshtein and M. Voloshin,
{\it Phys.  Lett. } {\bf {B249}} (1990) 261; 
{\it Phys. Rev.  } {\bf {D42}} (1990) 
180.

\bibitem{BNVrev}
For reviews see
M. P. Mattis, {\it Phys. Rept.} {\bf 214} (1992) 159;
V.A. Rubakov and M.E. Shaposhnikov, {\em Usp. Fiz. Nauk} {\bf  166} 
(1996) 493 [{\it Phys. Usp.} {\bf 39} (1996 ) 461],  [hep-ph/9603208].  

\bibitem{sphaleron}
R. Dashen, B. Hasslacher and A. Neveu,
{\it Phys. Rev. } {\bf D10} (1974) 4138,\\
F.R. Klinkhamer and  N.S. Manton,  {\it Phys. Rev.} {\bf D30} (1984) 
2212, \\
L.G. Yaffe, Phys. Rev. D40 (1989) 3463; \\
F. Klinkhamer, {\em Sphalerons and Energy Barriers in the 
Weinberg-Salam
Model},
in Proc. XXV Int. Conf. on High Energy Physics,
Singapore, 1990, Ed. 
K.K.~Phua and Y. Yamaguchi (World Scientific, 
Singapore, 1991) p. 913.

\bibitem{ZMSV}
V.  Zakharov, {\it Nucl. Phys.} {\bf B353} (1991) 683;\\
M.  Maggiore and  M. Shifman, {\it Nucl. Phys.}
{\bf B371} (1992) 177; {\it Phys. Rev.} {\bf D46} (1992) 
3550;\\
G. Veneziano, {\it Mod. Phys. Lett.} {\bf A7} (1992) 1661.

\bibitem{CSASKI}
C. Cs\'{a}ki and W. Skiba, {\it Phys. Rev.}  {\bf D58} (1998) 045008.

\bibitem{ORAI}
L. O'Raifeartaigh, {\it Nucl. Phys.} {\bf B96} (1975) 331.

\bibitem{FGP}
S. Ferrara, L. Girardello, and F. Palumbo,
{\it Phys. Rev. } {\bf D20} (1979) 403.

\bibitem{FIL}
P. Fayet and J. Iliopoulos,  {\it Phys. Lett.} {\bf B51} (1974) 461.

\bibitem{tHooft2}
 G. 't Hooft,  in  {\it Recent Developments in Gauge Theories},
 Eds. G. 't Hooft {\em et al.},  (Plenum Press, New York, 1980).

\bibitem{ARV}
 D. Amati, G.C. Rossi, and G. Veneziano,  {\it Nucl. Phys.} {\bf B249} 
(1985 ) 1.

\bibitem{31}
I. Affleck, M. Dine and N. Seiberg,  {\it Phys. Lett.} {\bf B137} (1984) 187. 

\bibitem{DMS}
 G. Dotti, A. Manohar,  and W. Skiba,  {\it Nucl. Phys.}
{\bf B531} (1998) 507.

\bibitem{Veldhuis}
T. ter Veldhuis, {\it Phys. Lett.} {\bf B367} (1996) 157. 

\bibitem{PST}
E. Poppitz, Y. Shadmi and S. Trivedi,
{\it Nucl. Phys.} {\bf B480} (1996) 125;\\
{\it Phys. Lett.} {\bf B388} (1996)  561.

\bibitem{KIST}
K. Intriligator and  S. Thomas, {\it Nucl. Phys.} {\bf B473} (1996) 
121.

\bibitem{CSS}
 C. Csaki, W. Skiba, and M. Schmaltz, {\it Nucl. Phys.}
{\bf B487} (1997) 128.

\bibitem{Mura}
H. Murayama, {\it Phys. Lett.} {\bf  B355} (1995 ) 187.

\bibitem{MB}
A. Belavin and A. Migdal, {\it Pis'ma ZhETF} {\bf 19} (1974) 317 
[{\it JETP Lett.}
{\bf 19} (1974) 181]; {\em  Scale Invariance and Bootstrap in the 
Non-Abelian Gauge 
Theories}, Landau Institute  Preprint-74-0894,  1974
(unpublished).

\bibitem{BZ}
T. Banks and  A. Zaks, {\it Nucl. Phys.} {\bf B196} (1982) 189.

\bibitem{BCI}
J. Brodie, P. Cho, and  K. Intriligator, {\it Phys. Lett.}
{\bf B429} (1998) 319.

\bibitem{CSSe}
 C. Csaki, M. Schmaltz, and W. Skiba, {\it Phys. Rev.}
{\bf D55} (1997) 7840.

\bibitem{Shir}
Y. Shirman, {\it Phys. Lett.} {\bf B389} (1996) 287.

\bibitem{CLP}
Z. Chacko, M. Luty and E. Pont\'{o}n, {\it J. High Energy Phys.}
{\bf 9812} (1998) 016
[hep-th/9810253].

\bibitem{KSVZ}
J. Kim, {\it Phys. Rev. Lett.} {\bf 43} (1979) 103;\\
M. Shifman, A. Vainshtein, and V. Zakharov, {\it Nucl. Phys. } {\bf
B166}  (1980) 493. 

\bibitem{KKS}
I. Kogan, A. Kovner, and  M.  Shifman, {\it Phys. Rev. } {\bf D57}
(1998) 5195.

\bibitem{EWD}
E. Witten,
{\it Nucl. Phys.} {\bf B507} (1997) 658.

\end{thebibliography}
\end{document}